\documentclass[twocolumn,tighten,twocolappendix]{aastex631}

\pdfoutput=1 
\usepackage{amsmath,amstext,longtable}
\usepackage[T1]{fontenc}
\usepackage{apjfonts} 
\usepackage{CJKutf8}
\usepackage[figure,figure*]{hypcap}
\received{08 Oct 2023}
\revised{30 Nov 2023}
\accepted{03 Dec 2023}
\submitjournal{The Astrophysical Journal Supplement Series}
\shortauthors{Kirkpatrick et al.}
\shorttitle{20-pc Mass Function}

\usepackage{color,amsmath,longtable}

\begin{document}

\title{The Initial Mass Function Based on the Full-sky 20-pc Census of $\sim$3,600 Stars and Brown Dwarfs}

\correspondingauthor{J.\ Davy Kirkpatrick}
\email{davy@ipac.caltech.edu}

\author[0000-0003-4269-260X]{J.\ Davy Kirkpatrick}
\affiliation{IPAC, Mail Code 100-22, California Institute of Technology, 1200 E. California Blvd., Pasadena, CA 91125, USA}
\affiliation{Backyard Worlds: Planet 9}


\author[0000-0001-7519-1700]{Federico Marocco}
\affiliation{IPAC, Mail Code 100-22, California Institute of Technology, 1200 E. California Blvd., Pasadena, CA 91125, USA}
\affiliation{Backyard Worlds: Planet 9}

\author[0000-0001-5072-4574]{Christopher R.\ Gelino}
\affiliation{NASA Exoplanet Science Institute, Mail Code 100-22, California Institute of Technology, 770 S. Wilson Avenue, Pasadena, CA 91125, USA}

\author[0000-0001-9778-7054]{Yadukrishna Raghu}
\affiliation{Washington High School, 38442 Fremont Blvd., Fremont, CA 94536, USA}
\affiliation{Backyard Worlds: Planet 9}

\author[0000-0001-6251-0573]{Jacqueline K.\ Faherty}
\affiliation{Department of Astrophysics, American Museum of Natural History, Central Park West at 79th Street, New York, NY 10024, USA}
\affiliation{Backyard Worlds: Planet 9}

\author[0000-0001-8170-7072]{Daniella C.\ Bardalez Gagliuffi}
\affiliation{Department of Astrophysics, American Museum of Natural History, Central Park West at 79th Street, New York, NY 10024, USA}
\affiliation{Backyard Worlds: Planet 9}

\author[0000-0003-1785-5550]{Steven D.\ Schurr}
\affiliation{IPAC, Mail Code 100-22, California Institute of Technology, 1200 E. California Blvd., Pasadena, CA 91125, USA}

\author{Kevin Apps}
\affiliation{Independent scholar, UK}


\author[0000-0002-6294-5937]{Adam C.\ Schneider}
\affiliation{United States Naval Observatory, Flagstaff Station, 10391 West Naval Observatory Road, Flagstaff, AZ 86005, USA}
\affiliation{Backyard Worlds: Planet 9}

\author[0000-0002-1125-7384]{Aaron M. Meisner}
\affiliation{NSF's National Optical-Infrared Astronomy Research Laboratory, 950 N. Cherry Ave., Tucson, AZ 85719, USA}
\affiliation{Backyard Worlds: Planet 9}

\author[0000-0002-2387-5489]{Marc J.\ Kuchner}
\affiliation{NASA Goddard Space Flight Center, Exoplanets and Stellar Astrophysics Laboratory, Code 667, Greenbelt, MD 20771, USA}
\affiliation{Backyard Worlds: Planet 9}


\author[0000-0001-7896-5791]{Dan Caselden}
\affiliation{Department of Astrophysics, American Museum of Natural History, Central Park West at 79th Street, New York, NY 10024, USA}
\affiliation{Backyard Worlds: Planet 9}

\author[0000-0002-4424-4766]{R.\ L.\ Smart}
\affiliation{Istituto Nazionale di Astrofisica, Osservatorio Astrofisico di Torino, Strada Osservatorio 20, I-10025 Pino Torinese, Italy}

\author[0000-0003-2478-0120]{S.\ L.\ Casewell}
\affiliation{School of Physics and Astronomy, University of Leicester, University Road, Leicester LE1 7RH, UK}

\author[0000-0002-9090-9191]{Roberto Raddi}
\affiliation{Universitat Polit{\`e}cnica de Catalunya, Departament de F{\'i}sica, c/ Esteve Terrades 5, 08860 Castelldefels, Spain}

\author[0000-0002-3239-5989]{Aurora Kesseli}
\affiliation{IPAC, Mail Code 100-22, California Institute of Technology, 1200 E. California Blvd., Pasadena, CA 91125, USA}


\author[0000-0003-4714-3829]{Nikolaj Stevnbak Andersen}
\affiliation{Sygehus Lillebalt, Department of Cardiology, Kolding, Denmark}
\affiliation{Backyard Worlds: Planet 9}

\author{Edoardo Antonini}
\affiliation{Backyard Worlds: Planet 9}

\author{Paul Beaulieu}
\affiliation{Backyard Worlds: Planet 9}

\author[0000-0003-2235-761X]{Thomas P.\ Bickle}
\affiliation {School of Physical Sciences, The Open University, Milton Keynes, MK7 6AA, UK} 
\affiliation{Backyard Worlds: Planet 9}

\author[0009-0000-5790-7488]{Martin Bilsing}
\affiliation{Backyard Worlds: Planet 9}

\author{Raymond Chieng}
\affiliation{Backyard Worlds: Planet 9}

\author[0000-0002-7630-1243]{Guillaume Colin}
\affiliation{Backyard Worlds: Planet 9}

\author{Sam Deen}
\affiliation{Backyard Worlds: Planet 9}

\author{Alexandru Dereveanco}
\affiliation{Backyard Worlds: Planet 9}

\author[0000-0002-2993-9869]{Katharina Doll}
\affiliation{Backyard Worlds: Planet 9}

\author[0000-0002-4143-2550]{Hugo A.\ Durantini Luca}
\affiliation{Backyard Worlds: Planet 9}

\author{Anya Frazer}
\affiliation{Backyard Worlds: Planet 9}

\author{Jean Marc Gantier}
\affiliation{Backyard Worlds: Planet 9}

\author[0000-0002-8960-4964]{L\'eopold Gramaize}
\affiliation{Backyard Worlds: Planet 9}

\author{Kristin Grant}
\affiliation{Backyard Worlds: Planet 9}

\author[0000-0002-7389-2092]{Leslie K.\ Hamlet}
\affiliation{Backyard Worlds: Planet 9}

\author[0009-0004-9088-7510]{\begin{CJK*}{UTF8}{} Hiro Higashimura ({\CJKfamily{min}東村滉}) \end{CJK*}}
\affiliation{Earl of March Intermediate School, 4 The Pkwy, Kanata, ON K2K 1Y4, Canada}

\author[0000-0001-8343-0820]{Michiharu Hyogo}
\affiliation{Meisei University, 2-1-1 Hodokubo, Hino, Tokyo 191-0042, Japan}
\affiliation{Backyard Worlds: Planet 9}

\author[0000-0002-4175-295X]{Peter A.\ Ja{\l}owiczor}
\affiliation{Backyard Worlds: Planet 9}

\author[0000-0003-3743-3320]{Alexander Jonkeren}
\affiliation{Backyard Worlds: Planet 9}

\author[0000-0003-4905-1370]{Martin Kabatnik}
\affiliation{Backyard Worlds: Planet 9}

\author[0000-0001-8662-1622]{Frank Kiwy}
\affiliation{Backyard Worlds: Planet 9}

\author{David W.\ Martin}
\affiliation{Backyard Worlds: Planet 9}

\author[0009-0000-8800-3174]{Marianne N.\ Michaels}
\affiliation{Backyard Worlds: Planet 9}

\author{William Pendrill}
\affiliation{Backyard Worlds: Planet 9}

\author{Celso Pessanha Machado}
\affiliation{Backyard Worlds: Planet 9}

\author[0000-0001-9692-7908]{Benjamin Pumphrey}
\affiliation{Backyard Worlds: Planet 9}

\author[0000-0003-4083-9962]{Austin Rothermich}
\affiliation{Physics Department, University of Central Florida, 4000 Central Florida Boulevard, Orlando, FL 32816, USA}
\affiliation{Backyard Worlds: Planet 9}

\author{Rebekah Russwurm}
\affiliation{Backyard Worlds: Planet 9}

\author[0000-0003-4864-5484]{Arttu Sainio}
\affiliation{Backyard Worlds: Planet 9}

\author{John Sanchez}
\affiliation{Backyard Worlds: Planet 9}

\author{Fyodor Theo Sapelkin-Tambling}
\affiliation{Backyard Worlds: Planet 9}

\author[0000-0002-7587-7195]{J\"org Sch\"umann}
\affiliation{Backyard Worlds: Planet 9}

\author{Karl Selg-Mann}
\affiliation{Backyard Worlds: Planet 9}

\author{Harshdeep Singh}
\affiliation{Backyard Worlds: Planet 9}

\author{Andres Stenner}
\affiliation{Backyard Worlds: Planet 9}

\author[0000-0003-3162-3350]{\begin{CJK*}{UTF8}{} Guoyou Sun ({\CJKfamily{gbsn}孙国佑}) \end{CJK*}}
\affiliation{Xingming Observatory, Mt.\ Nanshan, Urumqi, 830011, Xinjiang, PR China}
\affiliation{Backyard Worlds: Planet 9}

\author{Christopher Tanner}
\affiliation{Backyard Worlds: Planet 9}

\author[0000-0001-5284-9231]{Melina Th\'evenot}
\affiliation{Backyard Worlds: Planet 9}

\author{Maurizio Ventura}
\affiliation{Backyard Worlds: Planet 9}

\author{Nikita V.\ Voloshin}
\affiliation{Backyard Worlds: Planet 9}

\author{Jim Walla}
\affiliation{Backyard Worlds: Planet 9}

\author{Zbigniew W{\k e}dracki}
\affiliation{Backyard Worlds: Planet 9}


\author{Jose I.\ Adorno}
\affiliation{Department of Physics, University of Miami, Coral Gables, FL 33124, USA}
\affiliation{Department of Astrophysics, American Museum of Natural History, Central Park West at 79th Street, New York, NY 10024, USA}

\author[0000-0003-2094-9128]{Christian Aganze}
\affiliation{Department of Physics, Stanford University, Stanford CA 94305, USA}

\author[0000-0003-0580-7244]{Katelyn N.\ Allers}
\affiliation{Department of Physics and Astronomy, Bucknell University, Lewisburg, PA 17837, USA}

\author[0000-0002-5253-0383]{Hunter Brooks}
\affiliation{Department of Astronomy and Planetary Science, Northern Arizona University, Flagstaff, AZ 86011, USA}
\affiliation{Backyard Worlds: Planet 9}

\author[0000-0002-6523-9536]{Adam J.\ Burgasser}
\affiliation{Department of Astronomy \& Astrophysics, University of California San Diego, 9500 Gilman Drive, La Jolla, CA 92093, USA}

\author[0000-0002-2682-0790]{Emily Calamari}
\affiliation{Department of Astrophysics, American Museum of Natural History, Central Park West at 79th Street, New York, NY 10024, USA}
\affiliation{The Graduate Center, City University of New York, New York, NY 10016, USA}

\author[0000-0002-7898-7664]{Thomas Connor}
\affiliation{Jet Propulsion Laboratory, California Institute of Technology, 4800 Oak Grove Drive, Pasadena, CA 91109, USA}
\affiliation{Center for Astrophysics $\vert$\ Harvard\ \&\ Smithsonian, 60 Garden St., Cambridge, MA 02138, USA}

\author[0000-0003-4142-1082]{Edgardo Costa}
\affiliation{Universidad de Chile, Casilla 36-D, Santiago, Chile}

\author{Peter R.\ Eisenhardt}
\affiliation{Jet Propulsion Laboratory, California Institute of Technology, MS 169-237, 4800 Oak Grove Drive, Pasadena, CA 91109, USA}

\author[0000-0002-2592-9612]{Jonathan Gagn\'e}
\affiliation{Institute for Research on Exoplanets, Universit\'e de Montr\'eal, Montr\'eal, Canada}

\author[0000-0003-0398-639X]{Roman Gerasimov}
\affiliation{Department of Physics \& Astronomy, University of Notre Dame, Notre Dame, IN 46556, USA}

\author[0000-0003-4636-6676]{Eileen C.\ Gonzales}
\altaffiliation{51 Pegasi b Fellow}
\affiliation{Department of Physics and Astronomy, San Francisco State University, 1600 Holloway Avenue, San Francisco, CA 94132, USA}
\affiliation{Department of Astronomy and Carl Sagan Institute, Cornell University, 122 Sciences Drive, Ithaca, NY 14853, USA}

\author[0000-0002-5370-7494]{Chih-Chun Hsu}
\affiliation{Center for Astrophysics and Space Science, University of California San Diego, La Jolla, CA 92093, USA}
\affiliation{Center for Interdisciplinary Exploration and Research in Astrophysics (CIERA), Northwestern University, 1800 Sherman, Evanston, IL, 60201, USA}

\author[0000-0003-2102-3159]{Rocio Kiman}
\affiliation{Department of Astronomy, California Institute of Technology, Pasadena, CA 91125, USA}

\author{Guodong Li}
\affiliation{National Astronomical Observatories, Chinese Academy of Sciences,
Beijing 100012, China}
\affiliation{University of Chinese Academy of Sciences, Beijing 100049, China}

\author[0000-0002-5024-0075]{Ryan Low}
\affiliation{Department of Physics and Astronomy, University of Kansas, Lawrence, KS 66046, USA}

\author[0000-0003-2008-1488]{Eric Mamajek}
\affiliation{Jet Propulsion Laboratory, California Institute of Technology, MS 321-100, 4800 Oak Grove Drive, Pasadena, CA 91109, USA}

\author{Blake M.\ Pantoja}
\affiliation{Departamento de Astronom{\'i}a, Universidad de Chile, Camino al Observatorio, Cerro Cal{\'a}n, Santiago, Chile}

\author[0000-0001-9482-7794]{Mark Popinchalk}
\affiliation{Department of Astrophysics, American Museum of Natural History, Central Park West at 79th St., New York, NY 10024, USA}
\affiliation{Department of Physics and Astronomy, Hunter College, City University of New York, 695 Park Avenue, New York, NY 10065, USA}
\affiliation{The Graduate Center, City University of New York, New York, NY 10016, USA}

\author[0000-0002-5376-3883]{Jon M.\ Rees}
\affiliation{Lick Observatory, P.O.\ Box 85, Mount Hamilton, CA 95140, USA}

\author[0000-0003-2686-9241]{Daniel Stern}
\affiliation{Jet Propulsion Laboratory, California Institute of Technology, MS 169-237, 4800 Oak Grove Drive, Pasadena, CA 91109, USA}

\author[0000-0002-2011-4924]{Genaro Su\'arez}
\affiliation{Department of Astrophysics, American Museum of Natural History, Central Park West at 79th Street, New York, NY 10024, USA}

\author[0000-0002-9807-5435]{Christopher Theissen}
\affiliation{Department of Astronomy \& Astrophysics, University of California San Diego, 9500 Gilman Drive, La Jolla, CA 92093, USA}

\author[0000-0002-9390-9672]{Chao-Wei Tsai}
\affiliation{National Astronomical Observatories, Chinese Academy of Sciences,
Beijing 100012, China}

\author[0000-0003-0489-1528]{Johanna M. Vos}
\affiliation{School of Physics, Trinity College Dublin, The University of Dublin, Dublin 2, Ireland}
\affiliation{Department of Astrophysics, American Museum of Natural History, Central Park West at 79th Street, New York, NY 10024, USA}

\author{David Zurek}
\affiliation{Department of Astrophysics, American Museum of Natural History, Central Park West at 79th Street, New York, NY 10024, USA}

\author{The Backyard Worlds: Planet 9 Collaboration}

\begin{abstract}

A complete accounting of nearby objects -- from the highest-mass white dwarf progenitors down to low-mass brown dwarfs -- is now possible, thanks to an almost complete set of trigonometric parallax determinations from Gaia, ground-based surveys, and Spitzer follow-up. We create a census of objects within a Sun-centered sphere of 20-pc radius and check published literature to decompose each binary or higher-order system into its separate components. The result is a volume-limited census of $\sim$3,600 {\it individual} star formation products useful in measuring the initial mass function across the stellar ($<8 M_\odot$) and substellar ($\gtrsim 5 M_{Jup}$) regimes. Comparing our resulting initial mass function to previous measurements shows good agreement above 0.8$M_\odot$ and a divergence at lower masses. Our 20-pc space densities are best fit with a quadripartite power law, $\xi(M) = dN/dM \propto M^{-\alpha}$ with long-established values of $\alpha = 2.3$ at high masses ($0.55 < M < 8.00 M_\odot$) and $\alpha = 1.3$ at intermediate masses ($0.22 < M < 0.55 M_\odot$), but at lower masses we find $\alpha = 0.25$ for $0.05 < M <0.22 M_\odot$ and $\alpha = 0.6$ for $0.01 < M < 0.05 M_\odot$. This implies that the rate of production as a function of decreasing mass diminishes in the low-mass star/high-mass brown dwarf regime before increasing again in the low-mass brown dwarf regime. Correcting for completeness, we find a star to brown dwarf number ratio of, currently, 4:1, and an average mass per object of 0.41 $M_\odot$.

\end{abstract}


\section{Introduction}

The concept of the initial mass function is one of the most fundamental paradigms in astronomy. It embodies the observational evidence for how the universe turns gas into stars and provides an empirical framework on which to test and inform the underlying theory. The initial mass function has far-reaching influence, from providing the cornerstone for galaxy formation scenarios across all cosmic epochs to determining which stellar and substellar populations we see in our own solar neighborhood. 

Debate continues on whether the initial mass function is variable with time or dependent on environment, but its description over most of the range of stellar masses in the Milky Way is well determined. \cite{bastian2010} conclude that the initial mass function is universal for hydrogen-burning stars, at least within the measurement errors of most current observations, and \cite{andersen2008} specifically conclude that there is no strong evidence for environment-specific effects at masses above $\sim30M_{Jup}$. However, far less is known about the mass function at the low-mass end. Knowledge in this area tells us the creation ratio between stars and brown dwarfs and enlightens us on whether planetary mass objects formed via star formation are common compared to those formed via protoplanetary disks.
 
In this paper, we use recent advances in our knowledge of the nearby stellar census to explore in unprecedented detail the field initial mass function. Gaia has helped refine the nearby census down to spectral types of mid-/late-L out to 20 pc (\citealt{smart2020}). For colder spectral types, the WISE mission, together with follow-up parallaxes measured by Spitzer, has filled out this census down to early-Y dwarfs (\citealt{kirkpatrick2019,kirkpatrick2021}), with the help of many other ground-based endeavors (e.g., \citealt{best2021}). Our understanding of the low-mass end is dominated by solivagant L, T, and Y dwarfs, but much less is known about the frequency with which these low-mass objects exist as companions to hotter objects in the census. We rectify that gap in our understanding by building a complete census of all objects within 20 pc of the Sun and splitting those systems into their individual components.

In Section~\ref{sec:creating_the_census} we use previous nearby star lists, additions from Gaia, and published or newly discovered objects lacking Gaia astrometry to construct the census of objects in the 20-pc volume. In Section~\ref{sec:20pc_census} we discuss the format of the compiled census, which includes data on nomenclature, astrometry, spectral types, photometry, radial velocities, multiplicity, masses, and effective temperatures. In Section~\ref{sec:mass_methods} we discuss the methods used to directly measure masses. In Section~\ref{sec:multiples} we discuss the fact that some objects in our sample have strong evidence for multiplicity but generally lack sufficient evidence to characterize the mass of the subcomponents, which is a source of uncertainty in our final analysis. In Section~\ref{sec:mass_estimates} we discuss mass estimation for white dwarf progenitors, giants/subgiants, brown dwarfs, young stars, low metallicity stars (subdwarfs), and normal main sequence stars and discuss what role objects labeled as exoplanets play in our analysis. In Section~\ref{sec:further_analysis} we perform analysis of the brown dwarf initial mass function, and then we mate that to the stellar initial mass function. In Section~\ref{sec:discussion} we discuss the resulting initial mass function over the entire mass range by comparing our fit of the functional form to other estimates in the literature, and in Section~\ref{sec:conclusions} we summarize our conclusions. Auxiliary data and analyses are found in the Appendices. In Appendix~\ref{sec:appendix_phot_spec_astrom} we present photometric, spectroscopic, and astrometric follow-up used to further characterize 20-pc census members and candidates, and in Appendix~\ref{sec:appendix_proximas} we present a list of the "proximal" systems for each constellation.

\section{Creating the 20-pc Census\label{sec:creating_the_census}}

\subsection{Building the list of 20-pc systems\label{sec:building_the_list_of_20pc_systems}}

Our starter list for compiling the census of 20-pc systems was the Preliminary Version of the Third Catalog of Nearby Stars (CNS3; \citealt{gliese1991}), which represents the sum knowledge, prior to large-area digital surveys, of stars believed to lie within 25 pc of the Sun. We took all objects in CNS3 and cross-identified them with the Gaia Early Data Release 3 (eDR3; \citealt{gaia2020}) to provide updated parallaxes. Objects with parallax values $<$50 mas were removed from further consideration, and those with values $\ge$50 mas or lacking a Gaia eDR3 parallax were retained. Two objects listed in the CNS3 as possibly being within 20 pc had no parallax in Gaia DR2, Gaia eDR3, or the literature. These were added to a list, shown in Table~\ref{tab:poss_20pc_members_hi_mass}, of potential 20-pc members to consider further. Other additions to this list are discussed in Section~\ref{sec:missing_M_dwarfs}.

\startlongtable
\begin{deluxetable*}{lcrrccccc}
\tabletypesize{\scriptsize}
\tablecaption{Stars Lacking Trigonometric Parallaxes but Possibly within 20 pc\label{tab:poss_20pc_members_hi_mass}}
\tablehead{
\colhead{Name} &
\colhead{Approx. J2000 Coords} &
\colhead{J2000 RA} &           
\colhead{J2000 Dec} &           
\colhead{Sp.\ Ty.} &
\colhead{Sp.\ Ty.} &
\colhead{$d_{est}$ Lit.} &
\colhead{$d_{est}$ Lit.} &
\colhead{$d_{est}$ Adopt.\tablenotemark{a}} \\
\colhead{} &
\colhead{(hhmm$\pm$ddmm)} &
\colhead{(deg)} &
\colhead{(deg)} &
\colhead{} &
\colhead{Ref.} &
\colhead{(pc)} &
\colhead{Ref.} &
\colhead{(pc)} \\
\colhead{(1)} &
\colhead{(2)} &           
\colhead{(3)} &           
\colhead{(4)} &
\colhead{(5)} &
\colhead{(6)} &
\colhead{(7)} &
\colhead{(8)} &
\colhead{(9)} \\
}
\startdata
EGGR 285 AB                 & 0037$-$2053&    9.352994& $-$20.895999& DA3+M3.5& 1,2& $\sim$16    & 1 & 51-63\\
2MASS J02133021$-$4654505 AB& 0213$-$4654&   33.376130& $-$46.914003& M3.5+M3.5& 3& 19.0$\pm$4.4& 2 & $\ge$29\\
2MASS J03323578+2843554 ABC & 0332+2843  &   53.149418&    28.731716& M4+M6$\gamma$+L0$\gamma$& 
                                                                                4,5& 15.8$\pm$3.1& 2 & 55\\
TYC 2885-494-1              & 0401+4254  &   60.291356&    42.908987& \nodata & \nodata& 17.7$\pm$3.5& 2 & $\ge$20\\
PM J04248+5339 E            & 0424+5339  &   66.220239&    53.663644& M4      & 6& 19.0$\pm$3.2& 2 & $\ge$23\\
LP 780-23 AB                & 0640$-$1627&  100.036317& $-$16.456009& M2.5    & 7& $\sim$20.0& 3 & $\ge$20\\
PM J06574+7405              & 0657+7405  &  104.357470&    74.090588& M4      & 8& 17.0$\pm$3.2& 2 & $\ge$21\\
2MASS J07543412+0832252     & 0754+0832  &  118.641273&     8.540408& M2.5    & 9& 17.8$\pm$3.3& 2 & $\ge$24\\
PM J07591+1719              & 0759+1719  &  119.779474&    17.329659& M4-5    &10& 19.0$\pm$3.7& 2 & $\ge$23\\
LP 617-21 AB                & 1315$-$0249&  198.827584&  $-$2.831640& M3.5+M4.5&11& $\sim$18.7& 3 & $\ge$22\\
GSC 03466-00805 AB          & 1341+4854  &  205.365957&    48.912086& M3      &12& 19.2$\pm$3.6& 2 & $\ge$26\\
LP 386-49 AB                & 1625+2601  &  246.383899&    26.027218& M3      & 8& $\sim$15& 1 & $\ge$21\\
LTT 8875                    & 2208$-$0824&  332.135825&  $-$8.415613& M2.5    & 14& $\sim$19.3& 3 & $\ge$27\\
L 166-44                    & 2234$-$6107&  338.521173& $-$61.128008& M4.5    & 15& $\sim$18.9& 3 & $\ge$22\\
LP 822-37 AB                & 2311$-$1701&  347.991624& $-$17.032996& M4      & 15,16& $\sim$18.8& 3 & $\ge$18\\
\enddata
\tablenotetext{a}{This is the adopted distance estimate. See per-object notes below for details.}
\tablecomments{0037$-$2053: \cite{farihi2006} estimate independent distances of 63 pc for the white dwarf and 51 pc for the M dwarf, placing the system well outside of 20 pc.}
\tablecomments{0213$-$4654: There is a single Gaia eDR3 entry for this source with $G = 13.12{\pm}0.01$ mag and $G_{Bp} - G_{Rp} = 2.83{\pm}0.01$ mag, the latter suggesting an M4 dwarf (\citealt{kiman2019}). \cite{kiman2019} find that $M_G \approx 10.4$ mag for an M3.5 dwarf or $M_G \approx 10.8$ mag for an M4. If the Gaia source represents joint photometry of the system, then the implied distance is $\ge$41 pc; if the Gaia source represents only one component, then the implied distance is $\ge$29 pc. In either case, this system appears to be outside of 20 pc.}
\tablecomments{0332+2843: This young system, a likely member of the $\beta$ Pic Moving Group, has a distance estimate of 55$\pm$4 pc from \cite{malo2014}, placing it well outside of 20 pc.}
\tablecomments{0401+4254: Gaia eDR3 measures $G = 10.70{\pm}0.01$ mag and $G_{Bp} - G_{Rp} = 2.19{\pm}0.01$ mag. The color suggests a value of $M_G \approx 9.2$ mag (type $\sim$M1.5-M2; \citealt{kiman2019}), implying a distance of 20 pc if the object is single. Given that the object has no five-parameter astrometric solution in Gaia eDR3, it is likely a multiple system, which would push this distance estimate even larger.}
\tablecomments{0424+5339: Gaia eDR3 measures $G = 13.12{\pm}0.01$ mag and $G_{Bp} - G_{Rp} = 2.86{\pm}0.01$ mag. The color suggests a value of $M_G \approx 11.3$ mag (type $\sim$M4-M4.5), implying a distance of 23 pc if the object is single.}
\tablecomments{0640$-$1627: \cite{winters2015} derived the $\sim$20 pc distance estimate under the assumption that this object was single. Gaia eDR3 splits this into two nearly equal-magnitude components, pushing the distance estimate beyond 20 pc.}
\tablecomments{0657+7405: Gaia eDR3 measures $G = 12.37{\pm}0.01$ mag and $G_{Bp} - G_{Rp} = 2.70{\pm}0.01$ mag. The color suggests a value of $M_G \approx 10.8$ mag (type $\sim$M4), implying a distance of 21 pc if the object is single. (A previously overlooked measurement of $\varomega_{abs}=37.8{\pm}4.1$ mas from \citealt{finch2016b} places this object at $\sim$26.5 pc.)}
\tablecomments{0754+0832: Gaia eDR3 measures $G = 11.79{\pm}0.01$ mag and $G_{Bp} - G_{Rp} = 2.46{\pm}0.01$ mag. The color suggests a value of $M_G \approx 10.0$ mag (type $\sim$M2.5-M3), implying a distance of 24 pc if the object is single.}
\tablecomments{0759+1719: Gaia eDR3 measures $G = 12.71{\pm}0.01$ mag and $G_{Bp} - G_{Rp} = 2.80{\pm}0.01$ mag. The color suggests a value of $M_G \approx 10.9$ mag (type $\sim$M3.5), implying a distance of 23 pc if the object is single. If the absolute magnitude is even fainter, as the \cite{bowler2019} spectral type suggest, this moves the single-object estimate within 20 pc.}
\tablecomments{1315$-$0249: There is a single Gaia eDR3 entry for this source with $G = 12.92{\pm}0.01$ mag and $G_{Bp} - G_{Rp} = 2.83{\pm}0.01$ mag. The color implies $M_G = 11.2$ mag (M4-M4.5). If the Gaia source represents only the primary, then the implied distance is 22 pc. If the Gaia magnitude is a joint magnitude, the implied distance is even larger. In either case, this system appears to be outside of 20 pc.}
\tablecomments{1341+4854: There is a single Gaia eDR3 entry for this source with $G = 12.44{\pm}0.01$ mag and $G_{Bp} - G_{Rp} = 2.64{\pm}0.01$ mag (M3.5-M4). The measured, joint spectral type of the system implies $M_G = 10.0-10.4$ mag. (The $\Delta{i^\prime} = 0.5$ mag of the binary measured by \citealt{lamman2020}, would imply M components separated by only a half spectral subclass, so we assume an absolute magnitude range encompassing M3-M3.5.) If the Gaia source represents only the primary, then the implied distance is $\ge$26 pc. Other assumptions push this value larger, so this system is assumed to lie beyond 20 pc.}
\tablecomments{1625+2601: There is a single Gaia eDR3 entry for this source with $G = 11.59{\pm}0.01$ mag and $G_{Bp} - G_{Rp} = 2.52{\pm}0.01$ mag (M3). The $\Delta{i^\prime} = 0.5$ mag of the binary measured by \cite{lamman2020}, would imply M components separated by only a half spectral subclass, so we assume an absolute magnitude range encompassing M2.5-M3, or $M_G = 9.7-10.0$ mag. If the Gaia source represents only the primary, then the implied distance is $\ge$21 pc. Other assumption push this value larger, so this system is assumed to lie beyond 20 pc. (A previously overlooked measurement of $\varomega_{abs}=39.7{\pm}7.2$ mas from \citealt{finch2016b} places this object at $\sim$25.2 pc.)}
\tablecomments{2208$-$0824: \cite{huber2016} estimate a distance of 31.1 pc using reduced proper motion and colors covering a wide wavelength baseline. \cite{scholz2005} estimate a distance of 27.5 pc based on the 2MASS $J$-band magnitude and spectral type.}
\tablecomments{2234$-$6107: Gaia eDR3 measures $G = 13.61{\pm}0.01$ mag and $G_{Bp} - G_{Rp} = 3.04{\pm}0.01$ mag. The color suggests a value of $M_G \approx 11.9$ mag (type $\sim$M5), implying a distance of 22 pc if the object is single.}
\tablecomments{2311$-$1701: \cite{reid2007} estimate a distance of 10.0 pc and \cite{scholz2005} estimate 17.4 pc, assuming the object is single in both cases. Gaia eDR3 splits this into two sources with $G = 12.61{\pm}0.01$ mag and $G = 13.25{\pm}0.01$ mag, having colors of $G_{Bp} - G_{Rp} = 2.85{\pm}0.01$ mag and $G_{Bp} - G_{Rp} = 2.84{\pm}0.01$ mag, respectively. The colors suggest a value of $M_G \approx 11.3$ mag (type $\sim$M4.5) for both components, implying a distance of 18-25 pc.}
\tablerefs{References for Sp.\ Ty.: (1) \citealt{koester2009}, (2) \citealt{farihi2006}, (3) \citealt{bergfors2016}, (4) \citealt{malo2014}, (5) \citealt{calissendorff2020}, (6) \citealt{terrien2015}, (7) \citealt{jeffers2018}, (8) \citealt{lepine2013}, (9) \citealt{alonso-floriano2015}, (10) \citealt{bowler2019}, (11) \citealt{janson2012}, (12) \citealt{rajpurohit2020}, (14) \citealt{scholz2005}, (15) \citealt{rajpurohit2013}, (16) \citealt{reid2007}.}
\tablerefs{References for $d_{est}$: (1) \citealt{gliese1991}, (2) \citealt{finch2014}, (3) \citealt{winters2015}}
\end{deluxetable*}

As the next step, we searched the SIMBAD Astronomical Database (\citealt{wenger2000}) for all objects with reported parallaxes $\ge$50 mas that were not already included above. We crossmatched these against Gaia eDR3, again retaining those with values $\ge$50 mas or lacking a Gaia eDR3 parallax and removing from further consideration those objects with Gaia parallax values $<$50 mas.  

Next, we created an independent list of 20-pc members by selecting objects with Gaia eDR3 parallax values $\ge$50 mas. This list was vetted by a group of Backyard Worlds: Planet 9 (hereafter, Backyard Worlds; \citealt{kuchner2017}\footnote{\url {https://www.zooniverse.org/projects/marckuchner/backyard-worlds-planet-9}}) citizen scientists to produce a list of bona fide 20-pc members alongside a list of potential 20-pc members that lacked independent verification of proximity, such as displaying unmistakable proper motion in archival imagery. Although most objects in the first Gaia-selected list were already in the master census discussed above, this Gaia selection nonetheless added another $\sim$60 discoveries to the total, as well as another $\sim$70 objects needing further scrutiny.

With this revised master census in hand, we checked against several other online sources and published papers to ensure that no objects had inadvertently been dropped. We consulted the lists of 10-pc objects produced by \cite{reyle2021}\footnote{See also \url{https://gucds.inaf.it/GCNS/The10pcSample .}} and the Research Consortium on Nearby Stars (RECONS)\footnote{This 01 Nov 2020 list is available at \url {http://recons.org/publishedpi.2020.1101}. Note that LHS 225AB, which is noted by RECONS to fall within 20 pc, is confirmed to fall outside 20 pc by Gaia eDR3.}, but this did not add any new objects. We also searched the Gaia Catalog of Nearby Stars (GCNS) published by \cite{smart2020}, but this likewise did not indicate any missing objects. For white dwarfs specifically, we further checked recent lists by \cite{sion2014}, \cite{mccook2016}, \cite{hollands2018}\footnote{\cite{hollands2018} suggest that WD 1443+256 is within 20 pc and that the Gaia DR2 parallax of 1.44$\pm$0.55 mas is in error, but the Gaia DR3 parallax seems to confirm that the object is truly distant ($\varomega_{abs}=1.43{\pm}0.04$ mas). Two other objects in \cite{hollands2018}, WD 0454+620 and WD 2140+078, are also shown to be outside of the 20-pc sample by Gaia DR3.}, \cite{mccleery2020}, \cite{gentile2021}, and \cite{obrien2023} and also found no omissions.

With the release of Gaia DR3 (\citealt{gaia2022}), we performed final checks of our list. The astrometry in DR3 is identical to that in eDR3 except for binary and higher order systems in which the astrometric and/or spectroscopic data could be used to establish physical parameters for individual components. Specifically, we found 55 objects within 20 pc that had revised astrometry. For two of these -- HD 64606 and NLTT 25223 -- the revised DR3 parallaxes place them outside the 20-pc volume. These objects were dropped from our list, and we updated the Gaia astrometry for the other 53. We also checked each of the non-single star lists accompanying the DR3 release to search for systems in which the revised astrometry may have pushed a distance closer than 20 pc. We found one such object -- Ross 59 -- which we added to our list.

Roughly three quarters of our resulting master census is comprised of objects with parallaxes in Gaia DR3. The other quarter is missing from DR3. Some of these objects, such as Sirius, are too bright for Gaia astrometry, whereas others, such as very faint brown dwarfs, are undetected by Gaia. Most of the rest are missing because they are likely in multiple systems for which the Gaia five-parameter astrometric solution has still not converged to a publishable solution.

Given that even Gaia DR3 has limitations for nearby multiple systems and very faint brown dwarfs, we have consulted additional publications to check for other possible 20-pc members that we may have missed in our checks above. Given that earlier type stars are likely bright enough to have been identified prior to 1991, these missing objects fall into two categories: (1) nearby M dwarfs -- which constitute the majority of stars in the solar neighborhood -- discovered since the 1991 update of CNS3, and (2) newly discovered L, T, and Y dwarfs. These additional checks are discussed in the subsections below.

\subsubsection{Other published M dwarfs\label{sec:missing_M_dwarfs}}

To better complete the M dwarf list, we first consulted the all-sky compilation of \cite{finch2014}, who used the US Naval Observatory fourth CCD Astrograph Catalog (UCAC4; \citealt{zacharias2013}) in concert with the American Association of Variable Star Observers (AAVSO) Photometric All-Sky Survey (APASS\footnote{\url{https://www.aavso.org/apass}}) and Two Micron All-Sky Survey (2MASS; \citealt{skrutskie2006}) to identify objects within 25 pc of the Sun. The methodology used a suite of color to absolute magnitude relations to provide distance estimates for detections, although this was supplemented with proper motion detection in order to further distinguish nearby stars from background sources. We took this list (their tables 5 and 6) and selected those candidates having \cite{finch2014} estimated distances $\le$20 pc and, if available, other published distance estimates $\le$20 pc from their table 6. This resulted in 267 objects not already in our master census created above. Of these, 251 had parallaxes in Gaia eDR3 (or Gaia DR2, if parallaxes were lacking in eDR3) placing them outside of 20 pc. Of the remaining 16 objects, seven were found to have other published parallaxes or additional distance estimates placing them beyond 20 pc. The final nine possible additions are listed in Table~\ref{tab:poss_20pc_members_hi_mass} for further scrutiny.

Second, we cross-checked our master table against a volume-complete subsample of 0.1-0.3 $M_\odot$ M dwarfs within 15 pc of the Sun (\citealt{winters2021}) whose parallax data were pulled from both Gaia DR2 and the literature. We found that all of the host stars in those systems were already included in our master list.

Third, we combed through The Solar Neighborhood series of papers by RECONS -- specifically papers I (\citealt{henry1994}) through XLIX (\citealt{vrijmoet2022}) -- to identify all objects verified or suspected to fall within 20 pc of the Sun. Our earlier checks had identified all of the confirmed 20-pc objects, but there were, however, a small number of nearby candidates from \cite{winters2015} that still lack a trigonometric parallax from any source. These were also added to Table~\ref{tab:poss_20pc_members_hi_mass}. 

As discussed in the footnotes to Table~\ref{tab:poss_20pc_members_hi_mass}, we have used available photometry and spectroscopy to update the distance estimates for these objects. After additional scrutiny, we find that only one  of these -- LP 822-37 AB -- likely falls within 20 pc, so it has been added to our master census.

\subsubsection{M, L, T, and Y dwarf discoveries from Backyard Worlds\label{sec:BYW_discoveries}}

Since the recent publication of our 20-pc L, T, and Y dwarf census (\citealt{kirkpatrick2021}), new nearby low-mass stars and brown dwarfs have continued to be recognized via discovery and/or additional follow-up. Examples are a new parallax confirming the nearby nature of the extreme T subdwarf WISEA J181006.18$-$101000.5 (\citealt{lodieu2022}), the discovery and confirming parallax of the late-T dwarf VVV J165507.19$-$421755.5 (\citealt{schapera2022}), and the discovery and established physical companionship of the possible Y dwarf companion to Ross 19 (\citealt{schneider2021}). Some other 20-pc suspects, such as CWISE J061741.79+194512.8 AB (\citealt{humphreys2023}), have also been shown to fall outside the 20-pc volume after additional follow-up. Still other candidates -- other isolated field brown dwarfs identified by the Backyard Worlds team -- may yet prove to be new members of the 20-pc census.

To assess the status of each of these, we list in Table~\ref{tab:poss_20pc_members_MLTY} all newer M, L, T, and Y dwarf discoveries that had initial distance estimates of $<$25 pc. To obtain more informed distance estimates of these candidates, in addition to providing additional data on other objects previously believed to be in the 20-pc census, we have searched photometric archives for additional data longward of 1 $\mu$m (along with Gaia magnitudes in the case of brighter sources) and have performed other photometric\footnote{All photometry in this paper is reported on the Vega system.}, spectroscopic, or astrometric follow-up on selected targets. Our own 1.25 $\mu$m and 1.65 $\mu$m follow-up and reductions, along with our reduction of archival data at 3.6 $\mu$m 4.5 $\mu$m, are described in section~\ref{sec:appendix_photometry}. Our optical and near-infrared spectroscopic follow-up is discussed in Section~\ref{sec:appendix_spectroscopy}. Additional parallactic measurements are described in Section~\ref{sec:appendix_astrometry}.


\begin{deluxetable*}{lll}
\tabletypesize{\scriptsize}
\tablecaption{Potentially New M, L, T, and Y Dwarf Members of the 20-pc Census\label{tab:poss_20pc_members_MLTY}}
\tablehead{
\colhead{Column} &
\colhead{Description} &
\colhead{Example Entry} \\
\colhead{(1)} &
\colhead{(2)} &
\colhead{(3)} \\
}
\startdata
Name         & Object's discovery designation with J2000 coordinates & CWISE J180308.71$-$361332.1\\
DiscoveryRef & Discovery reference\tablenotemark{a}& Q\\
SpOp         & Optical spectral type\tablenotemark{b}& \nodata\\
SpIR         & Near-infrared spectral type\tablenotemark{b}& \nodata\\
SpRf         & Reference for the spectral type\tablenotemark{c}& -\\
G            & $G$-band magnitude from Gaia DR3 (mag)& \nodata\\
G\_RP        & $G_{RP}$-band magnitude from Gaia DR3 (mag)& \nodata\\
W1           & W1 magnitude from the WISE catalog indicated by the object designation (mag)\tablenotemark{d}& 19.048\\
W1err        & Uncertainty in W1 (mag)& null\\
W2           & W2 magnitude from the WISE catalog indicated by the object designation (mag)& 14.948\\
W2err        & Uncertainty in W2 (mag)& 0.029\\
ch1          & Spitzer/IRAC ch1 magnitude (mag)& \nodata\\
ch1err       & Uncertainty in ch1 (mag)& \nodata\\
ch2          & Spitzer/IRAC ch2 magnitude (mag)& \nodata\\
ch2err       & Uncertainty in ch2 (mag)& \nodata\\
S            & Reference for the Spitzer photometry\tablenotemark{e}& -\\
JMKO         & $J$-band magnitude on the Mauna Kea Observatories filter system (mag)\tablenotemark{f}& 18.44\\
Jerr         & Uncertainty in JMKO (mag)& 0.16\\
H            & $H$-band magnitude on either the Mauna Kea Observatories or 2MASS filter system (mag)& \nodata\\
Herr         & Uncertainty in H (mag)& \nodata\\
Ph           & Reference for J and H photometry\tablenotemark{g}& v-\\
DateObs      & UT date of observation for any JMKO or H values reported for the first time in this paper& \nodata\\
pmra         & Proper motion in Right Ascension from CatWISE2020 (arcsec yr$^{-1}$)& 0.25550\\
pmrerr       & Uncertainty in pmra (arcsec yr$^{-1}$)& 0.0431\\
pmdec        & Proper motion in Declination from CatWISE2020 (arcsec yr$^{-1}$)& 0.01887\\
pmderr       & Uncertainty in pmdec (arcsec yr$^{-1}$)& 0.0454\\
pmtot        & Total proper motion (mas yr$^{-1}$)& 256.2\\
pmerr        & Uncertainty in pmtot (mas yr$^{-1}$)& 62.6\\
pmsg         & Significance of the CatWISE2020 proper motion measurement (pmtot/pmerr)& 4.1\\
d\_J         & Distance estimate based on the M$_J$ vs $J-$W2 relation (pc)& \nodata\\
d\_H         & Distance estimate based on the M$_H$ vs spectral type relation (pc)& \nodata\\
d\_ch2       & Distance estimate based on the M$_{ch2}$ vs ch1$-$ch2 relation (pc)& \nodata\\
d\_W2        & Distance estimate based on the M$_{W2}$ vs W1$-$W2 relation (pc)& $<$8.93\\
d\_G         & Distance estimate based on the M$_G$ vs $G-J$ relation (pc)& \nodata\\
d\_GRP       & Distance estimate based on the M$_{GRP}$ vs $G_{RP}-J$ relation (pc)& \nodata\\
d\_adp       & Adopted distance (pc)& 16.9\\
M            & Method used to determine d\_adp\tablenotemark{h}& 6\\
Res          & Result\tablenotemark{i}& in\\
JW2          & $J-$W2 color (mag)& 3.49\\
ch12         & ch1$-$ch2 color (mag)& \nodata\\
W12          & W1$-$W2 color (mag)& $>$4.10\\
GJ           & $G-J$ color (mag)& \nodata\\
GRPJ         & $G_{RP}-J$ color (mag)& \nodata\\
SpW          & Spectral type suggested by the W1$-$W2 color& \nodata\\
SpS          & Spectral type suggested by the ch1$-$ch2 color & \nodata\\
SpJW         & Spectral type suggested by the $J-$W2 color& \nodata\\
Note         & Additional notes for this object& $J-$W2 suggests $\sim$T8-8.5 and\\
& & implies $d{\approx}16.9$ pc; motion\\
& & confirmed in VVV $J$-band\\
& & images\\
\enddata
\tablecomments{This table describes the columns available in the full, online table.}
\tablenotetext{a}{Alphabetic characters refer to discoveries in this paper by Backyard Worlds citizen scientists, and numeric characters refer to published literature references. Both an alphabetic and a numeric code are listed in cases for which a citizen scientist re-discovered a published object that we felt required a fresh look --
     A = Nikolaj Stevnbak Andersen,      
     B = Paul Beaulieu,      
     C = Guillaume Colin,               
     D = Dan Caselden,                    
     E = Andres Stenner,       
     F = Guoyou Sun,
     G = Sam Goodman,                   
     H = Leslie K. Hamlet,
     I = Nikita V. Voloshin,
     J = J\"org Sch\"umann,                  
     K = Martin Kabatnik,                
     L = L\'eopold Gramaize,                  
     M = David W. Martin,         
     N = Karl Selg-Mann,
     O = Frank Kiwy,
     P = William Pendrill,
     Q = Tom Bickle,
     R = Austin Rothermich,
     S = Arttu Sainio,
     T = Melina Th\'evenot,
     U = Alexandru Dereveanco,             
     V = Christopher Tanner,             
     W = Jim Walla,                        
     X = Alexander Jonkeren,         
     Y = Benjamin Pumphrey,
     Z = Zbigniew W{\k e}dracki,
     a = Hiro Higashimura,
     b = John Sanchez,
     c = mar.bil,
     1 = \cite{meisner2020a}, 
     2 = \cite{meisner2020b},
     3 = \cite{schneider2016},
     4 = \cite{schneider2017}, 
     5 = \cite{schneider2020},
     6 = \cite{schneider2021}, 
     7 = \cite{schneider2022}, 
     8 = \cite{zhang2019},
     9 = \cite{kellogg2017},
     10 = \cite{best2020},
     11 = \cite{martin2018},
     12 = \cite{kirkpatrick2021},
     13 = \cite{faherty2020},
     14 = \cite{bardalez2020},
     15 = \cite{luhman2014},
     16 = \cite{kota2022},
     17 = \cite{schapera2022},
     18 = \cite{humphreys2023},
     19 = Rothermich et al. (2022, in prep.),
     20 = \cite{skrzypek2016}.}
\end{deluxetable*}

\begin{deluxetable*}{lll}
\tabletypesize{\scriptsize}
\tablenum{2}
\tablecaption{{\it (continued)}}
\tablehead{
\colhead{Column} &
\colhead{Description} &
\colhead{Example Entry} \\
\colhead{(1)} &
\colhead{(2)} &
\colhead{(3)} \\
}
\startdata
..............................&
...........................................................................................................................................&
................................\\
\enddata
\tablenotetext{b}{Codes are $-$5.0=M5, 0.0=L0, 5.0=L5, 10.0=T0, 15.0=T5, 20.0=Y0, etc.}
\tablenotetext{c}{The first character is for the optical type and second character for the near-infrared type --
   A = \cite{schneider2017},
   E = \cite{martin2018},
   F = \cite{faherty2020},
   H = \cite{humphreys2023},
   I = \cite{kirkpatrick2021},
   J = \cite{kirkpatrick2016},
   K = this paper,
   L = \cite{kellogg2017}, 
   M = \cite{meisner2020b},
   R = Rothermich et al. (2022, in prep.),
   S = \cite{schapera2022},
   X = \cite{schneider2020},
   Y = \cite{schneider2022},
   Z = \cite{zhang2019}.}
\tablenotetext{d}{Values lacking uncertainties are 2-$\sigma$ brightness upper limits.}
\tablenotetext{e}{Codes are --   
   a = \cite{meisner2020a},
   b = \cite{meisner2020b},
   K = This paper.}
\tablenotetext{f}{Values lacking uncertainties are 5-$\sigma$ brightness upper limits.}
\tablenotetext{g}{Two-character code for the reference to JMKO and H, respectively. Note that $H$-band magnitudes are generally included only for sources with measured spectral types --
   2 = 2MASS All-Sky Point Source Catalog,
   G = Gemini/FLAMINGOS-2 (this paper),
   K = Keck/MOSFIRE (this paper),
   P = Palomar/WIRC (this paper),
   S = \cite{schneider2021},
   U = UHS,
   u = ULAS or UGPS,
   V = VHS,
   v = VVV,
   Z = VVV photometry from \cite{schapera2022}.}
\tablenotetext{h}{Codes are --
   1 = The average of d\_J, d\_H, d\_ch2 is used,
   2 = d\_W2 alone is used,
   3 = d\_G alone is used,
   4 = d\_GRP alone is used,
   5 = The distance is determined from the Gaia DR3 parallax,
   6 = See Note for details.}
\tablenotetext{i}{Codes are --
   in = Object assumed to be located within 20 pc of the Sun and included in the 20-pc census,
   out = Object assumed to be located beyond 20 pc and not included in the 20-pc census.}
\end{deluxetable*}

Using this set of compiled data, we have recomputed distance estimates, as listed in Table~\ref{tab:poss_20pc_members_MLTY}. Column d$_J$ is the distance estimate derived by comparing the measured $J_{MKO}$ magnitude to the predicted $M_{JMKO}$ magnitude derived from the M$_{JMKO}$ vs.\ $J_{MKO}-$W2 relation of \cite{kirkpatrick2021}\footnote{Note that for this distance estimate and others that follow, we consider WISE W2 and Spitzer/IRAC ch2 photometry to be interchangeable, as shown in figure 15 of \citealt{kirkpatrick2021}, meaning that we can use the $J_{MKO}-$ch2 color in the published relation as a proxy for the $J_{MKO}-$W2 color.}. This estimate is valid only for objects with $J_{MKO}-{\rm W2} \ge 4.0$ mag, as smaller values may lead to non-unique solutions for M$_{JMKO}$ (figure 20a of \citealt{kirkpatrick2021}). Column d$_H$ is the distance estimate derived by comparing the measured $H$ magnitude to the predicted $M_{H}$ magnitude derived from the M$_H$ vs.\ spectral type relation of \cite{kirkpatrick2021}\footnote{Note that the MKO- and 2MASS-based $H$-band filters are essentially identical, so we use $H_{MKO}$ and $H_{2MASS}$ magnitudes interchangeably, as further discussed in section 3.1 of \citealt{kirkpatrick2011}.}. The published relation is restricted to types of L0 and later. Column d$_{ch2}$ is the distance estimate derived by comparing the measured ch2 magnitude to the predicted $M_{ch2}$ magnitude derived from the M$_{ch2}$ vs.\ ch1$-$ch2 relation of \cite{kirkpatrick2021}. This estimate is valid only for objects with $0.3 \le {\rm ch1}-{\rm ch2} \le 3.7$ mag, as shown in Figure 18c of \cite{kirkpatrick2021}. Method 1 (M = 1 in the table) takes the average of these three independent distance measurements -- or as many of these as can be derived -- as the adopted distance.

Column d$_{W2}$ is the distance estimate derived by comparing the measured W2 magnitude to the predicted $M_{W2}$ magnitude derived from the M$_{W2}$ (M$_{ch2}$) vs.\ W1$-$W2 relation of \cite{kirkpatrick2021}. This estimate is valid only for objects with $1.0 \le {\rm W1}-{\rm W2} \le 4.5$ mag, as shown in figure 19c of \cite{kirkpatrick2021}. Method 2 (M = 2 in the table) takes this estimate as the adopted distance. 

Column d$_{G}$ is the distance estimate derived by comparing the measured Gaia G magnitude to the predicted $M_{G}$ magnitude derived from an M$_{G}$ vs.\ G$-J$ relation derived specifically for this paper. This estimate is valid only for objects with $1.5 \le {\rm G}-J \le 5.0$ mag. Method 3 (M = 3 in the table) uses this as the adopted distance. Method 4 (M = 4 in the table) is exactly the same as Method 3 except that its distance estimate, d$_{GRP}$ uses the Gaia G$_{RP}$ magnitude instead of G and uses an M$_{GRP}$ vs.\ G$_{RP}-J$ relation also derived specifically for this paper. Method 5 (M = 5 in the table) uses the Gaia DR3 parallax, if available, to establish the distance.

When none of the five estimation methods above apply, we use combinations of colors to solve for degeneracies among possible spectral type or absolute magnitude solutions, as discussed in the notes to the table. For a very small number of objects, the adopted distance is left blank, as no estimate will be possible until additional follow-up is acquired. 

Finally, as another arbiter of proximity to the Sun, Table~\ref{tab:poss_20pc_members_MLTY} lists the measured CatWISE2020 proper motions (\citealt{marocco2021, eisenhardt2020}) and how significantly those measurements differ from zero. Also, using the color vs.\ spectral type relations given in \cite{kirkpatrick2021}, we have estimated spectral type based on the W1$-$W2 color (SpW), ch1$-$ch2 color (SpS), and $J-$W2 color (SpJW), where the valid color ranges are $0.4 \le {\rm W1}-{\rm W2} \le 4.0$ mag, $0.3 \le {\rm ch1}-{\rm ch2} \le 3.0$ mag, and $4.0 \le J-{\rm W2} \le 8.5$ mag.

Of the 211 candidate objects in the table, 44 have adopted distance estimates placing them closer than 20 pc (res = in, as listed in the table). Although we have tentatively added these 44 objects to the 20-pc census, obtaining parallaxes of all objects in Table~\ref{tab:poss_20pc_members_MLTY} still believed to be within 25 pc would be desirable to more carefully determine which are the true 20-pc members.

\subsection{20-pc stars with newly discovered companions\label{sec:new_companions}}

\begin{figure*}
\includegraphics[scale=0.45,angle=0]{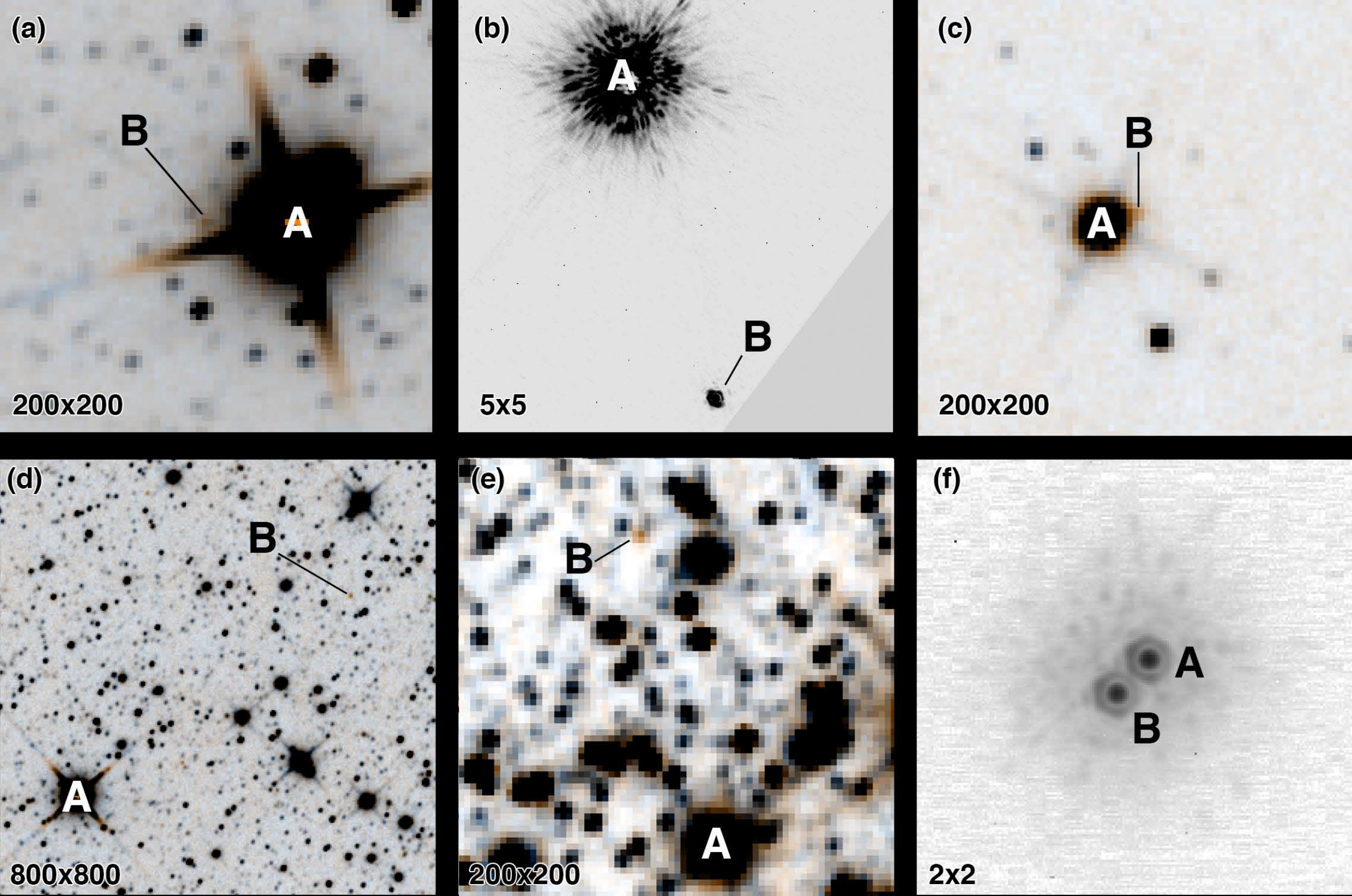}
\caption{Images illustrating the six new multiple systems discussed in the text. Each image has north up and east to the left, and the size of each (in arcsec) is noted in the legend. (a) HD 13579 and CWISER J021550.96+674017.2, (b) HD 17230 and Gaia DR3 25488745411919488, (c) G 43-23 and WISEU J100241.49+145914.9, (d) HD 170573 and CWISE J183207.94$-$540943.3, (e) G 155-42 and CWISE J184803.45$-$143232.3, and (f) 2MASS J19253089+0938235 A and B. All panels show WISE W1+W2 images from WiseView, except for panel (b), which shows a Keck/NIRC2 $K_p$-band image, and panel (f), which shows a Keck/NIRC2 $K$-band image.
\label{fig:companions}}
\end{figure*}

While assembling our nearby census, we discovered a small number of objects that fall in close proximity to other, higher mass stars in the list. These known stars and their possible companions are discussed further below and are illustrated in Figure~\ref{fig:companions}.

{\it HD 13579 (0215+6740), a K2 dwarf (\citealt{bidelman1985}) at 18.6 pc (Gaia DR3)}: The motion object CWISER J021550.96+674017.2 from Table~\ref{tab:poss_20pc_members_MLTY} was discovered by D.\ Caselden during a targeted search for companions to known 20-pc stars using multi-epoch imaging data from WISE (Figure~\ref{fig:companions}a). Follow-up $J_{MKO}$-band photometry from Keck/MOSFIRE (see Section~\ref{sec:appendix_photometry}) shows that current location of the companion coincides with a background source, rendering the $J-$W2 = 0.78$\pm$0.02 mag color useless as a gauge of spectral type. The measurement of W1$-$W2 = 0.49$\pm$0.02 mag from the CatWISE2020 Reject Catalog (\citealt{marocco2021}) is also contaminated, as the WISE imaging sequence shown in WiseView\footnote{\url{http://byw.tools/wiseview}} (\citealt{caselden2018}) indicates that this is a much redder source. The motion measurement from the CatWISE2020 Reject Catalog, 392$\pm$11 mas yr$^{-1}$ in RA and $-$102$\pm$10 mas yr$^{-1}$ in Dec, is also contaminated by background sources but shows a magnitude and direction roughly similar to the values for the 41$^{\prime\prime}$-separated K2 star HD 13579 (518.178$\pm$0.012 mas yr$^{-1}$ in RA and $-$305.636$\pm$0.014 mas yr$^{-1}$ in Dec; Gaia DR3). Using the WISE W2 epochal positions from the unTimely Catalog (\citealt{meisner2023-untimely}), a linear least-squares fit results in motions of 698$\pm$35 mas yr$^{-1}$ in RA and $-$500$\pm$71 mas yr$^{-1}$ in Dec, which are discrepant from the primary's motion values by 5.1$\sigma$ and 2.7$\sigma$ in RA and Dec, respectively. Curiously, the CatWISE2020 and unTimely motions bracket the Gaia motion values of the primary despite the fact that both the CatWISE2020 and unTimely measurements are WISE-based and are affected by the same background contaminants. Implanting a fake source into the WiseView image sequence with the same W2 magnitude as CWISER J021550.96+674017.2 but with the Gaia-measured motions of HD 13579 provides an excellent match to observed motion of CWISER J021550.96+674017.2 itself, but suggests that the CatWISE2020 value of W2 = 13.84$\pm$0.01 mag may be too bright. Given the close separation between the CWISER source and HD 13579 and motions that appear similar, we tentatively denote these as a physical pair with an apparent separation of 760 AU. If associated, the distance to HD 13579 implies a spectral type of $>$T4.5 for CWISER J021550.96+674017.2 based on the CatWISE W2 magnitude's possibly being biased too bright.

{\it HD 17230 (0246+1146), a K6 dwarf (\citealt{gray2003}) at 16.2 pc (Gaia DR3)}: K.\ Apps (see Section~\ref{sec:apps_catalog}) notes that there is a fainter star, Gaia DR3 25488745411919488 (G = 15.62 mag, $\Delta{G}$ = 7.51 mag), 3$\farcs$6 south of HD 17230 that has no parallax or proper motion solution in Gaia DR3. A search of the Keck Observatory Archive\footnote{\url {https://koa.ipac.caltech.edu}} by C.\ Gelino reveals two epochs of observations of HD 17230 with Keck/NIRC2 behind the adaptive optics system (\citealt{wizinowich2000}).  Raw images in the $K_p$ and $J$ filters with HD 17230 under a coronagraph (PI: J.\ Crepp; Program ID: C182N2) clearly show a star located $\sim$3.7$\arcsec$ from HD 17230 at a position angle of $\sim$195$^\circ$. This observation, taken on 2011 Aug 30 UT, can be compared to another taken on 2014 Oct 13 UT (PI: J.\ Crepp; Program ID: N100N2) in the narrow-band $K$-continuum. Given the substantial proper motion of HD 17230 of 263.88$\pm$0.03 mas yr$^{-1}$ in RA and $-211.58{\pm}0.03$ mas yr$^{-1}$ in Dec (Table~\ref{tab:20pc_census}), the fainter star should fall at a separation of $3{\farcs}4$ and position angle of $211^\circ$ if it were a background source. However, the second epoch shows the secondary at nearly the same separation and position angle as the first epoch, proving that the two stars are a common motion pair. This conclusion is further bolstered by the 2016-epoch Gaia DR3 positions, that place the fainter star at a separation of $3{\farcs}64$ and position angle of  $195.3^\circ$ from HD 17230. (Figure~\ref{fig:companions}b shows a first-epoch coronagraphic image in which the A component is seen only via its scattered light.) Using the Gaia DR3 parallax for the primary, this implies $M_G = 14.58$ mag (spectral type $\sim$M8) for the secondary. However, the true type might be somewhat later than this, as the lack of an astrometric solution in Gaia DR3 may mean that this companion is itself a multiple system. This companion, at an apparent physical separation of 59 AU, may be responsible for the radial velocity acceleration seen for HD 17230 over a decades-long timespan by \cite{rosenthal2021}.

{\it G 43-23 (1002+1149), an M4 dwarf (\citealt{reid1995}) at 17.9 pc (Gaia DR3)}: The motion object WISEU J100241.49+145914.9 from Table~\ref{tab:poss_20pc_members_MLTY} was discovered by D.\ Caselden during a targeted search for companions to known 20-pc stars using multi-epoch imaging data from WISE (Figure~\ref{fig:companions}c). This object lies only 15$\farcs$6 away from G 43-23, which has astrometry from Gaia DR3 of $\varomega_{abs} = 56.01{\pm}0.11$ mas, $\mu_\alpha = 157.02{\pm}0.11$ mas yr$^{-1}$, and $\mu_\delta = -235.65{\pm}0.10$ mas yr$^{-1}$. WISEU J100241.49+145914.9 itself is not listed in either the CatWISE2020 Catalog or the CatWISE2020 Reject Table, but a linear least-squares fit to its epochal unTimely positions (\citealt{meisner2023-untimely}) in W2 gives motions of $164{\pm}56$ mas yr$^{-1}$ in RA and $-233{\pm}43$ mas yr$^{-1}$ in Dec, nearly identical to the Gaia motions for the primary. Implanting a W2 = 14.55 mag source with the motions of G 43-23 into the WISE image sequence of WiseView (\citealt{caselden2018}) makes for a convincing doppelg{\"a}nger to WISEU J100241.49+145914.9 itself. The WISEU source's UHS detection at $J_{MKO} = 18.18{\pm}0.05$ mag results in a color of $J-$W2 = 3.63$\pm$0.11 mag, suggesting a type of $\sim$T8.5 and a distance of $\sim$14.3 pc, which is slightly closer than the 17.9 pc distance measured for G 43-23. Nonetheless, given the proximity of the two objects to each other and their nearly identical motions, we consider this to be a physical pair at an apparent physical separation of 280 AU. 

{\it HD 170573 (1833$-$5415), a K4.5 dwarf (\citealt{gray2006}) at 19.1 pc (Gaia DR3)}: The T7 dwarf CWISE J183207.94$-$540943.3 was discovered by G.\ Colin and B.\ Pumphrey and first published in \cite{kirkpatrick2021}, where Spitzer astrometric monitoring gave $\varomega_{abs} = 57.0{\pm}4.3$ mas, $\mu_\alpha = -129.1{\pm}11.6$ mas yr$^{-1}$, and $\mu_\delta = -172.1{\pm}9.7$ mas yr$^{-1}$. In assembling the full 20-pc census for this paper, it was noted that this object lies 10$\farcm$3 from the K4.5 dwarf HD 170573 (Figure~\ref{fig:companions}d), which has Gaia DR3 astrometric values of $\varomega_{abs} = 52.29{\pm}0.02$ mas, $\mu_\alpha = -121.05{\pm}0.02$ mas yr$^{-1}$, and $\mu_\delta = -142.04{\pm}0.02$ mas yr$^{-1}$. Until more accurate astrometry for the T dwarf becomes available, we will consider this pair to be physically associated because these values are only 1.1$\sigma$, 0.7$\sigma$, and 3.1$\sigma$ different for $\varomega_{abs}$, $\mu_\alpha$, and $\mu_\delta$, respectively. If a true binary, the projected separation is 11,800 AU.

{\it G 155-42 (1848$-$1434), an M3 dwarf (\citealt{gaidos2014}) at 17.1 pc (Gaia DR3)}: The motion object CWISE J184803.45$-$143232.3 from Table~\ref{tab:poss_20pc_members_MLTY} was discovered by S.\ Goodman while searching for unpublished motion objects in WISE imaging data. While assembling the 20-pc census for this paper, it was noted that this source falls 2$\farcm$45 away from G 155-42 (Figure~\ref{fig:companions}e). The CatWISE2020 Catalog (\citealt{marocco2021}) lists motions for CWISE J184803.45$-$143232.3 of $\mu_\alpha = -145{\pm}33$ mas yr$^{-1}$ and $\mu_\delta = -104{\pm}37$ mas yr$^{-1}$. A linear least-squares fit to the WISE W2 epochal positions from the unTimely Catalog (\citealt{meisner2022}) results in motions of $\mu_\alpha = -181{\pm}22$ mas yr$^{-1}$ and $\mu_\delta = -158{\pm}31$ mas yr$^{-1}$. The Gaia DR3 astrometry for G 155-42 is $\varomega_{abs} = 58.60{\pm}0.02$ mas, $\mu_\alpha = -236.45{\pm}0.02$ mas yr$^{-1}$, and $\mu_\delta = -237.26{\pm}0.02$ mas yr$^{-1}$. The measured motion values between the two sources differ by 2.8$\sigma$ and 3.6$\sigma$ in RA and Dec, respectively, for the CatWISE2020 motion of the potential secondary and by 2.5$\sigma$ and 2.6$\sigma$ for the unTimely motion. The $J-$W2 color of CWISE J184803.45$-$143232.3 from Table~\ref{tab:poss_20pc_members_MLTY} suggests a $\sim$T7.5 dwarf at a distance of $\sim$15.7 pc, which is sufficiently close to the 17.1 pc distance of G 155-42 that we tentatively consider them to be a physical pair with apparent physical separation of 2500 AU, pending improved astrometry for the secondary.

{\it 2MASS J19253089+0938235, an M8 dwarf (\citealt{west2015}) at 17.0 pc (Gaia DR3)}: C.\ Gelino finds two epochs of Keck/NIRC2 data for this object in the Keck Observatory Archive.  The first epoch (2019 May 22 UT; PI: Bond; Program ID: H299) shows two objects separated by 194 mas at a position angle of 146$^\circ$ and magnitude difference of ${\Delta}K$=0.29 mag.  Two objects are still present in the second epoch (2020 Jun 2 UT; PI: Mawet; Program ID: C249) but with a separation of 199 mas and position angle of 137$^\circ$ (Figure~\ref{fig:companions}f).  We conclude that 2MASS J19253089+0938235 is a closely-separated binary showing orbital motion because the pair shows measurably different separations and position angles but the astrometry of the second object is inconsistent with the motion of a background star, which would have exhibited a relative motion of approximately $-$80 mas in RA and +240 mas in Dec. Using a UKIDSS Galactic Plane Survey DR11PLUS star visible in the field and located at J2000 RA = 291.3801844 deg and Dec= +9.6377532 deg, we find $K$=10.53$\pm$0.03 mag for  2MASS J19253089+0938235A (the northwest component) and $K$=10.82$\pm$0.03 mag for 2MASS J19253089+0938235B (the southeast component). This object has been flagged as a possible member of the AB Dor Moving Group (\citealt{gagne2018b}). 

\subsection{Checks against the Fifth Catalog of Nearby Stars}

After we had completed our accounting of the 20-pc census, we were presented with an additional opportunity to further check for omissions or subtractions. \cite{golovin2022} recently published the Fifth Catalog of Nearby Stars (CNS5), a compilation of all stars and brown dwarfs within 25 pc of the Sun. Within the CNS5, there are 3,002 objects with parallaxes of 50 mas or greater, whereas our list has 3,588 individual objects that meet this criterion\footnote{Part of this discrepancy is due to the fact that the CNS5 has some entries whose components are not split into individual sources. Specifically, fifty entries are listed as double stars, seven as triples, and one as a quadruple. Even if these are split out as individual components, that still leaves a difference between the two lists of 519 objects.}. For the purposes of checking the completeness of our own census, we find that only twenty-two of these CNS5 objects were not included in our list. These are given in Table~\ref{tab:cns5_comparison}. Five of these are Gaia discoveries with relatively large Gaia parallax uncertainties. We show in section~\ref{sec:appendix_spectroscopy} that three of these are background objects based on their spectra, and we assume that the other two, given their even larger parallactic errors, are also background objects. Another fifteen have preferred parallaxes that place them beyond 20 pc, and these preferred parallaxes are either revised values in Gaia DR3 or published parallaxes (or new parallaxes discussed in Section~\ref{sec:appendix_astrometry}) with smaller uncertainties than those quoted in CNS5\footnote{We consider SIPS J1256$-$1257 B to be outside of the 20-pc volume because its primary star, SIPS J1256$-$1257 A, has a smaller parallax uncertainty and falls outside 20 pc. The same is true of Ross 776, based on the parallactic measurement for Ross 826, with which it shares common proper motion. For 2MASS J13585269+3747137 and 2MASSI J2249091+320549, we suspect that the CNS5 parallax values and uncertainties come from the same source as our values, \cite{best2020}, but have been rounded; however, the CNS5 does not cite individual references for its parallax entries, so we are not able to confirm this.}. The remaining two objects in Table~\ref{tab:cns5_comparison} deserve special note. The first, 2MASSI J0639559$-$741844, has a CNS5 parallax with a 16\% uncertainty, so we consider our spectrophotometric distance estimate, which places the object beyond 20 pc, to be preferable. (See \citealt{kirkpatrick2021} for a discussion on the credibility of parallaxes when the uncertainties exceed 12.5\%.) The second, APMPM J2330$-$4737 B, is a bit of a mystery, as we can find no corroborating evidence in the literature that it exists, and this is why it is not included in our census. In conclusion, our comparison to the CNS5 results in no new additions to our list.

\begin{deluxetable*}{lcccl}
\tabletypesize{\scriptsize}
\tablecaption{CNS5 Objects with $\varomega_{abs} \ge 50$ mas Not Included in Our 20-pc Census\label{tab:cns5_comparison}}
\tablehead{
\colhead{Object name} &
\colhead{CNS5 RA Dec (J2000)} &
\colhead{CNS5 $\varomega_{abs}$} &
\colhead{Our $\varomega_{abs}$} &
\colhead{Our reference} \\
\colhead{} &
\colhead{(hhmmss.ss$\pm$ddmmss.s)} &
\colhead{(mas)} &
\colhead{(mas)} &
\colhead{} \\
\colhead{(1)} &
\colhead{(2)} &
\colhead{(3)} &
\colhead{(4)} &
\colhead{(5)} 
}
\startdata
G 39-9                      &    04 22 34.31   +39 00 34.0&  50.03$\pm$0.03&  49.97$\pm$0.03   & Gaia DR3 \\
2MASS J05160945$-$0445499   &    05 16 09.41 $-$04 45 50.4&  54.00$\pm$4.00&  47.83$\pm$2.85   & Section 6.2 of \cite{kirkpatrick2021} \\
2MASSI J0639559$-$741844    &    06 39 55.99 $-$74 18 44.6&  51.00$\pm$8.00&  [46.1]           & Table 10 of \cite{kirkpatrick2021} \\
WISEA J064313.95+163143.6   &    06 43 13.99   +16 31 44.0&  50.05$\pm$0.27&  49.97$\pm$0.25   & Gaia DR3 \\
HD 64606 AC                 &    07 54 33.92 $-$01 24 45.2&  50.74$\pm$0.58&  48.55$\pm$0.13   & Gaia DR3 Non-single star lists \\
2MASS J08583467+3256275     &    08 58 34.32   +32 56 26.5&  50.30$\pm$3.70&  40.9$\pm$3.6     & \cite{best2020} \\
NLTT 25223                  &    10 45 14.83   +49 41 26.6&  53.13$\pm$0.42&  43.31$\pm$0.10   & Gaia DR3 Non-single star lists \\
CD$-$45 7872                &    12 35 58.50 $-$45 56 14.6&  52.67$\pm$3.05&  48.21$\pm$0.60   & Gaia DR2 \\
SIPS J1256$-$1257 B         &    12 56 01.85 $-$12 57 24.8&  52.00$\pm$3.00&  47.27$\pm$0.47   & Gaia DR3 (SIPS J1256$-$1257 A) \\
Kelu-1 AB                   &    13 05 39.80 $-$25 41 06.1&  53.80$\pm$0.70&  49.05$\pm$0.72   & Gaia DR3 \\
LP 220-13                   &    13 56 40.80   +43 42 59.8&  50.00$\pm$0.60&  46.30$\pm$0.58   & Gaia DR3 \\
2MASS J13585269+3747137     &    13 58 52.73   +37 47 12.8&  50.00$\pm$3.00&  49.6$\pm$3.1     & \cite{best2020} \\
Gaia DR3 6305165514134625024&    14 59 54.40 $-$18 32 15.9& 174.04$\pm$1.83&  background object& Table A2 \\
Gaia DR3 6013647666939138688&    15 29 22.77 $-$35 52 20.1&  56.76$\pm$0.97&  background object& Table A2 \\
SDSS J163022.92+081822.0    &    16 30 22.97   +08 18 22.3&  55.80$\pm$3.40&  41.76$\pm$2.79   & Table A3 \\
Gaia DR3 4118195139455558016&    17 38 53.15 $-$20 53 56.2&  53.16$\pm$2.33&  spurious parallax?  & Gaia DR3 \\
Gaia DR3 4062783361232757632&    17 59 55.76 $-$27 38 17.1&  59.45$\pm$1.18&  spurious parallax?  & Gaia DR3 \\
Gaia DR3 4479498508613790464&    18 39 31.62   +09 01 43.1& 121.98$\pm$0.93&  background object& Table A2 \\
Ross 776                    &    21 16 06.06   +29 51 51.5&  50.79$\pm$0.46&  49.91$\pm$0.02   & Gaia DR3 (Ross 826) \\
2MASSI J2249091+320549      &    22 49 10.08   +32 05 46.3&  50.00$\pm$3.00&  49.7$\pm$3.2     & \cite{best2020} \\
APMPM J2330$-$4737 B        &    23 30 15.28 $-$47 37 00.7&  73.67$\pm$0.08&  ---              & companion doesn't exist? \\
2MASS J23312378$-$4718274   &    23 31 23.92 $-$47 18 28.6&  56.50$\pm$7.50&  48.99$\pm$4.21   & Table A3 \\
\enddata
\end{deluxetable*}

\section{The 20-pc Census\label{sec:20pc_census}}

Our final 20-pc census is presented in Table~\ref{tab:20pc_census}. The content of this table is described in more detail in the subsections that follow.

\subsection{Nomenclature\label{sec:nomenclature}}

Not all researchers refer to the same star by the same name, so having a list of aliases is needed. As we entered each object into the census, we searched SIMBAD for alternative names. The name listed under the heading "DefaultName" in Table~\ref{tab:20pc_census} is the one that appeared as the default name in SIMBAD\footnote{Note that SIMBAD sometimes conflates system names and individual names. For example, there is a single record combining the system Ross 614 [AB] and the individual component Ross 614 A, although the component Ross 614 B has a separate record. Correcting these associations is beyond the scope of this paper.} when our initial search was performed.  For all of these objects, a deep dive into the literature is required to establish the current knowledge of multiplicity, spectral type, etc., so we also list alternative names to aid the literature search. Table~\ref{tab:20pc_census} therefore lists common names (e.g., Sirius), Bayer and Flamsteed designations, and designations from the HR, HD, BD, CD, and CPD catalogs. Table~\ref{tab:20pc_census} also lists designations from proper motion catalogs (Wolf, Ross, L, LP, G, LHS, LFT, NLTT, LTT, LSPM, SCR, UPM, APMPM, LEHPM, WT, SIPS, PM, and PM J), white dwarf catalogs (WD, LAWD, EGGR), all-sky photometric catalogs (2MASS, WISE), all-sky astrometric catalogs (Gaia, HIC, HIP, TYC, UCAC4, TIC), along with a few other miscellaneous catalogs that also have high usage (GJ, V*, Karmn, **). The field "VarType" is filled with the type of variability seen, if the object is a known variable star; this information was taken from the General Catalog of Variable Stars\footnote{\cite{samus2017} and \url{https://heasarc.gsfc.nasa.gov/W3Browse/all/gcvs.html}}. The references from which these designations were drawn are also listed in Table~\ref{tab:20pc_census} and serve as an homage to the many researchers who have helped advance our knowledge of the nearby census. 

A few common names ("CommonName"), not listed in SIMBAD, have been added from the list of star names\footnote{\url {https://www.iau.org/public/themes/naming\_stars/}} approved by the International Astronomical Union (IAU) and from \cite{allen1899}, along with certain double star names from the Washington Double Star (WDS) Catalog\footnote{\url {http://www.astro.gsu.edu/wds/} and \url{https://vizier.cds.unistra.fr/viz-bin/VizieR?-source=B/wds}}. The origin of these names is given under the heading "NamesRef", which is populated at the upper level for each system (i.e, on rows having integral values of "SystemHierarchy"). For 2MASS names, we supplemented the SIMBAD listings with the list of Gliese-2MASS crossmatches provided by \cite{stauffer2010}. For objects having no 2MASS-associated name in either of these lists, we searched the 2MASS Point Source Catalog (\citealt{cutri2003}) directly. In a few cases, SIMBAD listed more than one name with the "2MASS J" prefix, and for these we also checked the 2MASS Point Source Catalog directly to remove the incorrect association.

\clearpage

\startlongtable
\begin{deluxetable*}{llll}
\tabletypesize{\scriptsize}
\tablecaption{The 20-pc Census\label{tab:20pc_census}}
\tablehead{
\colhead{Column} &
\colhead{Description} &
\colhead{Sec} &
\colhead{Example Entry} \\
\colhead{(1)} &
\colhead{(2)} &
\colhead{(3)} &
\colhead{(4)} \\
}
\startdata
DefaultName & Default name in SIMBAD & \ref{sec:nomenclature} &nu Phe\\
\#CompsOnThisRow & Number of known components for this row & \ref{sec:multiplicity_and_exoplanets} & 1\\
AdoptedInitialMass & Adopted initial mass of this component ($M_\odot$) & \ref{sec:further_analysis} & 1.150\\
AdoptedInitialMassErr & Uncertainty in adopted initial mass ($M_\odot$) & \ref{sec:further_analysis} & 0.159\\
AdoptedInitialMassNote & Origin of adopted initial mass & \ref{sec:further_analysis} & TIC \\
Mass & Directly measured mass ($M_\odot$) & \ref{sec:mass_methods} & \nodata\\
MassErr & Uncertainty in directly measured mass ($M_\odot$) & \ref{sec:mass_methods} & \nodata\\
MassMethod & Method for direct mass measurement & \ref{sec:mass_methods} & \nodata\\
MassRef & Reference for direct mass measurement & \ref{sec:mass_methods} & \nodata\\
EstMassLit & Estimated mass from the literature ($M_\odot$) & \ref{sec:further_analysis} & \nodata\\
EstMassLitErr & Uncertainty on estimated mass from the literature ($M_\odot$) & \ref{sec:further_analysis} & \nodata\\
EstMassLitMethod & Method used for this estimated mass determination & \ref{sec:further_analysis} & \nodata\\
EstMassLitRef & Reference for this estimated mass & \ref{sec:further_analysis} & \nodata\\
EstMassTIC & Estimated mass from the TESS Input Catalog ($M_\odot$) & \ref{sec:mass_estimates_main_sequence}, \ref{sec:further_analysis} & 1.150\\
EstMassTICErr & Uncertainty on TESS Input Catalog estimated mass ($M_\odot$) & \ref{sec:mass_estimates_main_sequence}, \ref{sec:further_analysis} & 0.159\\
EstMassSH & Estimated mass from StarHorse ($M_\odot$) & \ref{sec:mass_estimates_main_sequence}, \ref{sec:further_analysis} & \nodata\\
EstMassSHErr & Uncertainty on StarHorse estimated mass ($M_\odot$) & \ref{sec:mass_estimates_main_sequence}, \ref{sec:further_analysis} & \nodata\\
EstMassMKs & Estimated mass from $M_{Ks}$ relation ($M_\odot$) & \ref{sec:mass_estimates_main_sequence}, \ref{sec:further_analysis} & \nodata\\
EstMassMKsErr & Uncertainty in $M_{Ks}$ relation estimated mass ($M_\odot$) & \ref{sec:mass_estimates_main_sequence}, \ref{sec:further_analysis} & \nodata\\
EstMassMG & Estimated mass from $M_G$ relation ($M_\odot$) & \ref{sec:mass_estimates_main_sequence}, \ref{sec:further_analysis} & \nodata\\
EstMassMGErr & Uncertainty in $M_G$ relation estimated mass ($M_\odot$) & \ref{sec:mass_estimates_main_sequence}, \ref{sec:further_analysis} & \nodata\\
Teff & Effective temperature, for L, T, Y dwarfs only (K) & \ref{sec:analysis_brown_dwarfs} & \nodata\\
Teff\_unc & Uncertainty in effective temperature (K) & \ref{sec:analysis_brown_dwarfs} & \nodata\\
\#Planets & No.\ of known exoplanets in NASA Exoplanet Archive & \ref{sec:multiplicity_and_exoplanets} & \nodata\\
RUWE & Gaia EDR3 renormalized unit weight error & \ref{sec:RUWE_LUWE} & 1.381\\
LUWE\_binary? & Possible binary flagged via local unit weight error & \ref{sec:RUWE_LUWE} & \nodata\\
Accelerator? & Accelerator flagged by \cite{brandt2021} or \cite{khovritchev2015} & \ref{sec:accelerators_lacking_other_info} & \nodata\\
EstMassAt3AU & \cite{kervella2022} mass estimate of companion if it is at 3 AU ($M_\odot$) & \ref{sec:accelerators_lacking_other_info} & \nodata\\
EstMassAt30AU & \cite{kervella2022} mass estimate of companion if it is at 30 AU ($M_\odot$) & \ref{sec:accelerators_lacking_other_info} & \nodata\\
SystemHierarchy & System hierarchy value & \ref{sec:multiplicity_and_exoplanets} & 158\\
\#CompsInThisSystem & No.\ of components, if this is top level of system & \ref{sec:multiplicity_and_exoplanets} & 1\\
SystemCode & System hierarchy value collapsed into an 8-digit integer & \ref{sec:multiplicity_and_exoplanets} & 01580000\\
CommonName & Common name & \ref{sec:nomenclature} & \nodata\\
Bayer/Flamsteed & Bayer or Flamsteed designation & \ref{sec:nomenclature} & *nu. Phe\\
HR & Bright Star Catalogue ("Harvard Revised") designation  & \ref{sec:nomenclature} & HR 370\\
HD & Henry Draper Catalogue designation & \ref{sec:nomenclature} & HD 7570\\
BD & Bonner Durchmusterung designation & \ref{sec:nomenclature} & \nodata\\
CD & Cordoba Durchmusterung designation & \ref{sec:nomenclature} & CD-46 346\\
CPD & Cape Photographic Durchmusterung designation & \ref{sec:nomenclature} & CPD-46 127\\
Wolf & Wolf motion survey designation & \ref{sec:nomenclature} & \nodata\\ 
Ross &   Ross motion survey designation & \ref{sec:nomenclature} & \nodata\\ 
L & 	 Bruce Proper Motion designation (``Luyten'', south) & \ref{sec:nomenclature} & \nodata\\ 
LP & 	 Luyten Palomar designation (north)  & \ref{sec:nomenclature} & \nodata\\ 
G & 	 Giclas motion survey designation & \ref{sec:nomenclature} & \nodata\\ 
LHS & 	 Luyten Half Second designation & \ref{sec:nomenclature} & LHS 1220\\   
LFT & 	 Luyten Five Tenths designation & \ref{sec:nomenclature} & LFT 119\\    
NLTT & 	 New Luyten Two Tenths & \ref{sec:nomenclature} & NLTT 4186\\ 
LTT & 	 Luyten Two Tenths designation & \ref{sec:nomenclature} & LTT 696\\ 
LSPM & 	 Lepine+Shara Proper Motion designation    & \ref{sec:nomenclature} & \nodata\\ 
SCR & 	 SuperCOSMOS+RECONS designation & \ref{sec:nomenclature} & \nodata\\ 
UPM & 	 UCAC3 Proper Motion designation & \ref{sec:nomenclature} & \nodata\\ 
APMPM &  Automated Plate Measurer Proper Motion designation & \ref{sec:nomenclature} & \nodata\\ 
LEHPM &  Liverpool-Edinburgh High Proper Motion designation & \ref{sec:nomenclature} & \nodata\\ 
WT & 	 Wroblewski+Torres motion survey designation & \ref{sec:nomenclature} & \nodata\\ 
SIPS & 	 Southern Infrared Proper motion Survey designation & \ref{sec:nomenclature} & \nodata\\ 
PM & 	 Proper Motion (B1950) survey designation & \ref{sec:nomenclature} & PM 01129-4548\\ 
PM J & 	 Proper Motion (J2000) survey designation & \ref{sec:nomenclature} & \nodata\\  	 
WD & White Dwarf designation & \ref{sec:nomenclature} & \nodata\\
LAWD & Luyten Atlas of White Dwarfs designation & \ref{sec:nomenclature} & \nodata\\
EGGR & Eggen+Greenstein designation & \ref{sec:nomenclature} & \nodata\\
2MASS & Two Micron All Sky Survey designation & \ref{sec:nomenclature} & 2MASS J01151112-4531540\\
WISE & Wide-field Infrared Survey Explorer designation & \ref{sec:nomenclature} & WISE J011511.83-453152.2\\
Gaia & Gaia designation & \ref{sec:nomenclature} & Gaia EDR3 4934923028038871296\\
HIC & Hipparcos Input Catalogue designation & \ref{sec:nomenclature} & HIC 5862\\
HIP & Hipparcos Catalogue designation & \ref{sec:nomenclature} & HIP 5862\\
TYC & Tycho-2 Catalog designation & \ref{sec:nomenclature} & TYC 8033-1232-1\\
UCAC4 & Fourth USNO CCD Astrograph Catalog designation & \ref{sec:nomenclature} & \nodata\\
TIC & TESS Input Catalog designation  & \ref{sec:nomenclature} & TIC 229092427\\
GJ & Gliese+Jahrei{\ss} nearby star catalog designation & \ref{sec:nomenclature} & GJ 55\\
V* & Variable Star designation & \ref{sec:nomenclature} & \nodata\\
VarType &  Type of variability seen, if column V* filled& \ref{sec:nomenclature} & \nodata\\
Karmn & CARMENES designation & \ref{sec:nomenclature} & \nodata\\
** & Multiple system designation & \ref{sec:nomenclature} & \nodata\\
NamesRef & Reference(s) for designations & \ref{sec:nomenclature} & SIMBAD\\
SexagesimalRA & Default J2000 right ascension, usually from SIMBAD & \ref{sec:astrometry} & 01 15 11.1214282378\\
SexagesimalDec & Default J2000 declination, usually from SIMBAD & \ref{sec:astrometry} & -45 31 53.992580679\\
RA & Decimal J2000 right ascension, if precision astrometry exists (deg) & \ref{sec:astrometry} & 18.80055895\\
RA\_unc & Uncertainty on decimal J2000 right ascension (mas) & \ref{sec:astrometry}& 0.0412\\
Dec & Decimal J2000 declination, if precision astrometry exists (deg) & \ref{sec:astrometry} & -45.53087326\\
Dec\_unc & Uncertainty on decimal J2000 declination (mas) & \ref{sec:astrometry} & 0.0499\\
Epoch & Epoch to which the decimal RA and Dec values above refer (yr) & \ref{sec:astrometry} & 2016.0\\
Parallax & Absolute parallax (mas) & \ref{sec:astrometry} & 65.527\\
Parallax\_unc & Uncertainty in the absolute parallax (mas) & \ref{sec:astrometry} & 0.0704\\
PMRA & Proper motion in right ascension (mas yr$^{-1}$) & \ref{sec:astrometry} & 665.086\\
PMRA\_unc & Uncertainty in PMRA (mas yr$^{-1}$) & \ref{sec:astrometry} & 0.052\\
PMDec & Proper motion in declination (mas yr$^{-1}$) & \ref{sec:astrometry} & 178.07\\
PMDec\_unc & Uncertainty in PMDec (mas yr$^{-1}$) & \ref{sec:astrometry} & 0.064\\
PlxPMRef & Reference for the parallax and proper motion values& \ref{sec:astrometry} & Gaia EDR3\\
Constellation & Constellation in which this object falls & \ref{sec:astrometry} & Phe\\
SpecTypeOpt & Published spectral type in the optical & \ref{sec:spectral_types} & F9 V Fe+0.4\\
SpTOpt\_indx & Machine-readable code for optical spectral type & \ref{sec:spectral_types} & 19.0\\
SpTOpt\_ref & Reference for the optical spectral type & \ref{sec:spectral_types} & Gray2006\\
SpecTypeNIR & Published spectral type in the near-infrared & \ref{sec:spectral_types} & \nodata\\
SpTNIR\_indx & Machine-readable code for the near-infrared type & \ref{sec:spectral_types} & \nodata\\
SpTNIR\_ref & Reference for the near-infrared spectral type & \ref{sec:spectral_types} & \nodata\\
Gaia\_RV & Radial velocity from Gaia DR3 (km s$^{-1}$) & \ref{sec:radial_velocities} & 11.90\\
Gaia\_RV\_unc & Uncertainty in Gaia\_RV (km s$^{-1}$) & \ref{sec:radial_velocities} & 0.12\\
G & $G$-band magnitude from Gaia eDR3 (mag) & \ref{sec:photometry} & 4.828\\
G\_unc & Uncertainty in $G$, as provided by VizieR (mag) & \ref{sec:photometry} & 0.003\\
G\_BP & $G_{BP}$-band magnitude from Gaia eDR3 (mag) & \ref{sec:photometry}& 5.108\\
G\_BP\_unc & Uncertainty in $G_{BP}$, as provided by VizieR (mag) & \ref{sec:photometry}& 0.003\\
G\_RP & $G_{RP}$-band magnitude from Gaia eDR3 (mag) & \ref{sec:photometry}& 4.380\\
G\_RP\_unc & Uncertainty in $G_{RP}$, as provided by VizieR (mag) & \ref{sec:photometry}& 0.004\\
JMKO & $J$-band photometry on the MKO system (mag)& \ref{sec:photometry}& \nodata\\
JMKOerr & Uncertainty in JMKO (mag) & \ref{sec:photometry}& \nodata\\
J2MASS & $J$-band photometry on the 2MASS system (mag)& \ref{sec:photometry}& 4.094\\
J2MASSerr & Uncertainty in J2MASS (mag) & \ref{sec:photometry}& 0.346\\
H & $H$-band photometry on the MKO system (mag)& \ref{sec:photometry}& 3.719\\
Herr & Uncertainty in H (mag) & \ref{sec:photometry}& 0.268\\
K & $K$-band photometry (mag)& \ref{sec:photometry}& \nodata\\
Kerr & Uncertainty in K (mag) & \ref{sec:photometry}& \nodata\\
Ks & $K_s$-band photometry (mag)& \ref{sec:photometry}& 3.782\\
Kserr & Uncertainty in Ks (mag) & \ref{sec:photometry}& 0.268\\
JHK\_ref & References for JMKO, J2MASS, H, K, and Ks& \ref{sec:photometry}& -22-2\\
2MASS\_contam? & Note if the 2MASS photometry is contaminated& \ref{sec:photometry}& \nodata\\
W1 & W1 photometry from WISE (mag)& \ref{sec:photometry}& 3.714\\
W1err & Uncertainty in W1 (mag) & \ref{sec:photometry}& 0.117\\
W2 & W2 photometry from WISE (mag)& \ref{sec:photometry}& 3.082\\
W2err & Uncertainty in W2 (mag) & \ref{sec:photometry}& 0.060\\
W3 & W3 photometry from WISE (mag)& \ref{sec:photometry}& 3.689\\
W3err & Uncertainty in W3 (mag) & \ref{sec:photometry}& 0.014\\
W4 & W4 photometry from WISE (mag)& \ref{sec:photometry}& 3.609\\
W4err & Uncertainty in W4 (mag) & \ref{sec:photometry}& 0.023\\
WISEphot\_ref & References for W1, W2, W3, and W4& \ref{sec:photometry}& WWWW\\
WISE\_contam? & Note if the WISE photometry is contaminated& \ref{sec:photometry}& \nodata\\
GeneralNotes & Special notes on this system/component& \nodata& \nodata\\
\enddata
\tablecomments{This summary table describes the columns available in the full, online table. This table is also available at the NASA Exoplanet Archive, {\url  https://exoplanetarchive.ipac.caltech.edu/docs/20pcCensus.html}.}
\tablerefs{References for mass measurements and estimates, astrometry, spectral types, and general notes --
Aberasturi2014=\cite{aberasturi2014},
Aberasturi2014b=\cite{aberasturi2014b},
Abt1965=\cite{abt1965},
Abt1970=\cite{abt1970},
Abt1976=\cite{abt1976},
Abt1981=\cite{abt1981},
Abt2006=\cite{abt2006},
Abt2017=\cite{abt2017},
Affer2005=\cite{affer2005},
Agati2015=\cite{agati2015},
Akeson2021=\cite{akeson2021},
Albert2011=\cite{albert2011},
Allen2000=\cite{allen2000},
Allen2012=\cite{allen2012},
AllendePrieto1999=\cite{allendeprieto1999},
Allers2013=\cite{allers2013},
Alonso-Floriano2015=\cite{alonso-floriano2015},
Andrade2019=\cite{andrade2019},
Artigau2010=\cite{artigau2010},
Azulay2015=\cite{azulay2015},
Azulay2017=\cite{azulay2017},
Bach2009=\cite{bach2009},
Bagnulo2020=\cite{bagnulo2020},
Baines2012=\cite{baines2012},
Baines2018=\cite{baines2018},
Bakos2006=\cite{bakos2006},
Balega1984=\cite{balega1984},
Balega2004=\cite{balega2004},
Balega2013=\cite{balega2013},
BardalezGagliuffi2014=\cite{bardalez2014},
BardalezGagliuffi2019=\cite{bardalez2019},
BardalezGagliuffi2020=\cite{bardalez2020},
Baroch2018=\cite{baroch2018},
Baroch2021=\cite{baroch2021},
Barry2012=\cite{barry2012},
Bartlett2017=\cite{bartlett2017},
Batten1992=\cite{batten1992},
Bazot2011=\cite{bazot2011},
Bazot2018=\cite{bazot2018},
Beamin2013=\cite{beamin2013},
Beavers1985=\cite{beavers1985},
Beichman2011=\cite{beichman2011},
Benedict2001=\cite{benedict2001},
Benedict2016=\cite{benedict2016},
Berdnikov2008=\cite{berdnikov2008},
Bergfors2010=\cite{bergfors2010},
Bergfors2016=\cite{bergfors2016}
Bernat2010=\cite{bernat2010}
Bernkopf2012=\cite{bernkopf2012},
Berski2016=\cite{berski2016},
Best2013=\cite{best2013},
Best2015=\cite{best2015},
Best2020=\cite{best2020},
Best2021=\cite{best2021},
Beuzit2004=\cite{beuzit2004},
Bidelman1980=\cite{bidelman1980},
Bidelman1985=\cite{bidelman1985},
Bihain2013=\cite{bihain2013},
Biller2022=\cite{biller2022},
Bochanski2005=\cite{bochanski2005},
Bonavita2020=\cite{bonavita2020},
Boden1999=\cite{boden1999},
Bond2017=\cite{bond2017},
Bond2018=\cite{bond2018},
Bond2020=\cite{bond2020},
Bonfils2005=\cite{bonfils2005},
Bonfils2013=\cite{bonfils2013},
Bonnefoy2014=\cite{bonnefoy2014},
Bonnefoy2018=\cite{bonnefoy2018},
Borgniet2019=\cite{borgniet2019},
Bouy2003=\cite{bouy2003},
Bouy2004=\cite{bouy2004},
Bouy2005=\cite{bouy2005},
Bowler2015a=\cite{bowler2015a},
Bowler2015b=\cite{bowler2015b},
Bowler2019=\cite{bowler2019},
Boyajian2012=\cite{boyajian2012},
Brandao2011=\cite{brandao2011},
Brandt2014=\cite{brandt2014},
Brandt2019=\cite{brandt2019},
Brandt2020=\cite{brandt2020},
Brandt2021=\cite{brandt2021},
BrandtG2021=\cite{brandtg2021},
Breakiron1974=\cite{breakiron1974},
Brewer2016=\cite{brewer2016},
Bruntt2010=\cite{bruntt2010},
Burgasser2003=\cite{burgasser2003},
Burgasser2004=\cite{burgasser2004},
Burgasser2006=\cite{burgasser2006},
Burgasser2007=\cite{burgasser2007},
Burgasser2008=\cite{burgasser2008},
Burgasser2008b=\cite{burgasser2008b},
Burgasser2010a=\cite{burgasser2010a},
Burgasser2010b=\cite{burgasser2010b},
Burgasser2011=\cite{burgasser2011},
Burgasser2013=\cite{burgasser2013},
Burgasser2015a=\cite{burgasser2015a},
Burgasser2015b=\cite{burgasser2015b},
Burningham2010=\cite{burningham2010},
Burningham2011=\cite{burningham2011},
Burningham2013=\cite{burningham2013},
Butler2017=\cite{butler2017},
Bychkov2013=\cite{bychkov2013},
Calissendorff2023=\cite{calissendorff2023},
Cannon1993=\cite{henrydraperextension},
CardonaGuillen2021=\cite{cardonaguillen2021},
Carrier2005=\cite{carrier2005},
Castro2013=\cite{castro2013},
Catalan2008b=\cite{catalan2008b},
CatWISE2020=\cite{marocco2021},
Chauvin2007=\cite{chauvin2007},
Che2011=\cite{che2011},
Chen2022=\cite{chen2022},
Chini2014=\cite{chini2014},
Chiu2006=\cite{chiu2006},
Chontos2021=\cite{chontos2021},
Christian2001=\cite{christian2001},
Christian2003=\cite{christian2003},
Cifuentes2020=\cite{cifuentes2020},
Clark2022=\cite{clark2022},
Climent2019=\cite{climent2019},
Close2007=\cite{close2007},
Compton2019=\cite{compton2019},
Corbally1984=\cite{corbally1984},
CortesContreras2017=\cite{cortescontreras2017},
Cowley1967=\cite{cowley1967},
Cowley1976=\cite{cowley1976},
Creevey2012=\cite{creevey2012},
Crifo2005=\cite{crifo2005},
Cruz2002=\cite{cruz2002},
Cruz2003=\cite{cruz2003},
Cruz2007=\cite{cruz2007},
Cruz2009=\cite{cruz2009},
Cushing2005=\cite{cushing2005},
Cushing2011=\cite{cushing2011},
Cushing2014=\cite{cushing2014},
Cushing2016=\cite{cushing2016},
Cvetkovic2010=\cite{cvetkovic2010},
Cvetkovic2011=\cite{cvetkovic2011},
Dahn1988=\cite{dahn1988},
Dahn2002=\cite{dahn2002},
Dahn2017=\cite{dahn2017},
Dalba2021=\cite{dalba2021},
Damasso2020=\cite{damasso2020},
David2015=\cite{david2015},
Davison2014=\cite{davison2014},
Deacon2012a=\cite{deacon2012a},
Deacon2012b=\cite{deacon2012b},
Deacon2017=\cite{deacon2017},
Deeg2008=\cite{deeg2008},
Delfosse1999a=\cite{delfosse1999a},
Delfosse1999b=\cite{delfosse1999b},
Delrez2021=\cite{delrez2021},
Diaz2007=\cite{diaz2007},
Dieterich2012=\cite{dieterich2012},
Dieterich2014=\cite{dieterich2014},
Dieterich2018=\cite{dieterich2018},
Dieterich2021=\cite{dieterich2021},
DiFolco2004=\cite{difolco2004},
Dittmann2014=\cite{dittmann2014},
Docobo2006=\cite{docobo2006},
Docobo2019=\cite{docobo2019},
DOrazi2017=\cite{dorazi2017},
DosSantos2017=\cite{dossantos2017},
Downes2006=\cite{downes2006},
Ducati2011=\cite{ducati2011},
Dupuy2010=\cite{dupuy2010},
Dupuy2012=\cite{dupuy2012},
Dupuy2016=\cite{dupuy2016},
Dupuy2017=\cite{dupuy2017},
Dupuy2019=\cite{dupuy2019},
Duquennoy1988=\cite{duquennoy1988},
Duquennoy1991=\cite{duquennoy1991},
Durkan2018=\cite{durkan2018},
Edwards1976=\cite{edwards1976},
Eggleton2008=\cite{eggleton2008},
Endl2006=\cite{endl2006},
Evans1961=\cite{evans1961},
Evans1964=\cite{evans1964},
Fabricius2000=\cite{fabricius2000},
Faherty2012=\cite{faherty2012},
Faherty2016=\cite{faherty2016},
Faherty2018=\cite{faherty2018},
Fan2000=\cite{fan2000},
Feng2021=\cite{feng2021},
Finch2016=\cite{finch2016},
Finch2018=\cite{finch2018},
Fischer2014=\cite{fischer2014},
Forveille1999=\cite{forveille1999},
Forveille2004=\cite{forveille2004},
Forveille2005=\cite{forveille2005},
Fouque2018=\cite{fouque2018},
Fuhrmann2008=\cite{fuhrmann2008},
Fuhrmann2011=\cite{fuhrmann2011},
Fuhrmann2012=\cite{fuhrmann2012},
Fuhrmann2012b=\cite{fuhrmann2012b},
Fuhrmann2016=\cite{fuhrmann2016},
Fuhrmann2017=\cite{fuhrmann2017},
}
\end{deluxetable*}

\clearpage

\startlongtable
\begin{deluxetable*}{llll}
\tabletypesize{\scriptsize}
\tablenum{4}
\tablecaption{{\it (continued)}}
\tablehead{
\colhead{Column} &
\colhead{Description} &
\colhead{Sec} &
\colhead{Example Entry} \\
\colhead{(1)} &
\colhead{(2)} &
\colhead{(3)} &
\colhead{(4)} \\
}
\startdata
.................................& ...............................................................................................& ...........& ..............................................\\
\enddata
\tablerefs{
Gagne2015=\cite{gagne2015},
GaiaDR2=\cite{gaia2016}, \cite{gaia2018}
GaiaEDR3=\cite{gaia2016}, \cite{gaia2021}
GaiaDR3-NSS=\cite{gaiaDR3-NSS},
Gaidos2014=\cite{gaidos2014},
Garcia2017=\cite{garcia2017},
Gardner2021=\cite{gardner2021},
Gatewood2003=\cite{gatewood2003},
Geballe2002=\cite{geballe2002},
GentileFusillo2019=\cite{gentilefusillo2019},
Giammichele2012=\cite{giammichele2012},
Gigoyan2012=\cite{gigoyan2012},
Gizis1997=\cite{gizis1997},
Gizis1997b=\cite{gizis1997b},
Gizis2000=\cite{gizis2000},
Gizis2000b=\cite{gizis2000b},
Gizis2002=\cite{gizis2002},
Gizis2002b=\cite{gizis2002b},
Gizis2011=\cite{gizis2011},
Gizis2015=\cite{gizis2015},
Gliese1991=\cite{gliese1991},
Gomes2013=\cite{gomes2013},
Goldin2006=\cite{goldin2006},
Goldin2007=\cite{goldin2007},
Goto2002=\cite{goto2002},
Grandjean2020=\cite{grandjean2020},
Gray2001=\cite{gray2001},
Gray2003=\cite{gray2003},
Gray2006=\cite{gray2006},
Greco2019=\cite{greco2019},
Griffin2004=\cite{griffin2004},
Griffin2010=\cite{griffin2010},
Guenther2003=\cite{guenther2003},
Guzik2016=\cite{guzik2016},
Halbwachs2000=\cite{halbwachs2000},
Halbwachs2012=\cite{halbwachs2012},
Halbwachs2018=\cite{halbwachs2018},
Hambaryan2004=\cite{hambaryan2004},
Hansen2022=\cite{hansen2022},
Harrington1993=\cite{harrington1993},
Hartkopf1994=\cite{hartkopf1994},
Hartkopf2012=\cite{hartkopf2012},
Hatzes2012=\cite{hatzes2012},
Hawley1997=\cite{hawley1997},
Hawley2002=\cite{hawley2002},
Heintz1986=\cite{heintz1986},
Heintz1990=\cite{heintz1990},
Heintz1993=\cite{heintz1993},
Heintz1994=\cite{heintz1994},
Helminiak2009=\cite{helminiak2009},
Helminiak2012=\cite{helminiak2012},
Henry1994=\cite{henry1994},
Henry1999=\cite{henry1999},
Henry2002=\cite{henry2002},
Henry2004=\cite{henry2004},
Henry2006=\cite{henry2006},
Henry2018=\cite{henry2018},
HenryDraperExtension=\cite{henrydraperextension},
Herbig1977=\cite{herbig1977},
Hinkley2011=\cite{hinkley2011},
Hipparcos=\cite{vanleeuwen2007},
Hollands2018=\cite{hollands2018},
Horch2011=\cite{horch2011},
Horch2012=\cite{horch2012},
Horch2017=\cite{horch2017},
Holberg2002=\cite{holberg2002},
Houk1982=\cite{houk1982},
Houk1988=\cite{houk1988},
Huber2009=\cite{huber2009},
Hussein2022=\cite{hussein2022},
Hsu2021=\cite{hsu2021},
Ireland2008=\cite{ireland2008},
Jackson1955=\cite{jackson1955},
Jahreiss2001=\cite{jahreiss2001},
Jahreiss2008=\cite{jahreiss2008},
Janson2012=\cite{janson2012},
Janson2014a=\cite{janson2014a},
Janson2014b=\cite{janson2014b},
Jao2003=\cite{jao2003},
Jao2008=\cite{jao2008},
Jao2011=\cite{jao2011},
Jao2014=\cite{jao2014},
Jeffers2018=\cite{jeffers2018},
Jeffers2020=\cite{jeffers2020},
Jeffries1993=\cite{jeffries1993},
Jodar2013=\cite{jodar2013},
Kallinger2010=\cite{kallinger2010},
Kallinger2019=\cite{kallinger2019},
Karovicova2022=\cite{karovicova2022},
Kasper2007=\cite{kasper2007},
Katoh2013=\cite{katoh2013},
Katoh2021=\cite{katoh2021},
Keenan1989=\cite{keenan1989},
Kellogg2015=\cite{kellogg2015},
Kendall2004=\cite{kendall2004},
Kendall2007=\cite{kendall2007},
Kennedy2012=\cite{kennedy2012},
Kervella2016=\cite{kervella2016},
Kervella2016b=\cite{kervella2016b},
Kervella2019=\cite{kervella2019},
Kervella2022=\cite{kervella2022},
Kesseli2019=\cite{kesseli2019},
Khovritchev2015=\cite{khovritchev2015},
Kilic2020=\cite{kilic2020},
King2010=\cite{king2010},
Kirkpatrick1991=\cite{kirkpatrick1991},
Kirkpatrick1994=\cite{kirkpatrick1994},
Kirkpatrick1995=\cite{kirkpatrick1995},
Kirkpatrick1997=\cite{kirkpatrick1997},
Kirkpatrick1999=\cite{kirkpatrick1999},
Kirkpatrick2000=\cite{kirkpatrick2000},
Kirkpatrick2001=\cite{kirkpatrick2001},
Kirkpatrick2008=\cite{kirkpatrick2008},
Kirkpatrick2010=\cite{kirkpatrick2010},
Kirkpatrick2011=\cite{kirkpatrick2011},
Kirkpatrick2012=\cite{kirkpatrick2012},
Kirkpatrick2013=\cite{kirkpatrick2013},
Kirkpatrick2014=\cite{kirkpatrick2014},
Kirkpatrick2016=\cite{kirkpatrick2016},
Kirkpatrick2019=\cite{kirkpatrick2019},
Kirkpatrick2021a=\cite{kirkpatrick2021},
Kirkpatrick2021b=\cite{kirkpatrick2021b},
Kirkpatrick2024=This paper,
Kiyaeva2001=\cite{kiyaeva2001},
Kluter2018=\cite{kluter2018},
Kniazev2013=\cite{kniazev2013},
Kochukhov2009=\cite{kochukhov2009},
Kochukhov2019=\cite{kochukhov2019},
Koen2017=\cite{koen2017},
Koizumi2021=\cite{koizumi2021},
Konopacky2010=\cite{konopacky2010},
Kraus2011=\cite{kraus2011},
Kuerster2008=\cite{kuerster2008},
Kuzuhara2013=\cite{kuzuhara2013},
Lacour2021=\cite{lacour2021},
Lamman2020=\cite{lamman2020},
Laugier2019=\cite{laugier2019},
Law2006=\cite{law2006},
Lazorenko2018=\cite{lazorenko2018},
Lee1984=\cite{lee1984},
Leggett2012=\cite{leggett2012},
Leinert2000=\cite{leinert2000},
Lepine2002=\cite{lepine2002},
Lepine2003=\cite{lepine2003},
Lepine2009=\cite{lepine2009},
Lepine2013=\cite{lepine2013},
Li2012=\cite{li2012},
Li2019=\cite{li2019},
Liebert2003=\cite{liebert2003},
Liebert2006=\cite{liebert2006},
Liebert2013=\cite{liebert2013},
Limoges2015=\cite{limoges2015},
Lindegren1997=\cite{lindegren1997},
Lindegren2018=\cite{lindegren2018},
Lindegren2021=\cite{lindegren2021},
Liu2002=\cite{liu2002},
Liu2005=\cite{liu2005},
Liu2010=\cite{liu2010},
Liu2012=\cite{liu2012},
Liu2016=\cite{liu2016},
Lloyd1994=\cite{lloyd1994},
Lodieu2005=\cite{lodieu2005},
Lodieu2007=\cite{lodieu2007},
Lodieu2012=\cite{lodieu2012},
Lodieu2022=\cite{lodieu2022},
Looper2007=\cite{looper2007},
Looper2008=\cite{looper2008},
LopezMorales2007=\cite{lopez2007},
Loth1998=\cite{loth1998},
Loutrel2011=\cite{loutrel2011},
Low2021=\cite{low2021},
Lowrance2002=\cite{lowrance2002},
Luck2017=\cite{luck2017},
Luhman2012=\cite{luhman2012},
Luhman2013=\cite{luhman2013},
Luhman2014=\cite{luhman2014},
Luhman2014b=\cite{luhman2014b},
Lurie2014=\cite{lurie2014},
Mace2013a=\cite{mace2013a},
Mace2013b=\cite{mace2013b},
Mace2018=\cite{mace2018},
Makarov2007=\cite{makarov2007},
Malkov2012=\cite{malkov2012},
Malkov2006=\cite{malkov2006},
Malo2014b=\cite{malo2014b},
Malogolovets2007=\cite{malogolovets2007},
Mamajek2012=\cite{mamajek2012},
Mamajek2018=\cite{mamajek2018},
Manjavacas2013=\cite{manjavacas2013},
Mann2019=\cite{mann2019},
Mariotti1990=\cite{mariotti1990},
Marocco2010=\cite{marocco2010},
Marocco2013=\cite{marocco2013},
Martin1995=\cite{martin1995},
Martin1998=\cite{martin1998},
Martin1998b=\cite{martin1998b},
Martin2018=\cite{martin2018},
Martinache2007=\cite{martinache2007},
Martinache2009=\cite{martinache2009},
Mason2009=\cite{mason2009},
Mason2017=\cite{mason2017},
Mason2018=\cite{mason2018},
Mason2018b=\cite{mason2018b},
McCleery2020=\cite{mccleery2020},
McCook2016=\cite{mccook2016},
Meisner2020a=\cite{meisner2020a},
Meisner2020b=\cite{meisner2020b},
Mendez2021=\cite{mendez2021},
Merc2021=\cite{merc2021},
Metcalfe2021=\cite{metcalfe2021},
Mitrofanova2020=\cite{mitrofanova2020},
Mitrofanova2021=\cite{mitrofanova2021},
Monnier2007=\cite{monnier2007},
Monnier2012=\cite{monnier2012},
Montagnier2006=\cite{montagnier2006},
}
\end{deluxetable*}

\startlongtable
\begin{deluxetable*}{llll}
\tabletypesize{\scriptsize}
\tablenum{4}
\tablecaption{{\it (continued)}}
\tablehead{
\colhead{Column} &
\colhead{Description} &
\colhead{Sec} &
\colhead{Example Entry} \\
\colhead{(1)} &
\colhead{(2)} &
\colhead{(3)} &
\colhead{(4)} \\
}
\startdata
.................................& ...............................................................................................& ...........& ..............................................\\
\enddata
\tablerefs{
Montes2006=\cite{montes2006},
Montes2007=\cite{montes2007},
Morales2009=\cite{morales2009},
Morbey1987=\cite{morbey1987},
Morgan1973=\cite{morgan1973},
Mosser2008=\cite{mosser2008},
Mugrauer2022=\cite{mugrauer2022},
Murray1986=\cite{murray1986},
Murray2011=\cite{murray2011},
Muzic2012=\cite{muzic2012},
Newton2014=\cite{newton2014},
Neuhauser2007=\cite{neuhauser2007},
Nidever2002=\cite{nidever2002},
Nielsen2019=\cite{nielsen2019},
Nilsson2017=\cite{nilsson2017},
Nordstrom2004=\cite{nordstrom2004},
OBrien2023=\cite{obrien2023},
Pan1990=\cite{pan1990},
Pepe2011=\cite{pepe2011},
Peretti2019=\cite{peretti2019},
Pettersen2006=\cite{pettersen2006},
Phan-Bao2006=\cite{phan-bao2006},
Phan-Bao2008=\cite{phan-bao2008},
Phan-Bao2017=\cite{phan-bao2017},
Piccotti2020=\cite{piccotti2020},
Pinamonti2022=\cite{pinamonti2022},
Pineda2016=\cite{pineda2016},
Pinfield2008=\cite{pinfield2008},
Pinfield2012=\cite{pinfield2012},
Pinfield2014a=\cite{pinfield2014a},
Pinfield2014b=\cite{pinfield2014b},
Pourbaix2000=\cite{pourbaix2000},
Pourbaix2004=\cite{pourbaix2004},
Poveda2009=\cite{poveda2009},
Pravdo2006=\cite{pravdo2006},
Raghavan2010=\cite{raghavan2010},
Rajpurohit2013=\cite{rajpurohit2013},
Ramm2005=\cite{ramm2005},
Ramm2016=\cite{ramm2016},
Ramm2021=\cite{ramm2021},
Rebassa-Mansergas2017=\cite{rebassa-mansergas2017},
Reffert2011=\cite{reffert2011},
Reid1990=\cite{reid1990},
Reid1995=\cite{reid1995},
Reid2000=\cite{reid2000},
Reid2001=\cite{reid2001},
Reid2003=\cite{reid2003},
Reid2004=\cite{reid2004},
Reid2005=\cite{reid2005},
Reid2006=\cite{reid2006},
Reid2007=\cite{reid2007},
Reid2008a=\cite{reid2008a},
Reid2008b=\cite{reid2008b},
Reiners2012=\cite{reiners2012},
Ren2013=\cite{ren2013},
Reuyl1943=\cite{reuyl1943},
Reyle2006=\cite{reyle2006},
Reyle2018=\cite{reyle2018},
Riaz2006=\cite{riaz2006},
Ribas2003=\cite{ribas2003},
Rice1998=\cite{rice1998},
Riedel2010=\cite{riedel2010},
Riedel2014=\cite{riedel2014},
Riedel2017=\cite{riedel2017},
Riedel2018=\cite{riedel2018},
Robert2016=\cite{robert2016},
Roberts2011=\cite{roberts2011},
Rodet2018=\cite{rodet2018},
Rodigas2011=\cite{rodigas2011},
Rodler2012=\cite{rodler2012},
Rodriguez2015=\cite{rodriguez2015},
Rosenthal2021=\cite{rosenthal2021},
Sahlmann2015=\cite{sahlmann2015},
Sahlmann2021=\cite{sahlmann2021},
Sahu2017=\cite{sahu2017},
Salama2021=\cite{salama2021},
Salama2022=\cite{salama2022},
Salim2003=\cite{salim2003},
Samus2003=\cite{samus2003},
Samus2017=\cite{samus2017},
Santos2003=\cite{santos2003},
SBC9=\cite{pourbaix2004},
SBC7=\cite{batten1978},
Schapera2022=\cite{schapera2022},
Schlieder2014=\cite{schlieder2014},
Schmidt2007=\cite{schmidt2007},
Schmidt2010=\cite{schmidt2010},
Schneider2014=\cite{schneider2014},
Schneider2015=\cite{schneider2015},
Schneider2017=\cite{schneider2017},
Schneider2019=\cite{schneider2019},
Schneider2020=\cite{schneider2020},
Schneider2021=\cite{schneider2021},
Scholz2002=\cite{scholz2002},
Scholz2002b=\cite{scholz2002b},
Scholz2003=\cite{scholz2003},
Scholz2004=\cite{scholz2004},
Scholz2005=\cite{scholz2005},
Scholz2014=\cite{scholz2014},
Schuster1979=\cite{schuster1979},
Schweitzer2019=\cite{schweitzer2019},
Segransan2000=\cite{segransan2000},
Seifahrt2008=\cite{seifahrt2008},
Shan2017=\cite{shan2017},
Shkolnik2009=\cite{shkolnik2009},
Shkolnik2010=\cite{shkolnik2010},
Shkolnik2012=\cite{shkolnik2012},
Silverstein2022=\cite{silverstein2022},
Sion2014=\cite{sion2014},
Skemer2016=\cite{skemer2016},
Skiff2013=\cite{skiff2013},
Skrutskie2006=\cite{skrutskie2006},
Skuljan2004=\cite{skuljan2004},
Smart-priv-comm=R.\ L.\ Smart (priv.\ comm.),
Smart2013=\cite{smart2013},
Smart2018=\cite{smart2018},
Smith2014=\cite{smith2014},
Soderhjelm1999=\cite{soderhjelm1999},
Soriano2010=\cite{soriano2010},
Sperauskas2019=\cite{sperauskas2019},
Stassun2016=\cite{stassun2016},
Stassun2019=\cite{stassun2019},
Stauffer2010=\cite{stauffer2010},
Stelzer2003=\cite{stelzer2003},
Stepanov2020=\cite{stepanov2020},
Stephenson1967=\cite{stephenson1967},
Stephenson1975=\cite{stephenson1975},
Strassmeier1990=\cite{strassmeier1990},
Struve1954=\cite{struve1954},
Struve1955=\cite{struve1955},
Subasavage2017=\cite{subasavage2017},
Takeda2007=\cite{takeda2007},
Tamazian2006=\cite{tamazian2006},
Teixeira2009=\cite{teixeira2009},
Terrien2015=\cite{terrien2015},
Thevenin2005=\cite{thevenin2005},
Thompson2013=\cite{thompson2013},
Thorstensen2003=\cite{thorstensen2003},
Tinney2003=\cite{tinney2003},
Tinney2014=\cite{tinney2014},
Tinney2018=\cite{tinney2018},
Tokovinin1997=\cite{tokovinin1997},
Tokovinin2005=\cite{tokovinin2005},
Tokovinin2006=\cite{tokovinin2006},
Tokovinin2008=\cite{tokovinin2008},
Tokovinin2012=\cite{tokovinin2012},
Tokovinin2014a=\cite{tokovinin2014a},
Tokovinin2014b=\cite{tokovinin2014b},
Tokovinin2016=\cite{tokovinin2016},
Tokovinin2017=\cite{tokovinin2017},
Tokovinin2019=\cite{tokovinin2019},
Tokovinin2020=\cite{tokovinin2020},
Tokovinin2021=\cite{tokovinin2021},
Tokovinin2021b=\cite{tokovinin2021b},
Toonen2017=\cite{toonen2017},
Torres2002=\cite{torres2002},
Torres2006=\cite{torres2006},
Torres2015=\cite{torres2015},
Torres2022=\cite{torres2022},
Tremblay2020=\cite{tremblay2020},
Trifonov2020=\cite{trifonov2020},
Upgren1972=\cite{upgren1972},
Valenti2005=\cite{valenti2005},
vanAltena1995=\cite{vanaltena1995},
vanBelle2007=\cite{vanbelle2007},
vanBiesbroeck1974=\cite{vanbiesbroeck1974},
vandeKamp1971=\cite{vandekamp1971},
vonBraun2011=\cite{vonbraun2011},
Volk2003=\cite{volk2003},
Vrba2004=\cite{vrba2004},
Vrijmoet2020=\cite{vrijmoet2020},
Vrijmoet2022=\cite{vrijmoet2022},
WandDuong2015=\cite{wardduong2015},
WDS=\url{http://www.astro.gsu.edu/wds/},
West2008=\cite{west2008},
West2011=\cite{west2011},
West2015=\cite{west2015},
Willmarth2016=\cite{willmarth2016},
Wilson1950=\cite{wilson1950},
Wilson2003=\cite{wilson2003},
Wilson2017=\cite{wilson2017},
Winters2011=\cite{winters2011},
Winters2017=\cite{winters2017},
Winters2018=\cite{winters2018},
Winters2019=\cite{winters2019},
Winters2019b=\cite{winters2019b},
Winters2020=\cite{winters2020},
Winters2021=\cite{winters2021},
Wittenmeyer2006=\cite{wittenmyer2006},
Wittenmeyer2011=\cite{wittenmyer2011},
Wittrock2017=\cite{wittrock2017},
Woitas2000=\cite{woitas2000},
Wright2013=\cite{wright2013},
Wright2018=\cite{wright2018},
Wyatt2007=\cite{wyatt2007},
Xia2019=\cite{xia2019},
Zasche2009=\cite{zasche2009},
Zeng2022=\cite{zeng2022},
Zhang2020=\cite{zhang2020},
Zhuchkov2012=\cite{zhuchkov2012},
Zurlo2018=\cite{zurlo2018}.
}
\end{deluxetable*}

\startlongtable
\begin{deluxetable*}{llll}
\tabletypesize{\scriptsize}
\tablenum{4}
\tablecaption{{\it (continued)}}
\tablehead{
\colhead{Column} &
\colhead{Description} &
\colhead{Sec} &
\colhead{Example Entry} \\
\colhead{(1)} &
\colhead{(2)} &
\colhead{(3)} &
\colhead{(4)} \\
}
\startdata
.................................& ...............................................................................................& ...........& ..............................................\\
\enddata
\tablecomments{Designation references:
{\bf Bayer} = \citealt{bayer1603};
{\bf Flamsteed} = Stellarum Inerrantium Catalogus Britannicus (Flamsteed 1725) \url {http://pbarbier.com/flamsteed/flamsteed.html},  
\url {http://www.ianridpath.com/ startales/flamsteed.html}, \citealt{lalande1783},
\citealt{hoffleit1991};
{\bf HR} = \citealt{schlesinger1930},
\citealt{schlesinger1940},
\citealt{hoffleit1964},
\citealt{hoffleit1982},
\citealt{hoffleit1983},
\citealt{hoffleit1991};
{\bf HD} = \citealt{cannon1918a},
\citealt{cannon1918b},
\citealt{cannon1919a},
\citealt{cannon1919b},
\citealt{cannon1920},
\citealt{cannon1921},
\citealt{cannon1922},
\citealt{cannon1923},
\citealt{cannon1924},
\citealt{cannon1925a},
\citealt{cannon1925b},
\citealt{cannon1927},
\citealt{cannon1928},
\citealt{cannon1931},
\citealt{cannon1936},
\citealt{cannon1949};
{\bf BD} = \citealt{schonfeld1886},
\citealt{argelander1903};
{\bf CD} = \citealt{thome1890},
\citealt{thome1892},
\citealt{thome1894},
\citealt{thome1900},
\citealt{thome1914},
\citealt{perrine1932};
{\bf CPD} = \citealt{gill1896},
\citealt{gill1897},
\citealt{gill1900},
\citealt{innes1903};
{\bf Wolf} = \citealt{wolf1919a},
\citealt{wolf1919b},
\citealt{wolf1919c},
\citealt{wolf1919d},
\citealt{wolf1919e},
\citealt{wolf1920a},
\citealt{wolf1920b},
\citealt{wolf1920c},
\citealt{wolf1920d},
\citealt{wolf1920e},
\citealt{wolf1920f},
\citealt{wolf1921a},
\citealt{wolf1921b},
\citealt{wolf1921c},
\citealt{wolf1921d},
\citealt{wolf1921e},
\citealt{wolf1921f},
\citealt{wolf1922},
\citealt{wolf1923a},
\citealt{wolf1923b},
\citealt{wolf1926a},
\citealt{wolf1924a},
\citealt{wolf1924b},
\citealt{wolf1924c},
\citealt{wolf1923c},
\citealt{wolf1924d},
\citealt{wolf1924e},
\citealt{wolf1924f},
\citealt{wolf1925a},
\citealt{wolf1925b},
\citealt{wolf1925c},
\citealt{wolf1925d},
\citealt{wolf1925e},
\citealt{wolf1925f},
\citealt{wolf1926b},
\citealt{wolf1926c},
\citealt{wolf1927a},
\citealt{wolf1927b},
\citealt{wolf1929},
\citealt{wolf1931};
{\bf Ross} = \citealt{ross1925},
\citealt{ross1926a},
\citealt{ross1926b},
\citealt{ross1926c},
\citealt{ross1927},
\citealt{ross1928},
\citealt{ross1929},
\citealt{ross1930},
\citealt{ross1931},
\citealt{ross1937},
\citealt{ross1939a},
\citealt{ross1939b};
{\bf L} = \citealt{luyten1963} (for which the ``L'' numbers were obtained from the NLTT Catalogue);
{\bf LP} = \citealt{luyten1970a} (specifically, volumes 1-9, 11-17, and 23-24),
\citealt{luyten1970b},
\citealt{luyten1970c},
\citealt{luyten1971},
\citealt{luyten1972a}, 
\citealt{luyten1972b},
\citealt{luyten1972c}, 
\citealt{luyten1972d}, 
\citealt{luyten1973a}, 
\citealt{luyten1973b}, 
\citealt{luyten1974a}, 
\citealt{luyten1974b}, 
\citealt{luyten1975a}, 
\citealt{luyten1975b}, 
\citealt{luyten1975c}, 
\citealt{luyten1976a}, 
\citealt{luyten1976b}, 
\citealt{luyten1976c}, 
\citealt{luyten1980a}, 
\citealt{luyten1980b}, 
\citealt{luyten1981},       
\citealt{luyten1982}, 
\citealt{luyten1983}, 
\citealt{luyten1985a},        
\citealt{luyten1985b}, 
\citealt{luyten1985c}, 
\citealt{luyten1987a},        
\citealt{luyten1987b};
{\bf G} = \citealt{giclas1971},
\citealt{giclas1978};
{\bf LHS} = \citealt{luyten1979a};
{\bf LFT} = \citealt{luyten1955};
{\bf NLTT} = \citealt{luyten1979b},
\citealt{luyten1979c},
\citealt{luyten1980c},
\citealt{luyten1980d};
{\bf LTT} = \citealt{luyten1957},
\citealt{luyten1961},
\citealt{luyten1962};
{\bf LSPM} = \citealt{lepine2005},
{\bf SCR} = \citealt{hambly2004},
\citealt{henry2004},
\citealt{subasavage2005a},
\citealt{subasavage2005b},
\citealt{finch2007},
\citealt{winters2011},
\citealt{boyd2011};
{\bf UPM} = \citealt{finch2010},
\citealt{finch2012};
{\bf APMPM} = \citealt{gizis1997c},
\citealt{scholz1999},
\citealt{schweitzer1999},
\citealt{scholz2000},
\citealt{reyle2002};
{\bf LEHPM} = \citealt{pokorny2003},
\citealt{pokorny2004};
{\bf WT} = \citealt{wroblewski1989},
\citealt{wroblewski1991},
\citealt{wroblewski1994},
\citealt{wroblewski1996},
\citealt{wroblewski1997},
\citealt{wroblewski1999},
\citealt{wroblewski2001};
{\bf SIPS} = \citealt{deacon2005},
\citealt{deacon2007};
{\bf PM} = \citealt{eggen1979},
\citealt{eggen1980};
{\bf PM J} = \citealt{lepine2005b},
\citealt{lepine2008},
\citealt{schlieder2010},
\citealt{lepine2011},
\citealt{schlieder2012},
\citealt{lepine2013};
{\bf WD} = \citealt{mccook1999};
{\bf LAWD} = \citealt{luyten1949};
{\bf EGGR} = \citealt{eggen1965a},
\citealt{eggen1965b},
\citealt{eggen1967},
\citealt{greenstein1969},
\citealt{greenstein1970},
\citealt{greenstein1974},
\citealt{greenstein1975},
\citealt{greenstein1976},
\citealt{greenstein1977},
\citealt{greenstein1979},
\citealt{greenstein1980},
\citealt{greenstein1984};
{\bf 2MASS} = \citealt{cutri2003};
{\bf WISE} = \citealt{cutri2012},
\citealt{cutri2013},
\citealt{eisenhardt2020},
\citealt{marocco2021};
{\bf Gaia} = \citealt{gaia2018},
\citealt{gaia2021};
{\bf HIC} = \citealt{turon1993};
{\bf HIP} = \citealt{ESA1997},
\citealt{vanleeuwen2007};
{\bf TYC} = \citealt{hog2000};
{\bf UCAC4} = \citealt{zacharias2013};
{\bf TIC} = \citealt{stassun2019};
{\bf GJ} = \citealt{gliese1969},
\citealt{woolley1970},
\citealt{gliese1979};
{\bf V*} = \citealt{samus2004};
{\bf Karmn} = \citealt{alonso-floriano2015},
\citealt{cortescontreras2017};
{\bf **} = This designation is unique to SIMBAD (\citealt{wenger2000}).
}
\tablecomments{Reference for photometry is a five-character code referring to the source of the $J_{MKO}$, $J_{2MASS}$, $H$, $K_{MKO}$, and $Ks_{2MASS}$ magnitudes, respectively, with these characters as the individual codes --
2 = 2MASS (\citealt{skrutskie2006}),
a = \cite{meisner2020a},
A = \cite{meisner2020b},
b = \cite{bardalez2020} for which the {\it HST} F125W magnitude limit for WISE 0830+2837 is used as its value for $J_{MKO}$,
B = Bigelow/2MASS from \cite{kirkpatrick2021},
c = \cite{boccaletti2003},
D = Database of Ultracool Parallaxes as of 2020 April: \cite{dupuy2012} and \cite{dupuy2013} and \cite{liu2016}, 
e = \cite{martin2018},
E = \cite{mcelwain2006}, 
f = \cite{faherty2012},  
F = \cite{freed2003}, 
g = \cite{mamajek2018},
G = Gemini-South/FLAMINGOS2 from \cite{kirkpatrick2021},
h = \cite{pinfield2014b}, 
H = \cite{pinfield2014a},
i = \cite{ireland2008},
I = \cite{dupuy2019},
j = \cite{janson2011},
J = \cite{faherty2014},
k = \cite{kirkpatrick2019}, 
K = \cite{kirkpatrick2011},
m = \cite{mace2013a},
M = Magellan/PANIC from \cite{kirkpatrick2021},
p = PAIRITEL from \cite{kirkpatrick2021},
P = Palomar/WIRC from \cite{kirkpatrick2021},
Q = \cite{deacon2017},
r = \cite{deacon2012b},
s = \cite{schneider2015},
S = SOAR/OSIRIS from \cite{kirkpatrick2021},
t = \cite{tinney2014},
T = \cite{thompson2013},
u = ULAS or UGPS or UGCS (\citealt{lawrence2007}),
U = UHS (\citealt{dye2018}),
v = VVV (\citealt{minniti2010}),
V = VHS (\citealt{mcmahon2013}),
w = \cite{wright2013},
W = \cite{best2020},
x = see Table~\ref{tab:poss_20pc_members_MLTY} in this paper,
z = \cite{meisner2023},
Z = \cite{schapera2022}.
}
\end{deluxetable*}

\clearpage

\pagebreak

\subsection{Astrometry\label{sec:astrometry}}

For each object in the 20-pc census, we list approximate sexagesimal Right Ascension (RA) and Declination (Dec) coordinates at equinox J2000, given under "SexagesimalRA" and "SexagesimalDec" in Table~\ref{tab:20pc_census}. For close multiple systems, the positions of the two objects may be identical, as these coordinates are meant to provide only a crude position for matching the system across catalogs. For more precise coordinates, we also provide RA and Dec in decimal degrees ("RA" and "Dec") at the yearly epoch provided in the "Epoch" column, along with the coordinate uncertainties ("RA\_unc" and "Dec\_unc"). Also listed are the absolute parallax ("Parallax") and its uncertainty ("Parallax\_unc") and the (usually) absolute proper motions and their uncertainties in RA and Dec ("PMRA", "PMDec", "PMRA\_unc", and "PMDEC\_unc"). The reference for these decimal coordinates, parallax, and motion measurements is also given ("PlxPMRef"). Note that for some multiple systems, this more precise astrometry may exist only for the composite system or primary and not for each individual component. (As asterisk in the "PlxPMRef" column indicates that the parallax and motion values for another object in the system are used in lieu of actual measurements for this component.) Furthermore, for some recent brown dwarf discoveries, only positions and proper motions are given, as parallaxes have not yet been measured.

As a final note on positions, we provide the constellation in which each object is located ("Constellation"), based on the VizieR tool\footnote{\url {http://vizier.cfa.harvard.edu/viz-bin/VizieR?-source=VI/42}} that uses the constellation boundaries provided by \cite{roman1987}. This column can be used to determine the nearest object in each constellation, as further explored in Appendix~\ref{sec:appendix_proximas}.

\subsection{Spectral types\label{sec:spectral_types}}

For higher mass stars -- typically those with types earlier than mid-M -- our primary sources for spectral types were the NStars papers by \cite{gray2003} and \cite{gray2006}. This was done to assure that as many of our referenced types as possible were classified against a homogeneous system of standards, in this case, the MKK System of \cite{morgan1943}. This system was subsequently updated to the MK system of \cite{johnson1953}, which itself was expanded and updated by \cite{morgan1973} (the revised MK system), \cite{keenan1976}, and \cite{morgan1978}. (See \citealt{hearnshaw2014} more a more detailed history.) 

Classification for objects of later type has followed the precepts of the MK System, thereby pushing this homogeneity into the late-M (\citealt{boeshaar1976}, \citealt{boeshaar1985}, \citealt{kirkpatrick1991}, \citealt{kirkpatrick2010}), L (\citealt{kirkpatrick1999}, \citealt{kirkpatrick2010}), T (\citealt{burgasser2006}, \citealt{kirkpatrick2010}), and Y (\citealt{cushing2011}, \citealt{kirkpatrick2012}) dwarf sequences. Classification is dependent upon the wavelength range over which the typing is done, so Table~\ref{tab:20pc_census} specifies whether the spectral type was obtained in the visible to photographic near-infrared region ($<$1 $\mu$m; "SpecTypeOpt") or the classical near-infrared region (1-2.5 $\mu$m; "SpecTypeNIR"). References for the spectral types can be found under "SpTOpt\_ref" and "SpTNIR\_ref". For ease of plotting, the spectral types have been converted into a numerical code, with the luminosity type (if listed) ignored. The scale\footnote{There are no O- or B-type stars within the 20-pc volume. The closest O star is $\zeta$ Oph (O9.5 V, d = 112 pc; \citealt{howarth2014, vanleeuwen2007}), and the closest B star is $\alpha$ Leo (B8 IVn, d = 24.3 pc; \citealt{fuhrmann2011b, vanbelle2009, vanleeuwen2007}).} is set so that 0=A0, 10=F0, 20=G0, 30=K0, 40=M0, 50=L0, 60=T0, and 70=Y0; a type of L8.5 would thus be encoded as 58.5. These codes can be found under "SpTOpt\_indx" and "SpTNIR\_indx". Note that the original MK classification system's standards jump from K5 to K7 to M0 in the late-K sequence, although K6 standards were eventually added in the late 1980s (\citealt{keenan1988,keenan1989}). As a result, there are very few objects with codes of $\sim$36 or $\sim$38-39. 

For white dwarfs, Table~\ref{tab:20pc_census} uses types primarily taken from the compilations of \cite{sion2014} and \cite{mccook2016}, with post-2016 discoveries taken from more recent literature or from Section~\ref{sec:appendix_spectroscopy}. The use of these references assures that all white dwarfs are on the spectroscopic classification system proposed by \cite{liebert1994}. All white dwarf classifications have been assigned based on optical spectra, and the corresponding optical spectral index, "SpTOpt\_indx", is coded to be the \cite{liebert1994} temperature index + 100. That is, our index is set so that DA2=102, DAZ5.8=105.8, DA9.2=109.2, DZ12.6 =112.6, etc. For any white dwarf lacking a temperature index, our spectral index has been arbitrarily assigned a code of 100, as a temperature index of 0.0 cannot exist (and no white dwarf in Table~\ref{tab:20pc_census} has a temperature index lower than 2.0).

\subsection{Photometry\label{sec:photometry}}

Table~\ref{tab:20pc_census} provides photometry in several systems that have hemispheric or all-sky coverage. As discussed in section~\ref{sec:color-magnitude_diagrams}, objects in the 20-pc census span a vast dynamic range in absolute luminosity, amounting to over twenty-nine magnitudes (a difference of $5{\times}10^{11}$ in brightness) in $J$-band alone. Thus, special care must be taken when choosing photometry for Table~\ref{tab:20pc_census}.

For the traditional "visible" wavelength regime, Gaia eDR3 magnitudes and uncertainties at $G$, $G_{BP}$, and $G_{RP}$ are listed ("G", "G\_BP", "G\_RP", "G\_unc", "G\_BP\_unc", and "G\_RP\_unc"). The brightest reported $G$-band magnitude is $\sim$2 mag (\citealt{gaia2021}), and objects with $G < 8$ mag, $G_{BP} \lesssim 4$ mag, and $G_{RP} \lesssim 4$ mag have residual saturation effects, as detailed in \cite{gaia2021} and \cite{riello2021}. At the faint end, Gaia is complete to $G \approx$ 20 mag (depending upon source crowding and galactic latitude; \citealt{gaia2021}), which means that the more distant late-L dwarfs in the 20-pc census, along with most of the T dwarfs and all of the Y dwarfs, are too faint for Gaia photometric measurements (figure 26 of \citealt{smart2020}; see also figure 2 of \citealt{theissen2018}).

For the traditional near-infrared wavelength regime, $J$, $H$, and $K$ magnitudes are provided, with the caveat that there are two main filter systems in use: the 2MASS filter system\footnote{\url {https://irsa.ipac.caltech.edu/data/2MASS/docs/releases/allsky/doc/ sec6\_4b.html}} and the MKO filter system (\citealt{tokunaga2002}). The $H$-band filter is almost identical between the two, but the $J$ and $K$ filters are quite different. As a result, we provide five separate entries to cover the possibilities -- $J_{MKO}$, $J_{2MASS}$, $H$, $K$, and $K_s$ -- along with their uncertainties ("JMKO", "J2MASS", "H", "K", "Ks", "JMKOerr", "J2MASSerr", "Herr", "Kerr", and "Kserr"). The references for this photometry are given in the "JHK\_ref" column. The  $J_{2MASS}$ entries mostly come from 2MASS, whereas the $J_{MKO}$ entries come mostly from surveys based at the United Kingdom Infrared Telescope (UKIRT; e.g., the UKIRT Hemisphere Survey, UHS -- \citealt{mcmahon2013}) and the Visible and Infrared Survey Telescope for Astronomy (VISTA; e.g. the VISTA Hemisphere Survey, VHS --  \citealt{dye2018}). The $K_{MKO}$ entries come primarily from UKIRT-based surveys, whereas the $K_s$ entries come from both 2MASS and VISTA-based surveys. The $H$ entries are pulled from all three sets of surveys.  2MASS provides the only reliable photometry at the bright end of our sample, albeit with large uncertainties, and extends to a S/N = 10 limit of $J = 15.8$ mag, $H = 15.1$ mag, and $K_s = 14.3$ mag at its faint end\footnote{\url {https://irsa.ipac.caltech.edu/data/2MASS/docs/releases/allsky/doc/ sec2\_2.html}}. UKIRT and VISTA provide reliable photometry between their bright limit ($J \approx 12$ mag for UHS and $\sim$11.5-12.5 mag in $J$, $H$, and $K_s$ for VHS; \citealt{dye2018}, \citealt{gonzalez-fernandez2018}) and their detection limit ($J \approx 19$ mag for UHS and $J \approx 20$ mag, $H \approx 19$ mag and $K_s \approx 18$ mag for VHS; \citealt{dye2018}, \citealt{gonzalez-fernandez2018}) and provide higher angular resolution than 2MASS. We have therefore favored 2MASS photometry for near-infrared magnitudes brighter than $\sim$12 mag and UKIRT/VISTA for fainter magnitudes. For objects even fainter than the UKIRT/VISTA limits, or for objects in areas not yet covered by the public UKIRT and VISTA releases, we have pulled objects from the literature or from Appendix~\ref{sec:appendix_photometry}.

At longer near-infrared wavelengths and extending into the near mid-infrared, we also provide WISE-based W1 (3.4 $\mu$m), W2 (4.6 $\mu$m), W3 (12 $\mu$m), and W4 (22 $\mu$m) magnitudes and their uncertainties ("W1", "W2", "W3", "W4", "W1err", "W2err", "W3err", and "W4err"). The reference for the WISE photometry is given in the "WISEphot\_ref" column. For W1 and W2, magnitudes brighter than W1 $\approx$ 8 mag and W2 $\approx$ 7 mag were pulled from the WISE All-sky Source Catalog and fainter magnitudes were pulled from the AllWISE Source Catalog, in accordance with the suggestion made in the AllWISE Explanatory Supplement\footnote{\url {https://wise2.ipac.caltech.edu/docs/release/allwise/expsup/sec2\_1.html}}. Photometry at W3 and W4 was pulled from the WISE All-sky Release. The only exceptions to the above are objects that were not detected in either of these releases and are instead found only in the CatWISE2020 Catalog. For these sources, only W1 and W2 photometry is presented, as CatWISE2020 has no W3 or W4 photometry.

In the case of 2MASS and WISE photometry, we further provide columns "2MASS\_contam?" and "WISE\_contam?". A "yes" in these columns indicates that the associated photometry is likely compromised by another nearby object or artifact, as judged via our by-eye assessments of the multi-epoch WiseView image blinks (\citealt{caselden2018}), as the poorer image scales of 2MASS (pixel scale of 1$\arcsec$) and WISE (pixel size of 1$\farcs$375) translate to a higher likelihood of source blending.

\subsection{Radial velocities\label{sec:radial_velocities}}

Gaia DR3 provides all-sky radial velocities for stars with $G_{RVS} \lesssim 14$ mag (\citealt{katz2022}) and effective temperatures as high as 14,500 K (\citealt{blomme2022}). These radial velocities and their uncertainties are also listed in Table~\ref{tab:20pc_census} ("GaiaRV" and "GaiaRV\_unc").

\subsection{Multiplicity}

Even after all systems within the 20-pc volume have been noted, one difficult step remains: correctly determining, based on current knowledge, the multiplicity of each system so that each individual component can be correctly accounted for in the mass distribution. We took a multi-pronged approach at tackling this problem, as described below.

\subsubsection{The "Stellar Ambassadors" program}

The first approach was to crowdsource the initial reconnaissance of the literature. With the help of the citizen scientist super users of the Backyard Worlds: Planet 9 project, we set up a program whereby volunteers could sign up to investigate the multiplicity of randomly selected 20-pc systems. To make this more enjoyable, the following mission statement was provided:

\begin{quote}
    
{\it "Our science-fictional Earth Coalition is currently laying the groundwork to explore all of the `worlds' within 20 parsecs (65 light years) of the Sun. Scientists on the Earth Coalition's Board of Advisors have a list of host `suns' within this volume of space, but the details in that list are a bit spotty. The Coalition is seeking to flesh out these details using our Stellar Ambassadors program.}

{\it "If you choose to become a Stellar Ambassador, your role will be to represent planet Earth to a small number of stellar systems within 20 parsecs. As we reach out for the first time to each of these stellar neighbors, you will be Earth's representative to them. But you need to be knowledgeable of the systems for which you're responsible, and that will involve your gaining knowledge of each system you're assigned. (By `system', we're referring to a host star and any of its companions -- other stars, brown dwarfs, or exoplanets -- in orbit around it.)"}

\end{quote} 

Each volunteer was tasked with determining (a) the number of stars, brown dwarfs, and exoplanets in each system, (b) the spectral types of the (sub)stellar components, and (c) the masses of each component, if the masses have been measured. Each Stellar Ambassador was initially assigned a set of $\sim$12 systems, and additional sets would be assigned if the Ambassador wished to analyze more. Importantly, participants were asked to track the reference material that they used for their data collection, regardless of whether they started with SIMBAD, VizieR, Wikipedia, or some other encyclopedic compendium. In total, twenty-one super users participated in the program, which allowed us to cover 56\% of the systems with primaries earlier than L0. (All 20-pc objects with primaries later than this had already been scrutinized in \citealt{kirkpatrick2021}.) These efforts were coordinated in weekly and bi-weekly calls with the volunteers.

The product of this exercise was, as expected, an inhomogeneous set of results, as individual Ambassadors concentrated on different portions of the exercise or used entirely different methodologies in their workflows. Nonetheless, it was these varied approaches that enabled us to determine the references on which it would be the most lucrative to focus our early attention. For example, despite the varied approaches, many of the same references kept appearing again and again in the Ambassadors' reports. These repeating references underscored the vast groundwork laid by exoplanet-finding searches in characterizing potential host stars, as well as the breadth of methods used to measure masses of stars within the solar neighborhood, a topic explored more fully in Section~\ref{sec:mass_methods}. The references that arose from the Stellar Ambassador program were the first resources we used to populate Table~\ref{tab:20pc_census}
with information on multiplicity, mass measurements, and mass estimates.

\subsubsection{In-depth literature checks}

After this first reconnaissance of the oft-referenced literature, our second approach was the inevitable deep-dive into the literature for each individual object. For this, we used the extensive per-object references compiled by SIMBAD. We concentrated on literature with high-resolution imaging and radial velocity monitoring, in order to judge the multiplicity of each system. We also looked for paper titles that referenced mass measurements and variability (such as eclipsing binaries, RS CVn variables, etc.). Because of time constraints, we were not able to review each reference in detail, but a paper well stocked with results would often allow us to populate Table~\ref{tab:20pc_census} with information for many systems at once, which sped up the process for objects further down the list. We also relied heavily on the Washington Double Star Catalog\footnote{\url {https://vizier.cds.unistra.fr/viz-bin/VizieR?-source=B/wds}} and the Ninth Catalog of Spectroscopic Binaries (\citealt{pourbaix2009}), although the former reference lists both confirmed and possible companions that themselves must be studied individually to gauge true companionship. 

\subsubsection{The Apps catalog\label{sec:apps_catalog}}

After our in-depth literature checks were completed, we became aware of an unpublished list of objects within 30 pc of the Sun that (now co-author) K.\ Apps has been carefully curating since 2009. A comparison of the Apps catalog to our list revealed twenty-eight objects, mainly companions, that have been disproved via published literature but that our list still included. These have now been removed from Table~\ref{tab:20pc_census}. The comparison to the Apps list also revealed another twenty-three objects, almost all of which are the second components in spectroscopic binaries or companions revealed by high-resolution imaging, whose discovery literature we had missed. These objects have now been added to Table~\ref{tab:20pc_census}.

\subsubsection{Multiplicity parameters and exoplanets\label{sec:multiplicity_and_exoplanets}}

\begin{figure*}
\includegraphics[scale=0.70,angle=0]{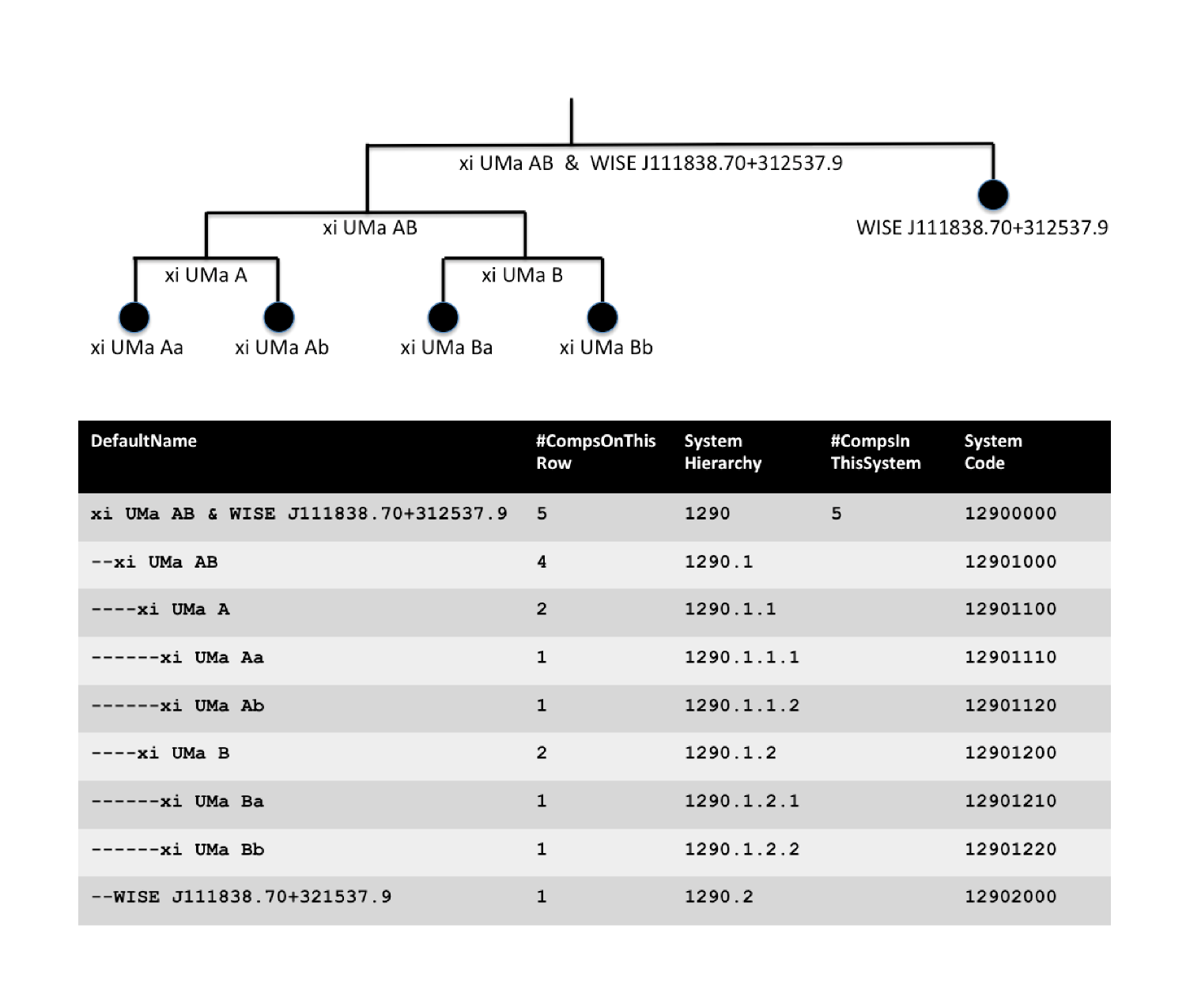}
\caption{Mobile diagram for the $\xi$ UMa system along with sample columns from Table~\ref{tab:20pc_census}. The mobile diagram at top shows a stylized representation of this quintuple system, illustrating the pair of close doubles ($\xi$ UMa A and $\xi$ UMa B) and their distant common proper motion companion (WISE J111838.70+312537.9). The table at bottom shows the nine rows for this system, representing the nine vertices (with labels) in the mobile diagram. Table~\ref{tab:20pc_census} entries for DefaultName, \#CompsOnThisRow, SystemHierarchy, \#CompsInThisSystem, and SystemCode are shown for illustration.
\label{fig:mobile_diagram}}
\end{figure*}

To encapsulate knowledge from the multiplicity checks above, we include several additional columns in Table~\ref{tab:20pc_census} and split the components of each system into separate rows. An example for one system is illustrated in the mobile diagram (see \citealt{evans1968}) of Figure~\ref{fig:mobile_diagram}. In the column "DefaultName", the entry for the system as a whole appears as "{\tt xi UMa AB \& WISE 111838.70+312537.9}". The names of the first subdivision in the mobile diagram of this multiple are denoted by a double hyphen at the beginning of the name, which in this case are "{\tt ---xi UMa AB}" and "{\tt ---WISE 111838.70+321537.9}". Further hierarchical branches are denoted by four hyphens (e.g., "{\tt ------xi UMa A}"), six hyphens (e.g., "{\tt ---------xi UMa Aa}"), etc. The column "\#CompsOnThisRow" refers to the number of known components on that row of the table. To select only individual objects in the census, for example, one can downselect only those rows for which "\#CompsOnThisRow" equals 1. There is also a "SystemHierarchy" column, giving a code for each division within the system. This is comprised of a four-digit integer (e.g., "{\tt 1297}") that uniquely identifies the system, followed by decimal subdivisions (e.g., "{\tt 1297.1}" and "{\tt 1297.2}") to identify subcomponents. For subcomponents that are themselves binaries, further decimal subdivisions (e.g., "{\tt 1297.1.1}" and "{\tt 1297.1.2}") are assigned, etc. Table~\ref{tab:20pc_census} also lists a column called "\#CompsInThisSystem" that gives the total number of components in the system. This field is populated only for the top level of each system (those rows having no decimal subdivisions in the "SystemHierarchy" column) and can be summed to find the total number of individual components in the table. Additionally, Table~\ref{tab:20pc_census} includes a column called "SystemCode" that collapses the "SystemHierarchy" format into an eight-digit integer comprised of the four-digit system identifier followed by four additional digits representing any other subdivisions of the "SystemHierarchy" code, but with the decimals removed (e.g., "{\tt 12971210}"). Note that when lower subdivisions are lacking, those digits are filled with zeroes. This "SystemCode" column is useful if the user prefers to sort the systems in Table~\ref{tab:20pc_census} into their mobile diagrams rather than keeping the table's default ordering, which sorts by RA. 

Note that our accounting of components above includes only those stellar and brown dwarf members of the system, but not any of the known exoplanets. For the latter, we also include a column in Table~\ref{tab:20pc_census} named "\#Planets" that reports the number of exoplanets listed in the NASA Exoplanet Archive\footnote{\url {https://exoplanetarchive.ipac.caltech.edu/}} as of 01 Sep 2022. To match objects from Table~\ref{tab:20pc_census} to objects in this archive, we used the Transiting Exoplanet Survey Satellite (TESS; \citealt{ricker2015}) Input Catalog (TIC; \citealt{stassun2019}) designations. It should be noted that, whereas we use a formation-based definition for brown dwarfs in this paper, the NASA Exoplanet Archive uses a mass-based definition for exoplanet vs.\ brown dwarf and sets the dividing line, somewhat arbitrarily, at 30 M$_{Jup}$\footnote{\url {https://exoplanetarchive.ipac.caltech.edu/docs/exoplanet\_criteria.html}}. As a result, there will be some double counting of objects, as these may appear in both the substellar and exoplanet lists. We will return to this point in Section~\ref{sec:exoplanets}.

\subsection{Mass parameters and effective temperature}

The final parameters in Table~\ref{tab:20pc_census} relate to our need to assign masses to all individual objects within the 20-pc census. In Section~\ref{sec:mass_methods}, we discuss the various methods for which masses can be directly measured. For objects whose masses must, instead, be estimated, Section~\ref{sec:mass_estimates} provides additional discussion. Stars with measured accelerations (see Section~\ref{sec:proper_motion_anomaly}) are further discussed in Section~\ref{sec:accelerators_lacking_other_info}, and objects whose Gaia astrometry suggests hidden companions are discussed in Section~\ref{sec:RUWE_LUWE}. 

Mass estimation techniques work well for hydrogen-burning stars because there is a direct mapping from color, temperature, and spectral type to mass on the main sequence. These same techniques fail for brown dwarfs because color, temperature, and spectral type vary with age, and the age of a brown dwarf is generally an unmeasurable quantity. Estimating the masses for brown dwarfs, therefore, requires a different tack, one that we approach statistically through their distribution of effective temperatures, as further discussed in Section~\ref{sec:brown_dwarfs}.

\section{Masses from Direct Measurement\label{sec:mass_methods}}

There are many ways of measuring stellar masses. Some methods (1) measure mass directly using only observational data, (2) lean lightly on theoretical assumptions when a full suite of needed observational data is not available, (3) derive masses by comparing available data to an empirical data grid for stars with directly measured masses, and (4) compare observables to theoretical models. Examples of these third and fourth groups are methodologies such as The Cannon (\citealt{ness2016}) and StarHorse (\citealt{queiroz2018}). However, the aim of this section is to establish nearby fiducial objects for which masses have been (semi-)directly measured, in order to establish our own empirical grid (method 3) to estimate masses for the remainder of the 20-pc census. Toward this goal, we use the next two subsections to discuss methods 1 and 2 as they have been applied to nearby objects. Table~\ref{tab:20pc_census} includes directly measured masses for objects that have such values ("Mass" and "Mass err") along with the technique used for the measurement ("Mass method") and its citation ("Mass reference").

\subsection{Multiple systems} 

Mass measurements can be made for objects in binary or multiple systems, once sufficient information has been collected to define the orbits. For compact objects, mass can also be deduced from the gravitational redshift; observationally, this can only be done in multiple systems, as it requires at least one additional, non-compact, co-moving object with which to disentangle the part of the redshift due to radial velocity. More specifics are given below.

\subsubsection{Visual binaries}

For a visual binary whose orbit can be observed, the ratio of the masses is just
\begin{equation}
\label{eqn:VB_mass_ratio}
   \frac{M_1}{M_2} = \frac{a_2}{a_1},
\end{equation}
where $M_1$ and $M_2$ are the masses of the two objects and $a_1$ and $a_2$ are the (physical, not apparent) semi-major axes of their respective orbits. The total mass of the system, $M_1 + M_2$, can be derived from the equation
\begin{equation}
\label{eqn:VB_total_mass}
   M_1 + M_2 = \frac{4\pi^2(a_1 + a_2)^3}{GP^2{\cos^3{i}}},
\end{equation}
where $G$ is the gravitational constant, $P$ is the orbital period, and $i$ is the inclination of the orbit on the plane of the sky. The distance to the system must also be measured so that $a_1$ and $a_2$ are in physical, not angular, units, and the inclination can be deduced from the difference between the offset of the center of mass and the focus of the projected ellipse (\citealt{carroll1996}). Individual masses can be measured by combining equations~\ref{eqn:VB_mass_ratio} and \ref{eqn:VB_total_mass}. A list of visual (and other) multiple systems can be found in the Washington Double Star Catalog\footnote{\url{https://vizier.cds.unistra.fr/viz-bin/VizieR?-source=B/wds}}.

\subsubsection{Spectroscopic binaries with eclipses}

For spectroscopic binaries in which the radial velocities of both stars can be measured (SB2s), the ratio of the masses is just
\begin{equation}
    \frac{M_1}{M_2} = \frac{v_2}{v_1},
\end{equation}
where $v_1$ and $v_2$ are the maximum velocity amplitudes with respect to the mean radial velocity curves of the system. The sum of the masses can be obtained via the equation
\begin{equation}
    M_1 + M_2 = \frac{P(v_1 + v_2)^3}{2{\pi}G\sin^3{i}}.
\end{equation}
The inclination cannot be determined unless the SB2 is also an eclipsing system, in which case the nearly edge-on orientation means that $i \approx 90^\circ$, allowing for a mass determination for both components.

There is a class of eclipsing single-lined spectroscopic binary (SB1) systems for which masses can also be derived (\citealt{stassun2017}; \citealt{stevens2018}). These are systems with a single stellar host and a transiting exoplanet. Because the combined light of the system is almost entirely that of the host star, available all-sky data sets can provide photometry across a wide swath of the electromagnetic spectrum -- from the ultraviolet to the near mid-infrared -- so that the star's apparent bolometric luminosity can be measured. Accurate parallaxes from Gaia provide the distances needed to convert this to absolute bolometric luminosity. These photometric points span either side of the flux peak in these objects, so they also provide a semi-empirical measurement of effective temperature, as well. The radius of the host star, $R$, can then be derived from the Stefan-Boltzmann Law
\begin{equation}
\label{eqn:stefan-boltzmann}
    R = \sqrt{\frac{L_{\rm bol}}{4\pi\sigma{T_{\rm eff}}^4}},
\end{equation}
where $L_{\rm bol}$ is its bolometric luminosity, $T_{\rm eff}$ is its effective temperature, and $\sigma$ is the Stefan-Boltzmann constant. In the limit where the mass and radius of the exoplanet are far smaller than those of the host star, the density of the host star, $\rho$, can be calculated directly from observable quantities using the equation
\begin{equation}
    \rho = \frac{3{\pi}}{GP^2}{a_{\rm n}}^3,
\end{equation}
where $a_{\rm n}$ is the "normalized" semi-major axis (see \citealt{sandford2017} for details) and $P$ is the orbital period, both of which can be measured from the transit light curve. (This simplified form assumes a circular orbit. More generalized forms of this equation can be found in \citealt{seager2003}.) The stellar mass, $M$, then follows from
\begin{equation}
    M = \frac{4}{3}{\pi}R^3\rho.
\end{equation}

A list of SB1 and SB2 systems (see \citealt{pourbaix2009}) can be found at the Centre de Donn{\'e}es astronomiques de Strasbourg\footnote{\url{cdsarc.u-strasbg.fr/ftp/cats/B/sb9}}. A list of 158 detached eclipsing binaries with well measured stellar properties is given in \cite{stassun2016}. 

\subsubsection{Astrometric binaries\label{sec:astrometric_binaries}}

Astrometric binaries are those systems in which the presence of an unseen companion can be inferred from the non-linear motion of the primary, once its parallactic motion is accounted for. A careful mapping of the astrometric orbit results in the following measurement
\begin{equation}
\label{eqn:astrometric_binary}
    \frac{M_2}{(M_1 + M_2)^{2/3}} = \frac{r_{ap}}{(1+e)} \Bigl(\frac{2\pi}{P\sqrt{G}}\Bigr)^{2/3} 
\end{equation}
where $M_1$ is the mass of the luminous component, $M_2$ is the mass of the invisible component, $r_{ap}$ is the orbital separation of the luminous component at apastron, $e$ is the eccentricity of the orbit, $P$ is the orbital period, and $G$ is the gravitational constant (\citealt{andrews2019}).

It is possible to measure individual masses in astrometric binaries if the right conditions are met. We consider here an astrometric binary in which the secondary contributes little or no light to the system, as would be the case in a system comprised of a main sequence star and a black hole, neutron star, cold brown dwarf, or exoplanet companion. In this case, the light of the system comes almost entirely from the primary star, so an analysis of its broad-wavelength spectrum or spectral energy distribution built from broad-wavelength photometry can be used to deduce, with the help of empirical relations, its mass, $M_1$. Then the mass of the companion, $M_2$, can be measured using Equation~\ref{eqn:astrometric_binary}. Gaia will produce orbits of hundreds of thousands of such astrometric binaries over its anticipated lifetime (\citealt{halbwachs2023}).

\subsubsection{Binaries with acceleration (aka proper motion anomaly)\label{sec:proper_motion_anomaly}}

\begin{figure}
\includegraphics[scale=0.325,angle=0]{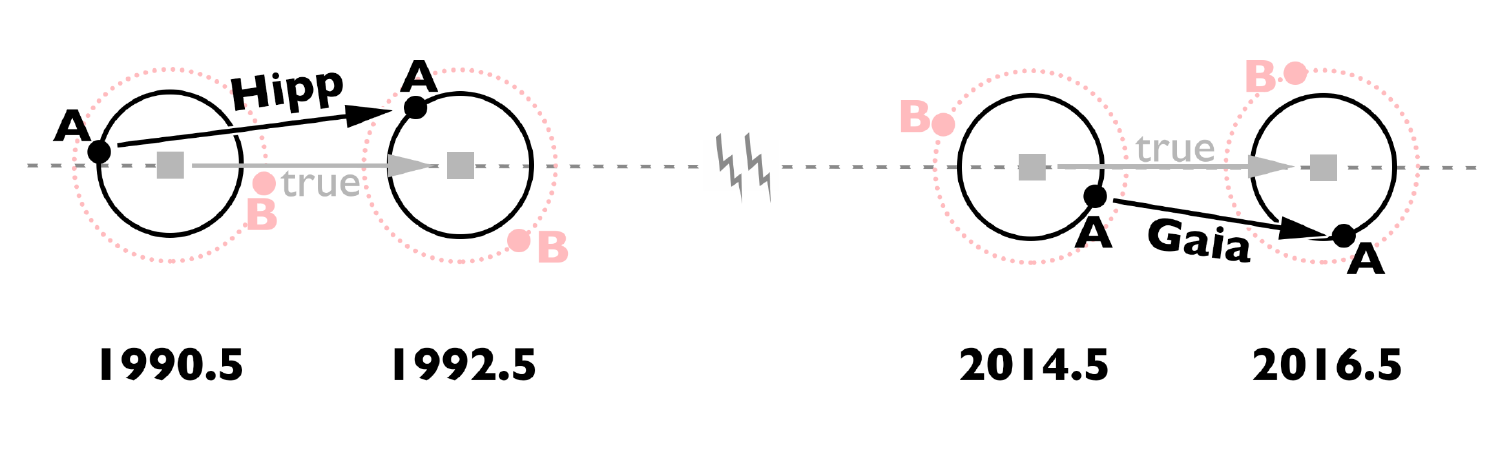}
\caption{Schematic diagram demonstrating the concept of proper motion anomaly. A binary star system, comprised of an A component (solid black orbit and black points) and a lower-mass B component (dotted pink orbit and pink points) is shown at four separate times corresponding to approximate start and end dates of Hipparcos (left pair) and Gaia DR2 (right pair). The center of mass (grey squares) moves from left to right over time, and the true proper motion of the system over the Hipparcos and Gaia timeframes is represented by the two grey arrows. Assuming that the A component dominates the light of the system, neither Hipparcos nor Gaia will measure this true motion because the center of light will move with component A as the stars orbit their barycenter. The black arrow at left thus shows the proper motion that would be measured by Hipparcos, and the black arrow at right shows the motion measured by Gaia DR2. The disagreement between these two independent measurements is termed "proper motion anomaly" and provides evidence that the system has an unseen component. (For simplicity, we have removed parallactic motion by showing only those points at the same parallax factor, as depicted by the time stamps at the bottom of the figure.)
\label{fig:pm_anomaly}}
\end{figure}

Proper motion measurements at two different epochs have the capability of identifying hidden companions if those two motion values differ significantly from one another. (This would be labeled as an astrometric binary, see Section~\ref{sec:astrometric_binaries}, once additional astrometric epochs are obtained.) The reason is that the proper motion of the system's photocenter will deviate from a straight line unless both components contribute equally to the light output. This methodology was first used by \cite{bessel1844} to deduce hidden companions to Sirius and Procyon. An illustration of the effect, which is known both as "proper motion anomaly" and as "acceleration", is shown in Figure~\ref{fig:pm_anomaly}. This procedure has seen a recent revival now that high quality Hipparcos motions from the early 1990s and high quality Gaia DR2 motions from the mid-2010s can be compared.

The lack of agreement between the motion measurements is sufficient to identify a hidden companion, and only a few other measurements are needed to derive the companion's mass. This can be computed from the following equation from \cite{brandt2019}
\begin{equation}
    M = \frac{{s^2}({a_{PM}}^2+{a_{RV}}^2)^{\frac{3}{2}}}{G({\varomega}a_{PM})^2}\label{eqn:accleration},
\end{equation}
where $s$ is the projected separation between the companion and host star, $a_{PM}$ is the host star's acceleration on the plane of the sky, $a_{RV}$ is the host star's acceleration along the line of sight, and $\varomega$ is the parallax of the system. (See Equation~\ref{eqn:kervella} for a different treatment.) This equation holds only if all measurements can be approximated to refer to the same orbital epoch. Otherwise, as detailed in \cite{brandt2019}, more complex orbital fitting is required.

\subsubsection{Compact objects with gravitational redshifts}

Finally, gravitational redshift can be used to measure the mass if the surface gravity can also be determined. Within the 20-pc sample, this is realistically measurable only in {\bf relatively} massive compact objects {\bf like} white dwarfs\footnote{Brown dwarfs, like white dwarfs, are electron degenerate but are less compact and less massive, so their gravitational redshifts are only $\sim$0.5 km s$^{-1}$ for the most massive examples. Although this is comparable to the effect seen for the Sun, $\sim$0.6 km s$^{-1}$, it is one hundred times smaller than the effect seen for a 0.8 M$_\odot$ white dwarf ($\sim$50 km s$^{-1}$).}. The observed velocity shift, $v_{gr}$, due to gravitational redshift is given by
\begin{equation}
   v_{gr} = \frac{GM}{Rc},
\end{equation}
where $c$ is the speed of light (e.g., \citealt{chandra2020}). Because the star's mass is related to its surface gravity, $g$, via the equation
\begin{equation}
\label{eqn:surface_gravity}
    g = \frac{GM}{R^2},
\end{equation}
the mass can be computed from
\begin{equation}
   M = \frac{c^2{v_{gr}}^2}{Gg}.
\end{equation}
The surface gravity can be measured from the white dwarf's spectrum by comparing to model atmospheres. In practice, though, this method cannot be applied to single white dwarfs because the gravitational redshift is not separable from the radial velocity. If the star is part of a co-moving multiple system or is a member of a young cluster or association, however, then the degeneracy between the radial velocity component and gravitational redshift component can be broken.

\subsection{Single objects} 

Researchers have employed several methods that are capable of measuring the masses of {\it individual} objects. These techniques -- gravitational lensing, asteroseismology, and surface convection monitoring (aka "flickering") -- are described below.

\subsubsection{Gravitational lensing}

Gravitational lensing occurs when a mass moves very close to the line of sight between an observer and a background object. The mass of the intervening object acts as a lens that alters the apparent position of the background source as seen by the observer (\citealt{gaudi2012}) and is potentially measurable for any object. The two temporarily generated images of the background source have a morphology that is azimuthally asymmetric, and this manifests itself observationally as a shift in the centroid. The astrometric shift of the photocenter is given by 
\begin{equation}
    {\boldsymbol \delta}(t) = {\bf u}(t)\frac{\theta_E}{u(t)^2 + 2}
\end{equation}
where $\theta_E$ is the angular Einstein radius, which can be expressed as 
\begin{equation}
    {\theta_E}^2 = \frac{4GM_l}{c^2}({D_l}^{-1} - {D_s}^{-1}) 
\end{equation}
(\citealt{walker1995}, \citealt{lu2016}). Here, $u$ and $\bf{u}$ represent the scalar and vector time-dependent lens-source separation in the plane of the sky normalized to $\theta_E$, $M_l$ is the mass of the lens, and $D_l$ and $D_s$ are the distances to the lens and source, respectively. When the distances to the lens and source are known, the monitoring of the astrometric shift as a function of time enables a measurement of the mass of the lens. These equations show that closer lenses produce larger astrometric signals, which makes this a valuable technique for measuring the masses of nearby objects, the main limitation being that such encounters of a lens and a background source happen only rarely and very accurate astrometry is needed to predict such encounters a priori. This technique has so far been successfully applied to only two objects in the 20-pc sample (\citealt{sahu2017}, \citealt{zurlo2018}) but promises to become more valuable as more accurate Gaia parallaxes and proper motions become available for stars all across the Milky Way.

\subsubsection{Asteroseismology}

Asteroseismology is the study of oscillations in stellar atmospheres, and the characteristics of these oscillations can be used to deduce a star's physical parameters. Any star having a mechanism that can drive oscillations -- such as surface convection, pulsations, tidal effects in a close binary, or opacity effects (the $\kappa$-mechanism) -- can potentially have its mass measured. Equation 52 in \cite{aerts2021} can be rewritten to show that the stellar mass, $M$, can be determined from these oscillations using the relation 
\begin{equation}
    M \sim \frac{{\nu_{max}}^3{T_{\rm eff}}^{\frac{3}{2}}}{\Delta\nu^{4}}.
\end{equation} 
(In the absence of a definitive theoretical model for convection, the scaling of this relation is based on observations of the Sun, as described in \citealt{kjeldsen1995}.) Here, $\Delta{\nu}$ is the large frequency separation, $\nu_{max}$ is the frequency of maximum power, and ${T_{\rm eff}}$ is the effective temperature.  The quantity $\nu$ can also be thought of as the inverse of twice the sound travel time between the stellar center and the stellar surface (Eq.\ 39 of \citealt{aerts2021}). Figures 4 and 10 of \cite{aerts2021} graphically demonstrate how $\nu$ and $\nu_{max}$ are measured in practice. The effective temperature, ${T_{\rm eff}}$, is obtained by comparing broad-wavelength spectroscopy of the star to model atmospheres. 

\subsubsection{Surface convection monitoring ("flickering")}

The full asteroseismology treatment above requires high-quality data over a sufficient time baseline with which to resolve the individual oscillation modes. However, variations in surface convection alone require less exquisite data and can be used to measure the mass, if certain ancillary quantities have also been well measured (\citealt{stassun2018}). The needed quantity is $\nu_{max}$ from above, which has been shown to depend on the star's gravity, $g$, and effective temperature, $T_{\rm eff}$, through the relation 
\begin{equation}
    g = \frac{\nu_{max}\sqrt{T_{\rm eff}}}{{C}} 
\end{equation}
(\citealt{brown1991}), where $C$ is a normalization constant obtained by calibrating to stars with gravity measurements independently determined via asteroseismology (\citealt{kallinger2016}). 
The effective temperature is, as above, obtained by comparing broad-wavelength spectroscopy to model atmospheres. The mass can then be measured via Equation~\ref{eqn:surface_gravity}, where the star's radius can be measured directly via interferometry or through the Stefan-Boltzmann Law in Equation~\ref{eqn:stefan-boltzmann}. The bolometric luminosity can be measured from the aforementioned broad-wavelength spectrum along with an accurate trigonometric parallax.

\section{Multiples Lacking Sufficient Data for Mass Determination\label{sec:multiples}}

There are some systems for which acceleration has been measured or whose astrometry indicates the presence of multiple components but for which insufficient data exist to compute the masses of the individual objects. Such systems are important to note because their mass accounting is still incomplete. This serves as an additional source of uncertainty in our mass function analysis.

\subsection{Multiples known only through limited acceleration data\label{sec:accelerators_lacking_other_info}}

Currently, there are many accelerating objects within the 20-pc census that lack the additional data needed for companion mass computations via Equation~\ref{eqn:accleration}. We nonetheless still note these as binaries in Table~\ref{tab:20pc_census}, and we split out those cases here for individual discussion.

\cite{khovritchev2015} have identified likely accelerators by comparing the proper motion measured between the first and second Palomar Observatory Sky Surveys (POSS-I and POSS-II; \citealt{minkowski1963, reid1991, lasker1998}) to a motion derived using first-epoch data from other sky surveys (2MASS, SDSS, WISE) and their own second-epoch follow-up astrometry. With these two independent measurements, they can compare a long-baseline motion over 50 yr to one derived more instantaneously, over only $\sim$10 yr. \cite{brandt2021} have similarly intercompared the near-instantaneous Hipparcos-measured proper motion from the early 1990's, the near-instantaneous Gaia-measured motion from the mid- to late-2010's, and a long-baseline motion constructed from the Hipparcos-to-Gaia baseline. Both sets of authors have identified objects with significant motion discrepancies and labeled these as likely binaries. These objects are noted in Table~\ref{tab:20pc_census} using the column labeled "Accelerator?".

\subsubsection{Accelerators from POSS vs.\ recent motion comparison\label{sec:accelerators}}

The \cite{khovritchev2015} list of $\sim$2400 objects covers only a portion of the northern sky ($30^\circ < {\rm Dec} < 70^\circ$) for bright ($V < 17$ mag), high motion ($\mu > 300$ mas yr$^{-1}$) stars. Within 20 pc of the Sun, nine such accelerators are identified, only two of which -- BD+66 34 and G 96-29 (Capella HL) -- were already identified as known multiples in Table~\ref{tab:20pc_census}. The other seven are listed in Table~\ref{tab:accelerator_bright_khovritchev}.

\begin{deluxetable}{lll}
\tabletypesize{\scriptsize}
\tablecaption{New 20-pc Accelerators from the \cite{khovritchev2015} Sample\label{tab:accelerator_bright_khovritchev}}
\tablehead{
\colhead{J2000 RA \& Dec} &
\colhead{Name} &
\colhead{Sp.\ Type} \\
\colhead{(1)} &
\colhead{(2)} &
\colhead{(3)} \\
}
\startdata
00 38 59.04 +30 36 58.4&  Wolf 1056& M2.5\\
00 57 02.69 +45 05 09.8&  G 172-30 & M3 \\
01 03 19.84 +62 21 55.8&  Wolf 47  & M5 V \\
01 38 21.62 +57 13 57.0&  Ross 10  & M2.5 \\
06 01 11.05 +59 35 49.9&  G 192-13 & M3.5 V\\
19 08 29.93 +32 16 51.6&  G 207-19 & M3.5 \\
23 07 29.92 +68 40 05.2&  G 241-45 & M3 \\
\enddata
\end{deluxetable}

We note that none of these seven objects is identified as a high-significance accelerator in the \cite{brandt2021} reference discussed in the following subsection. This is because the \cite{brandt2021} Hipparcos-to-Gaia accelerations could not be computed for these seven stars, as none are in the Hipparcos Catalog\footnote{Ross 10 has a Hipparcos designation in SIMBAD (HIP 7635) but does not appear in \cite{vanleeuwen2007}.}. To further explore the underlying data for these \cite{khovritchev2015} accelerators, we have produced finder charts that show all seven in the POSS-I, POSS-II, 2MASS, and WISE images. A few of these appear to be blended with a background object at one of the POSS epochs. The most notably affected are G 172-30, which is blended at POSS-I with an object fainter by $\Delta$G = 5.8 mag; Wolf 47\footnote{The primary in this system, BD+61 195, is 295$\arcsec$ away from Wolf 47 itself and would not be responsible for any acceleration.}, which is blended at POSS-I with an object fainter by $\Delta$G = 6.4 mag; and G 192-13, which is blended at POSS-II with an object fainter by $\Delta$G = 6.2 mag. (Ross 10 moves past a star of near-equal magnitude in all of the images, the possible blending being worst at the POSS-II and 2MASS epochs.) This having been noted, whether or not objects with these magnitude differences could perturb the POSS measurements enough to affect the 50-yr proper motion measurements is not clear.  Future releases from a longer baseline Gaia data set should determine whether the accelerations seen for these seven objects are real.

\subsubsection{Accelerators from Hipparcos vs.\ Gaia comparisons}

The \cite{brandt2021} list of $\sim$115,000 objects covers the entire sky for objects in common to Hipparcos and Gaia eDR3 ($G \lesssim 11$ mag). This list also gives the computed $\chi^2$ value between the two proper motions measured with the best precision, which is usually the Gaia-specific and Hipparcos-to-Gaia measurements. We conservatively set a false alarm rate of $Q = e^{-\chi^2/2} < 0.1\%$, corresponding to $\chi^2 > 13.8$, to select high-confidence accelerators for analysis here. Using this criterion produces $\sim$33,750 objects, of which 194 fall within the 20-pc census. These 194 are denoted in Table~\ref{tab:20pc_census} with a "yes" in the "Accelerator?" column.

\cite{kervella2022} have also produced a catalog of possible accelerators based on a comparison of the short-baseline Gaia-specific motions and the long-baseline Hipparcos-to-Gaia motions. As this list is based on the same underlying data as the list produced by \cite{brandt2021}, many of the same accelerators are flagged by both teams. Under the assumption that the companion mass is much less than that of the primary and that the (circular) orbit is perpendicular to the line of sight, \cite{kervella2022} have further used the proper motion measures to estimate the mass of the hidden companion using the equation
\begin{equation}
    m =  \Bigl(4740.470\frac{\Delta\mu}{\varomega}\Bigr)\sqrt{\frac{rM}{G}},
    \label{eqn:kervella}
\end{equation} 
where $m$ is the companion mass, $M$ is the primary mass, $G$ is the gravitational constant, $r$ is the orbital radius, $\Delta\mu$ (the difference in motion measurements) is in units of mas yr$^{-1}$, and $\varomega$ is in units of mas. The constant of 4740.470 is used to convert $\Delta\mu/\varomega$ into units of m s$^{-1}$ (\citealt{kervella2019}). Companion masses are estimated using estimated primary masses generally from isochrone fitting for the brightest stars or from an absolute $K$-band relation for the fainter stars, as described further in \cite{kervella2022}. Companion masses are dependent upon the unknown value of the separation between components, so \cite{kervella2022} constructed estimates for assumed separations of 3, 5, 10, and 30 AU. In Table~\ref{tab:20pc_census} we include the extrema of these mass estimates in columns labeled "EstMassAt3AU" and "EstMassAt30AU" for all objects tagged as accelerators. (In a small number of cases, a \citealt{brandt2021} accelerator was not deemed to be an accelerator by \citealt{kervella2022}, so these estimates are not given.)

\startlongtable
\begin{deluxetable*}{lllDccl}
\tabletypesize{\scriptsize}
\tablecaption{20-pc Accelerators in Known Close Binary/Multiple Systems\label{tab:accelerator_binaries}}
\tablehead{
\colhead{J2000 RA \& Dec} &
\colhead{Name} &
\colhead{Sp.\ Type} &
\multicolumn2c{$\chi^2$} &
\colhead{$M_{est}$ @ 3 AU} &
\colhead{$M_{est}$ @ 30 AU} &
\colhead{Note} \\
\colhead{} &
\colhead{} &
\colhead{} &
\multicolumn2c{} &
\colhead{($M_\odot$)} &
\colhead{($M_\odot$)} &
\colhead{} \\
\colhead{(1)} &
\colhead{(2)} &
\colhead{(3)} &
\multicolumn2c{(4)} &
\colhead{(5)} &
\colhead{(6)} &
\colhead{(7)} \\
}
\decimals
\startdata
00 49 26.76 $-$23 12 44.9&  HD 4747 AB      &G9 V          & 12980    & 0.0338&   0.1532& \nodata\\   
00 50 33.25   +24 49 00.2&  FT Psc AB       &M3 V kee\tablenotemark{a}      &   768.3  & null  &   null  & \nodata\\   
00 58 27.94 $-$27 51 25.4&  CD$-$28 302 AB  &M3 V          &    24.25 & 0.0078&   0.0772& \nodata\\   
01 41 47.13   +42 36 48.2&  HD 10307 AB     &G1 V          &  2540    & 0.1971&   0.6338& \nodata\\   
01 55 57.46 $-$51 36 32.0&  $\chi$ Eri AB   &G9 IV         &   569.3  & 0.0970&   0.1717& \nodata\\   
02 05 04.88 $-$17 36 52.7&  BD$-$18 359 AB  &M3            &  6920    & 0.0925&   0.6897& \nodata\\   
02 10 25.92 $-$50 49 25.5&  HD 13445 AB     &K1 V          & 78930    & 0.0957&   0.4335&    1 known exoplanet\\
02 19 10.08 $-$36 46 41.2&  L 440-30 AB     &M2.5+ V       &    90.10 & 0.0034&   0.0273& \nodata\\   
02 36 04.90   +06 53 12.4&  HD 16160 AB     &K3 V          &  6078    & 0.0942&   0.4769& \nodata\\   
02 45 06.20 $-$18 34 21.4&  $\tau^1$ Eri AB &F6 V          &  4366    & 0.1935&   0.5594& \nodata\\   
02 46 17.28   +11 46 30.9&  HD 17230 AB     &K6 V          &    60.88 & 0.0022&   0.0125& \nodata\\     
03 01 51.39 $-$16 35 36.0&  BD$-$17 588 ABC &M2.5 V        &   215.2  & 0.0147&   0.1728&    2 known exoplanets \\
03 48 01.70   +68 40 38.8&  G 221-24 AB     &K6 V + M2 V   &  1011    & 0.1764&   1.2275& \nodata\\   
05 08 35.04 $-$18 10 19.4&  L 737-9 AB      &M3.5 V        &    41.46 & 0.0024&   0.0169& \nodata\\   
05 19 12.66 $-$03 04 25.7&  HD 34673 AB     &K3 V          &  2500    & 0.0128&   0.0739& \nodata\\   
05 22 37.48   +02 36 11.6&  HD 35112 AB     &K2.5 V        &192800    & 0.0806&   1.0007& \nodata\\   
05 28 44.87 $-$65 26 55.2&  AB Dor ACaCb    &K2 V k        &  1173    & 0.1355&   0.6137& \nodata\\     
05 32 14.66   +09 49 14.9&  Ross 42 AB      &M4            &    17.34 & 0.0010&   0.0079& \nodata\\   
05 54 22.96   +20 16 34.5&  $\chi^1$ Ori AB &G0 V CH-0.3   &   430.1  & 0.0578&   0.2097& \nodata\\   
06 10 34.61 $-$21 51 52.7&  Gl 229 AB       &M1 V          & 13040    & 0.0078&   0.0544&    2 known exoplanets \\
06 17 16.13   +05 05 59.9&  HD 43587 AaAb   &G0 V          & 22570    & 0.2482&   0.7983& \nodata\\   
06 26 10.25   +18 45 24.8&  HD 45088 AaAb   &K3 V k        & 19680    & 0.0712&   0.3606& \nodata\\   
06 36 18.29 $-$40 00 23.6&  CD$-$39 2700 AB &K8 V k        & 14140    & 0.2114&   1.4834& \nodata\\   
07 16 19.77   +27 08 33.1&  G 109-55 AB     &M2.5 V        &   243.4  & 0.0240&   0.1787& \nodata\\   
07 19 31.27   +32 49 48.3&  BD+33 1505 AB   &M0 V          &  1301    & 0.0174&   0.1160& \nodata\\   
07 20 07.37   +21 58 56.3&  $\delta$ Gem AaAb  &F2 V kF0mF0   & 36.94 & 0.0322&   0.0612& \nodata\\   
07 28 51.36 $-$30 14 49.3&  CD$-$29 4446 AB &M2            &  1377    & 0.1433&   0.8673& \nodata\\   
07 36 07.07 $-$03 06 38.7&  BD$-$02 2198 AB &M1 V          & 18520    & 0.1057&   0.7205& \nodata\\   
08 31 37.57   +19 23 39.4&  CU Cnc AaAbAc   &M4            &  2767    & 0.0466&   0.3328& \nodata\\   
08 36 25.47   +67 17 41.8&  BD+67 552 AB    &M0.5          &  9935    & 0.1336&   0.8255& \nodata\\   
08 39 07.90 $-$22 39 42.8&  HD 73752 AaAbB  &G5 IV         &  2804    & 0.2452&   0.8217& \nodata\\   
08 42 44.53   +09 33 24.1&  BD+10 1857 AaAb &M0            &   186.0  & 0.3145&   1.9750& \nodata\\    
08 57 04.68   +11 38 48.8&  BD+12 1944 AB   &M1.5          & 25170    & 0.0973&   0.6882& \nodata\\    
09 14 53.65   +04 26 34.2&  HD 79555 AB     &K3+ V         & 22350    & 0.1772&   1.0191&    also in Table~\ref{tab:LUWE_binaries} \\
09 29 08.93 $-$02 46 08.2&  $\tau^1$ Hya AaAb  &F5.5 IV-V  &  2351    & 0.3076&   0.7993& \nodata\\    
09 32 51.43   +51 40 38.3&  $\theta$ UMa AB &F5.5 IV-V     &   518.5  & 0.0727&   0.1540& \nodata\\   
09 35 39.50   +35 48 36.5&  11 LMi AaAb     &G8+ V         &   496.1  & 0.0091&   0.0364& \nodata\\    
09 45 40.07 $-$39 02 26.5&  L 462-119 AB    &M2.5 V        &  1697    & 0.1046&   0.8348& \nodata\\   
09 53 11.78 $-$03 41 24.4&  BD$-$02 3000 AB &M2            &    80.34 & 0.0052&   0.0351& \nodata\\   
11 11 33.15 $-$14 59 28.9&  HD 97233 AB     &K5 V (k)      &   471.7  & 0.1108&   0.7535& \nodata\\   
11 21 26.67 $-$20 27 14.0&  HD 98712 A      &K6 V ke       &  6618    & 0.1439&   0.8376& \nodata\\      
12 00 44.46 $-$10 26 46.1&  HD 104304 AB    &G8 IV         & 57500    & 0.2325&   0.8631& \nodata\\   
12 23 33.20   +67 11 18.5&  G 237-64 AB     &M2.5          &  1767    & null  &   null  & \nodata\\   
12 28 57.59   +08 25 31.1&  Wolf 414 AB     &M3.5 V + M5 V &  2059    & 0.1223&   1.0562& \nodata\\   
12 44 14.55   +51 45 33.4&  HD 110833 AaAbB &K3            &    95.44 & 0.0307&   0.1438& \nodata\\   
13 00 46.56   +12 22 32.7&  BD+13 2618 AB   &M1.5          &  3393    & 0.0530&   0.3667&    1 known exoplanet \\  
13 19 33.59   +35 06 36.6&  BD+35 2436 AaAb &M1            &    79.89 & 0.0151&   0.0948& \nodata\\     
13 47 15.74   +17 27 24.8&  $\tau$ Boo AB   &F7 IV-V       &  7745    & 0.1133&   0.2944&    1 known exoplanet \\  
13 52 35.85 $-$50 55 18.1&  HD 120780 AaAb  &K2 V          &   479.5  & 0.0332&   0.1594& \nodata\\   
14 03 32.34   +10 47 12.3&  HD 122742 AB    &G6 V          &   640.6  & 0.0597&   0.2446& \nodata\\   
14 54 29.24   +16 06 03.8&  BD+16 2708 ABaBb    &M3 V      &  2764    & 0.0089&   0.0654& \nodata\\     
15 41 16.57   +75 59 34.0&  Ross 1057 AB    &M3.5          &  1780    & 0.0576&   0.6208& \nodata\\   
15 44 01.82   +02 30 54.6&  $\psi$ Ser ABaBb&G5 V          &  4965    & null  &   null  & \nodata\\     
16 05 40.48 $-$20 27 00.1&  HD 144253 AB    &K3 V          &   189.3  & 0.0439&   0.2222& \nodata\\   
16 28 28.14 $-$70 05 03.8&  $\zeta$ TrA AaAb    &F9 V      &    47.58 & 0.0110&   0.0369& \nodata\\   
17 09 31.54   +43 40 52.8&  G 203-47 AB     &M3.5 V + wd   &    51.01 & 0.0072&   0.0830& \nodata\\   
17 19 03.84 $-$46 38 10.4&  41 Ara A        &G9 V          &   832.7  & 0.0160&   0.0689& \nodata\\   
17 30 11.20 $-$51 38 13.1&  CD$-$51 10924 AB&M0 V          &   270.2  & 0.0044&   0.0274&    4 known exoplanets \\ 
17 34 59.62   +61 52 28.2&  26 Dra AB       &G0 IV-V       & 11960    & 0.3382&   1.1333& \nodata\\      
17 46 14.42 $-$32 06 08.4&  CD$-$32 13298 AaAb  &M3 V      &    27.02 & 0.0080&   0.0597& \nodata\\      
17 46 27.55   +27 43 14.6&  $\mu^1$ Her AaAb   &G5 IV      & 12600    & 0.1266&   0.4130& \nodata\\      
18 07 01.59   +30 33 43.6&  b Her AB        &F9 V          & 66470    & 0.0973&   1.0649& \nodata\\  
18 10 26.15 $-$62 00 08.0&  $\iota$ Pav AB  &G0 V          &   192.8  & 0.0356&   0.1192& \nodata\\   
18 57 01.64   +32 54 04.7&  HD 176051 AB    &F9 V          & 38160    & null  &   null  & \nodata\\
19 23 34.01   +33 13 19.1&  HD 182488 AB    &G9+ V         &  1700    & 0.0085&   0.0349& \nodata\\   
19 31 07.97   +58 35 09.6&  HD 184467 AB    &K2 V          &    27.50 & 0.0314&   0.1422& \nodata\\      
19 54 17.74 $-$23 56 27.9&  HD 188088 AaAb  &K2 IV (k)     &    15.46 & 0.0011&   0.0050& \nodata\\   
20 04 06.22   +17 04 12.7&  15 Sge AB       &G0 V          & 31560    & 0.0890&   0.3162& \nodata\\    
20 05 09.78   +38 28 42.6&  HD 190771 AB    &G2 V          & 12760    & 0.0639&   0.2319& \nodata\\   
20 10 19.57 $-$20 29 36.4&  HD 191391 AB    &K6 V k        & 38130    & 0.0819&   0.4895& \nodata\\   
20 44 21.95   +19 44 58.7&  HD 352860 AB    &M0.5 V        &  1583    & 0.1875&   1.2533& \nodata\\   
20 56 48.54 $-$04 50 49.1&  Ross 193 AaAb   &M3            &    72.54 & 0.0050&   0.0395& \nodata\\   
21 00 05.39   +40 04 12.7&  BD+40 883 AaAbB &M2            & 46070    & 0.1350&   0.8783& \nodata\\     
21 19 45.63 $-$26 21 10.4&  HD 202940 AaAbB &G7 V          &   600.4  & 0.0172&   0.1989& \nodata\\   
21 49 05.76 $-$72 06 09.1&  CD$-$72 1700 AB &M1            & 19020    & null  &   null  & \nodata\\   
22 07 00.67   +25 20 42.4&  $\iota$ Peg AaAb    &F5 V      &    66.25 & 0.0205&   0.0533& \nodata\\   
22 18 15.61 $-$53 37 37.5&  HD 211415 AB    &G0 V          &   612.3  & 0.0132&   0.0478& \nodata\\   
22 36 09.69 $-$00 50 29.8&  HD 214100 AB    &M1 V          & 27200    & 0.0672&   0.4472& \nodata\\   
22 38 45.57 $-$20 37 16.1&  FK Aqr AaAb     &M2            &    47.50 & 0.0014&   0.0091& \nodata\\   
23 01 51.54 $-$03 50 55.4&  HD 217580 AB    &K2.5 V        &   232.6  & 0.0591&   0.2966& \nodata\\    
23 39 20.91   +77 37 56.5&  $\gamma$ Cep AB &K1 III        &  4771    & 0.2376&   0.4206&    1 known exoplanet \\
23 52 25.41   +75 32 40.4&  HD 223778 AaAbB &K3 V          &124300    & 0.1261&   0.5909& \nodata\\      
23 55 39.78 $-$06 08 33.4&  BD$-$06 6318 AB &M2.5 V k      &  2137    & 0.0497&   0.3530& \nodata\\   
\enddata
\tablenotetext{a}{Some optical spectral types for late-K and M dwarfs include information about chromospheric activity. An "e" generally indicates that H$\alpha$ emission is present. Other values include "(k)" for slight emission reversals or infilling of the \ion{Ca}{2} H and K lines, "k" for emission reversals in \ion{Ca}{2} H and K that do not rise to the level of the local continuum, "ke" for such emission that rises above the local continuum level, and "kee" for strong emission in \ion{Ca}{2} H and K along with H$\beta$ and possibly H$\gamma$ and H$\delta$ (\citealt{gray2003}). Because chromospheric activity is variable, these emission-line classification suffixes pertain only to the epoch of spectroscopic observation.}
\end{deluxetable*}

\begin{deluxetable*}{lllDcccc}
\tabletypesize{\scriptsize}
\tablecaption{20-pc Accelerators whose only Close Companions are Known Exoplanets\label{tab:accelerator_exoplanets}}
\tablehead{
\colhead{J2000 RA \& Dec} &
\colhead{Name} &
\colhead{Sp.\ Type} &
\multicolumn2c{$\chi^2$} &
\colhead{$M_{est}$ @ 3 AU} &
\colhead{$M_{est}$ @ 30 AU} &
\colhead{\# of} &
\colhead{Distance to Stellar}\\
\colhead{} &
\colhead{} &
\colhead{} &
\multicolumn2c{} &
\colhead{($M_\odot$)} &
\colhead{($M_\odot$)} &
\colhead{Known Exoplanets} &
\colhead{Companion (AU)} \\
\colhead{(1)} &
\colhead{(2)} &
\colhead{(3)} &
\multicolumn2c{(4)} &
\colhead{(5)} &
\colhead{(6)} &
\colhead{(7)} &
\colhead{(8)} \\
}
\decimals
\startdata
00 16 12.68 $-$79 51 04.2&  HD 1237 A       &G8.5 V (k) &    26.86 &    0.0019&   0.0080&   1&      70\\ 
00 18 22.88   +44 01 22.6&  GX And          &M1.5 V     &  2456    &    0.0013&   0.0114&   2&     122\\
03 32 55.84 $-$09 27 29.7&  $\epsilon$ Eri  &K2 V (k)   &    33.89 &    0.0013&   0.0062&   1& \nodata\\
04 52 05.73   +06 28 35.6&  Wolf 1539       &M3.5       &    35.10 &    0.0014&   0.0137&   1& \nodata\\
05 11 40.59 $-$45 01 06.4&  Kapteyn's Star  &sdM1 p     &    29.58 &    0.0001&   0.0007&   1& \nodata\\
05 37 09.89 $-$80 28 08.8&  $\pi$ Men       &G0 V       &    60.98 &    0.0073&   0.0244&   3& \nodata\\
07 54 10.88 $-$25 18 11.4&  CD$-$24 6144    &M0         &    18.89 &    0.0007&   0.0043&   2&    7110\\
09 14 24.68   +52 41 10.9&  HD 79211        &K7 V       &    27.48 &    0.0059&   0.0401&   1&     108\\ 
10 08 43.14   +34 14 32.1&  HD 87883        &K2.5 V     &   713.6  &    0.0087&   0.0408&   1& \nodata\\
16 10 24.32   +43 49 03.5&  14 Her          &K0 IV-V    &  1009    &    0.0126&   0.0502&   2& \nodata\\
16 12 41.78 $-$18 52 31.8&  LP 804-27       &M3 V       &   331.0  &    0.0076&   0.0562&   1& \nodata\\
20 03 37.41   +29 53 48.5&  HD 190360       &G7 IV-V    &    14.83 &    0.0016&   0.0054&   2&    2847\\  
21 33 33.98 $-$49 00 32.4&  HD 204961       &M2         &   278.3  &    0.0008&   0.0062&   2& \nodata\\
22 03 21.65 $-$56 47 09.5&  $\epsilon$ Ind A&K4 V (k)   &   287.5  &    0.0030&   0.0157&   1&    1464\\ 
\enddata
\end{deluxetable*}

We divide the resulting list of 20-pc accelerators into three subgroups. The first, listed in Table~\ref{tab:accelerator_binaries}, comprises eighty-three objects in known close binary and multiple systems. For all of these, the host star is known to have a close-in companion that Gaia eDR3 fails to detect or provide a full astrometric solution for, and these companions range in mass from the substellar regime into the planetary regime. For a host star at a distance of 10 pc, its Hipparcos-to-Gaia acceleration can be detected if the companion has a separation below a few$\times$100 AU (Figure 12 from \citealt{kervella2022}). Companions at this separation range can also be detected with high-resolution imaging techniques or via radial velocity monitoring, and some have independently measured masses. As one example, the companion in the 19.5-yr spectroscopic binary HD 10307 AB has a measured dynamical mass from \cite{torres2022} of 0.254$\pm$0.019 M$_\odot$, and that system has $a = 7.7$ AU, $i = 100^\circ$ and $e = 0.44$. The \cite{kervella2022} companion mass estimates of 0.20 M$_\odot$ at 3 AU and 0.63 M$_\odot$ at 30 AU bracket the dynamically measured values well, as the assumptions used were reasonable for this system. As another example, the companion to the 1.35-yr spectroscopic binary HD 184467 AB has a measured dynamical mass of 0.868$\pm$0.025 M$_\odot$ (\citealt{piccotti2020}), and the system has $a = 0.7$ AU, $i = 145^\circ$ and $e = 0.34$ (\citealt{arenou2000}). The \cite{kervella2022} companion mass estimate of 0.03 M$_\odot$ at 3 AU compares unfavorably to the measured value possibly because of the \cite{kervella2022} assumption that the secondary mass is much less than that of the primary. This demonstrates that, although the \cite{kervella2022} companion mass estimates listed in Table~\ref{tab:accelerator_binaries} provide a guide as to whether the companion causing the acceleration is already known or is a still hidden member, additional astrometric data is needed before the masses can be reliably measured. As can be seen from the full entries in Table~\ref{tab:20pc_census}, many objects in Table~\ref{tab:accelerator_binaries} are triples, so it is also unclear {\it how many} objects are contributing to the measured acceleration.

The second list, shown in Table~\ref{tab:accelerator_exoplanets}, gives fourteen objects known to host exoplanets but lacking any "close" stellar or substellar companions. Here we define "close" to mean within $\sim$50 AU. Six of these objects, as listed in the final column of the table, have more widely separated stellar companions at apparent separations of $\gtrsim$70 AU. The \cite{kervella2022} mass estimates for all fourteen of these objects are quite low and, for assumed separations of a few AU, correspond to masses traditionally thought of as being in the planetary range. Thus, the accelerations for these objects are likely caused by the known exoplanet(s) in the system.  \cite{kervella2022} provides additional analysis on the stars $\epsilon$ Eri, Kapteyn's Star, $\epsilon$ Ind A, and $\pi$ Men, while noting that Kapteyn's Star has no significant proper motion anomaly as measured by them.

The final list, shown in Table~\ref{tab:accelerator_new_ones}, has ninety-seven objects whose closest known companions are resolved by Gaia eDR3 or have no known companions at all. For many of these, the nearest known companion falls close enough to the accelerator star ($\lesssim$100 AU; Figure 12 of \citealt{kervella2022}) that it may be the object causing the acceleration. Examples are CD$-$44 3045 A,  VV Lyn Aa, CD$-$36 6589 A, Ross 52 A, BD+45 2247 A, and Wolf 1225 A. Objects for which the nearest known companion lies beyond this separation or for which no companions are currently known are the hosts most likely to harbor new additions to the 20-pc census. Examples of stars with likely hidden companions are G 32-7, CD$-$22 526, HD 13579, LP 837-53, HD 43162 A, HD 52698, G 250-34, BD$-$17 3088, $\mu$ Vir, $\beta$ TrA, and $\theta$ Cyg.

Tables~\ref{tab:accelerator_binaries}-\ref{tab:accelerator_new_ones} highlight that the accounting of all components within the 20-pc census is still incomplete, as there is overwhelming evidence of additional, tightly separated companions. As only $<$200 of the $\sim$3,000 Gaia-detected primaries show such evidence, it is tempting to conclude that our tally of higher mass (non-exoplanet) companions is nearing completion. We caution, however, that our criteria for selecting accelerators was set very conservatively and that many real accelerators likely exist with a measured significance below our cutoff value. As the time baseline of Gaia observations is extended, accelerations will be increasingly sensitive to longer-period companions that, for higher (non-exoplanet) masses, are potentially verifiable with direct imaging techniques. Furthermore, Gaia observations over this same extended time baseline will remove the need to compare to the shallower Hipparcos data, enabling acceleration data for lower-mass primaries between the Hipparcos and Gaia limits ($11 \lesssim G \lesssim 21$ mag). Finally, less than a third of all systems in the 20-pc census of Table~\ref{tab:20pc_census} have both a Hipparcos entry and a Gaia DR3 astrometric solution, so many objects within our sample volume are unavailable for similar acceleration analysis.

\startlongtable
\begin{deluxetable*}{lllDccccl}
\tabletypesize{\scriptsize}
\tablecaption{20-pc Accelerators with More Distant (or No Known) Companions\label{tab:accelerator_new_ones}}
\tablehead{
\colhead{J2000 RA \& Dec} &
\colhead{Name} &
\colhead{Sp.\ Type} &
\multicolumn2c{$\chi^2$} &
\colhead{$M_{est}$ @ 3 AU} &
\colhead{$M_{est}$ @ 30 AU} &
\colhead{\# of} &
\colhead{Dist.\ to Next} &
\colhead{Note} \\
\colhead{} &
\colhead{} &
\colhead{} &
\multicolumn2c{} &
\colhead{($M_\odot$)} &
\colhead{($M_\odot$)} &
\colhead{Components} &
\colhead{Nearest Known} &
\colhead{} \\
\colhead{} &
\colhead{} &
\colhead{} &
\multicolumn2c{} &
\colhead{} &
\colhead{} &
\colhead{in System} &
\colhead{Member (AU)} &
\colhead{} \\
\colhead{(1)} &
\colhead{(2)} &
\colhead{(3)} &
\multicolumn2c{(4)} &
\colhead{(5)} &
\colhead{(6)} &
\colhead{(7)} &
\colhead{(8)} &
\colhead{(9)} \\
}
\decimals
\startdata
00 05 41.02   +45 48 43.6&  HD 38 A         &K6 V           &  6209    &0.0219&   0.1406& 2&   70& \nodata\\   
00 16 14.63   +19 51 37.5&  G 32-7          &M4             &    14.42 &0.0082&   0.0677& 3&  387& \nodata\\   
00 45 48.29 $-$41 54 33.1&  HD 4378 A       &K5             &    64.43 &0.0021&   0.0120& 2&   75& \nodata\\   
00 49 06.29   +57 48 54.6&  $\eta$ Cas A    &F9 V           &   659.4  &0.0220&   0.0780& 2&   79& \nodata\\   
00 49 09.90   +05 23 19.0&  Wolf 28         &DZ7.4          &  1621    &0.0007&   0.0044& 1& \nodata& \nodata\\   
01 03 14.15   +20 05 52.3&  G 33-35 A       &M1.5           &    28.44 &0.0040&   0.0300& 2&   40& \nodata\\   
01 32 26.20 $-$21 54 18.4&  CD$-$22 526     &M1.5 V (k)     &   151.0  &0.0039&   0.0284& 1& \nodata& \nodata\\   
01 39 47.56 $-$56 11 47.2&  p Eri B         &K2 V           &  1077    &0.0184&   0.0864& 2&   94& \nodata\\   
02 15 42.55   +67 40 20.3&  HD 13579        &K2             &    58.54 &0.0041&   0.0184& 2&  736  & \nodata\\
02 37 52.79 $-$58 45 11.1&  L 174-28        &M3 V           &    14.88 &0.0010&   0.0111& 1& \nodata& \nodata\\   
03 12 04.53 $-$28 59 15.4&  $\alpha$ For A  &F6 V           &  2451    &0.0692&   0.1861& 3&   75& \nodata\\   
03 16 13.83   +58 10 02.5&  Ross 370 A      &M2             &   130.2  &0.0058&   0.0441& 2&   69& \nodata\\   
03 23 35.26 $-$40 04 35.0&  HD 21175 A      &K1 V           & 69960    &0.0950&   0.4497& 2&   46& \nodata\\   
03 48 01.03   +68 40 22.4&  HD 23189        &K2 V           &    41.23 &0.0036&   0.0222& 3&  303& \nodata\\   
03 57 28.70 $-$01 09 34.1&  HD 24916 A      &K4 V           &    62.16 &null  &   null  & 3&  168& \nodata\\   
04 31 11.51   +58 58 37.5&  G 175-34        &M4.5 V         &   208.4  &0.0018&   0.0203& 2&   57& \nodata\\   
04 53 31.20 $-$55 51 37.1&  CD$-$56 1032 A  &M3 V           &   117.4  &0.0032&   0.0265& 2&   83& \nodata\\   
05 03 23.90   +53 07 42.5&  BD+52 911 A     &M0.5           &    82.34 &0.0023&   0.0167& 2&   78& \nodata\\   
05 45 48.28   +62 14 12.4&  BD+62 780       &M0             &    94.82 &0.0013&   0.0078& 1& \nodata& \nodata\\   
05 55 43.21 $-$26 51 23.4&  LP 837-53       &M2.5 V         &    21.48 &0.0010&   0.0070& 1& \nodata& \nodata\\   
06 13 45.30 $-$23 51 43.0&  HD 43162 A      &G6.5 V         &    16.04 &0.0013&   0.0053& 4&  408& \nodata\\      
06 33 43.28 $-$75 37 48.0&  L 32-9          &M3             &    25.61 &0.0013&   0.0106& 2&  192& \nodata\\   
06 37 11.23 $-$50 02 17.7&  CD$-$49 2340 A  &K8 V (k)       &  7109    &0.0262&   0.1645& 2&   35& \nodata\\   
06 57 46.63 $-$44 17 28.2&  CD$-$44 3045 B  &M3             & 21590    &0.0283&   0.2796& 2&   19& \nodata\\   
07 01 13.73 $-$25 56 55.5&  HD 52698        &K1 V (k)       &   734.1  &0.0491&   0.2224& 1& \nodata&  Also in Table~\ref{tab:LUWE_binaries} \\
07 07 50.42   +67 12 04.9&  G 250-34\tablenotemark{a}
                                            &M1.5           & 16320    &0.1070&   0.7770& 1& \nodata& \nodata\\   
07 31 57.71   +36 13 10.1&  VV Lyn Aa       &M3             & 48470    &0.0862&   0.6179& 3&   19& \nodata\\      
07 57 57.78 $-$00 48 51.9&  HD 65277 A      &K3+ V          &   645.3  &0.0081&   0.0409& 2&   95& \nodata\\   
08 08 13.19   +21 06 18.2&  BD+21 1764 A    &K7 V           &   140.8  &0.0058&   0.0359& 4&  190& \nodata\\   
08 50 42.30   +07 51 52.5&  BD+08 2131 A    &K5 V           &   206.6  &null  &   null  & 2&   21& \nodata\\      
09 01 17.48   +15 15 56.8&  HD 77175 A      &K5             &   210.8  &0.0062&   0.0361& 2&   93& \nodata\\      
09 06 45.35 $-$08 48 24.6&  BD$-$08 2582    &M0             &   143.1  &0.0007&   0.0042& 2&  123& \nodata\\      
09 14 22.77   +52 41 11.8&  HD 79210        &M0 V           &   178.2  &0.0053&   0.0344& 2&  108& \nodata\\      
09 43 55.61   +26 58 08.4&  Ross 93         &M3.5           &    23.32 &0.0023&   0.0211& 1& \nodata& \nodata\\   
10 12 08.15 $-$18 37 04.1&  BD$-$17 3088    &M0             &   267.3  &0.0561&   0.3563& 2& 7225& \nodata\\      
10 31 24.22   +45 31 33.8&  BD+46 1635 A    &K7 V           &   111.8  &0.0026&   0.0151& 2&   74& \nodata\\      
10 41 09.30 $-$36 53 43.7&  CD$-$36 6589 A  &M0.5 V         & 101600   &0.1453&   0.9567& 2&   13& \nodata\\      
10 41 51.83 $-$36 38 00.1&  CD$-$35 6662    &M0 V (k)       &    31.36 &0.0045&   0.0306& 2&  305& \nodata\\      
11 05 28.58   +43 31 36.3&  BD+44 2051 A    &M1 V           &    25.09 &0.0004&   0.0040& 2&  156& \nodata\\      
11 11 19.48   +43 25 02.4&  G 122-2 A       &M2.5 V         &   140.0  &0.0178&   0.1259& 2&   60& \nodata\\      
11 15 11.90   +73 28 30.7&  HD 97584 A      &K3             &    47.01 &0.0014&   0.0072& 2&   94& \nodata\\      
11 34 29.49 $-$32 49 52.8&  20 Crt A        &K0- V          &   257.4  &0.0039&   0.0175& 2&  146& \nodata\\      
11 45 34.44 $-$20 21 12.4&  LP 793-33       &M2.5 V         &   618.2  &0.0245&   0.1946& 2&  295& \nodata\\      
11 45 42.92 $-$64 50 29.5&  LAWD 37         &DQ6.4          &    37.90 &0.0006&   0.0040& 1& \nodata& \nodata\\   
12 08 24.82 $-$24 43 44.0&  $\alpha$ Crv A  &F1 V           &    34.53 &0.0166&   0.0431& 2&   47& \nodata\\      
12 23 00.16   +64 01 51.0&  Ross 690        &M3             &   125.9  &0.0039&   0.0288& 1& \nodata& \nodata\\   
12 41 06.49   +15 22 36.0&  HD 110315 A     &K4.5 V         & 21830    &0.0394&   0.2098& 2&   32& \nodata\\      
12 41 39.63 $-$01 26 57.9&  $\gamma$ Vir A  &F2 V           &   551.7  &0.1854&   0.4774& 2&   28& \nodata\\      
13 02 20.69 $-$26 47 13.6&  HD 113194       &K5 V (k)       &    82.76 &0.0092&   0.0528& 1& \nodata& Also in Table~\ref{tab:LUWE_binaries}\\   
13 06 15.40   +20 43 45.3&  BD+21 2486 A    &K7             &  4329    &0.0282&   0.1623& 3&   31& \nodata\\   
13 14 15.14 $-$59 06 11.7&  HD 114837 A     &F6 V Fe-0.4    &    76.39 &0.0117&   0.0327& 2&   84& \nodata\\      
13 16 51.05   +17 01 01.8&  HD 115404 A     &K2.5 V (k)     &  2205    &0.0089&   0.0588& 2&   85& \nodata\\      
13 20 58.05   +34 16 44.2&  BD+35 2439      &M1.5           &   199.1  &0.0042&   0.0301& 1& \nodata& \nodata\\   
13 23 32.78   +29 14 15.0&  HD 116495 A     &M0 V           &   213.5  &0.0072&   0.0414& 2&   30& \nodata\\      
13 28 21.08 $-$02 21 37.1&  Ross 486 A      &M3 V           &    20.64 &0.0015&   0.0118& 2&  112& \nodata\\      
13 32 44.60   +16 48 39.1&  G 63-36 A       &M2.5 V         &    51.28 &0.0258&   0.1893& 2&   45& \nodata\\      
13 47 42.16 $-$32 25 48.1&  HD 120036 A     &K6.5 V (k)     &   173.5  &0.0071&   0.0431& 2&  135& \nodata\\      
13 49 04.00   +26 58 47.8&  HD 120476 A     &K3.5 V         & 63800    &0.0441&   0.3214& 2&   39& \nodata\\      
13 55 02.56 $-$29 05 25.9&  HD 121271 A     &M0 V k         & 172300   &0.0536&   0.3033& 2&   38& \nodata\\      
14 19 00.90 $-$25 48 55.5&  HD 125276 A     &F9 V Fe-1.5 CH-0.7  & 904.8 & 0.0170&   0.0616& 2&  64& \nodata\\      
14 42 21.58   +66 03 20.8&  G 239-25 A      &M2             &    62.92 &0.0056&   0.0497& 2&   25& \nodata\\      
14 43 03.62 $-$05 39 29.5&  $\mu$ Vir       &F2 V           &    23.50 &0.0384&   0.0746& 2&  793& \nodata\\      
14 51 23.39   +19 06 01.6&  $\xi$ Boo A     &G7 V           & 11500    &0.0689&   0.2820& 2&   38& \nodata\\      
14 53 51.40   +23 33 21.0&  Ross 52 A       &M3 V           &  6509    &0.1125&   0.9852& 2&   10& \nodata\\      
14 57 28.00 $-$21 24 55.7&  HD 131977       &K4 V           &   354.6  &0.0046&   0.0239& 4&  145& \nodata\\     
15 00 55.57   +45 25 34.6&  BD+45 2247 A    &M0.5           & 80070    &0.0951&   0.6269& 2&   23& \nodata\\      
15 47 29.10 $-$37 54 58.7&  HD 140901 A     &G7 IV-V        &    18.43 &0.0023&   0.0085& 2&  220& \nodata\\      
15 55 08.56 $-$63 25 50.6&  $\beta$ TrA     &F1 V           &    31.48 &0.0163&   0.0295& 1& \nodata& \nodata\\   
16 16 45.31   +67 15 22.5&  EW Dra          &M3             &    15.22 &0.0007&   0.0052& 2&  695& \nodata\\      
16 20 03.51 $-$37 31 44.4&  CD$-$37 10765 A &M2 V           &  2809    &0.0074&   0.0656& 2&   36& \nodata\\      
16 28 52.66   +18 24 50.6&  HD 148653 A     &K2 V           &  2069    &0.0163&   0.0827& 2&   46& \nodata\\      
16 55 25.22 $-$08 19 21.3&  Wolf 629        &M3.5 V         &    15.84 &0.0005&   0.0059& 5&  469& \nodata\\  
16 55 38.01 $-$32 04 03.7&  HD 152606       &K8 V k         &  1687    &0.0142&   0.0874& 1& \nodata& \nodata\\  
16 56 48.57 $-$39 05 38.2&  CD$-$38 11343 A &M3             &   550.3  &0.0184&   0.1312& 3&   49& \nodata\\      
16 57 53.18   +47 22 00.1&  HD 153557 A     &K3 V           &   875.6  &0.0102&   0.0514& 3&   91& \nodata\\      
17 15 20.98 $-$26 36 10.2&  36 Oph B        &K0 V           &   29.24  &0.0017&   0.0078& 3&   30& \nodata\\      
17 21 00.37 $-$21 06 46.6&  $\xi$ Oph A     &F2 V           &   82.92  &0.0248&   0.0646& 2&   71& \nodata\\      
17 35 13.62 $-$48 40 51.1&  CD$-$48 11837 A &M1.5 V         & 8601     &0.0174&   0.1332& 2&   45& \nodata\\      
17 46 34.23 $-$57 19 08.6&  L 205-128       &M4             &   37.12  &0.0001&   0.0012& 1& \nodata& \nodata\\   
17 57 48.50   +04 41 36.1&  Barnard's Star  &M4 V           &   74.89  &0.0000&   0.0005& 1& \nodata& \nodata\\  
18 42 46.70   +59 37 49.4&  HD 173739       &M3 V           &  723.0   &0.0020&   0.0209& 2&   41& \nodata\\      
18 42 46.89   +59 37 36.7&  HD 173740       &M3.5 V         &   91.38  &0.0026&   0.0309& 2&   41& \nodata\\   
18 57 30.59 $-$55 59 30.8&  HD 175224 A     &M1             &  412.4   &0.0066&   0.0378& 2&   33& \nodata\\      
19 36 26.53   +50 13 16.0&  $\theta$ Cyg    &F3+ V          &  103.1   &0.0277&   0.0648& 2& 2150&   Other component in Table~\ref{tab:LUWE_binaries} \\
19 45 49.75   +32 23 13.7&  G 125-30        &M1.5           &   25.44  &0.0044&   0.0353& 1& \nodata& \nodata\\   
19 46 23.93   +32 01 01.4&  HD 331161 A     &M0.5 V         &  420.0   &0.0068&   0.0472& 2&   79& \nodata\\      
20 02 34.16   +15 35 31.5&  HD 190067 A     &K0 V Fe-0.9    &  679.7   &0.0071&   0.0322& 2&   62& \nodata\\      
20 41 51.13 $-$32 26 06.7&  AT Mic A        &M4.5 V         & 20360    &0.0891&   0.5735& 3&   21& \nodata\\      
21 02 40.75   +45 53 05.2&  HD 200560 A     &K2.5 V         &  6003    &0.0230&   0.1106& 2&   52& \nodata\\      
21 06 53.94   +38 44 57.9&  61 Cyg A        &K5 V           &   317.4  &0.0068&   0.0386& 2&  111& \nodata\\      
21 06 55.26   +38 44 31.4&  61 Cyg B        &K7 V           & 16640    &0.0070&   0.0431& 2&  111& \nodata\\   
21 07 10.38 $-$13 55 22.5&  HD 200968 A     &G9.5 V (k)     &  4553    &0.0180&   0.0816& 2&   64& \nodata\\      
21 38 00.39   +27 43 25.4&  BD+27 4120 A    &M0.5+ V        &   359.2  &0.0266&   0.1860& 2&   44& \nodata\\      
22 14 31.41   +27 51 18.7&  G 188-49 A      &K7.5           &    20.55 &0.0023&   0.0157& 2&   59& \nodata\\      
22 23 29.13   +32 27 33.9&  Wolf 1225 A     &M3.5           &  5174    &0.1158&   0.7867& 2&   19& \nodata\\      
23 31 52.17   +19 56 14.1&  BD+19 5116 A    &M3.5 V         & 17570    &0.0172&   0.1549& 2&   34& \nodata\\      
23 39 37.39 $-$72 43 19.8&  HD 222237       &K3+ V          &   700.6  &0.0046&   0.0231& 1& \nodata& \nodata\\   
\enddata  
\tablenotetext{a}{A single epoch of Keck/NIRC2 data is available in the Keck Observatory Archive for this object.  The $K_s$- and $H$-band observations taken on 22 Jan 2021 (UT; PI: Crepp; Program ID: D313) show an object $0{\farcs}22$ from G 250-34 at a position angle of 328$^\circ$ and $\Delta{K_s} = 2.9$ mag. No other high-resolution observations were found in other archives to help confirm or refute this possible companion.}
\end{deluxetable*}

\clearpage

\subsection{Multiples with large RUWE values\label{sec:RUWE_LUWE}}

The Gaia Renormalized Unit Weight Error (RUWE) is a measure of the goodness of fit of the single-star astrometric model to the observed astrometry and is expected to be $\sim$1.0 if the fit is a good representation (\citealt{lindegren2021b}). This parameter is pulled from Gaia DR3 and is listed in the "RUWE" column of Table~\ref{tab:20pc_census}. Values significantly higher than unity can indicate either that the object is an unresolved, physical multiple (\citealt{penoyre2020}) or that some other effect is causing the photocenter to deviate from expectations. Two examples of the latter are a chance alignment with a marginally resolved background star or  single-star variability that confounds the RUWE renormalization itself (\citealt{belokurov2020}). The typically quoted value for selecting likely binaries using this statistic is RUWE $>$ 1.4 (e.g., \citealt{fabricius2021}), although \cite{stassun2021} have shown that values of 1.0 $<$ RUWE $<$ 1.4 are also highly predictive of unresolved multiplicity. While the RUWE normalization works well across the full population of Gaia-measured stars, \cite{penoyre2022b} note that it does a somewhat less adequate job when a selection of nearby (d $<$ 100 pc; the GCNS of \citealt{smart2020}) stars alone is analyzed. For that reason, they define a new statistic, which they term the Local Unit Weighted Error (LUWE), that improves upon RUWE for these nearer objects. 

Values of RUWE and LUWE change with each subsequent release of Gaia data, and there is valuable information contained within the differences. Later Gaia releases have data (and astrometric solutions) covering a longer timespan, so for unresolved multiple systems with periods roughly equal to or longer than the timespans of the data release, the RUWE (or LUWE) values may continue to run high or even become larger between Gaia DR2 and Gaia eDR3 simply because the photocentric displacement caused by orbital motion in an unresolved binary makes the single-object astrometric solution fit less well with an extended data set. (See \citealt{penoyre2022a} for additional discussion.) Conversely, unresolved binaries with shorter periods should improve and eventually get full astrometric solutions in the Gaia non-single star lists. 

With these observations in mind, \cite{penoyre2022b} devised a two-part criterion to select the most likely hidden multiples in the 100-pc sample: (1) LUWE$_{eDR3} > 2$ and (2) $\Delta{\rm LUWE} \equiv {\rm LUWE}_{eDR3} - {\rm LUWE}_{DR2} > - {\rm LUWE}_{eDR3}/3$. Within our 20-pc census, 104 objects meet these criteria, and these are the ones labeled with a "yes" in the Table~\ref{tab:20pc_census} column named "LUWE\_binary?". Of these, 73 are known from previous literature to be binary and were already labeled as such in our census. The other 31, listed in Table~\ref{tab:LUWE_binaries}, are newly identified multiples. Nine of these are part of higher-order multiples, as indicated by the notes in the table. For eight of these systems, Gaia has detections of both the new high-LUWE object and the other component (sometimes a double itself) with which it has physical companionship. The ninth system, however, is a new triple system for which Gaia detects only the new high-LUWE binary G 43-23 but not the common-proper-motion T dwarf companion, WISEU J100241.49+145914.9, discussed in Section~\ref{sec:new_companions}. Note that the LUWE criteria from \cite{penoyre2022b} are meant to be conservative, so other hidden binaries will exist with LUWE or $\Delta$LUWE values outside of the bounds noted above. 

\startlongtable
\begin{deluxetable}{lllcl}
\tabletypesize{\scriptsize}
\tablecaption{New 20-pc Multiple Systems Identified Through LUWE\label{tab:LUWE_binaries}}
\tablehead{
\colhead{J2000 RA} &
\colhead{J2000 Dec} &
\colhead{Name} &
\colhead{Note} &
\colhead{Sp.\ Type\tablenotemark{a}} \\
\colhead{(1)} &
\colhead{(2)} &
\colhead{(3)} &
\colhead{(4)} &
\colhead{(5)} \\
}
\startdata
01 46 29.35& $-$53 39 32.6&   2MASS J01462935$-$5339325&   1&   M4.5e    \\
04 34 45.33& $-$00 26 46.5&   G 82-33                  &    &   M4 V     \\
07 01 13.73& $-$25 56 55.5&   HD 52698\tablenotemark{b}                 &    &   K1 V (k) \\
07 08 07.01& $-$22 48 47.3&   LP 840-16                &    &   M2       \\
07 49 42.14& $-$03 20 34.0&   UCAC4 434-042012         &   2&   M3.5 V   \\
08 25 52.82&   +69 02 01.1&   LP 35-347                &    &   M5.5 V   \\
10 02 42.45&   +14 59 13.0&   G 43-23                  &   3&   M4 V     \\
10 14 53.12&   +21 23 46.0&   G 54-19                  &    &   M4.5 V   \\
10 39 45.41& $-$44 30 37.0&   TYC 7722-1583-1          &    &   M3       \\
11 45 34.44& $-$20 21 12.4&   LP 793-33                &   4&   M2.5 V   \\
13 02 20.69& $-$26 47 13.6&   HD 113194\tablenotemark{b}                &    &   K5 V (k) \\
13 30 40.95& $-$20 39 03.7&   UCAC4 347-066233         &    &   M4       \\
13 40 08.79&   +43 46 38.0&   Ross 1026                &    &   M3.5     \\
13 58 52.21&   +27 52 14.2&   UPM J1358+2752           &    &   \nodata  \\
14 17 22.10&   +45 25 46.0&   FBS 1415+456             &   5&   M5/6     \\
14 23 43.74&   +14 26 51.4&   LP 440-17                &    &   M7 (NIR) \\
14 24 18.70& $-$35 14 32.7&   2MASSI J1424187$-$351432 &    &   M6.5 V   \\
14 28 17.58&   +05 18 45.8&   G 65-53                  &   6&   M3.5     \\
15 10 16.82& $-$02 41 08.1&   TVLM 868-110639          &    &   M9 V     \\
15 18 31.46&   +20 36 28.2&   UCAC4 554-051865         &    &   M4.5 V   \\
15 52 06.55& $-$33 59 19.0&   UCAC3 113-186615         &    &   \nodata  \\
17 33 53.18&   +16 55 13.1&   LSPM J1733+1655          &    &   M5.5     \\
18 30 39.45& $-$03 56 18.9&   UCAC4 431-076686         &    &   M4.5 V   \\
18 43 12.51& $-$33 22 46.1&   CD$-$33 13497            &   7&   M1       \\
18 53 25.37&   +02 50 48.7&   G 141-46                 &    &   M2.5     \\
19 36 14.39&   +50 13 10.1&   UCAC3 281-150921         &   8&   M2/3     \\
20 12 59.94&   +01 12 58.3&   2MASS J20125995+0112584  &    &   M6       \\
20 33 36.67& $-$21 20 10.1&   2MASS J20333668$-$2120096&    &   M2       \\
22 10 13.19& $-$71 46 06.2&   PM J22102$-$7146         &    &   \nodata  \\
22 17 18.97& $-$08 48 12.3&   Wolf 1561 A              &   9&   M4 V     \\
22 24 24.65& $-$58 26 13.6&   UCAC3 64-480761          &    &   \nodata  \\
\enddata
\tablenotetext{a}{Spectral types are taken from Table~\ref{tab:20pc_census}. All are optical types except for LP 440-17, which is a near-infrared type.}
\tablenotetext{b}{This object is also in the accelerator list of Table~\ref{tab:accelerator_new_ones}.}
\tablecomments{
\begin{itemize}
\item (1) 0146$-$5339: This is the 35{\farcm}9-distant companion to the F9 dwarf q01 Eri. 
\item (2) 0749$-$0320: This is the 3{\farcm}9-distant companion to the M3.5 binary PM J07498$-$0317 AB.
\item (3) 1002+1459: This object also has a 15{\farcs}6-distant companion, WISEU J100241.49+145914.9, announced in this paper (Section~\ref{sec:new_companions}).
\item (4) 1145$-$2021: This is a physical system with the M5e star LP 793-34, 15{\farcs}2 distant.
\item (5) 1417+4525: This is the 59{\farcs}2-distant companion of the M0 star BD+46 1951.
\item (6) 1428+0518: This is a physical system with the M4 star G 65-54, 1{\farcm}0 distant.
\item (7) 1843$-$3322: This is a physical system with the M6 star CE 507, 15{\farcs}0 distant.
\item (8) 1936+5013:  This is the 1{\farcm}9-distant companion to the F3+ dwarf $\theta$ Cyg, which is listed amongst the accelerators in Table~\ref{tab:accelerator_new_ones}.
\item (9) 2217$-$0848: This is a physical system with the M5 dwarf binary Wolf 1561 BaBb, 7{\farcs}9 distant.
\end{itemize}
}
\end{deluxetable}

\subsection{Multiplicity (and oddities) identified through color-magnitude diagrams\label{sec:color-magnitude_diagrams}}

\begin{figure*}
\includegraphics[scale=0.825,angle=0]{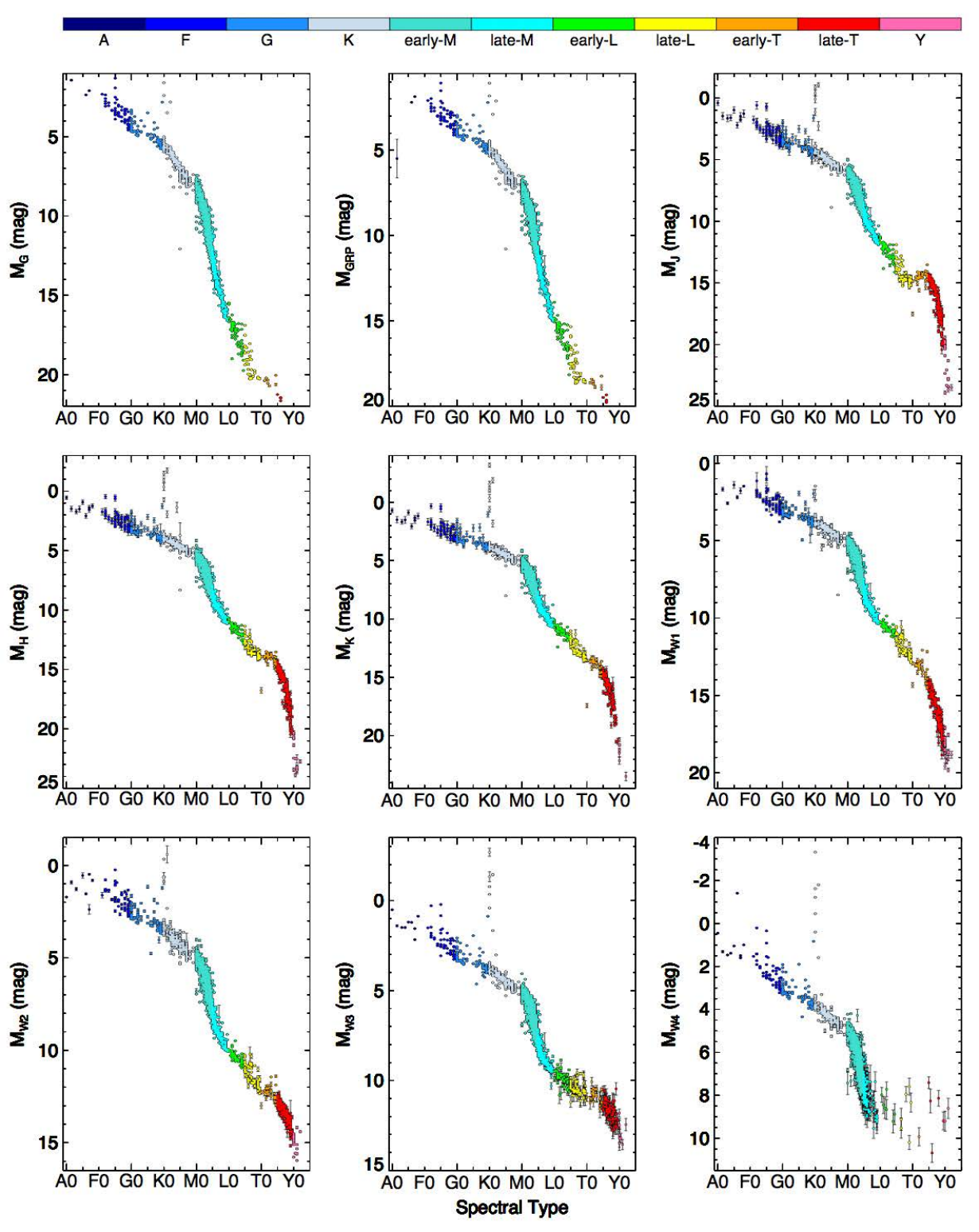}
\caption{Various absolute magnitudes plotted against spectral type for the 20-pc census. See text for details.
\label{fig:AbsMag_SpType}}
\end{figure*}

\begin{figure*}
\includegraphics[scale=0.775,angle=0]{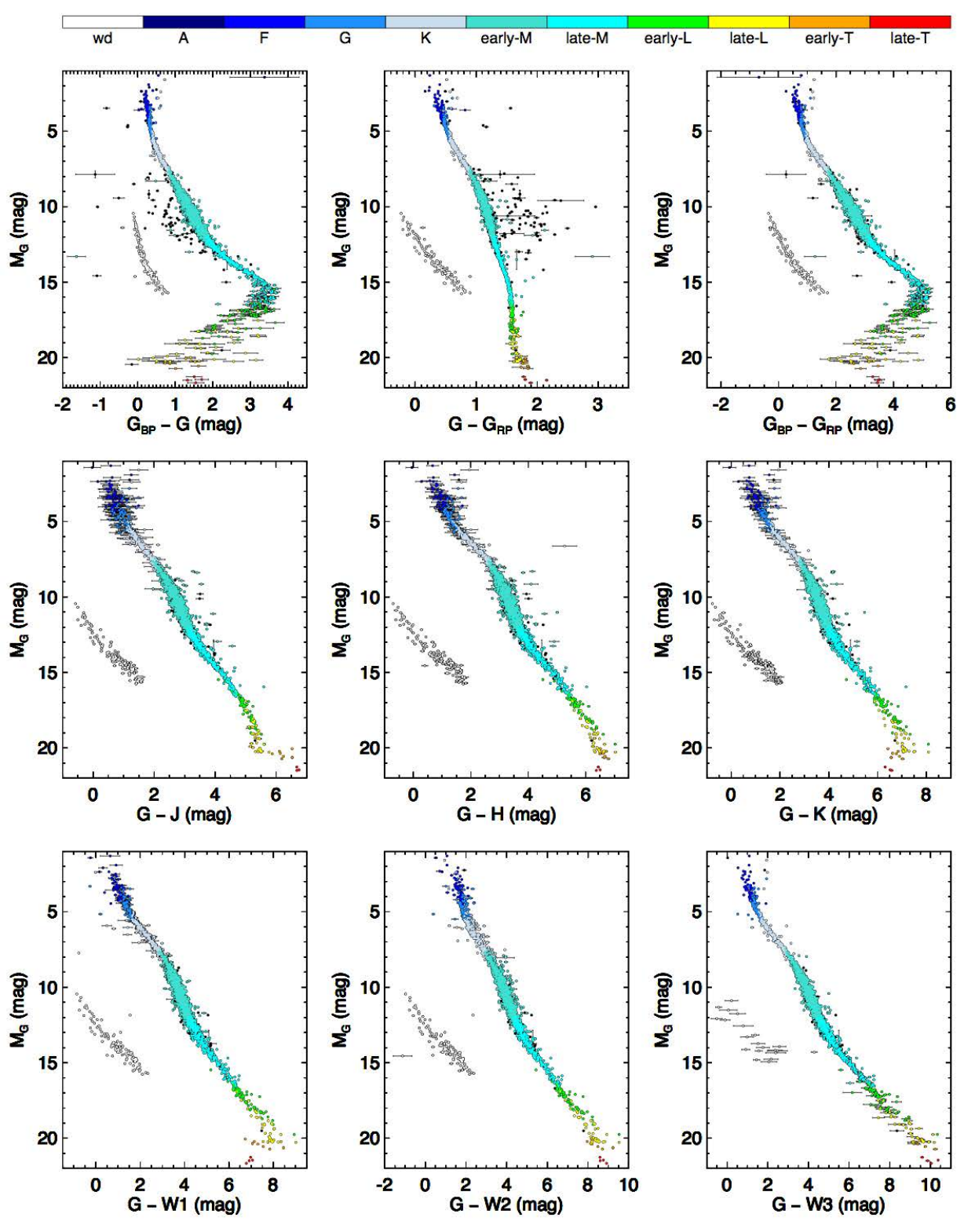}
\caption{Absolute $G$-band magnitude plotted against various colors for the 20-pc census. See text for details. The spray of mostly black points (i.e., objects with no measured spectral types) to the left of the main sequence in the $G_{BP}-G$ vs.\ $M_G$ diagram and to the right of the main sequence in the $G-G_{RP}$ vs.\ $M_G$ diagram represents components in close binaries near the Gaia resolution limit. The $G_{BP}$ and $G_{RP}$ magnitudes are calculated from the fluxes in a 3.5$\times$2.1 arcsec$^2$ field, whereas the $G$ magnitudes are calculated from a profile fit to a much higher-resolution image (section 8 of \citealt{evans2018}). For binaries just above the Gaia resolution limit, this means that per-component BP and RP fluxes will often include light from the other object, whereas the $G$ flux will not (\citealt{halbwachs2022}). This effect pushes such objects blueward in $G_{BP}-G$ color and redward in $G-G_{RP}$ color, as these diagrams show.
\label{fig:MG_color}}
\end{figure*}

\begin{figure*}
\includegraphics[scale=0.85,angle=0]{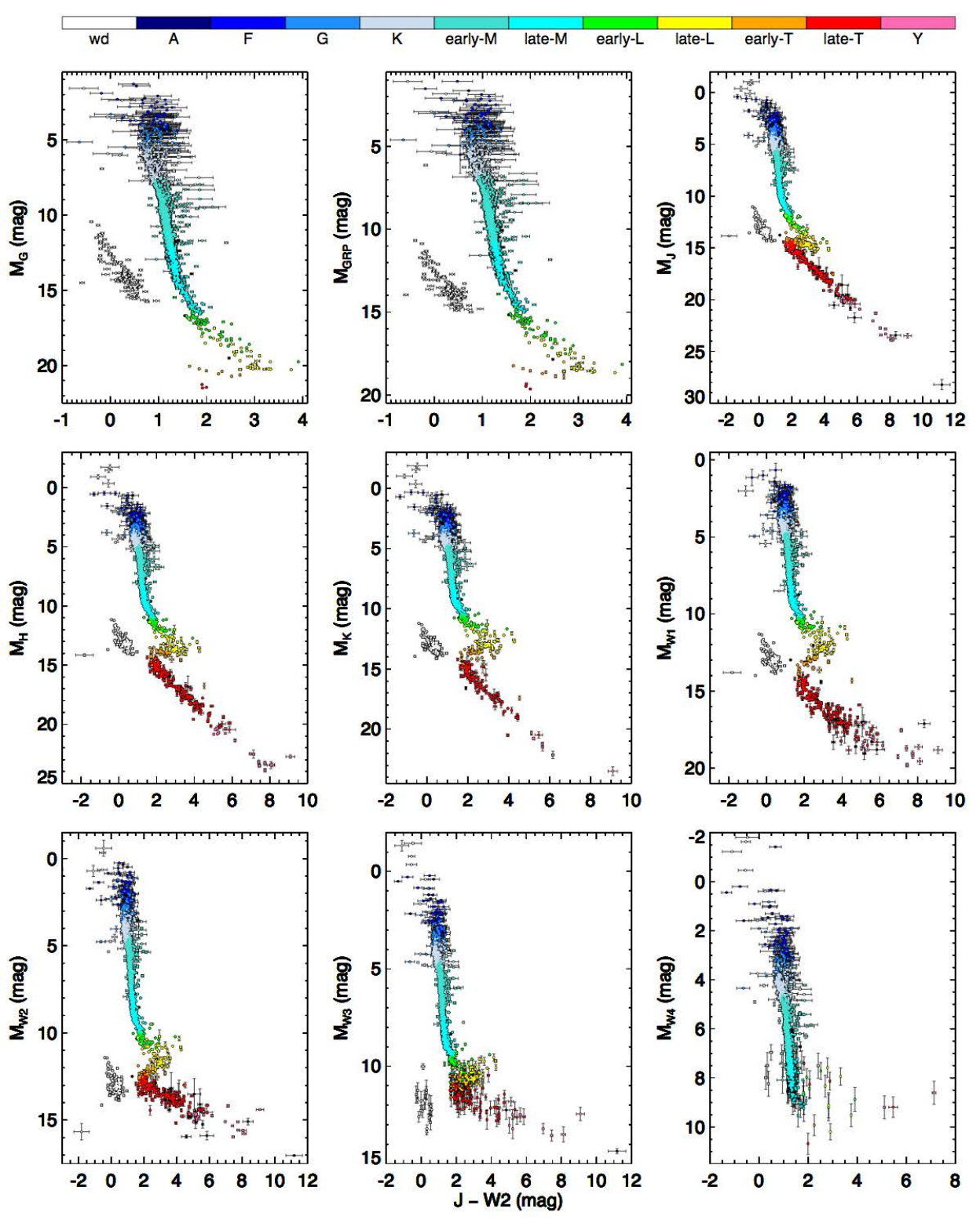}
\caption{Various absolute magnitudes plotted against $J-$W2 color for the 20-pc census. See text for details.
\label{fig:AbsMag_JW2}}
\end{figure*}

\begin{figure*}
\includegraphics[scale=0.85,angle=0]{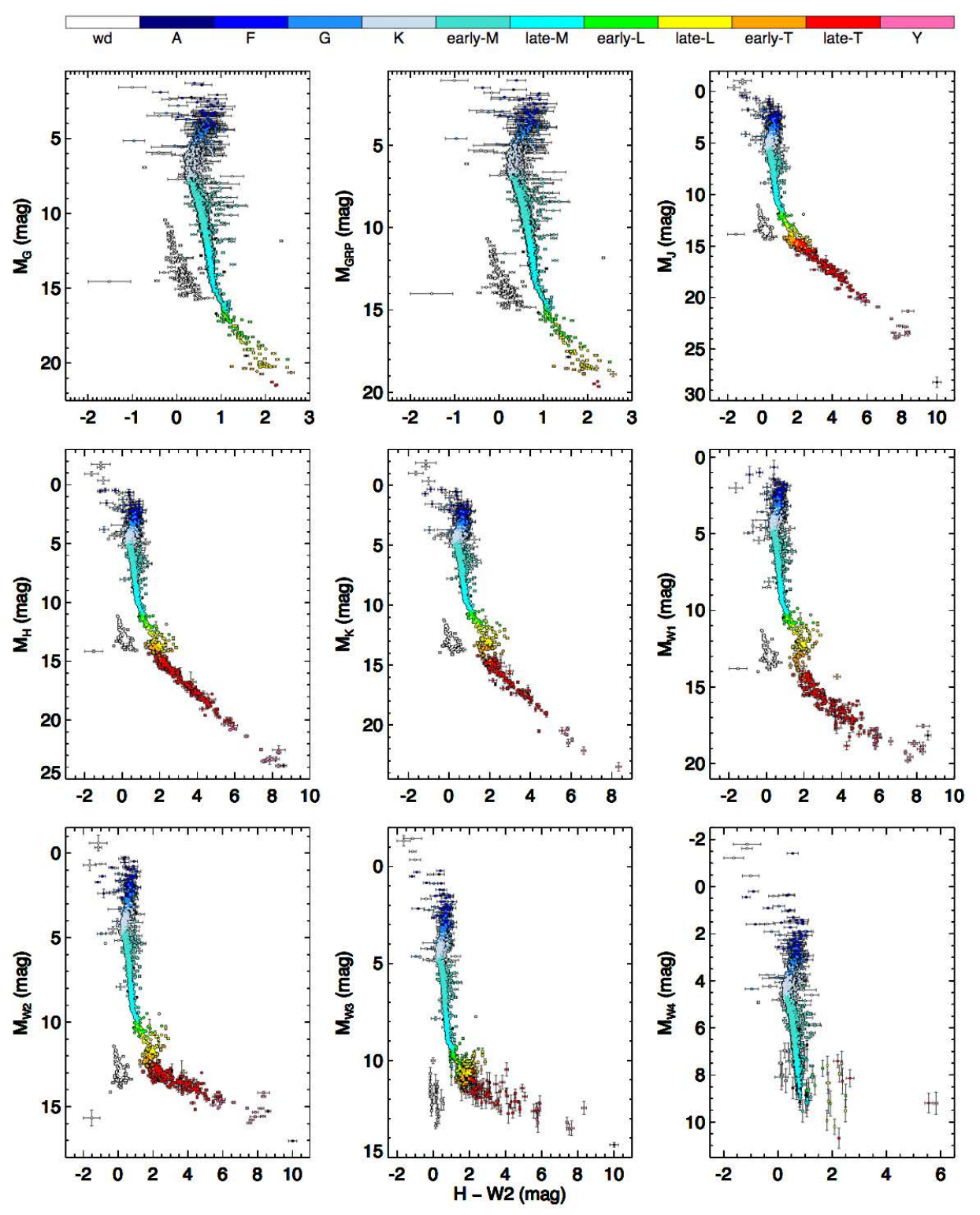}
\caption{Various absolute magnitudes plotted against $H-$W2 color for the 20-pc census. See text for details.
\label{fig:AbsMag_HW2}}
\end{figure*}

\begin{figure*}
\includegraphics[scale=0.425,angle=0]{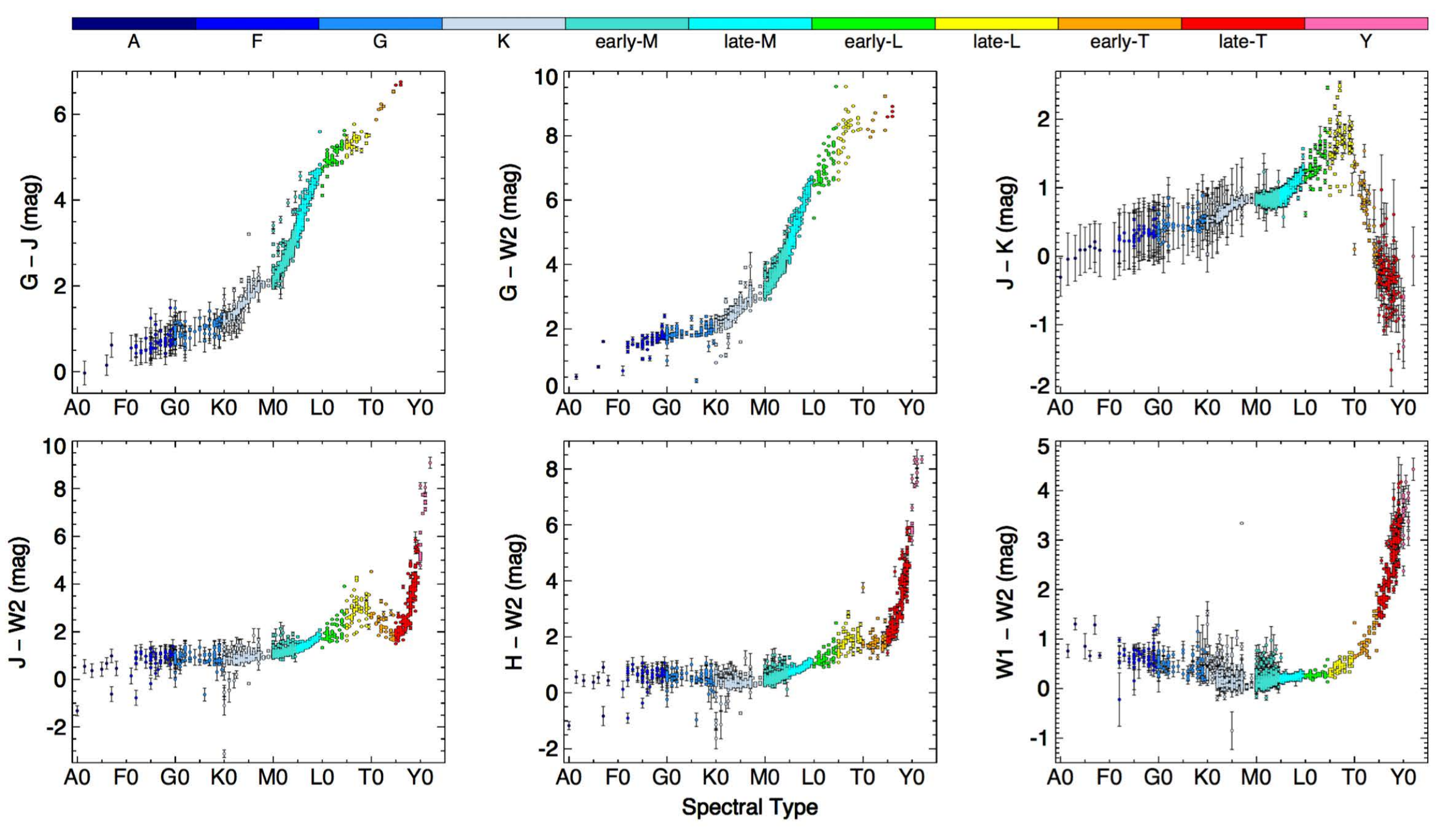}
\caption{Various colors plotted against spectral type for the 20-pc census. See text for details.
\label{fig:colors_SpType}}
\end{figure*}

\begin{figure*}
\includegraphics[scale=0.825,angle=0]{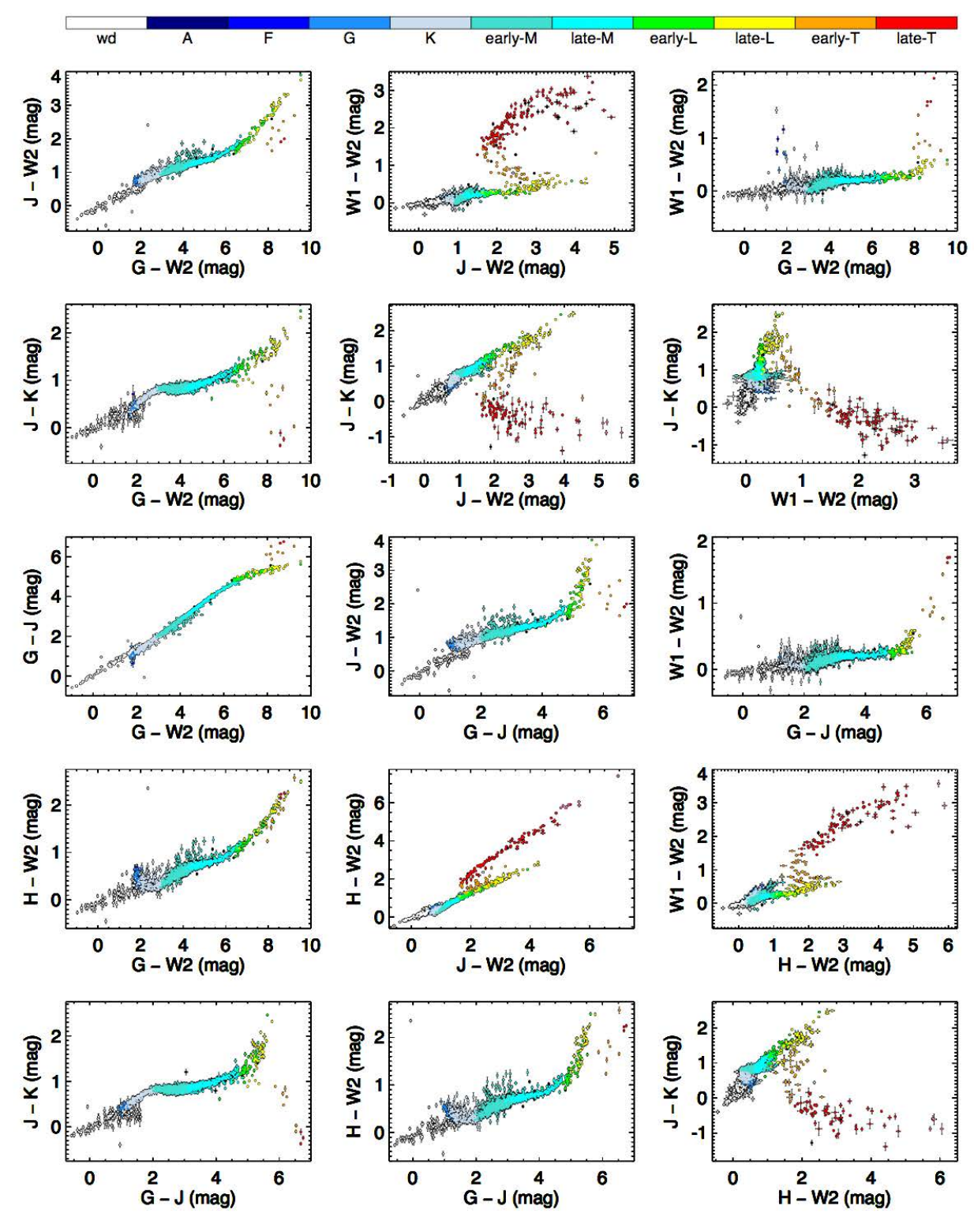}
\caption{Various color-color diagrams for the 20-pc census. See text for details.
\label{fig:color_color}}
\end{figure*}

In Figures~\ref{fig:AbsMag_SpType}-\ref{fig:color_color}, we show several color-type, color-color, and color-magnitude diagrams as a final method for identifying unresolved binaries. These diagrams also illustrate the rich diversity of colors and absolute magnitudes that objects within the 20-pc census possess.

Each plot shows photometry only for those objects believed to be single components ("\# Components this Row" = 1 in Table~\ref{tab:20pc_census}) and whose photometry is uncontaminated ("2MASS\_contam?" and/or "WISE\_contam?" not equal to "yes" in Table~\ref{tab:20pc_census}). Each object is color coded by its spectral type, as shown by the color bar in each figure\footnote{This color palette was chosen so that each color could also be differentiated by readers with deuteranopia, protanopia, tritanopia, or achromatopsia.}. Preference is given to the near-infrared spectral type if listed; otherwise, the optical spectral type is used. (It should be noted that for stars of type A through M, near-infrared classifications are given in Table~\ref{tab:20pc_census} only when no optical type is available, so this criterion is only relevant for the L, T, and Y dwarfs.) Each object is plotted as a solid black dot, the center of which is colored if the spectral type is known; that is, objects lacking a spectral type appear only as black dots. Furthermore, for plots that involve $J$ or $K$ bands, preference is given to MKO magnitudes; otherwise 2MASS magnitudes are used.

In Figure~\ref{fig:AbsMag_SpType}, only those objects with absolute magnitude uncertainties below 1.0 mag are shown, to keep the plots more legible. 
In Figure~\ref{fig:MG_color}, objects are shown only if their uncertainties in $M_G$ are below 1.0 mag and their color uncertainties are below 0.5 mag. 
In Figure~\ref{fig:AbsMag_JW2} (or \ref{fig:AbsMag_HW2}), objects are shown only if their absolute magnitude uncertainties are below 1.0 mag and their $J-$W2 (or $H-$W2) uncertainties are also below 1.0 mag.
In Figure~\ref{fig:colors_SpType}, objects are shown only if the color uncertainty is less than 0.5 mag for Gaia-based color plots or less than 1.0 mag for all other colors.
In Figure~\ref{fig:color_color}, points are shown only if their color uncertainties are generally less than 0.10-0.20 mag.

We have examined each of these diagrams in detail and have identified objects that fall significantly far from the common loci of main sequence stars or white dwarfs to warrant special attention. There are several classes of objects, however,  that we do not discuss in this section but address elsewhere: (1) Stars with bright magnitudes may be problematic and have quoted uncertainties insufficiently small to capture these problems. Given that these bright stars are generally well characterized already, we concentrate only on those not believed to be main sequence stars (category \#4 below). (2) L, T, and Y dwarfs have already been examined in detail via color-type, color-magnitude, and color-color diagrams in \cite{kirkpatrick2021}. (3)  Subgiant, giant, and bright giant stars are discussed in Section~\ref{sec:subgiants-giants}. (4) Low-metallicity (subdwarf) stars are discussed in Section~\ref{sec:subdwarfs}. (5) Young objects are discussed in Section~\ref{sec:young-objects}.

We begin with objects whose placement on these diagrams could potentially highlight a problem with their measured parallaxes. These are all objects that the Gaia survey is placing within the 20-pc volume for the first time. With the exception of the object at $11^h59^m-36^{\circ}34^{\prime}$, all of these objects have higher than normal Gaia parallax uncertainties as compared to objects of similar $G$ magnitude. We discuss each of these individually below:

\begin{itemize}

\item Gaia EDR3 4966072879648455296 (0229$-$3606): This object, whose spectral type has yet to be determined but whose Gaia eDR3/DR3 parallax is $50.66{\pm}0.61$ mas, falls near or just above the main sequence on most color-magnitude diagrams. Its apparent magnitudes are similar to 20-pc objects of the same color, so there is no reason to question its inclusion in our census. Its location on color-magnitude diagrams along with its high DR3 RUWE value of 5.438 indicate possible unresolved binarity.

\item Gaia EDR3 3330473222213987072 (0623+1018): This M3 dwarf (see Section ~\ref{sec:appendix_spectroscopy}) has a Gaia eDR3/DR3 parallax of $50.80{\pm}1.55$ mas. The derived $M_G$ value is $\sim$8 magnitudes fainter than that expected for an average M3 dwarf, and the $M_{W2}$ value is $\sim$9 magnitudes fainter. The Gaia parallax value for this object is clearly in error, so it has been removed from the 20-pc census.

\item Gaia EDR3 3460907947316392704 (1159$-$3634): This is an M9.5 dwarf with a Gaia eDR3/DR3 parallax of $50.10{\pm}0.18$ mas. Its apparent magnitudes fall within the range of other M9.5 dwarfs within the 20-pc census. In absolute magnitude, it falls above the main sequence by as much as a magnitude for objects of similar color, and its Gaia RUWE value is 1.482. This position on color-magnitude diagrams cannot be explained by binarity alone, but a slightly larger parallax in tandem would solve the discrepancy. In any event, there is no reason to exclude this object from Table~\ref{tab:20pc_census}.

\item Gaia EDR3 6025146733201615616 (1624$-$3212): This object, of unknown type, has a Gaia eDR3/DR3 parallax of $59.01{\pm}0.12$ mas. Its apparent magnitude falls in the range expected for objects of similar color within 20 pc. On plots of absolute magnitude vs.\ color, however, it appears anomalous. On the $M_G$ vs.\ $G_{BP}-G_{RP}$ plot, it falls 0.4 mag more luminous than objects of similar color; on the $M_G$ vs.\ $G-J$ plot, it is also more luminous, but by 2.4 mag. Whether these issues indicate a problem with the measured photometry, the measured astrometry, or both -- or whether the object has an unusual spectrum -- is currently unknown. This object is retained in Table~\ref{tab:20pc_census}.

\item Gaia DR2 4062191480304598656 (AB) (1736$-$2515): This object, also of unknown type, has a Gaia DR2 parallax of $60.24{\pm}0.83$ mas and is a known double (\citealt{vrijmoet2022}). Gaia DR3 lists two sources near this position, but neither have a parallax or proper motion measurement. The apparent magnitude of the DR2 source is at odds with the range expected for objects of similar color within 20 pc for many combinations of apparent magnitude vs.\ color, such as $G$ vs.\ $G-$W2, $J$ vs.\ $J-$W2, and $H$ vs.\ $J-K_s$. The object also has a very small Gaia-measured proper motion of $25.2{\pm}1.1$ mas yr$^{-1}$ and lies near the Galactic Center at $l,b = (2{\fdg}1,+3{\fdg}1)$. This object is most likely a background object with a faulty parallax, so we have removed it from the 20-pc census.

\item Gaia DR2 1795813379365971072 (2151+2328): No spectrum has been acquired of this object, and it appears to be a very close double in both Gaia DR2 and eDR3/DR3. However, only one of these components has a parallax measurement in DR2, and neither one does in DR3. This object lies well below and blueward of the main sequence on many apparent magnitude vs.\ color plots such as $J$ vs.\ $G-J$, W2 vs.\ $G_{BP}-G{RP}$, and $H$ vs.\ $J-$W2. The object also has a small Gaia DR2 motion of only $45.7{\pm}0.7$ mas yr$^{-1}$. This is likely a background source with a bogus parallax, so we have removed it from the 20-pc census.

\end{itemize}

The rest of our analysis deals with objects that are outliers for various other reasons. As discussed below, these reasons include possible unresolved binarity, unusual atmospheric composition, variability corrupting pan-epoch colors, and suspected typographical errors in published literature values.

\begin{itemize}

\item HD 1237 B (0016$-$7951): For its spectral type of M4, this object has $G_{BP} - G$ and $G_{BP} - G_{RP}$ colors much bluer than expected, while its $G - G_{RP}$ color is much redder than expected. No other separate photometry of the B component is given in Table~\ref{tab:20pc_census}. Given that the A component is eight magnitudes brighter in $G$ than the B component and lies only $4{\farcs}0$ away, we suspect a problem with the measured photometry of B that is not adequately reflected in its quoted uncertainties.

\item EGGR 246 (0041$-$2221): This is an oddly blue white dwarf in colors that use W1 or W2 magnitudes. The object is also blue relative to other white dwarfs in $J-H$ color, though normal in $G-J$. This carbon-bearing object has a peculiar spectral type, DQpec9.3, and is believed to have a mixed hydrogen-helium atmosphere. The known infrared flux deficit is thought to be caused by absorption by H$_2$ via collisions with neutral He  (\citealt{giammichele2012,bergeron1994,bergeron2022}).

\item LP 941-19 (0213$-$3345): Although this DA4.5 white dwarf has contaminated WISE photometry, it falls in an odd position on plots based only on Gaia photometry. Specifically, at its value of $M_G$ it falls $\sim$0.5 mag blueward of the white dwarf locus in $G_{BP}-G$ and $\sim$0.3 mag redward in $G-G_{RP}$. There is very little literature on this source, and our spectrum of it (Section~\ref{sec:appendix_spectroscopy}) is the first published. It is not yet clear if this spectrum differs markedly enough from other DA white dwarfs to account for the color discrepancies or whether the Gaia magnitudes themselves are at fault. 

\item HD 21209 A (0323$-$4959): The only oddity with this K dwarf is its anomalously blue W1$-$W2 color. The value in Table~\ref{tab:20pc_census}, which is from the WISE All-Sky Source Catalog, is W1$-$W2 = $-0.17{\pm}0.06$ mag. Although this is the preferred WISE catalog for sources of this brightness (W1 = 5.56 mag), the AllWISE Source Catalog gives a very similar color of W1$-$W2 = $=-0.15{\pm}0.13$ mag. This color may be due to the slightly subsolar metallicity of the object ([Fe/H]=$-0.44{\pm}0.19$, \citealt{soto2018}; $-0.41{\pm}0.04$, \citealt{sousa2008}; $-0.39{\pm}0.02$, \citealt{tsantaki2013}).

\item HD 23189 (0348+6840): This early-K dwarf is underluminous for its type at $M_G$, $M_J$, $M_H$, $M_{Ks}$, and $M_{W2}$. When colors formed from Gaia-based magnitudes are compared to the mean colors of objects of the same type, it appears normal, whereas the W1$-$W2 color is slightly bluer than normal. We suspect that the \cite{gray2003} type of K2 V is a typographical error, as independent assessments of the type from spectra, colors, and luminosity considerations (\citealt{adams1935, bidelman1985, mermilliod1987, stassun2019}) suggest a spectral type closer to K7.

\item 2MASS J05053461+4648017 (0505+4648): This is an M8 dwarf (see Section ~\ref{sec:appendix_spectroscopy}) with a Gaia DR3 parallax value of $56.84{\pm}0.60$ mas. The absolute values calculated with this parallax are similar to those of other known M8 dwarfs in the census, and a previous parallax of $69.5{\pm}4.7$ mas (\citealt{dittmann2014}) also places it within 20 pc. Its location on color-magnitude diagrams such as $M_G$ vs.\ $G-$W2 along with its high DR3 RUWE value of 4.823 indicate possible unresolved binarity.

\item DENIS J071807.3$-$350220 (0718$-$3502): On the $G_{BP}-G_{RP}$ and $G-J$ vs.\ various absolute magnitude diagrams, this object appears to be $\sim$0.7 mag above the locus of other objects of the same color. This is, therefore, likely a near equal-magnitude double. This object is also flagged as a possible binary in the Apps Catalog, again based on its position in color-magnitude diagrams.

\item SCR J0818$-$3110 (0818$-$3110): This DZ white dwarf is an outlier on the $G-J$ vs. W1$-$W2 diagram and, in fact, any diagram involving W1$-$W2 color. This issue has been hinted at previously in Figure 4 of \cite{kawka2021}, which shows that the best model fit to existing spectra and photometry fails to match the observed W1$-$W2 color. Although the effect is known, its reason has apparently not yet been established and may be caused by variability or missing opacity sources in the models.

\item UPM J0901$-$6526 (0901$-$6526): This K5 dwarf is an outlier on all plots showing spectral type but appears normal on color-color and color-magnitude plots. We suggest that the published type of K5 (\citealt{riaz2006}) results from a transcription error in the data for this star and that the actual spectral type is closer to M5.

\item APMPM J1251$-$2121 (1250$-$2121): This M6/6.5 dwarf has a Gaia DR3 parallax of $56.79{\pm}0.19$ mas. Its apparent magnitudes fall within the range expected for an M6 dwarf within the 20-pc volume, and a previous parallax measurement of $57.7{\pm}1.7$ mas (\citealt{winters2015}) is in agreement with the Gaia one. On the $M_G$ vs.\ $G_{BP}-G_{RP}$ diagram, it falls $\sim$0.5 mag above the main sequence, and on the $M_G$ vs.\ $G-$W2 diagram, it falls $\sim$0.7 mag above. This and the Gaia DR3 RUWE value of 2.888 suggest unresolved binarity. 

\item HD 113194 (1302$-$2647): Although this K5 dwarf has a Gaia DR3 parallax with a relatively large uncertainty ($56.94{\pm}0.19$ mas), the earlier Hipparcos parallax ($56.87{\pm}1.11$ mas) is in agreement. This object also has a high Gaia DR3 RUWE value, is listed as a high-LUWE object (see Section~\ref{sec:RUWE_LUWE}), and shows acceleration (see Section~\ref{sec:proper_motion_anomaly}), in addition to falling $\sim$0.5 mag above the main sequence on the $G$ vs.\ $G_{BP}-G_{RP}$ diagram. This object is almost certainly an unresolved binary. It is also considered to be binary in the Apps Catalog, based on its position on color-magnitude diagrams. The Gaia DR3 main catalog reports a radial velocity of $-17.56{\pm}7.24$ km s$^{-1}$ using seventeen observations over 920 days, along with an amplitude of radial velocity variations of 60.25 km s$^{-1}$, further supporting the hypothesis of binarity. The P-value for radial velocity constancy ({\tt rv\_chisq\_pvalue}) is also 0.0.

\item 2MASSW J1421314+182740 (1421+1827): This M9.5 dwarf has WISE photometry contaminated by a background source, but it appears unusual in non-WISE colors as well. Specifically, it has oddly blue $G_{BP}-G$ and $G_{BP}-G_{RP}$ colors compared to other objects of similar $M_G$. However, it appears normal for its $M_G$ in $G-G_{RP}$, $G-J$, $G-H$, and $J-K_s$. This may simply indicate an issue with the $G_{BP}$ magnitude that the formal uncertainty fails to adequately capture.

\item LP 222-65 (1516+3910): This mid-M dwarf lies consistently $\sim$0.6 mag above the main sequence relative to objects of the same color and spectral type on color-magnitude diagrams. This is an isolated object with no obvious problems with its photometry, so we believe this is an unresolved near-equal magnitude binary.

\item UCAC4 554-051865 (1518+2036): This mid-M dwarf 
is $\sim$0.6 mag more luminous than objects of similar color on the $M_G$ vs.\ $G_{BP}-G_{RP}$ and $M_G$ vs.\ $G_{BP}-G$ diagrams and has a large RUWE and LUWE value (see Section~\ref{sec:RUWE_LUWE}). It is likely an unresolved binary.

\item L 339-19 (1640$-$4559): This M3 dwarf shows anomalously red $G-$W3 and $J-$W3 colors for its absolute magnitude, and even more anomalously red $G-$W4 and $J-$W4 colors. A more careful look at the WISE images shows that the W3 detection is likely real, but the W4 detection likely is not. The W3 photometry from AllWISE (reported in Table~\ref{tab:20pc_census}) is 7.10$\pm$0.05 mag and that from WISE All-Sky is 6.57$\pm$0.04 mag. In $G-$W3 and using the AllWISE value, the object lies 1.0 mag redward of objects of the same absolute $G$ magnitude; using the WISE All-Sky value shows the object to lie 0.5 mag redward. Archival Spitzer/IRAC and Spitzer/MIPS photometry of this object exists in the GLIMPSE I Spring '07 Catalog (\citealt{benjamin2003}) and MIPSGAL Archive (\citealt{carey2009}) at IRSA: ch1 = 7.830$\pm$0.038, ch2 = 7.781$\pm$0.045, ch3 = 7.724$\pm$0.037 (5.8 $\mu$m), ch4 = 7.705$\pm$0.026 (8.0 $\mu$m), and [24$\mu$m] = 7.14$\pm$0.24 mag. Running these new data points and the tabulated Table~\ref{tab:20pc_census} photometry through the Virtual Observatory Spectral energy distribution Analyzer (VOSA\footnote{\url{http://svo2.cab.inta-csic.es/theory/vosa/}}; \citealt{bayo2008}) suggests not only that the W4 magnitude is in error but that the W3 magnitude is spuriously bright relative to the bracketing IRAC and MIPS data points. The spectral energy distribution is otherwise typical of that of an M3 dwarf. We therefore conclude that there is no infrared excess in this object.

\item UCAC4 317-104829 (1706$-$2643): This DAH white dwarf is normal in Gaia-only colors, colors formed using Gaia minus near-infrared magnitudes, and colors formed from $J$, $H$, and $K_s$ magnitudes. It is, however, oddly blue in W1$-$W2. We assume that this anomalous color may be intrinsic to the star and a result of its strong magnetic field, although it should be cautioned that this white dwarf is located against a busy region of the Galactic Plane and may suffer from contamination in its WISE photometry.

\item DENIS-P J1733423$-$165449 (1733$-$1654): This L1 dwarf has WISE photometry that is contaminated by background sources, but it also shows unusual colors in Gaia-only measurements. Gaia DR3 lists two other point sources within $2{\farcs}1$ of this object, so its Gaia photometry may be adversely affected in a way that the formal uncertainties fail to capture.

\item LSPM J1733+1655 (1733+1655): The Gaia DR3 parallax of $60.91{\pm}0.48$ mas has a relatively large uncertainly for its magnitude and is in disagreement with an earlier published value of $85.40{\pm}3.30$ mas by \cite{dittmann2014}. This mid-M dwarf is more luminous than objects of similar color by $\sim$1.6 mag on the $G$ vs.\ $G_{BP}-G$, $G_{BP}-G_{RP}$, $G-G_{RP}$, $G-J$, and $G-$W2 diagrams, if the Gaia DR3 parallax is used. This overluminosity decreases to $\sim$0.9 mag if the \cite{dittmann2014} parallax is used instead. This is a high-RUWE/LUWE object as well (Section~\ref{sec:RUWE_LUWE}), and so is likely an unresolved multiple system with problematic Gaia astrometry. \cite{clark2022} identify a candidate companion at separation $0{\farcs}14$ and position angle 101$^\circ$ at epoch 2017.3 and again at separation $0{\farcs}36$ and position angle 63$^\circ$ at epoch 2019.7. C.\ Gelino also finds a single epoch of Keck/NIRC2 data in the Keck Observatory Archive for LSPM J1733+1655. These are Br$\gamma$ and $J$-continuum  observations taken on 2015 Jul 10 UT (PI: Hansen; Program ID: U050N2), from which we measure a separation of $0{\farcs}11$ at position angle 248$^\circ$. If we assume all three of these measurements refer to the same star and it is a stationary background object, we derive motions of LSPM J1733+1655 of $-0.100 \arcsec {\rm yr}^{-1}$ in RA and $-0.051 \arcsec {\rm yr}^{-1}$ in Dec, which can be compared to the measured Gaia DR3 values of $-0.135 \arcsec {\rm yr}^{-1}$ in RA and $-0.130 \arcsec {\rm yr}^{-1}$ in Dec. The derived magnitude and direction of motion lead us to conclude that the background star hypothesis is sound. 

\item LP 388-55 A (1735+2634): This late-M dwarf is anomalously red, by 0.25 mag, in $G-G_{RP}$ color but looks normal in the $G_{BP}-G$ color compared to objects of similar $M_G$ magnitude. Curiously, all Gaia-based absolute magnitudes ($M_G$, $M_{GBP}$, and $M_{GRP}$) are consistent with the reported spectral type. The B component is an early-L that is not directly imaged by Gaia but may nonetheless be subtly affecting the Gaia magnitudes of the A component.

\item LP 44-334 A (1840+7240): This primary in a M6.5 dwarf system has a Gaia DR3 parallax ($52.78{\pm}0.09$ mas) with a relatively large uncertainty for its magnitude, but this value compares favorably to the earlier published value of $59.3{\pm}2.2$ mas by \cite{lepine2009}. The $G_{BP}-G$ color is too blue for its $M_G$ value, the $G_{BP}-G_{RP}$ color is normal, and the $G-G_{RP}$ color is too red. These issues are likely caused by the nearness of the B component, only $0{\farcs}8$ away, which is likely corrupting the photometry of the A component.

\item LP 867-15 (1842$-$2328): The colors for this M0 dwarf are more consistent with an M4 dwarf than with an M0. Pending spectroscopic verification, we assume that this object has been misclassified.

\item SCR J2012$-$5956 (2012$-$5956): This object, a DC9.9 white dwarf, falls below the white dwarf locus for most colors. It is very blue relative to other white dwarfs in $J-K_s$, $J-H$, and $H-K_s$ but looks like other white dwarfs in colors made with Gaia-only magnitudes. It is somewhat blue in $G-J$, $G-H$, and $G-K_s$ colors. As with EGGR 246 above, the infrared flux deficit is believed
to be caused by H$_2$-He collision-induced absorption (\citealt{giammichele2012}).

\item LEHPM 2-783 (2019$-$5816): This M6.5 dwarf is overluminous in all Gaia-based colors. (Many other colors are nearly degenerate with absolute magnitude or type in this spectral type range.) On both the $G_{BP}-G_{RP}$ vs.\ $M_H$ and the $G-$W2 vs. $M_{W2}$ plots, the overluminosity is $\sim$0.7 mag. \cite{ujjwal2020} mark this as a possible member of the $\beta$ Pic Moving Group, and \cite{riaz2006} note that it is a strong X-ray emitter with strong H$\alpha$ emission.

\item LP 12-90 (2322+7847): This mid-M dwarf lies above the main sequence by $\sim$0.75 mag on the $M_G$ vs.\ $G_{BP}-G$ plot. On many other plots of absolute magnitude vs.\ color, it lies similarly above (and redward of) the main sequence.  This could be another unresolved binary -- if confirmed, this would make its system with HD 220140 AB a quadruple -- but the primary in this system is a young, naked T Tauri star (\citealt{makarov2007}), meaning that its position may be solely due to its youth.

\item ZZ Psc (2328+0514): This white dwarf is anomalously red in colors involving WISE magnitudes -- so much so that it falls far from the white dwarf sequence itself. On a plot of $M_G$ vs.\ W1$-$W2, for example, it lies substantially redward of both the white dwarf locus and the main sequence. This object, also known as G 29-38, is known to have a debris disk around it, the first evidence of which was uncovered by \cite{zuckerman1987}. For an update on this object, see \cite{cunningham2022}.

\end{itemize}

\section{Masses from Estimation\label{sec:mass_estimates}}

Only a small fraction of objects within the 20-pc census has masses measurable by methods 1 or 2 described in the introduction of Section~\ref{sec:mass_methods}. For the rest, we must rely on methods 3 and 4 of that section, which depend on comparison to empirical trends or to theoretical models.

In the first subsection below, we discuss mass measurements for objects not on the main sequence -- namely, white dwarfs, giants/subgiants, and brown dwarfs. In the second subsection, we summarize mass estimation for main sequence stars. In the third subsection, we discuss other complications -- youth, subsolar metallicity, and formation scenario -- that may need to be considered when assigning accurate mass estimates for special objects.

\subsection{Individual objects not on the main sequence}

\subsubsection{White dwarfs\label{sec:wd_masses}}

Masses have been measured via one of the methods described in Section~\ref{sec:mass_methods} for a handful of white dwarfs in the 20-pc census, but 
these represent the end-state masses of the stellar remnants and are not suitable for analysis of the {\it initial} mass function.
Rather, what is needed are the initial masses {\it before} evolution off the main sequence. Techniques have been established that use the final mass of the remnant to estimate the initial mass of the progenitor.

For white dwarfs lacking a direct mass measurement, one can estimate the final mass of the white dwarf using one of the following two semi-empirical methods. The first is to use spectroscopic observations of the depth and width of the hydrogen Balmer, \ion{He}{1}, or \ion{He}{2} lines to establish, after comparison to atmospheric models, the $\log{(g)}$ and $T_{\rm eff}$ for each object. Further comparison of these two parameters to cooling models provides the remnant mass (e.g., \citealt{tremblay2011,genest-beaulieu2019,bergeron1992,finley1997}). Whereas this first method is applicable only to DA (hydrogen atmosphere) or DB (helium atmosphere) white dwarfs, an alternate method can be used both for objects lacking hydrogen lines as well as for objects lacking spectroscopic observations. In this second method, masses can still be estimated if an accurate parallax has been measured. Here, absolute fluxes across as wide a swath of wavelength space as possible are compared to model atmospheres to provide $\log{(g)}$ and $R$, from which the mass can be derived from equation~\ref{eqn:surface_gravity} (e.g., \citealt{bergeron2019, tremblay2019,giammichele2012,bergeron2001,koester1979}).

The next step is to convert this final mass into an initial mass using an initial-to-final mass relation (e.g., \citealt{weidemann2000}). The empirical form of this relationship has been established using white dwarfs that are members of open clusters of known age. As described above, spectroscopic observations of the Balmer lines in these stars can be compared to atmospheric models to derive $\log{(g)}$ and $T_{\rm eff}$ for each object. A comparison of these parameters to cooling models provides both the remnant mass as well as the cooling time since the object left the tip of the asymptotic giant branch. The known cluster age minus this cooling time gives the main sequence lifetime of the object, which can then be related back to an initial mass using theoretical evolutionary isochrones. This same technique can also be applied to white dwarfs in globular clusters. Due to their much older ages, these clusters can provide white dwarfs of lower final mass than those available in young open clusters. Because these globular clusters are much more distant, their white dwarfs are faint and more difficult to study, so old low-mass white dwarfs are still not well represented by cluster methods. 

This lack of low-mass examples can be partly mitigated by the use of old, wide binaries for which the second component can be age dated and the separation between components is large enough that no mass transfer has occurred during the system's evolution. Examples are wide subgiant + white dwarf binaries in which the system can be dated from its more recently evolved member (\citealt{barrientos2021}), wide F/G/K dwarf + white dwarf binaries in which the age of the main sequence star can be estimated from activity diagnostics (\citealt{catalan2008}, \citealt{zhao2012}), and white dwarf pairs in which comparison of the higher-mass white dwarf to known cluster white dwarfs can provide an age for the binary, and the difference in the cooling times for the white dwarf pair gives the main sequence lifetime of the lower-mass white dwarf (\citealt{andrews2015}). Using the results of these methods, the trend of final mass with initial mass can be fit. As can be seen from figure 9 of \cite{barrientos2021}, the relation shows considerable scatter at lower masses, as the age dating methods for individual systems are generally less robust than those from clusters. The relations we adopt here are the cluster-based tripartite parameterization found in equations 4-6 of \cite{cummings2018} and the quadripartite parameterization found in table 1 of \cite{elbadry2018}, 
The former relation is applicable to white dwarfs with $0.56 < M_{\rm final} < 1.24 M_\odot$ ($0.83 < M_{\rm initial} < 7.20 M_\odot$), and the latter relation, which is
based on nearby white dwarfs with accurate Gaia parallaxes, is applicable to white dwarfs with  $0.50 < M_{\rm final} < 1.37 M_\odot$ ($0.95 < M_{\rm initial} < 8.00 M_\odot$). 
We further note that neither the cluster nor field methods have yet extended the initial-to-final mass relation below final masses of $0.50 M_\odot$. (As discussed further below, white dwarfs with final masses below $0.45 M_\odot$ require binary interactions, as a single progenitor would imply an age older than the Universe; \citealt{marsh1995}).

Specifically, we apply the following methodology to assign {\it final} masses to white dwarfs in the 20-pc census. First, we use directly measured masses, whenever such measurements are available. For others, we use final mass estimates that are based on accurate parallaxes, high S/N spectra, and/or broad-wavelength data spanning the white dwarf's spectral energy distribution. For all other objects, we resort to the Gaia-centric estimates of \cite{gentile2021} and \cite{gentilefusillo2019}. These estimates use only a small fraction of the white dwarf's spectral energy distribution -- spanning the Gaia optical bandpasses -- and thus lead to separate solutions for hydrogen- vs.\ helium-atmosphere objects. When our own follow-up has determined the spectral type of the object, we use this information to break the degeneracy; otherwise, a hydrogen-atmosphere object is assumed, as noted in Table~\ref{tab:wd_masses}.

\startlongtable
\begin{deluxetable*}{lccccccccc}
\tabletypesize{\scriptsize}
\tablecaption{Mass Measurements and Estimates for White Dwarfs in the 20-pc Census\label{tab:wd_masses}}
\tablehead{
\colhead{Name} &
\colhead{Abbrev.} &
\colhead{WD} &
\colhead{Final} &
\colhead{Final} &
\colhead{Ref.} &
\colhead{Initial} &
\colhead{Initial} &
\colhead{Initial} &
\colhead{Initial} \\
\colhead{} &
\colhead{Coords} &
\colhead{Name\tablenotemark{a}} &
\colhead{Mass} &
\colhead{Mass} &
\colhead{} &
\colhead{Mass} &
\colhead{Mass} &
\colhead{Mass} &
\colhead{Mass} \\
\colhead{} &
\colhead{(J2000)} &
\colhead{} &
\colhead{($M_\odot$)} &
\colhead{Method} &
\colhead{} &
\colhead{(Cummings)\tablenotemark{b}} &
\colhead{(El-Badry)\tablenotemark{c}} &
\colhead{(Adopted)\tablenotemark{d}} &
\colhead{Method} \\
\colhead{} &
\colhead{} &
\colhead{} &
\colhead{} &
\colhead{} &
\colhead{} &
\colhead{($M_\odot$)} &
\colhead{($M_\odot$)} &
\colhead{($M_\odot$)} &
\colhead{} \\
\colhead{(1)} &
\colhead{(2)} &           
\colhead{(3)} &           
\colhead{(4)} &
\colhead{(5)} &
\colhead{(6)} &
\colhead{(7)} &
\colhead{(8)} &
\colhead{(9)} &
\colhead{(10)} \\
}
\startdata
LAWD 96&                   0002$-$4309& WD 2359$-$434  & 0.78$\pm$0.03  & Spec                   & 1 &3.2$\pm$1.5&    3.4$\pm$0.2&    3.3$\pm$1.5&      IFMR \\
LAWD 1&                    0002$-$3413& WD 0000$-$345  & 0.88$\pm$0.10  & Phot                   & 1 &3.8$\pm$1.3&    4.7$\pm$0.3&    4.3$\pm$1.3&      IFMR \\
LP 464-57&                 0007+1230  & WD 0004+122    & 0.57$\pm$0.15  & Phot                   & 2 &1.0$\pm$1.9&    1.7$\pm$0.1&    1.4$\pm$1.9&      IFMR \\
EGGR 381&                  0012+5025  & WD 0009+501    & 0.73$\pm$0.04  & Phot                   & 1 &2.9$\pm$1.4&    3.1$\pm$0.2&    3.0$\pm$1.5&      IFMR \\
L 50-73&                   0013$-$7149& WD 0011$-$721  & 0.59$\pm$0.00  & Phot                   & 1 &1.3$\pm$0.5&    1.9$\pm$0.1&    1.6$\pm$0.5&      IFMR \\
G 158-45&                  0014$-$1311& WD 0011$-$134  & 0.72$\pm$0.07  & Phot                   & 1 &2.9$\pm$1.5&    3.0$\pm$0.2&    3.0$\pm$1.5&      IFMR \\
EGGR 246&                  0041$-$2221& WD 0038$-$226  & 0.53$\pm$0.01  & Phot                   & 1 &\nodata    &    1.3$\pm$0.1&    1.3$\pm$0.1&      IFMR \\ 
Wolf 28&                   0049+0523  & (WD 0046+051)  & 0.68$\pm$0.02  & Phot                   & 1 &2.4$\pm$0.7&    2.8$\pm$0.2&    2.6$\pm$0.7&      IFMR \\
Wolf 1516&                 0118+1610  & WD 0115+159    & 0.69$\pm$0.04  & Phot                   & 1 &2.5$\pm$0.8&    2.9$\pm$0.2&    2.7$\pm$0.8&      IFMR \\
LP 991-16 A&               0124$-$4240& WD 0121$-$429  & 0.41$\pm$0.01\tablenotemark{e}  
                                                                        & Phot                   & 1 &\nodata    &    \nodata    &    1.9$\pm$0.9&      Ultra-low\\ 
LP 991-16 B&               0124$-$4240& \nodata        & \nodata        & \nodata                & 1 &\nodata    &    \nodata    &    1.9$\pm$0.9&      Ultra-low\\ 
EGGR 307&                  0125$-$2600& WD 0123$-$262  & 0.58$\pm$0.00  & Phot                   & 1 &1.1$\pm$0.4&    1.8$\pm$0.1&    1.5$\pm$0.4&      IFMR \\
LAWD 10 A&                 0137$-$0459& WD 0135$-$052.1& 0.47           & {\bf Orbit}            & 4 &\nodata    &    \nodata    &    1.0$\pm$0.1&      Low\\ 
LAWD 10 B&                 0137$-$0459& WD 0135$-$052.2& 0.52           & {\bf Orbit}            & 4 &\nodata    &    1.2$\pm$0.1&    1.2$\pm$0.1&      IFMR\\ 
L 88-59&                   0143$-$6718& WD 0141$-$675  & 0.48$\pm$0.06  & Spec                   & 1 &\nodata    &    \nodata    &    1.0$\pm$0.1&      Low\\
EGGR 268&                  0151+6425  & WD 0148+641    & 0.81$\pm$0.01  & {\bf GravRed}          & 5 &3.3$\pm$1.5&    3.5$\pm$0.2&    3.4$\pm$1.5&      IFMR \\
GD 279&                    0152+4700  & WD 0148+467    & 0.63$\pm$0.03  & Spec                   & 1 &1.8$\pm$0.6&    2.3$\pm$0.2&    2.1$\pm$0.7&      IFMR \\
$\chi$ Eri B  &            0155$-$5136& \nodata        & \nodata        & \nodata                & - &\nodata    &    \nodata    &    1.9$\pm$0.9&      Conjecture\\  
HD 13445 B&                0210$-$5049& WD 0208$-$510  & 0.597$\pm$0.010& {\bf Accel}            & 6 &1.4$\pm$0.5&    2.0$\pm$0.1&    1.7$\pm$0.5&      IFMR \\
EGGR 168&                  0211+3955  & WD 0208+396    & 0.48$\pm$0.10  & Spec                   & 7 &\nodata    &    \nodata    &    1.0$\pm$0.1&      Low\\     
LP 649-66&                 0212$-$0804& \nodata        & 0.53$\pm$0.02  & Gaia H-atm*            & 8 &\nodata    &    1.3$\pm$0.1&    1.3$\pm$0.1&      IFMR\\
LP 941-19&                 0213$-$3345& \nodata        & 0.37$\pm$0.05  & Gaia H-atm*            & 8 &\nodata    &    \nodata    &    1.9$\pm$0.9&      Ultra-low\\
EGGR 471&                  0232$-$1411& WD 0230$-$144  & 0.66$\pm$0.06  & Phot                   & 1 &2.1$\pm$0.9&    2.7$\pm$0.2&    2.4$\pm$1.0&      IFMR \\
LP 830-14&                 0235$-$2400& WD 0233$-$242  & 0.58$\pm$0.00  & Phot                   & 1 &1.1$\pm$0.4&    1.8$\pm$0.1&    1.5$\pm$0.4&      IFMR \\
EGGR 473&                  0248+5423  & WD 0245+541    & 0.73$\pm$0.03  & Phot                   & 1 &2.9$\pm$1.4&    3.1$\pm$0.2&    3.0$\pm$1.5&      IFMR \\
CPD$-$69 177&              0310$-$6836& WD 0310$-$688  & 0.67$\pm$0.03  & Spec                   & 1 &2.3$\pm$0.7&    2.8$\pm$0.2&    2.5$\pm$0.7&      IFMR \\
$\alpha$ For Bb &          0312$-$2859& \nodata        & \nodata        & \nodata                & - &\nodata    &    \nodata    &    1.9$\pm$0.9&      Conjecture\\    
EGGR 566&                  0325$-$0149& WD 0322$-$019  & 0.63$\pm$0.05  & Phot                   & 1 &1.8$\pm$0.8&    2.3$\pm$0.2&    2.1$\pm$0.8&      IFMR \\
Wolf 219&                  0344+1826  & WD 0341+182    & 0.57$\pm$0.06  & Phot                   & 1 &1.0$\pm$0.9&    1.7$\pm$0.1&    1.4$\pm$0.9&      IFMR \\
G 7-16&                    0400+0814  & WD 0357+081    & 0.61$\pm$0.06  & Phot                   & 1 &1.5$\pm$0.9&    2.1$\pm$0.2&    1.8$\pm$0.9&      IFMR \\
${\rm o^2}$ Eri B&         0415$-$0739& (WD 0413$-$077)& 0.573$\pm$0.018& {\bf Orbit}            & 9 &1.1$\pm$0.5&    1.7$\pm$0.1&    1.4$\pm$0.5&      IFMR \\
$\epsilon$ Ret B&          0416$-$5917& WD 0415$-$594  & 0.60$\pm$0.12  & Spec                   & 7 &1.4$\pm$1.6&    2.0$\pm$0.2&    1.7$\pm$1.6&      IFMR \\
EGGR 169&                  0425+1211  & WD 0423+120    & 0.65$\pm$0.04  & Phot                   & 1 &2.0$\pm$0.7&    2.6$\pm$0.2&    2.3$\pm$0.8&      IFMR \\
EGGR 180&                  0431+5858  & WD 0426+588    & 0.675$\pm$0.051& {\bf Lensing}          & 10&2.3$\pm$0.9&    2.8$\pm$0.2&    2.6$\pm$0.9&      IFMR \\
HD 283750 B&               0436+2709  & WD 0433+270    & 1.12$\pm$0.01  & {\bf GravRed}          & 5 &6.1$\pm$1.2&    6.5$\pm$0.4&    6.3$\pm$1.2&      IFMR \\
EGGR 41&                   0437$-$0849& WD 0435$-$088  & 0.53$\pm$0.02  & Phot                   & 1 &\nodata    &    1.3$\pm$0.1&    1.3$\pm$0.1&      IFMR \\
LP 777-1&                  0505$-$1722& WD 0503$-$174  & 0.53$\pm$0.01  & Spec                   & 23&\nodata    &    1.3$\pm$0.1&    1.3$\pm$0.1&      IFMR\\
V371 Ori B &               0533+0156  & \nodata        & 0.63$\pm$0.17  & Dyn+mod                & 18&1.8$\pm$2.2&    2.3$\pm$0.2&    2.1$\pm$2.2&      IFMR\\    
EGGR 248&                  0551$-$0010& WD 0548$-$001  & 0.69$\pm$0.03  & Phot                   & 1 &2.5$\pm$0.7&    2.9$\pm$0.2&    2.7$\pm$0.8&      IFMR \\
UCAC4 398-010797&          0554$-$1035& \nodata        & 0.68$\pm$0.01  & Gaia He-atm            &8 & 2.4$\pm$0.6&    2.8$\pm$0.2&    2.6$\pm$0.7&      IFMR \\
EGGR 45&                   0555$-$0410& WD 0552$-$041  & 0.82$\pm$0.01  & Phot                   & 1 &3.4$\pm$1.5&    3.7$\pm$0.2&    3.6$\pm$1.6&      IFMR \\
EGGR 290&                  0556+0521  & WD 0553+053    & 0.72$\pm$0.03  & Phot                   & 1 &2.9$\pm$1.4&    3.0$\pm$0.2&    3.0$\pm$1.4&      IFMR \\
G 249-36 B &               0605+6049  & \nodata        & 1.03$\pm$0.08  & Dyn+mod                &21 &5.2$\pm$1.3&    5.9$\pm$0.4&    5.6$\pm$1.4&      IFMR \\  
$\alpha$ CMa B&            0645$-$1643& WD 0642$-$166  & 1.017$\pm$0.025& {\bf GravRed}          & 11&5.1$\pm$1.1&    5.9$\pm$0.4&    5.5$\pm$1.1&      IFMR \\
EGGR 484&                  0647+0231  & WD 0644+025    & 0.85$\pm$0.15  & Spec                   & 7 &3.6$\pm$1.8&    4.2$\pm$0.3&    3.9$\pm$1.8&      IFMR \\
LAWD 23&                   0647+3730  & WD 0644+375    & 0.69$\pm$0.03  & Spec                   & 1 &2.5$\pm$0.7&    2.9$\pm$0.2&    2.7$\pm$0.8&      IFMR \\
L 454-9&                   0657$-$3909& WD 0655$-$390  & 0.59$\pm$0.00  & Phot                   & 1 &1.3$\pm$0.5&    1.9$\pm$0.1&    1.6$\pm$0.5&      IFMR \\
EGGR 485&                  0700+3157  & WD 0657+320    & 0.60$\pm$0.02  & Phot                   & 1 &1.4$\pm$0.5&    2.0$\pm$0.2&    1.7$\pm$0.6&      IFMR \\
SCR J0708$-$6706&          0708$-$6706& WD 0708$-$670  & 0.57$\pm$0.00  & Phot                   & 1 &1.0$\pm$0.4&    1.7$\pm$0.1&    1.4$\pm$0.4&      IFMR \\
EGGR 52 A&                 0730+4810  & WD 0727+482A   & 0.51$\pm$0.01  & Phot                   & 1 &\nodata    &    1.1$\pm$0.1&    1.1$\pm$0.1&      IFMR \\
EGGR 52 B&                 0730+4810  & WD 0727+482B   & 0.65$\pm$0.01  & Phot                   & 1 &2.0$\pm$0.6&    2.6$\pm$0.2&    2.3$\pm$0.6&      IFMR \\
EGGR 321&                  0733+6409  & WD 0728+642    & 0.58$\pm$0.00  & Phot                   & 1 &1.1$\pm$0.4&    1.8$\pm$0.1&    1.5$\pm$0.4&      IFMR \\
$\alpha$ CMi B&            0739+0513  & WD 0736+053    & 0.553$\pm$0.022& {\bf Orbit}            & 13&\nodata    &    1.5$\pm$0.1&    1.5$\pm$0.1&      IFMR \\
LAWD 25&                   0740$-$1724& WD 0738$-$172  & 1.11$\pm$0.05  & {\bf GravRed}          & 5 &6.0$\pm$1.2&    6.4$\pm$0.4&    6.2$\pm$1.3&      IFMR \\
VB 3&                      0745$-$3355& (WD 0743$-$336)& 0.55$\pm$0.01  & Phot                   & 1 &\nodata    &    1.5$\pm$0.1&    1.5$\pm$0.1&      IFMR \\
EGGR 426&                  0750+0711  & WD 0747+073.1  & 0.48$\pm$0.01  & Phot                   & 1 &\nodata    &    \nodata    &    1.0$\pm$0.1&      Low\\ 
EGGR 427&                  0750+0711  & WD 0747+073.2  & 0.56$\pm$0.01  & Phot                   & 1 &0.9$\pm$0.4&    1.6$\pm$0.1&    1.2$\pm$0.4&      IFMR \\
LAWD 26&                   0753$-$6747& WD 0752$-$676  & 0.73$\pm$0.06  & Phot                   & 1 &2.9$\pm$1.5&    3.1$\pm$0.2&    3.0$\pm$1.5&      IFMR \\
SCR J075$3$-2524&          0753$-$2524& WD 0751$-$252  & 0.52$\pm$0.01  & Gaia-H                 & 8 &\nodata    &    1.2$\pm$0.1&    1.2$\pm$0.1&      IFMR \\
L 97-3 A&                  0806$-$6618& WD 0806$-$661  & 0.58$\pm$0.03  & Phot                   & 1 &1.1$\pm$0.6&    1.8$\pm$0.1&    1.5$\pm$0.6&      IFMR \\
UPM J0812$-$3529&          0812$-$3529& (WD 0810$-$353)& 0.70$\pm$0.01  & Gaia H-atm*            & 8& 2.6$\pm$0.7&    2.9$\pm$0.2&    2.8$\pm$0.7&      IFMR \\
G 111-64&                  0814+4845  & WD 0810+489    & 0.57$\pm$0.00  & Phot                   & 1 &1.0$\pm$0.4&    1.7$\pm$0.1&    1.4$\pm$0.4&      IFMR \\
SCR J0818$-$3110&          0818$-$3110& WD 0816$-$310  & 0.57$\pm$0.00  & Phot                   & 1 &1.0$\pm$0.4&    1.7$\pm$0.1&    1.4$\pm$0.4&      IFMR \\
SCR J0821$-$6703&          0821$-$6703& WD 0821$-$669  & 0.66$\pm$0.01  & Phot                   & 1 &2.1$\pm$0.6&    2.7$\pm$0.2&    2.4$\pm$0.6&      IFMR \\
CD$-$32 5613&              0841$-$3256& WD 0839$-$327  & 0.45$\pm$0.05  & Spec                   & 7 &\nodata    &    \nodata    &    1.0$\pm$0.1&      Low\\
LP 726-1&                  0842$-$1347& WD 0840$-$136  & 0.57$\pm$0.00  & Phot                   & 1 &1.0$\pm$0.4&    1.7$\pm$0.1&    1.4$\pm$0.4&      IFMR \\
$\iota$ UMa Ab &           0859+4802  & \nodata        & 1.00$\pm$0.30  & Dyn+mag                & 19&4.9$\pm$3.0&    5.8$\pm$0.4&    5.3$\pm$3.0&      IFMR \\ 
LP 606-32&                 0859$-$0058& WD 0856$-$007  & 0.52$\pm$0.01  & Gaia He-atm            &8 &\nodata     &    1.2$\pm$0.1&    1.2$\pm$0.1&      IFMR \\
EGGR 250&                  0915+5325  & WD 0912+536    & 0.75$\pm$0.02  & Phot                   & 1 &3.0$\pm$1.5&    3.2$\pm$0.2&    3.1$\pm$1.5&      IFMR \\
EGGR 252&                  1001+1441  & WD 0959+149    & 0.70$\pm$0.01  & Gaia He-atm            &8 & 2.6$\pm$0.7&    2.9$\pm$0.2&    2.8$\pm$0.7&      IFMR \\
LP 315-42&                 1011+2845  & WD 1008+290    & 0.68$\pm$0.01  & Phot                   & 1 &2.4$\pm$0.6&    2.8$\pm$0.2&    2.6$\pm$0.7&      IFMR \\
WT 1759&                   1012$-$1843& WD 1009$-$184  & 0.59$\pm$0.02  & Phot                   & 1 &1.3$\pm$0.5&    1.9$\pm$0.1&    1.6$\pm$0.5&      IFMR \\
EGGR 350&                  1023+6327  & WD 1019+637    & 0.57$\pm$0.05  & Phot                   & 1 &1.0$\pm$0.8&    1.7$\pm$0.1&    1.4$\pm$0.8&      IFMR \\
LP 37-186&                 1037+7110  & WD 1033+714    & 0.58$\pm$0.00  & Phot                   & 1 &1.1$\pm$0.4&    1.8$\pm$0.1&    1.5$\pm$0.4&      IFMR \\
EGGR 535&                  1038$-$2040& WD 1036$-$204  & 0.60$\pm$0.01  & Phot                   & 1 &1.4$\pm$0.5&    2.0$\pm$0.2&    1.7$\pm$0.5&      IFMR \\
BD$-$18 3019 B&            1045$-$1906& WD 1043$-$188  & 0.53$\pm$0.11  & Phot                   & 1 &\nodata    &    1.3$\pm$0.1&    1.3$\pm$0.1&      IFMR \\
LAWD 34&                   1057$-$0731& WD 1055$-$072  & 0.85$\pm$0.04  & Phot                   & 1 &3.6$\pm$1.6&    4.2$\pm$0.3&    3.9$\pm$1.6&      IFMR \\
$\xi$ UMa Bb &             1118+3131  & \nodata        & \nodata        & \nodata                & - &\nodata    &    \nodata    &    1.9$\pm$0.9&      Conjecture\\  
SCR J1118$-$4721&          1118$-$4721& WD 1116$-$470  & 0.57$\pm$0.00  & Phot                   & 1 &1.0$\pm$0.4&    1.7$\pm$0.1&    1.4$\pm$0.4&      IFMR \\
Ross 627&                  1124+2121  & WD 1121+216    & 0.61$\pm$0.11  & Spec                   & 7 &1.5$\pm$1.5&    2.1$\pm$0.2&    1.8$\pm$1.5&      IFMR \\
20 Crt B&                  1134$-$3250& WD 1132$-$325  & 0.60$\pm$0.01  & Gaia He-atm            &8 & 1.4$\pm$0.5&    2.0$\pm$0.2&    1.7$\pm$0.5&      IFMR \\
GD 140&                    1137+2947  & WD 1134+300    & 0.97$\pm$0.03  & Spec                   & 1 &4.7$\pm$1.0&    5.6$\pm$0.4&    5.1$\pm$1.1&      IFMR \\
LAWD 37&                   1145$-$6450& (WD 1142$-$645)& 0.61$\pm$0.01  & Phot                   & 1 &1.5$\pm$0.5&    2.1$\pm$0.2&    1.8$\pm$0.5&      IFMR \\
SSSPM J1148$-$7458&        1147$-$7457& \nodata        & 0.488$\pm$0.003& Gaia He-atm            &8 &\nodata     &    \nodata    &    1.0$\pm$0.1&      Low\\
SDSS J115052.32+683116.1&  1150+6831  & WD 1148+687    & 0.69$\pm$0.04  & Spec                   & 2 &2.5$\pm$0.8&    2.9$\pm$0.2&    2.7$\pm$0.8&      IFMR \\
LP 852-7&                  1205$-$2333& WD 1202$-$232  & 0.59$\pm$0.03  & Spec                   & 1 &1.3$\pm$0.6&    1.9$\pm$0.1&    1.6$\pm$0.6&      IFMR \\
G 197-47&                  1211+5724  & WD 1208+576    & 0.56$\pm$0.09  & Phot                   & 1 &0.9$\pm$1.2&    1.6$\pm$0.1&    1.2$\pm$1.2&      IFMR \\
WG 21&                     1226$-$6612& WD 1223$-$659  & 0.45$\pm$0.02  & Spec                   & 1 &\nodata    &    \nodata    &    1.0$\pm$0.1&      Low\\
WG 22&                     1238$-$4948& WD 1236$-$495  & 1.13$\pm$0.14  & Spec                   & 7 &6.2$\pm$1.8&    6.5$\pm$0.4&    6.3$\pm$1.8&      IFMR \\
Wolf 457&                  1300+0328  & WD 1257+037    & 0.70$\pm$0.06  & Phot                   & 1 &2.6$\pm$1.0&    2.9$\pm$0.2&    2.8$\pm$1.0&      IFMR \\
EGGR 436&                  1308+8502  & WD 1309+853    & 0.71$\pm$0.02  & Phot                   & 1 &2.8$\pm$0.7&    3.0$\pm$0.2&    2.9$\pm$0.7&      IFMR \\
ER 8&                      1312$-$4728& WD 1310$-$472  & 0.63$\pm$0.04  & Phot                   & 1 &1.8$\pm$0.7&    2.3$\pm$0.2&    2.1$\pm$0.7&      IFMR \\
LP 854-50&                 1319$-$2147& WD 1316$-$215  & 0.99$\pm$0.01  & Gaia H-atm             & 8 &4.9$\pm$1.0&    5.7$\pm$0.4&    5.3$\pm$1.1&      IFMR \\
LAWD 45&                   1319$-$7823& WD 1315$-$781  & 0.69$\pm$0.02  & Phot                   & 1 &2.5$\pm$0.7&    2.9$\pm$0.2&    2.7$\pm$0.7&      IFMR \\
BD-07 3632&                1330$-$0834& WD 1327$-$083  & 0.50$\pm$0.06  & {\bf GravRed}          & 5 &\nodata    &    1.0$\pm$0.1&    1.0$\pm$0.1&      IFMR \\
Wolf 489&                  1336+0340  & (WD 1334+039)  & 0.54$\pm$0.03  & Phot                   & 1 &\nodata    &    1.4$\pm$0.1&    1.4$\pm$0.1&      IFMR \\
LSPM J1341+0500&           1341+0500  & (WD 1338+052)  & 0.58$\pm$0.15  & Phot                   & 2 &1.1$\pm$1.9&    1.8$\pm$0.1&    1.5$\pm$1.9&      IFMR \\
EGGR 438&                  1348+2334  & WD 1345+238    & 0.45$\pm$0.02  & Phot                   & 1 &\nodata    &    \nodata    &    1.0$\pm$0.1&      Low\\
PG 1350$-$090&             1353$-$0916& WD 1350$-$090  & 0.68$\pm$0.03  & Spec                   & 1 &2.4$\pm$0.7&    2.8$\pm$0.2&    2.6$\pm$0.7&      IFMR \\
VVV J141159.32$-$592045.7& 1411$-$5920& \nodata        & 0.66$\pm$0.01  & Gaia H-atm*            & 8& 2.1$\pm$0.6&    2.7$\pm$0.2&    2.4$\pm$0.6&      IFMR \\
LP 801-9&                  1447$-$1742& WD 1444$-$174  & 0.82$\pm$0.05  & Phot                   & 1 &3.4$\pm$1.6&    3.7$\pm$0.2&    3.6$\pm$1.6&      IFMR \\
G 137-24&                  1535+1247  & WD 1532+129    & 0.57$\pm$0.15  & Phot                   & 2 &1.0$\pm$1.9&    1.7$\pm$0.1&    1.4$\pm$1.9&      IFMR \\
HD 140901 B&               1547$-$3755& WD 1544$-$377  & 0.58$\pm$0.01  & {\bf GravRed}         & 5 & 1.1$\pm$0.5&    1.8$\pm$0.1&    1.5$\pm$0.5&      IFMR \\
CD$-$38 10980&             1623$-$3913& WD 1620$-$391  & 0.65$\pm$0.01  & {\bf GravRed}         & 5 & 2.0$\pm$0.6&    2.6$\pm$0.2&    2.3$\pm$0.6&      IFMR \\
Ross 640&                  1628+3646  & WD 1626+368    & 0.58$\pm$0.03  & Phot                   & 1 &1.1$\pm$0.6&    1.8$\pm$0.1&    1.5$\pm$0.6&      IFMR \\
G 138-38&                  1632+0851  & WD 1630+089    & 0.59$\pm$0.15  & Spec                   & 2 &1.3$\pm$1.9&    1.9$\pm$0.1&    1.6$\pm$1.9&      IFMR \\
EGGR 258&                  1634+5710  & WD 1633+572    & 0.57$\pm$0.04  & Phot                   & 1 &1.0$\pm$0.7&    1.7$\pm$0.1&    1.4$\pm$0.7&      IFMR \\
PG 1633+434&               1635+4317  & WD 1633+433    & 0.68$\pm$0.04  & Phot                   & 1 &2.4$\pm$0.8&    2.8$\pm$0.2&    2.6$\pm$0.8&      IFMR \\
DN Dra&                    1648+5903  & WD 1647+591    & 0.77$\pm$0.03  & {\bf Astero}          & 14& 3.1$\pm$1.5&    3.3$\pm$0.2&    3.2$\pm$1.5&      IFMR \\
UCAC4 317-104829&          1706$-$2643& (WD 1703$-$267)& 0.808$\pm$0.009& Gaia H-atm             & 8 &3.3$\pm$1.5&    3.5$\pm$0.2&    3.4$\pm$1.5&      IFMR \\
EGGR 494&                  1708+0257  & WD 1705+030    & 0.68$\pm$0.09  & Phot                   & 1 &2.4$\pm$1.3&    2.8$\pm$0.2&    2.6$\pm$1.3&      IFMR \\
G 203-47 B &               1709+4340  & WD 1708+437    & $>$0.50        & Dyn+mod                &22 &$>$1.4     &    $>$2.0     &    $>$1.4     &      IFMR \\      
PM J17476$-$5436&          1747$-$5436& \nodata        & 0.48$\pm$0.01  & Gaia He-atm            &8 & \nodata    &    \nodata    &    1.0$\pm$0.1&      Low\\
EGGR 372&                  1748+7052  & WD 1748+708    & 0.79$\pm$0.01  & Phot                   & 1 &3.2$\pm$1.5&    3.4$\pm$0.2&    3.3$\pm$1.5&      IFMR \\
EGGR 199&                  1749+8246  & WD 1756+827    & 0.55$\pm$0.13  & Spec                   & 7 &\nodata    &    1.5$\pm$0.1&    1.5$\pm$0.1&      IFMR \\
LSR J1817+1328&            1817+1328  & WD 1814+134    & 0.68$\pm$0.02  & Phot                   & 1 &2.4$\pm$0.7&    2.8$\pm$0.2&    2.6$\pm$0.7&      IFMR \\
G 227-28&                  1821+6101  & WD 1820+609    & 0.56$\pm$0.05  & Phot                   & 1 &0.9$\pm$0.8&    1.6$\pm$0.1&    1.2$\pm$0.8&      IFMR \\
EGGR 176&                  1824$-$1308& WD 1821$-$131  & 1.06$\pm$0.07  & Spec                   & 3 &5.5$\pm$1.3&    6.1$\pm$0.4&    5.8$\pm$1.3&      IFMR \\
UCAC4 508-079937&          1825+1135  & \nodata        & 0.51$\pm$0.01  & Gaia H-atm             & 8 &\nodata    &    1.1$\pm$0.1&    1.1$\pm$0.1&      IFMR \\
EGGR 374&                  1830+5447  & WD 1829+547    & 0.90$\pm$0.07  & Phot                   & 1 &4.0$\pm$1.1&    5.0$\pm$0.3&    4.5$\pm$1.2&      IFMR \\
LAWD 73&                   1900+7039  & WD 1900+705    & 0.93$\pm$0.02  & Phot                   & 1 &4.3$\pm$1.0&    5.3$\pm$0.3&    4.8$\pm$1.0&      IFMR \\
EGGR 375&                  1918+3843  & WD 1917+386    & 0.75$\pm$0.04  & Phot                   & 1 &3.0$\pm$1.5&    3.2$\pm$0.2&    3.1$\pm$1.5&      IFMR \\
LAWD 74&                   1920$-$0740& WD 1917$-$077  & 0.62$\pm$0.02  & Phot                   & 1 &1.6$\pm$0.6&    2.2$\pm$0.2&    1.9$\pm$0.6&      IFMR \\
UCAC4 482-095741&          1921+0613  & \nodata        & 0.68$\pm$0.01  & Gaia H-atm             & 8 &2.4$\pm$0.6&    2.8$\pm$0.2&    2.6$\pm$0.7&      IFMR \\
GD 219&                    1921+1440  & WD 1919+145    & 0.74$\pm$0.03  & Spec                   & 1 &3.0$\pm$1.4&    3.1$\pm$0.2&    3.1$\pm$1.5&      IFMR \\
PY Vul&                    1937+2743  & WD 1935+276    & 0.66$\pm$0.02  & {\bf Astero}           & 15&2.1$\pm$0.6&    2.7$\pm$0.2&    2.4$\pm$0.7&      IFMR \\
LAWD 79&                   1956-0102  & WD 1953$-$011  & 0.79$\pm$0.13  & Spec                   & 7 &3.2$\pm$1.7&    3.4$\pm$0.2&    3.3$\pm$1.7&      IFMR \\
Wolf 1130 B&               2005+5426  & WD 2003+542    & 1.24$\pm$0.17  & {\bf Orbit}            & 16&7.2$\pm$2.0&    7.2$\pm$0.4&    7.2$\pm$2.1&      IFMR \\
EGGR 498&                  2005$-$1056& WD 2002$-$110  & 0.72$\pm$0.01  & Phot                   & 1 &2.9$\pm$1.4&    3.0$\pm$0.2&    3.0$\pm$1.4&      IFMR \\
CD$-$30 17706&             2010$-$3013& WD 2007$-$303  & 0.60$\pm$0.02  & Spec                   & 1 &1.4$\pm$0.5&    2.0$\pm$0.2&    1.7$\pm$0.6&      IFMR \\
SCR J2012$-$5956&          2012$-$5956& WD 2008$-$600  & 0.44$\pm$0.01  & Phot                   & 1 &\nodata    &    \nodata    &    1.9$\pm$0.9&      Ultra-low\\
EC 20173$-$3036&           2020$-$3027& \nodata        & 0.75$\pm$0.01  & Gaia H-atm*            & 8& 3.0$\pm$1.5&    3.2$\pm$0.2&    3.1$\pm$1.5&      IFMR \\
HD 340611&                 2034+2503  & WD 2032+248    & 0.64$\pm$0.03  & Spec                   & 1 &1.9$\pm$0.7&    2.4$\pm$0.2&    2.2$\pm$0.7&      IFMR \\
EGGR 140&                  2044$-$6805& WD 2039$-$682  & 0.98$\pm$0.03  & Spec                   & 1 &4.8$\pm$1.0&    5.6$\pm$0.4&    5.2$\pm$1.1&      IFMR \\
EGGR 261&                  2049+3728  & WD 2047+372    & 0.81$\pm$0.03  & Spec                   & 1 &3.3$\pm$1.5&    3.5$\pm$0.2&    3.4$\pm$1.5&      IFMR \\
G 187-8 A &                2050+2630  & WD 2048+263    & 0.24$\pm$0.04\tablenotemark{e}  
                                                                        & Phot                   & 1 &\nodata    &    \nodata    &    1.9$\pm$0.9&      Ultra-low\\
G 187-8 B &                2050+2630  & \nodata        & \nodata        & \nodata                & 1 &\nodata    &    \nodata    &    1.9$\pm$0.9&      Ultra-low\\
UCAC4 325-215293&          2052$-$2504& (WD 2049$-$253)& 0.47$\pm$0.01  & Gaia He-atm            &8 & \nodata    &    \nodata    &    1.0$\pm$0.1&      Low\\
Ross 193 B&                2056$-$0450& WD 2054$-$050  & 0.49$\pm$0.01  & Spec                   & 23&\nodata    &    \nodata    &    1.0$\pm$0.1&      Low\\
WT 765&                    2101$-$4906& \nodata        & 0.53$\pm$0.01  & Gaia H-atm             & 8 &\nodata    &    1.3$\pm$0.1&    1.3$\pm$0.1&      IFMR \\
LAWD 83&                   2113$-$8149& WD 2105$-$820  & 0.78$\pm$0.10  & Spec                   & 7 &3.2$\pm$1.6&    3.4$\pm$0.2&    3.3$\pm$1.6&      IFMR \\
EGGR 378 &                 2118+5412  & WD 2117+539    & 0.56$\pm$0.03  & Spec                   & 1 &0.9$\pm$0.6&    1.6$\pm$0.1&    1.2$\pm$0.6&      IFMR \\
$\nu$ Oct B &              2141$-$7723& \nodata        & 0.55$\pm$0.05  & Dyn+mod                &20 &\nodata    &    1.5$\pm$0.1&    1.5$\pm$0.1&      IFMR \\   
L 570-26&                  2141$-$3300& WD 2138$-$332  & 0.70$\pm$0.02  & Phot                   & 1 &2.6$\pm$0.7&    2.9$\pm$0.2&    2.8$\pm$0.7&      IFMR \\
EGGR 148&                  2142+2059  & WD 2140+207    & 0.48$\pm$0.04  & Phot                   & 1 &\nodata    &    \nodata    &    1.0$\pm$0.1&      Low\\
PHL 1716&                  2143$-$0659& \nodata        & 0.87$\pm$0.01  & Gaia H-atm             & 8 &3.7$\pm$0.9&    4.5$\pm$0.3&    4.1$\pm$1.0&      IFMR \\
UCAC4 747-070768&          2151+5917  & (WD 2150+591)  & 0.57$\pm$0.01  & Gaia H-atm             & 8 &1.0$\pm$0.4&    1.7$\pm$0.1&    1.4$\pm$0.5&      IFMR \\
WG 39&                     2157$-$5100& WD 2154$-$512  & 0.60$\pm$0.04  & Phot                   & 1 &1.4$\pm$0.7&    2.0$\pm$0.2&    1.7$\pm$0.7&      IFMR \\
CD Oct&                    2204$-$7513& WD 2159$-$754  & 0.92$\pm$0.04  & Spec                   & 1 &4.2$\pm$1.0&    5.3$\pm$0.3&    4.7$\pm$1.1&      IFMR \\
WD 2211$-$392&             2214$-$3859& WD 2211$-$392  & 0.80$\pm$0.04  & Phot                   & 1 &3.3$\pm$1.5&    3.5$\pm$0.2&    3.4$\pm$1.5&      IFMR \\
SCR J2230$-$7515&          2230$-$7515& WD 2226$-$755  & 0.58$\pm$0.00  & Phot                   & 1 &1.1$\pm$0.4&    1.8$\pm$0.1&    1.5$\pm$0.4&      IFMR \\
SCR J2230$-$7513&          2230$-$7513& WD 2226$-$754  & 0.58$\pm$0.00  & Phot                   & 1 &1.1$\pm$0.4&    1.8$\pm$0.1&    1.5$\pm$0.4&      IFMR \\
EGGR 155&                  2249+2236  & WD 2246+223    & 1.11$\pm$0.21  & Spec                   & 7 &6.0$\pm$2.3&    6.4$\pm$0.4&    6.2$\pm$2.3&      IFMR \\
EGGR 283 A&                2251+2939  & WD 2248+293    & 0.35$\pm$0.07\tablenotemark{e} 
                                                                        & Phot                   & 1 &\nodata    &    \nodata    &    1.9$\pm$0.9&      Ultra-low\\
EGGR 283 B&                2251+2939  & \nodata        & \nodata        & \nodata                & 1 &\nodata    &    \nodata    &    1.9$\pm$0.9&      Ultra-low\\
EGGR 453&                  2253$-$0646& (WD 2251$-$070)& 0.58$\pm$0.03  & Phot                   & 1 &1.1$\pm$0.6&    1.8$\pm$0.1&    1.5$\pm$0.6&      IFMR \\
LSPM J2309+5506E&          2309+5506  & WD 2307+548    & 0.59$\pm$0.15  & Spec                   & 2 &1.3$\pm$1.9&    1.9$\pm$0.1&    1.6$\pm$1.9&      IFMR \\
ZZ Psc&                    2328+0514  & WD 2326+049    & 0.593$\pm$0.012& {\bf Astero}           & 15&1.3$\pm$0.5&    1.9$\pm$0.1&    1.6$\pm$0.5&      IFMR \\
GD 1212&                   2338$-$0741& WD 2336$-$079  & 0.62$\pm$0.03  & {\bf Astero}           & 17&1.6$\pm$0.6&    2.2$\pm$0.2&    1.9$\pm$0.7&      IFMR \\
LAWD 93&                   2343+3232  & WD 2341+322    & 0.65$\pm$0.11  & {\bf GravRed}          & 5 &2.0$\pm$1.5&    2.6$\pm$0.2&    2.3$\pm$1.5&      IFMR \\
\enddata
\tablenotetext{a}{White dwarf designations from \cite{mccook2016}. Those in parentheses are "WD" designations found in SIMBAD but not in \cite{mccook2016}.}
\tablenotetext{b}{The initial mass derived using the initial-to-final mass relation of \cite{cummings2018}. Uncertainties are derived by propagating the listed uncertainty for the final mass and the uncertainties listed for the coefficients of the initial-to-final mass relations in equations 4-6 of \cite{cummings2018}.}
\tablenotetext{c}{The initial mass derived using the initial-to-final mass relation of \cite{elbadry2018}. One-sigma uncertainties are chosen to match the initial mass 95.4\% (2-sigma) envelope shown in figure 3 of \cite{elbadry2018}.}
\tablenotetext{d}{This is our adopted initial mass. When both a \cite{cummings2018} estimate and an \cite{elbadry2018} estimate are available, this adopted mass in the unweighted average of those two.}
\tablenotetext{e}{Quoted mass estimate assumes the system is a single white dwarf.}
\tablerefs{References for final mass measurements and estimates: 
(1) \cite{giammichele2012},
(2) \cite{limoges2015},
(3) \cite{gianninas2011},
(4) \cite{maxted2002},
(5) \cite{silvestri2001},
(6) \cite{brandt2019},
(7) \cite{bedard2017},
(8) \cite{gentile2021},
(9) \cite{mason2017},
(10) \cite{sahu2017},
(11) \cite{joyce2018},
(12) \cite{gentilefusillo2019},
(13) \cite{liebert2013},
(14) \cite{romero2013},
(15) \cite{romero2012},
(16) \cite{mace2018},
(17) \cite{hermes2014},
(18) \cite{baroch2021},
(19) \cite{zhuchkov2012},
(20) \cite{ramm2021},
(21) \cite{winters2020},
(22) \cite{delfosse1999a},
(23) \cite{blouin2019}.
}
\tablecomments{Codes for mass methods: Direct measurements of final mass (in boldface) --
{\bf Accel} = dynamical acceleration,
{\bf Astero} = asteroseismology,
{\bf GravRed} = gravitational redshift,
{\bf Lensing} = gravitational lensing,
{\bf Orbit} = dynamical orbital analysis.
Estimates of final mass (in normal font) --
{\it Dyn+mag} = partial dynamical orbital analysis and absolute magnitude,
{\it Dyn+mod} = partial dynamical orbital analysis used along with modeling,
{\it Gaia H-atm} = Gaia-centric hydrogen-atmosphere solution,
{\it Gaia H-atm*} = assumes a Gaia-centric hydrogen-atmosphere solution even though no spectrum is available,
{\it Gaia He-atm} = Gaia-centric helium-atmosphere solution,
{\it Phot} = uses fit to the spectral energy distribution,
{\it Spec} = uses fit to the spectrum.
Method for estimating the initial mass --
IFMR = initial-to-final-mass relation,
Low = arbitrary assignment for low final masses $0.45 \le M < 0.56 M_\odot$,
Ultra-low = special assignment for ultra-low final masses $M < 0.45 M_\odot$ (see text for details), 
Conjecture = case-by-case handling (see text for details).
}
\end{deluxetable*}

There are a number of caveats that must be addressed before proceeding with the {\it initial} mass determinations, however: (1) The parameterization of the initial-to-final mass relation is heavily reliant on white dwarfs with hydrogen lines in their spectra, but roughly half of the white dwarfs in the 20-pc census have no hydrogen lines and are classified as DB (helium lines), DC (no lines), DQ (carbon bands), or DZ (metal lines). Are such objects expected to follow the same initial-to-final mass relation as the DA white dwarfs? (2) Several 20-pc white dwarfs have masses lower than the $0.50 M_\odot$ lower bound of current initial-to-final mass relations. How do we address this complication? (3) Given the age of the Milky Way, a white dwarf with mass below ${\sim}0.45 M_\odot$ should not yet exist (\citealt{fontaine2001}, \citealt{sun2018}), as there has been insufficient time for its single-star progenitor to have evolved off the main sequence. Nonetheless, a few 20-pc white dwarfs have final masses so low that they imply cooling ages longer than a Hubble time. How do we interpret this seeming contradiction? We address these issues below.

For issue (1), consider that within the 20-pc census, the most massive white dwarfs\footnote{One possible exception to this is Wolf 1130 B, which is believed to have a progenitor mass near $8 M_\odot$ and an oxygen-neon core (\citealt{mace2018}).} are akin to Sirius B and have initial mass estimates of $\sim$5 $M_\odot$. With this in mind, it is helpful to summarize some relevant facts about white dwarfs, as stated in the excellent review by \cite{fontaine2001}. The alpha process (carbon burning) occurs only in stars more massive than $\sim$8 $M_\odot$\footnote{This is close to the lower mass boundary of neutron star formation, which is believed to be $\sim8-11 M_\odot$ (\citealt{woosley2002})}, so all white dwarfs within 20 pc of the Sun will have stopped their thermonuclear burning at the triple-alpha process (helium burning). These white dwarfs will have a core made primarily of carbon and oxygen and will be compositionally stratified, with a helium-rich envelope and a separate hydrogen-rich envelope. These two layers are very thin, but their opacity regulates the core's energy output. As such, these thin layers play a critical role in determining white dwarf cooling times. 

Current observational evidence shows that white dwarfs evolve spectroscopically as they cool, sometimes appearing as hydrogen-atmosphere stars (DA) and sometimes as helium-atmosphere stars (DB and other classes\footnote{The other non-DA spectral classes of white dwarfs are also thought to be helium-atmosphere stars but with different temperatures and/or trace pollutants than normal DB stars, with DO stars (ionized helium present) at the hot end of the evolutionary sequence and DC, DQ, and DZ stars at the cool end (figure 4 of \citealt{fontaine2013}).}), meaning that some unknown process -- convection, diffusion, and/or mixing -- reorders the two outer envelopes over time. Specifically, the ratio of helium-atmosphere to hydrogen-atmosphere white dwarfs changes as a function of effective temperature, with relatively few helium-atmosphere stars being identified in the range $30,000 < T_{\rm eff} < 45,000$K (\citealt{shibahashi2007}) and relatively few helium-atmosphere stars again being found in the cooler range $5000 < T_{\rm eff} < 6000$K (\citealt{bergeron1997}). This implies that most nearby white dwarfs evolve as hydrogen-atmosphere stars, an assertion that \cite{fontaine2001} say is further bolstered by the fact that the observational data and theoretical predictions -- particularly with regards to the white dwarf luminosity function -- are in excellent agreement. We are therefore confident in using the initial-to-final mass relations (IFMR) to estimate the initial masses of 20-pc objects having final masses $>0.50 M_\odot$. These initial mass estimates are listed in Table~\ref{tab:wd_masses} and are noted with "IFMR" in the final column.

For issue (2), we note that of the 161 white dwarfs in Table~\ref{tab:wd_masses}, only twelve have $0.45 \le M < 0.50 M_\odot$. An object at the upper end of this range would have a predicted initial mass of $<0.9 M_\odot$ according to the initial-to-final mass relation of \cite{cummings2018} (their equation 4) or $\sim1.0{\pm}0.1 M_\odot$ according to the relation of \cite{elbadry2018} (their table 1 and their figure 3). Additionally, the lowest mass object to have evolved off the main sequence during the lifetime of the Universe (13.8 Gyr; \citealt{planck2020}) is predicted\footnote{Below masses of $\sim5 M_\odot$, objects of sub-solar metallicity are predicted to evolve more rapidly than those of solar metallicity (figure 5 of \citealt{mowlavi1998}; see also figure 11 of \citealt{mowlavi2012}). However, very few stars in the solar neighborhood have metallicities significantly below solar. See Section~\ref{sec:subdwarfs}.} to have an initial mass of $10^{-log(1.38)/2.5} = 0.9 M_\odot$ (equation 1.88 of \citealt{hansen1994}). Thus, we can consider all twelve of these objects to have initial masses of $\sim1.0{\pm}0.1 M_\odot$, and these low-mass estimates are noted by "Low" in the final column of Table~\ref{tab:wd_masses}. Non-evolved late-F and early-G dwarfs in this same mass range significantly outnumber (by $\sim5\times$) the lowest mass white dwarf progenitors, so this part of the white dwarf population makes only a small contribution to this mass slice of the initial mass function anyway.

For issue (3), a small number of white dwarfs in the census have quoted masses below ${\sim}0.45 M_\odot$. There are two scenarios that can explain the existence of such objects. The first is that unresolved white dwarf doubles will be misinterpreted as being overluminous because, at a fixed value of $T_{\rm eff}$, the Stefan-Boltzmann law will imply a falsely large radius, which results in an erroneously small mass. Of the ultra-low mass systems within 20 pc, three (LP 991-16, G 187-8, and EGGR 283) were suspected to be double degenerate systems by \cite{giammichele2012}, and those predictions appear to have been verified by Gaia DR3: LP 991-16 and G 187-8 both fall significantly above the white dwarf sequence on the Gaia-based color-magnitude diagram, and EGGR 283 is shown to be an astrometric double with an orbital period of 278.0 d. The masses quoted for these objects in Table~\ref{tab:wd_masses} are therefore biased low, and these white dwarfs likely have masses $>0.45 M_\odot$. A fourth system, LP 941-19, has a Gaia DR3 RUWE value of 1.817 and may yet prove to be a double star in its own right and thus also have a biased mass estimate. 

The second scenario is that common-envelope mergers or episodic mass loss can produce remnants with pure helium cores (\citealt{serenelli2001}), and for these the cooling times are much longer than those of white dwarfs with carbon-oxygen cores, particularly if the hydrogen envelope is massive enough for its own sustained burning (\citealt{alberts1996}, \citealt{sarna1999}). This scenario likely explains the low mass estimate for SCR J2012$-$5956, which has a Gaia DR3 RUWE value of 0.98 and is likely to be a single star.

As \cite{giammichele2012} explain, without knowing more about the companion objects in the double degenerate systems, determining new mass estimates for the individual objects is not straightforward. Whether or not the other ultra-low-mass systems are the products of a double-object merger or mass loss from a single star is also guesswork. For these seven white dwarf systems, we have estimated initial masses as follows. We know only that their initial masses likely fall between $\sim$0.9 and $\sim$8.0 $M_\odot$. Because the median initial mass of 20-pc white dwarfs with solid mass estimates is $1.9 M_\odot$, we arbitrarily assign each a mass of $1.9{\pm}0.9 M_\odot$ that, at 1-sigma, encompasses the initial mass range spanned by 72\% of the 20-pc white dwarf sample. These estimates are marked with "Ultra-low" in the last column of Table~\ref{tab:wd_masses}.

Finally, we note three objects in Table~\ref{tab:wd_masses} whose white dwarf natures are more speculative. Each of these objects is discussed individually below:

\begin{itemize}
    \item $\chi$ Eri B: Both \cite{fuhrmann2012} and \cite{fuhrmann2016} posit that the appreciable X-ray luminosity coming from this system emanates not from the G9 subgiant primary (\citealt{gray2006}), $\chi$ Eri A, but from the secondary, $\chi$ Eri B. Although this companion is seven magnitudes fainter at $V$ band, they stipulate that it could account for the anomalous X-ray flux if it were a white dwarf.

    \item $\alpha$ For Bb: The primary in this system, $\alpha$ For A, is an F6 dwarf (\citealt{gray2006}) and the secondary, $\alpha$ For B, is a G7 dwarf (\citealt{corbally1984}). This B component was found to be a 3.75-d radial velocity binary by \cite{fuhrmann2016}. No spectral lines are visible from the tertiary component, meaning that its mass would have to be below $0.35 M_\odot$ if it were an M dwarf. However, \cite{fuhrmann2016} find that the spectrum of $\alpha$ For Ba is enhanced in barium content relative to other dwarfs of similar spectral type and relative to its primary star, $\alpha$ For A. As barium is an abundant product of nucleosynthesis on the red giant branch, the authors speculate that $\alpha$ For Ba was polluted during the post-main sequence evolution of $\alpha$ For Bb, now a white dwarf.

    \item $\xi$ UMa Bb: This is part of a quintuple system. The primary, $\xi$ UMa A, is an F8.5: dwarf (\citealt{keenan1989}) and the secondary, $\xi$ UMa B, is a G2 dwarf (\citealt{keenan1989}). Both components are spectroscopic binaries. The A component is an RS CVn double (\citealt{samus2003}), and the B component is a well known double with a period of $\sim$4 d, as summarized in \cite{fuhrmann2008}. A distant, co-moving fifth member of this system, the T8.5 dwarf WISE J111838.70+312537.9, has also been identified; \citealt{wright2013}. As with $\alpha$ For Ba discussed above, $\xi$ UMa Ba has an enhanced barium abundance, leading \cite{fuhrmann2016} to speculate that $\xi$ UMa Bb is a white dwarf and the donor responsible for the extra barium content. 
\end{itemize}

For these three white dwarfs, we also arbitrarily assign each an initial mass of $1.9{\pm}0.9 M_\odot$, as done for the objects with ultra-low mass estimates. Estimates for these three white dwarfs are marked with "Conjecture" in the last column of Table~\ref{tab:wd_masses}.

\subsubsection{Giants and subgiants\label{sec:subgiants-giants}}

There are a number of objects in the 20-parsec census that have evolved off the main sequence but have not yet become white dwarfs. Table~\ref{tab:giants_subgiants} includes all objects in Table~\ref{tab:20pc_census} that have a luminosity class more luminous than V and/or fall in a locus on the absolute magnitude vs.\ color diagrams that identifies them as post-main sequence stars. 

Several of these have direct mass measurements either from orbital dynamics or asteroseismology. The rest have had their masses estimated from other methods, primarily via comparison of their placement on the HR diagram in relation to modeled evolutionary tracks or via fits of their spectra to atmospheric models. For some objects with IV-V or IV luminosity classes, other published spectral types indicate a V luminosity class or their placement on the HR diagram suggests a main sequence star. This is reflected in the mass estimates given in Table~\ref{tab:giants_subgiants}. 

\startlongtable
\begin{deluxetable*}{lccccl}
\tabletypesize{\scriptsize}
\tablecaption{Giants and Subgiants in the 20-pc Census\label{tab:giants_subgiants}}
\tablehead{
\colhead{Name} &
\colhead{Abbrev.\ Coords} &
\colhead{Sp.\ Type\tablenotemark{a}} &                          
\colhead{Mass\tablenotemark{b}} &                          
\colhead{Mass}&
\colhead{Method}\\
\colhead{} &
\colhead{J2000} &
\colhead{} &
\colhead{($M_\odot$)} &                          
\colhead{Ref.} &                          
\colhead{}\\
\colhead{(1)} &                          
\colhead{(2)} &
\colhead{(3)} &
\colhead{(4)} &
\colhead{(5)} &
\colhead{(6)}
}
\startdata
$\beta$ Cas      &0009+5908   &  F2 III            & 1.91$\pm$0.02     & 9 &  Interferometry + rapid-rotation models                        \\
$\alpha$ Tri A   &0153+2934   &  F6 IV             & 1.70              & 15&  Spectral fit to atmospheric models                                    \\
$\chi$ Eri A     &0155$-$5136 &  G9 IV             & 1.58              & 13&  Placement on evolutionary tracks                              \\
10 Tau           &0336+0024   &  F9 IV-V           & 1.139$\pm$0.016   & 23&  Interferometry + model isochrones                                    \\
$\delta$ Eri     &0343$-$0945 &  K1 III-IV         & 1.33$\pm$0.07     & 8 &  {\bf Asteroseismology}                                              \\
$\tau^6$ Eri     &0346$-$2314 &  F5 IV-V           & 1.44$\pm$0.13     & 13&  Placement on evolutionary tracks                              \\
$\epsilon$ Ret A &0416$-$5918 &  K2 III            & 1.48              & 25&  Spectral fit to atmospheric models                                    \\
HD 283750 Aa     &0436+2707   &  K3 IV ke          & 0.84$\pm$0.19     & 15&  Placement on evolutionary tracks                              \\
$\pi^3$ Ori      &0449+0657   &  F6 IV-V           & 1.283$\pm$0.006   & 23&  Interferometry + model isochrones                                    \\
$\alpha$ Aur Aa  &0516+4559   &  G1 III            & 2.569$\pm$0.007   & 27&  {\bf Orbital dynamics}                                              \\
$\alpha$ Aur Ab  &0516+4559   &  K0 III            & 2.483$\pm$0.007   & 27&  {\bf Orbital dynamics}                                              \\
$\xi$ Gem        &0645+1253   &  F5 IV-V           & 1.706$\pm$0.012   & 23&  Interferometry + model isochrones                              \\
HD 53143         &0659$-$6120 &  K0 IV-V (k)       & 1.0               & 22&  Placement on color-magnitude diagram                                \\
$\alpha$ Gem Aa  &0734+3153   &  A1.5 IV+          & 2.98              & 26&  Estimated from spectral type/color                            \\
$\beta$ Gem      &0745+2801   &  K0 III            & 1.91$\pm$0.09     & 16&  {\bf Asteroseismology}                                              \\
$\rho$ Pup\tablenotemark{c}
                 &0807$-$2418 &  F5II kF2II mF5II  & 1.9$\pm$0.1       & 1 &  Placement of general $\rho$ Pup class on evolutionary tracks     \\
HD 73752 Aa      &0839$-$2239 &  G5 IV             & 1.21              & 14&  Spectral fit to atmospheric models                                    \\
$\rho^1$ Cnc A   &0852+2819   &  K0 IV-V           & 0.91$\pm$0.02     & 28&  Placement on evolutionary tracks                              \\
10 UMa A         &0900+4146   &  F5 IV-V           & 1.396$\pm$0.002   & 4&  {\bf Orbital dynamics}                                              \\
HD 78366\tablenotemark{d}       
                 &0908+3352   &  G0 IV-V           & 1.08              & 15&  Spectral fit to atmospheric models                                    \\
$\tau^1$ Hya Aa  &0929$-$0246 &  F5.5 IV-V         & 1.20              & 11&  Placement on evolutionary tracks                              \\
$\theta$ UMa A   &0932+5140   &  F5.5 IV-V         & 1.506$\pm$0.095   & 23&  Interferometry + model isochrones                                    \\
15 LMi           &0948+4601   &  G0 IV-V           & 1.11+0.15         & 7 &  Spectral fit to atmospheric models                                    \\
$\delta$ Leo     &1114+2031   &  A5 IV(n)          & 2.061$\pm$0.006   & 23&  Interferometry + model isochrones                                    \\
$\beta$ Vir      &1150+0145   &  F8.5 IV-V         & 1.42$\pm$0.08     & 8 &  {\bf Asteroseismology}                                              \\
HD 104304 A      &1200$-$1026 &  G8 IV             & 0.98              & 25&  Spectral fit to atmospheric models                                    \\
e Vir Aa         &1316+0925   &  G0 IV             & 1.22              & 12&  Placement on evolutionary tracks                              \\
70 Vir           &1328+1346   &  G5 V\tablenotemark{f}& 1.14$\pm$0.08  & 31&  $L_{bol}$ + $T_{\rm eff}$ + spectrum-based log(g)\\
$\tau$ Boo A     &1347+1727   &  F7 IV-V           & 1.34              & 25&  Spectral fit to atmospheric models                                    \\
$\eta$ Boo Aa    &1354+1823   &  G0 IV             & 1.77$\pm$0.11     & 8 &  {\bf Asteroseismology}                                              \\
$\theta$ Cen     &1406$-$3622 &  K0 III            & 1.27              & 13&  Placement on evolutionary tracks                              \\
$\alpha$ Boo     &1415+1910   &  K0 III            & 0.80$\pm$0.20\tablenotemark{g}     & 19&  {\bf Asteroseismology}                                              \\
HD 125072        &1419$-$5922 &  K3 IV             & 0.88              & 25&  Spectral fit to atmospheric models                                    \\
$\alpha$ Cen B   &1439$-$6050 &  K2 IV C2+1**      & 0.909$\pm$0.003   & 30&  {\bf Orbital dynamics}                                           \\
HD 130948 A      &1450+2354   &  F9 IV-V           & 1.18$\pm$0.16     & 7 &  Spectral fit to atmospheric models                                    \\
$\lambda$ Ser    &1546+0721   &  G0 IV-V           & 1.15$\pm$0.15     & 7 &  Spectral fit to atmospheric models                                    \\
HD 140901 A      &1547$-$3754 &  G7 IV-V           & 0.99              & 25&  Spectral fit to atmospheric models                                    \\
14 Her           &1610+4349   &  K0 IV-V           & 0.73$\pm$0.10     & 7 &  Spectral fit to atmospheric models                                    \\
$\zeta$ Her A    &1641+3136   &  G2 IV             & 1.04$\pm$0.03     & 20&  {\bf Orbital dynamics}                                              \\
$\epsilon$ Sco A &1650$-$3417 &  K1 III            & 1.4$\pm$0.1       & 18&  {\bf Asteroseismology}                                              \\
HD 154088        &1704$-$2834 &  K0 IV-V           & 0.92              & 25&  Spectral fit to atmospheric models                                    \\
HD 158614 A      &1730$-$0103 &  G9- IV-V Hdel1    & 0.963$\pm$0.005   & 4 &  {\bf Orbital dynamics}                                              \\
$\alpha$ Oph A   &1734+1233   &  A5 IV nn          & 2.20$\pm$0.06     & 17&{\bf Orbital dynamics}                                              \\
26 Dra A         &1734+6152   &  G0 IV-V           & 1.06              & 10&  Orbital dynamics + astrophysical assumptions                  \\
$\mu$ Ara        &1744$-$5150 &  G3 IV-V           & 1.21$\pm$0.13     & 8 &  {\bf Asteroseismology}                                              \\
$\mu^1$ Her A    &1746+2743   &  G5 IV             & 1.10$^{+0.11}_{-0.06}$& 21&{\bf Asteroseismology}                                              \\
$\eta$ Ser       &1821$-$0253 &  K0 III-IV         & 1.45$\pm$0.21     & 8 &  {\bf Asteroseismology}                                              \\
110 Her          &1845+2032   &  F5.5 IV-V         & 1.422$\pm$0.009   & 23&  Interferometry + model isochrones                                    \\
b Aql            &1924+1156   &  G7 IV Hdel1       & 1.186$\pm$0.015   & 23&  Interferometry + model isochrones                                    \\
$\delta$ Aql Aa  &1925+0306   &  F1 IV-V(n)        & 1.45              & 11&  Placement on evolutionary tracks                              \\
HD 188088 Aa     &1954$-$2356 &  K2 IV (k)         & 0.85              & 29&  Orbital dynamics + other assumptions\tablenotemark{e}          \\
$\beta$ Aql A    &1955+0624   &  G9.5 IV           & 1.26$\pm$0.18     & 8 &  {\bf Asteroseismology}                                              \\
HD 190360        &2003+2953   &  G7 IV-V           & 0.92$\pm$0.12     & 7 &  Spectral fit to atmospheric models                                    \\
$\delta$ Pav     &2008$-$6610 &  G8 IV             & 1.07$\pm$0.13     & 8 &  {\bf Asteroseismology}                                              \\
$\eta$ Cep       &2045+6150   &  K0 IV             & 1.6               & 2 &  Placement on evolutionary tracks                              \\
$\nu$ Oct A      &2141$-$7723 &  K1 III            & 1.6$\pm$0.1       & 24&  Placement on evolutionary tracks                              \\
$\delta$ Cap Aa  &2147$-$1607 &  kA5hF0mF2 III     & 2.0               & 6 &  {\bf Orbital dynamics}                                              \\
$\gamma$ Cep A   &2339+7737   &  K1 III            & 1.294$\pm$0.081     & 5 &  {\bf Orbital dynamics}                              \\
\enddata
\tablerefs{References for the mass measurements and estimates: 
(1)  \citealt{abt2017},        
(2)  \citealt{affer2005},      
(3)  \citealt{allendeprieto1999},   
(4)  \citealt{andrade2019},      
(5)  \citealt{mugrauer2022},       
(6)  \citealt{batten1992},      
(7)  \citealt{brewer2016},     
(8)  \citealt{bruntt2010}.   
(9)  \citealt{che2011},         
(10) \citealt{cvetkovic2010},    
(11) \citealt{david2015},       
(12) \citealt{dorazi2017},      
(13) \citealt{fuhrmann2012},    
(14) \citealt{fuhrmann2011},
(15) \citealt{fuhrmann2008},  
(16) \citealt{hatzes2012},      
(17) \citealt{gardner2021},   
(18) \citealt{kallinger2019},   
(19) \citealt{kallinger2010},   
(20) \citealt{katoh2013},         
(21) \citealt{li2019},          
(22) \citealt{nielsen2019},      
(23) \citealt{boyajian2012},
(24) \citealt{ramm2021},    
(25) \citealt{takeda2007},     
(26) \citealt{tokovinin2008}, 
(27) \citealt{torres2015},      
(28) \citealt{vonbraun2011},
(29) \citealt{fekel2017},
(30) \citealt{akeson2021},
(31) \citealt{stassun2017}.
}
\tablenotetext{a}{References for Sp.\ Type can be found in Table~\ref{tab:20pc_census}.}
\tablenotetext{b}{Methods in bold involve direct mass measurements.}
\tablenotetext{c}{Luminosity class II suggests a more evolved state for this star than its placement on the HR diagram -- subgiant or giant -- attests. The cause for this "anomalous luminosity effect" is unknown but is a feature of the $\rho$ Puppis class of pulsators (\citealt{gray2009}).}
\tablenotetext{d}{\cite{fuhrmann2008} believes this object is young, not evolved.}
\tablenotetext{e}{\cite{fekel2017} calculates minimum masses ($M\sin^3{i}$) from orbital dynamics of 0.8463$\pm$0.0014 $M_\odot$ for HD 188088 A and 0.8316$\pm$0.014 $M_\odot$ for HD 188088 B. These authors believe that both components are normal K dwarfs, the subgiant classification likely resulting from the slight metal richness of these stars. Given that these minimum masses are close to the mass expected for this dwarf class, the inclination is suspected of being near $90^\circ$ despite the lack of eclipses in the system.}
\tablenotetext{f}{Some references classify this object as G4 V-IV (e.g., \citealt{strassmeier2018}), and its placement just above the main sequence indicates that it may just be moving into a later stage of evolution.}
\tablenotetext{g}{Evolutionary models suggest a mass of $1.08{\pm}0.06 M_\odot$, toward the upper end of the range deduced from asteroseismology, and a relatively old age of $7.1^{+1.5}_{-1.2}$ Gyr (\citealt{ramirez2011}). (An object with a mass of $\sim$0.90 $M_\odot$ has a main sequence lifetime exceeding a Hubble time and will not yet have evolved to a giant state; see section 1.7 of \citealt{hansen1994}.)}
\end{deluxetable*}

One curious observation from the absolute magnitude vs.\ spectral type diagrams of Figure~\ref{fig:AbsMag_SpType} is the vertical locus of evolved stars lying well above the main sequence but concentrated almost exclusively at a spectral class K0. On the $M_{W4}$ vs.\ spectral type plot, for example, we find fifteen objects (not counting the young, main sequence star $\beta$ Pic, whose unusual position is caused by its debris disk) that lie more than one magnitude above the main sequence for their spectral classes. These can be divided into a group of five objects ($\beta$ Cas, $\rho$ Pup, $\alpha$ Tri, $\eta$ Boo, and $\zeta$ Her) with classes between F2 and G2 and a group of ten objects ($\delta$ Eri, $\beta$ Aql A, $\eta$ Cep, $\gamma$ Cep A, $\nu$ Oct A, $\eta$ Ser, $\beta$ Gem, $\theta$ Cen, $\epsilon$ Sco A, and $\alpha$ Boo) with a very narrow range of types from G9.5 to K1. As we can see from Table~\ref{tab:giants_subgiants}, this first group of five objects has a mass range of 1.0-1.9 $M_\odot$ and falls in a locus in Figure~\ref{fig:AbsMag_SpType} indicating evolution off the main sequence and onto the evolutionary subgiant branch\footnote{Note that the assigning of an object's  subgiant or giant {\it luminosity class} via spectroscopic gravity diagnostics does not necessarily equate to its presumed {\it evolutionary} status as a subgiant or red giant branch star via its placement in color magnitude diagrams.}. The other group, of ten objects, is comprised of stars with an identical mass range (1.0-1.9 $M_\odot$) that are now ascending the red giant branch. Evolution along the subgiant branch is more rapid than the climb up the red giant branch, explaining the overabundance of red giants (Tables 3-4 of \citealt{iben1967}). The fact that these latter stars are concentrated so narrowly in spectral type is a consequence of the fact that at typical disk ages for masses in this same range, the red giant branch is confined narrowly to a temperature of $\sim$5000K (Figure 13 of \citealt{iben1967}, Figure 8 of \citealt{bressan2012}), which is the temperature that corresponds to early-K giant spectral classes (Figure 2 of \citealt{dyck1996}, Figure 4 of \citealt{richichi1999}, Table 10 of \citealt{heiter2015}). Objects above this mass range ($> 1.9 M_\odot$) are few in number in the 20-pc census ($<$3\% of the total; see Table~\ref{tab:20pc_census}) and evolve through their giant phases in no more than a few tens of Myr (Table 3 of \citealt{iben1967}); as a consequence of their rarity and rapid evolution, no such earlier type giants are seen.

\subsubsection{Brown dwarfs\label{sec:brown_dwarfs}}

Brown dwarfs follow no mass-luminosity relation because they constantly cool over time. If the age of the brown dwarf is known, this can be used to estimate the mass from evolutionary models, but age is a difficult parameter to measure for non-youthful disk objects. We therefore must resort to simulations to tease out information regarding the mass function. In \cite{kirkpatrick2019,kirkpatrick2021}, we took the empirical distribution of brown dwarf {\it effective temperatures} and compared that to various predicted temperature distributions modeled by taking the shape of the brown dwarf mass function, the value of its low-mass cutoff, and the underlying evolutionary model suite as free parameters. For the analysis of this paper, we will employ those same methods, using an updated suite of predictions by Raghu et al. (submitted).

Here, we compare the \cite{kirkpatrick2021} accounting of all 525 known 20-pc L, T, and Y dwarfs to that given in Table~\ref{tab:20pc_census}. Additions and subtractions to this tally are listed in Table~\ref{tab:brown_dwarf_additions_subtractions}. We find that eight objects have fallen out of the 20-pc sample, all because of new parallax measurements or revised distance estimates that place them outside of 20 pc. On the other hand, sixty-five objects are newly added. These additions include thirty-eight new discoveries (thirty-seven by the Backyard Worlds citizen science group, four of which are new companions), nine new companions recently announced in the literature, one new companion announced here but found in Gaia, three new published parallaxes with d$<$20 pc, twelve previously overlooked companions, and two previously overlooked objects (DENIS J065219.7-253450, presumably due to a transcription error, and SSSPM J1444-2019, whose subdwarf type had earlier been updated from late-M to early-L). To facilitate analysis on the revised $T_{\rm eff}$ distribution, we have listed in Table~\ref{tab:brown_dwarf_additions_subtractions} the estimated temperatures of each of the additions and subtractions. Further analysis can be found in Section~\ref{sec:further_analysis}.

\startlongtable
\begin{deluxetable*}{lllc}
\tabletypesize{\scriptsize}
\tablecaption{Additions to and Subtractions from the 20-pc L, T, and Y Census of \cite{kirkpatrick2021}\label{tab:brown_dwarf_additions_subtractions}}
\tablehead{
\colhead{Object} &
\colhead{Reason for Change} &
\colhead{Reference} &
\colhead{$T_{\rm eff}$ (K)\tablenotemark{a}}\\
\colhead{(1)} &
\colhead{(2)} &
\colhead{(3)} &
\colhead{(4)}
}
\startdata
\multicolumn{4}{c}{\bf Additions} \\
WISE J003110.04+574936.3 B & New companion& \cite{best2021} & 1275$\pm$200\tablenotemark{b}\\
CWISE J003507.81$-$153233.5& New discovery& This paper (Table~\ref{tab:poss_20pc_members_MLTY}) & 686$\pm$79\\
2MASSW J0036159+182110 B   & Overlooked companion& \cite{bernat2010} & 1125$\pm$79\\
CWISE J013343.58+803153.1  & New discovery& This paper (Table~\ref{tab:poss_20pc_members_MLTY}) & 1181$\pm$79\\
CWISE J014433.03$-$545545.5& New discovery& This paper (Table~\ref{tab:poss_20pc_members_MLTY}) & 751$\pm$79\\
CWISER J021550.96+674017.2 & New companion to HD 13579& This paper (Table~\ref{tab:poss_20pc_members_MLTY} and Section~\ref{sec:new_companions}) & 1125$\pm$79\\
CWISER J021612.11+423015.9 & New discovery& This paper (Table~\ref{tab:poss_20pc_members_MLTY}) & 460$\pm$79\\
L 440-30 B (0219$-$3646)   & Overlooked companion& \cite{kuerster2008} & 1613$\pm$134\\
Ross 19 B (0219+3518)      & New companion& \cite{schneider2021} & 460$\pm$79\\
CWISE J032600.46+421058.5  & New discovery& This paper (Table~\ref{tab:poss_20pc_members_MLTY}) & 624$\pm$79\\
WISE J033605.05$-$014350.4 B& New companion& \cite{calissendorff2023} & 325$\pm$79\tablenotemark{c}\\
Wolf 227B (0352+1701)       & Overlooked companion& \cite{winters2018}& 2000$\pm$81\tablenotemark{h}\\
CWISE J035856.18+480244.9  & New discovery& This paper (Table~\ref{tab:poss_20pc_members_MLTY}) & 511$\pm$79\\
L 375-2 B (0432$-$3947)   & New discovery& \cite{silverstein2022}& 1200$\pm$333\tablenotemark{i}\\
LP 775-31 B (0435$-$1606) & Overlooked companion& \cite{cortescontreras2017}& 1600$\pm$81\tablenotemark{j}\\
Wolf 230 C (0507+1758)    & Overlooked companion& \cite{winters2020}& 1200$\pm$333\tablenotemark{k}\\
CWISE J053046.20+440849.2  & New discovery& This paper (Table~\ref{tab:poss_20pc_members_MLTY}) & 1125$\pm$79\\
CWISE J060938.91+062513.2  & New discovery& This paper (Table~\ref{tab:poss_20pc_members_MLTY}) & 1420$\pm$134\\
DENIS J063001.4$-$184014 B & New companion& \cite{sahlmann2021} & 1420$\pm$134\\
DENIS J063001.4$-$184014 (C)& New companion& \cite{sahlmann2021} & 1420$\pm$134\\
WISEA J064750.85$-$154616.4 B& New companion& \cite{best2021} & 1275$\pm$200\tablenotemark{b}\\
DENIS J065219.7$-$253450   & Overlooked object& Gaia DR3 & 2196$\pm$88\\
PSO J103.0927+41.4601 B (0652+4127) & New companion& \cite{best2021} & 1190$\pm$100\tablenotemark{d}\\
CWISE J075227.38+053802.6  & New discovery& This paper (Table~\ref{tab:poss_20pc_members_MLTY}) & 1273$\pm$79\\
CWISE J075853.12$-$232645.8 & New discovery& This paper (Table~\ref{tab:poss_20pc_members_MLTY}) & 1209$\pm$79\\
L186-67 Ab (0822$-$5726)   & Overlooked companion& \cite{bergfors2010}& 2091$\pm$88\tablenotemark{l}\\
CWISE J083130.98+154018.4  & New discovery& This paper (Table~\ref{tab:poss_20pc_members_MLTY}) & 566$\pm$79\\
CWISE J092710.37$-$474155.5& New discovery& This paper (Table~\ref{tab:poss_20pc_members_MLTY}) & 624$\pm$79\\
LP 788-1 B (0931$-$1717)   & Overlooked companion& \cite{winters2017}, \cite{vrijmoet2020}& 1200$\pm$333\tablenotemark{m}\\
WISEU J100241.49+145914.9  & New companion to G 43-23& This paper (Table~\ref{tab:poss_20pc_members_MLTY} and Section~\ref{sec:new_companions}) & 624$\pm$79\\
CWISE J100521.10$-$691226.8& New discovery& This paper (Table~\ref{tab:poss_20pc_members_MLTY}) & 1190$\pm$79\\
CWISE J100628.98+105408.5  & New discovery& This paper (Table~\ref{tab:poss_20pc_members_MLTY}) & 566$\pm$79\\
CWISE J105349.12$-$460239.1& New discovery& This paper (Table~\ref{tab:poss_20pc_members_MLTY}) & 686$\pm$79\\
1RXS J121408.0-234516 B    & New companion& Gaia + this paper (Table~\ref{tab:20pc_census})& 1250$\pm$150\tablenotemark{n}\\
e Vir Ab (1316+0925)       & Overlooked companion& \cite{kuzuhara2013,bonnefoy2018} & 624$\pm$79\\
2MASSW J1326201$-$272937 B & New companion& \cite{best2021} & 1400$\pm$200\tablenotemark{e}\\
WT 460 B (1411$-$4132)     & Overlooked companion& \cite{montagnier2006} & 2096$\pm$134\\
SSSPM J1444$-$2019         & Revised spectral type& \cite{kirkpatrick2016} & 2207$\pm$88\\
DENIS-P J1454078$-$660447 B  & Overlooked companion&  \cite{vrijmoet2020}& 1100$\pm$250\tablenotemark{f}\\
WISEA J153429.75$-$104303.3& New parallax& \cite{kirkpatrick2021b} & 686$\pm$79\\
2MASS J15345325+1219495    & New parallax& Gaia DR3 & 1532$\pm$88\\
SCR J1546$-$5534 B         & New companion& \cite{vrijmoet2022}&  2085$\pm$88\tablenotemark{p}\\
CWISE J161546.07+671227.4  & New discovery& This paper (Table~\ref{tab:poss_20pc_members_MLTY}) & 1254$\pm$79\\
CWISE J163336.14$-$325305.3& New discovery& This paper (Table~\ref{tab:poss_20pc_members_MLTY}) & 819$\pm$79\\
VVV J165507.19$-$421755.5  & New discovery& \cite{schapera2022} and this paper (Table~\ref{tab:poss_20pc_members_MLTY}) & 1125$\pm$79\\
DENIS-P J170548.38$-$051645.7 B& Overlooked companion& \cite{dieterich2014} & 1838$\pm$134\\
CWISE J171221.50+495318.2  & New discovery& This paper (Table~\ref{tab:poss_20pc_members_MLTY}) & 624$\pm$79\\
CWISE J171338.81$-$183322.7& New discovery& This paper (Table~\ref{tab:poss_20pc_members_MLTY}) & 1227$\pm$79\\
CWISE J173830.94$-$773024.3& New discovery& This paper (Table~\ref{tab:poss_20pc_members_MLTY}) & 624$\pm$79\\
2MASS J17502484$-$0016151 B& Overlooked companion& \cite{henry2018} & 1100$\pm$250\tablenotemark{f}\\
CWISE J180308.71$-$361332.1& New discovery& This paper (Table~\ref{tab:poss_20pc_members_MLTY}) & 686$\pm$79\\
CWISE J181005.77$-$101001.2& New parallax & \cite{lodieu2022} & 800$\pm$100\tablenotemark{g}\\
CWISE J181125.34+665806.4  & New discovery + parallax & This paper (Table~\ref{tab:poss_20pc_members_MLTY} and Section~\ref{sec:white_bear}) & 412$\pm$79\\
CWISE J184803.45$-$143232.3& New companion to G 155-42& This paper (Table~\ref{tab:poss_20pc_members_MLTY} and Section~\ref{sec:new_companions}) & 686$\pm$79\\
CWISE J203438.09$-$462543.1& New discovery& This paper (Table~\ref{tab:poss_20pc_members_MLTY}) & 624$\pm$79\\
CWISE J210057.80$-$624555.4& New discovery& This paper (Table~\ref{tab:poss_20pc_members_MLTY}) & 1296$\pm$134\\
CWISE J215841.48+732842.8  & New discovery& This paper (Table~\ref{tab:poss_20pc_members_MLTY}) & 624$\pm$79\\
2MASSW J2224438$-$015852 B & New companion& \cite{best2021} & 1400$\pm$200\tablenotemark{e}\\
CWISE J222701.50+260450.0  & New discovery& This paper (Table~\ref{tab:poss_20pc_members_MLTY}) & 460$\pm$79\\
CWISE J223002.32+424655.3  & New discovery& This paper (Table~\ref{tab:poss_20pc_members_MLTY}) & 751$\pm$79\\
CWISE J224547.21$-$433341.5& New discovery& This paper (Table~\ref{tab:poss_20pc_members_MLTY}) & 686$\pm$79\\
CWISE J230930.81+145630.6  & New discovery& This paper (Table~\ref{tab:poss_20pc_members_MLTY}) & 686$\pm$79\\
CWISE J233817.04$-$732930.3& New discovery& This paper (Table~\ref{tab:poss_20pc_members_MLTY}) & 624$\pm$79\\
CWISE J233819.49$-$385421.2& New discovery& This paper (Table~\ref{tab:poss_20pc_members_MLTY}) & 1254$\pm$79\\
CWISE J235120.60$-$700026.2& New discovery& This paper (Table~\ref{tab:poss_20pc_members_MLTY}) & 460$\pm$79\\
\multicolumn{4}{c}{\bf Subtractions} \\
CWISE J061741.79+194512.8 A& Revised distance estimate& \cite{humphreys2023} & 1465$\pm$134\\
CWISE J061741.79+194512.8 B& Revised distance estimate& \cite{humphreys2023} &  686$\pm$79\\
Kelu-1 A (1305$-$2541)     & New parallax& Gaia DR3 & 1931$\pm$134\\
Kelu-1 B (1305$-$2541)     & New parallax& Gaia DR3 & 1750$\pm$134\\
2MASSI J1526140+204341     & New parallax& Gaia DR3 & 1518$\pm$157\\
SDSS J163022.92+081822.0   & New parallax& This paper (Section~\ref{sec:nparsec_astrometry}) & 970$\pm$88\\
2MASS J23174712$-$4838501  & New parallax& Gaia DR3 & 1537$\pm$197\\
2MASS J23312378$-$4718274  & New parallax& This paper (Section~\ref{sec:nparsec_astrometry}) & 1125$\pm$79\\
\enddata
\tablenotetext{a}{For the new additions, temperature values with uncertainties of $\pm$88K were determined via the $M_H$ vs.\ $T_{\rm eff}$ relation in table 13 of \cite{kirkpatrick2021} and those with uncertainties of $\pm$134K or $\pm$79K were determined with the (assumed) spectral type vs.\ $T_{\rm eff}$ relations in the same table, unless otherwise noted. For subtractions, the temperature values are taken from table 11 of \cite{kirkpatrick2021}.}
\tablenotetext{b}{No information is yet available on this companion, so the temperature estimate is set to cover the spectral type range from late-L to mid-T.}
\tablenotetext{c}{Temperature estimate is taken from \cite{calissendorff2023}, although the uncertainty has been inflated to match the typical uncertainties in \cite{kirkpatrick2021}.}
\tablenotetext{d}{No information is yet available on this companion, so the temperature estimate is set to cover the spectral type range from early-T to mid-T.}
\tablenotetext{e}{No information is yet available on this companion, so the temperature estimate is set to cover the spectral type range from mid-L to early-T.}
\tablenotetext{f}{Little information is available on this companion, so the temperature estimate is set to cover the spectral type range from mid-L to mid-T.}
\tablenotetext{g}{Temperature estimate is taken from \cite{lodieu2022}.}
\tablenotetext{h}{\cite{winters2018} conclude this is likely an early-L dwarf, so we set the temperature to cover the typical range for L0 to L4 dwarfs.}
\tablenotetext{i}{We assume a huge temperature range to encompass the full substellar regime, as \cite{silverstein2022} is able to provide only limited constraints on this companion.}
\tablenotetext{j}{The $\Delta{I}$ magnitude from \cite{cortescontreras2017} suggests a mid-L dwarf, so we estimate a temperature corresponding to this range.}
\tablenotetext{k}{We assume a huge temperature range to encompass the full substellar regime, as \cite{winters2020} is able to provide only an minimum mass of $\sim44 M_{Jup}$ for this companion.}
\tablenotetext{l}{We use the $\Delta{i^\prime}$ magnitude listed in \cite{bergfors2010} to estimate a type of L1, which we use for the temperature estimation.}
\tablenotetext{m}{We assume a huge temperature range to encompass the full substellar regime, as this object is only known as a likely brown dwarf (\citealt{vrijmoet2022}).}
\tablenotetext{n}{The $M_G$ value derived from Gaia DR3 data suggests an early-T dwarf, and we assign a temperature appropriate for late-L through mid-T given that no spectrum yet exists.}
\tablenotetext{p}{We base our temperature estimate on the estimated L1 type we derive in Table~\ref{tab:20pc_census}.}
\end{deluxetable*}

\subsection{Other complications}

\subsubsection{Youth\label{sec:young-objects}}

Will the estimation of masses for young objects be biased if those estimates use a relation based on much older stars? Evolutionary models suggest that below a mass of $\sim0.4 M_\odot$, the contraction of a star down to the main sequence follows a Hayashi track along which the star's effective temperature remains approximately fixed (section 16.2.5 of \citealt{stahler2004}). If a temperature-based metric is used for estimating the masses of such stars, then such estimates will be accurate. At higher masses, however, the descent along the Hayashi track will be interrupted when a radiative zone develops. The star then moves via a Henyey track along which the temperature slowly increases until the star reaches the main sequence. For stars with masses above $\sim0.4 M_\odot$, this evolution to the main sequence occurs within the first 100 Myr.


\begin{deluxetable*}{lllcc}
\tabletypesize{\scriptsize}
\tablecaption{20-pc Members of Young Associations and Moving Groups with Ages $<$ 100 Myr\label{tab:young_objects}}
\tablehead{
\colhead{Objects in System} &
\colhead{Coords.} &
\colhead{Spectral Type\tablenotemark{a}} &                          
\colhead{Membership} &
\colhead{Likelihood\tablenotemark{b}} \\
\colhead{} &                          
\colhead{J2000} &
\colhead{} &
\colhead{} &                          
\colhead{} \\
\colhead{(1)} &                          
\colhead{(2)} &
\colhead{(3)} &
\colhead{(4)} &                          
\colhead{(5)} 
}
\startdata
2MASSW J0045214+163445       & 0045+1634  & L2 $\beta$                   & Argus      & high \\
Ross 15                      & 0159+5831  & M4                           & Carina     & possible \\
$[$BHR2005$]$ 832-2          & 0311+0106  & M5.5                         & $\beta$ Pic& possible \\
LP 944-20                    & 0339$-$3525& M9                           & Argus      & possible \\
BD$-$21 1074 ABC             & 0506$-$2135& M1.5 V e, $[$M1$]$, $[$M2.5$]$& $\beta$ Pic& confirmed \\
PSO J076.7092+52.6087        & 0506+5236  & T4.5                         & Argus      & possible \\
V2689 Ori, PM J05366+1117    & 0536+1119  & M0.5 V ek, M4                & $\beta$ Pic& possible \\
$\beta$ Pic                  & 0547$-$5103& A6 V                         & $\beta$ Pic& confirmed \\
AP Col                       & 0604$-$3433& M4.5 V e                     & Argus      & confirmed \\
2MASS J06244595$-$4521548    & 0624$-$4521& L5                           & Argus      & high \\
LSPM J0714+3702              & 0714+3702  & M8                           & Argus      & possible \\
WISEPA J081958.05$-$033529.0 & 0819$-$0335& T4                           & $\beta$ Pic& high \\
G 161-71\tablenotemark{c}                     
                             & 0944$-$1220& M4.5 V                       & Argus      & high \\
TWA 22 AB                    & 1017$-$5354& M5:, M5.5:                   & $\beta$ Pic& confirmed \\
WISE J104915.57$-$531906.1 AB& 1049$-$5319& L7.5, T0.5:                  & Argus\tablenotemark{d}      & high \\
SDSS J121951.45+312849.4     & 1219+3128  & L9.5                         & Argus      & confirmed \\
G 164-47                     & 1309+2859  & M4 V                         & Carina     & possible \\
$\alpha$ Cir AB              & 1442$-$6458& A7 Vp SrCrEu, K5 V           & $\beta$ Pic& possible\tablenotemark{e} \\
2MASS J17534518$-$6559559    & 1753$-$6559& L4                           & Argus      & possible \\
WISE J180001.15$-$155927.2   & 1800$-$1559& L4.5                         & $\beta$ Pic& possible \\
UCAC3 152-281185, 
UCAC3 152-281176             & 1845$-$1409& M5 V e, M5 V e               & Argus      & possible \\
HD 182488 AB                 & 1923+3313  & G9+ V, T7:                   & Argus      & confirmed \\
LEHPM 2-1265 AB              & 2033$-$4903& $[$M5 composite$]$           & $\beta$ Pic& possible \\
AU Mic, AT Mic AB            & 2041$-$3226& M1, M4.5 V, M4 V             & $\beta$ Pic& confirmed \\
PSO J319.3102$-$29.6682      & 2117$-$2940& T0:                          & $\beta$ Pic& high \\
WISEPC J225540.74$-$311841.8 & 2255$-$3118& T8                           & $\beta$ Pic& high \\
HD 220140 AB, LP 12-90       & 2319+7900  & K2 V k, M4, M5: V            & Columba    & possible \\
G 190-28, G 190-27           & 2329+4128  & M3.5, M4.5                   & Columba    & possible \\
LP 704-15, LP 704-14         & 2357$-$1258& M3 V, M4 V                   & Argus      & possible \\
\enddata
\tablenotetext{a}{References for spectral type are listed in Table~\ref{tab:20pc_census} along with information on whether the types are based on optical or near-infrared data. Spectral types in brackets are estimates based on the absolute $G$-band magnitude.}
\tablenotetext{b}{Our notes translate to the following codes in the {\tt moca\_mtid} column in table {\tt moca\_membership\_types} of the MOCA database: confirmed = BF (bona fide member), high = HM (high-likelihood candidate member), and possible = CM (candidate member). Objects with MOCA codes of LM (low-likelihood candidate member), AM (ambiguous candidate member), and R (rejected candidate member) were ignored.}
\tablenotetext{c}{On the $G_{BP}-G_{RP}$ vs.\ $M_G$ and $G-J$ vs.\ $M_G$ diagrams of Figure~\ref{fig:MG_color}, this M4.5 dwarf appears to be $\sim$1.5 mag above the locus of other objects of the same color despite having a Gaia DR3 RUWE value of 1.114. This could alternatively be interpreted as this object's being $\sim$0.5 mag redder than the bulk of objects of similar absolute magnitude. In colors not involving Gaia magnitudes, no discrepancy is seen; this object falls at an M dwarf spectral type where many of these other colors ($J-K_s$, $J-$W2, W1$-$W2) are degenerate with spectral type over a large range, so such a discrepancy may not be noticeable. \cite{malo2014b} and \cite{ujjwal2020} identify this as a possible member of the Argus Association, so these discrepancies may be related to a young, active chromosphere.}
\tablenotetext{d}{Also identified as a member of the 500-Myr old Oceanus Group (\citealt{gagne2023}).}
\tablenotetext{e}{The possible $\beta$ Pic Moving Group association for $\alpha$ Cir A is listed for both components, as the "high"  likelihood Greater Scorpius-Centaurus association for $\alpha$ Cir B is highly suspect.}
\end{deluxetable*}

This means that objects in the 20-pc census that have masses above $\sim0.4 M_\odot$ and ages less than 100 Myr should have their mass estimates more carefully considered. The Montreal Open Clusters and Associations database (\url{https://mocadb.ca/}, Gagn\'e et al., in prep., \citealt{gagne2018}) is a compilation of known stellar associations, stellar streams, moving groups, and open clusters within 500 pc of the Sun. A search of this database on 2023 May 18 for objects within 20 pc of the Sun and likely belonging to one of these young groups yielded 217 systems. The only objects in this list with ages below 100 Myr are those believed to be members\footnote{Several 20-pc objects are tentatively associated with Greater Scorpius-Centaurus (age $\approx$ 15 Myr) according to \cite{kerr2021}, but closer investigation of kinematics, color-magnitude diagrams, lithium abundances, rotation periods, and non-detections at X-ray wavelengths suggests that these are much older objects.} of the $\beta$ Pic Moving Group ($\sim$26 Myr), the Columba Association ($\sim$42 Myr), the Argus Association ($\sim$45 Myr), the Carina Assocation ($\sim$45 Myr), and the Octans-Near Association ($\sim$55 Myr)\footnote{These age estimates are taken directly from the table {\tt calc\_association\_properties} in the Montreal Open Clusters and Associations database.}. These are listed in Table~\ref{tab:young_objects}.

Because a mass of $0.4 M_\odot$ corresponds to a spectral type of M2.5-M3 (table 7 of \citealt{mann2019}), we can use spectral type to identify which of the young 20-pc objects are the ones whose mass estimations may need special handling. The only objects in Table~\ref{tab:young_objects} with spectral types earlier than this are 
BD$-$21 1074 ABC, 
V2689 Ori, 
$\beta$ Pic, 
$\alpha$ Cir AB, 
HD 182488 A, 
AU Mic, 
and HD 220140 A. Three of these are early M dwarfs for which the brief jog along the Henyey track before reaching the main sequence covers such a small range in temperature that their mass estimates should not be unduly affected. 

This leaves only four individual objects to consider, and two of these have dynamical mass measurements already. For $\beta$ Pic, \cite{lacour2021} used astrometry of the exoplanet system to derive the mass of the host star (1.75$\pm$0.03 $M_\odot$) using only a uniform mass prior between 1.4 and 2.0 $M_\odot$ on $\beta$ Pic itself. HD 182488 A has a loosely constrained dynamical mass measurement from \cite{brandt2019} of $0.94^{+0.17}_{-0.27} M_\odot$. 
This leaves only two young systems with possibly skewed mass estimates, and this represents such a small percentage of 20-pc stars with types earlier than M0 ($<1\%$) that no bias will be imparted on the overall derived mass distribution.

We acknowledge that our understanding of  young moving groups near the Sun is still evolving. Our Sun is currently moving through three groups -- the $\beta$ Pic Moving Group, the AB Dor Moving Group, and the recently identified (but older) Oceanus Group (\citealt{gagne2023}) -- but it remains unlikely that many new early-M and hotter dwarfs within 20 pc will be associated with any newly recognized groups. Such young objects would have already revealed themselves through, for example, high chromospheric activity.

Young {\it brown dwarfs}, on the other hand, require their own special handling. For brown dwarfs, we deduce the form of the mass function via the empirical temperature distribution. It has been well established, however, that young brown dwarfs follow a different spectral type (or color) to $T_{\rm eff}$ relation than their older counterparts (\citealt{faherty2016}). Corrections to the temperature estimates for these objects were already established for the brown dwarf portion of the 20-pc census in \cite{kirkpatrick2021}, and none of the new brown dwarfs discussed in Table~\ref{tab:brown_dwarf_additions_subtractions} are known to be youthful themselves. Therefore, no additional work is required here.

\subsubsection{Non-solar metallicity\label{sec:subdwarfs}}

Objects with non-solar metallicity raise two concerns. The first is that metal-poor objects may belong to the Galactic halo population and could skew our calculation of the nearby mass function, which concentrates on the Galactic disk. The second is that these objects, even if true disk members, may be sufficiently metal poor that standard solar-metallicity relations will not adequately predict their masses. Are either of these concerns justified?

A number of objects in Table~\ref{tab:20pc_census} have spectroscopic classifications indicating subsolar metallicity. For objects earlier than early-M, these classifications can generally be identified via the iron index, "Fe\#", which attempts to encode the abundance of metals relative to hydrogen in the spectrum if the spectrum does not match the standards of solar-metallicity (\citealt{gray2009}). Underabundances are encoded as negative numbers. For objects of spectral type late-K and later, metal poor spectral types (\citealt{gizis1997,lepine2003,kirkpatrick2005,lepine2007,burgasser2007c,zhang2017}) are usually denoted with prefixes of sd (subdwarf), esd (extreme subdwarf), or usd (ultra subdwarf). Table~\ref{tab:low_metallicity} lists all objects in the 20-pc census that have one of these low-metallicity classifications. 

\startlongtable
\begin{deluxetable*}{lccccc}
\tabletypesize{\scriptsize}
\tablecaption{20-pc Objects with Low-metallicity and/or Halo Kinematics\label{tab:low_metallicity}}
\tablehead{
\colhead{Name} &
\colhead{Coords} &
\colhead{SpType\tablenotemark{a}} &
\colhead{Radial Vel.} &
\colhead{RV Ref.} &
\colhead{Kinem.} \\
\colhead{} &
\colhead{J2000} &
\colhead{} &
\colhead{(km s$^{-1}$)} &
\colhead{} &
\colhead{Group} \\
\colhead{(1)} &                          
\colhead{(2)} &
\colhead{(3)} &
\colhead{(4)} &
\colhead{(5)} &
\colhead{(6)}
}
\startdata
85 Peg         &  0002+2704   &  G5 V Fe-1                &  $-$35.57$\pm$0.35& 1& thin disk  \\    
6 Cet	       &  0011$-$1528 &  F8 V Fe-0.8 CH-0.5       &     14.95$\pm$0.13& 3& thin disk  \\  
HD 4391        &  0045$-$4733 &  G5 V Fe-0.8              &  $-$10.92$\pm$0.12& 3& thin disk  \\
$\mu$ Cas      &  0108+5455   &  K1 V Fe-2                &  $-$97.09$\pm$0.25& 3& thick disk \\    
LP 410-38      &  0230+1648   &  sdM6e                    &     49.56$\pm$6.65& 3& thin disk\\
LP 651-7       &  0246$-$0459 &  M6 V                     &     33.28$\pm$2.32& 3& halo \\
BD+33 529      &  0252+3423   &  (sd)K5 V ([Fe/H]=$-$0.63)&  $-$49.89$\pm$0.15& 3& thick disk \\
G 174-25       &  0258+5014   &  sdM3                     &  $-$30.24$\pm$0.45& 3& thin disk\\
LP 994-33      &  0302$-$3950 &  sdM5                     &  \nodata          & -& thick disk?\\
Ross 578       &  0338$-$1129 &  d/sdM2                   &  $-$84.88$\pm$0.56& 3& halo \\
HD 25329       &  0403+3516   &  K3 Vp Fe-1.7             &  $-$25.57$\pm$0.13& 3& halo       \\
$\zeta$ Dor    &  0505$-$5728 &  F9 V Fe-0.5              &   $-$1.45$\pm$0.12& 3& thin disk  \\
Kapteyn's Star &  0511$-$4501 & sdM1p                     &    245.05$\pm$0.13& 3& halo\\
EGGR 290       &  0556+0521   & DAP8.7                    & $-$414.02$\pm$10.41&3& (RV in error)\tablenotemark{h}\\
2MASS J06453153$-$6646120& 0645$-$6646& sdL8              & \nodata           & -& thick disk?\\
YY Gem         &  0734+3152   &  M0.5 Ve Fe-2             &     32.66         & 2& thin disk  \\
212 Pup        &  0752$-$3442 &  F5 V Fe-0.5              &     28.03$\pm$0.16& 3& thin disk  \\
HD 65583       &  0800+2912   &  K0 V Fe-1.3              &     14.79$\pm$0.12& 3& thick disk \\    
UPM J0812$-$3529& 0812$-$3529 &  DC                       & $-$373.74$\pm$8.18& 3& (RV in error)\tablenotemark{h}\\
$\alpha$ Cha   &  0818$-$7655 &  F5 V Fe-0.8              &  $-$12.60$\pm$0.12& 3& thin disk  \\
WISE J083337.83+005214.2 & 0833+0052  & (sd)T9            & \nodata           & -& thick disk?\\
$\psi$ Vel     &  0930$-$4028 &  F3 V Fe-0.7              &      8.80$\pm$1.80& 4& thin disk  \\
L 750-42       &  0943$-$1747 & sdM3                      &     97.36$\pm$0.58& 3& thick disk\\
HD 88230       &  1011+4927   &  K6e V Fe-1               &  $-$26.48$\pm$0.12& 3& thin disk  \\
CWISE J105512.11+544328.3\tablenotemark{c}
                         & 1055+5443  & [sdT8]            & \nodata           & -& thick disk?\\
HD 103095      &  1152+3743   &  K1 V Fe-1.5              &  $-$98.05$\pm$0.12& 3& halo       \\   
10 CVn	       &  1244+3916   &  F9 V Fe-0.3              &     80.49$\pm$0.12& 3& thick disk \\ 
HD 114837      &  1314$-$5906 &  F6 V Fe-0.4              &  $-$63.57$\pm$0.12& 3& thin disk  \\
SDSS J141624.08+134826.7\tablenotemark{b} 
                         & 1416+1348  & sdL7              & $-$87$\pm$33      & 7& thin disk\\
ULAS J141623.94+134836.3\tablenotemark{b} 
                         & 1416+1348  & (sd)T7.5          & $-$87$\pm$33\tablenotemark{g}      & 7& thin disk\\
HD 125276      &  1419$-$2548 &  F9 V Fe-1.5 CH-0.7       &  $-$22.28$\pm$0.13& 3& thin disk  \\
WISE J142320.84+011638.0\tablenotemark{d}
                         & 1423+0116  & sdT8              & $-$19.21$\pm$0.14\tablenotemark{g} & 3& thin disk\\
$\sigma$ Boo   &  1434+2944   &  F4 V kF2 mF1 (metal weak)&      0.75$\pm$0.12& 3& thin disk  \\
SSSPM J1444$-$2019       & 1444$-$2019& sdL0              & \nodata           & -& halo\\
WISEA J153429.75$-$104303.3\tablenotemark{f}
                         & 1534$-$1043& [esdT/Y?]         & \nodata           & -& halo\\
$\chi$ Her     &  1552+4227   &  G0 V Fe-0.8...           &  $-$55.99$\pm$0.12& 3& thin disk  \\
HD 144579      &  1604+3909   &  K0 V Fe-1.2              &  $-$59.44$\pm$0.12& 3& thin disk  \\
HD 145417      &  1613$-$5734 &  K3 V Fe-1.7              &      8.68$\pm$0.13& 3& thick disk \\
b Her          &  1807+3033   &  F9 V metal-weak          &   $-$0.38$\pm$0.13& 3& thin disk  \\
CWISE J181005.77$-$101001.2&   1810$-$1010& esdT0:      & \nodata           & -& thick disk?\\
$\chi$ Dra     &  1821+7243   &  F7 V (metal-weak)        &     31.90$\pm$0.14& 5& thin disk  \\
HD 190067      &  2002+1535   &  K0 V Fe-0.9              &     20.37$\pm$0.12& 3& thin disk  \\
WISE J200520.38+542433.9\tablenotemark{e} 
                         & 2005+5424  & sdT8              & $-$107.6\tablenotemark{g}          & 8& thick disk\\
Ross 769       &  2104$-$1657&   M1 V                     &  $-$10.89$\pm$0.27& 3& halo \\
$\gamma$ Pav   &  2126$-$6521 &  F9 V Fe-1.4 CH-0.7       &  $-$29.78$\pm$0.12& 3& thin disk  \\
G 188-49 B     &  2214+2751   &  \nodata                  & $-$351.15$\pm$8.18& 3& (RV in error)\tablenotemark{h}\\
53 Aqr         &  2226$-$1644 &  G1 V + G5 V Fe-0.8 CH-1  &      2.28$\pm$0.15& 6& thin disk  \\
\enddata
\tablerefs{The references for radial velocity are --
(1) \citealt{gaia2018},
(2) \citealt{fouque2018},
(3) \citealt{gaia2022},
(4) \citealt{holmberg2007},
(5) \citealt{pourbaix2004},
(6) \citealt{maldonado2010},
(7) \citealt{abazajian2009},
(8) \citealt{gizis1997}.
}
\tablenotetext{a}{References for spectral type are listed in Table~\ref{tab:20pc_census} along with information on whether the types are based on optical or near-infrared data. Spectral types sometimes encode information about metallicity. For the coldest objects, this is generally done via a prefix such as sd (subdwarf), esd (extreme subdwarf), or usd (ultra subdwarf), with variations such as d/sd (indicating a spectral morphology intermediate between a normal, solar metallicity dwarf and a subdwarf) also possible (figure 11 of \citealt{kirkpatrick2005}). For hotter stars, the metallicity is included in one of two ways. A suffix type such as "mF1" for an F4 V star would indicate that the metal lines better match that of an F1 standard, despite the fact that the hydrogen line morphology matches the F4 standard. An alternative way of expressing this is to subtract the metal-line best-match subtype from the H-line best-match subtype and to convert that to a metal index; for example, if the iron lines best matched an F1 dwarf although the hydrogen lines best matched a F9 dwarf ($\Delta = +8$ subtypes), the metallicity index for iron would be expressed as Fe $= -0.13{\Delta} - 0.26$, or more compactly as Fe$-$1.3. See chapter 6 of \cite{gray2009} for more details.}
\tablenotetext{b}{These two objects form a common-proper-motion pair.}
\tablenotetext{c}{This object does not yet have spectroscopic observations, so its status as a subdwarf is assumed based on its location on color-magnitude diagrams.}
\tablenotetext{d}{This is the companion to HD 126053, which is typed as a G1.5 V.}
\tablenotetext{e}{This is the wide companion in the $\xi$ Ursae Majoris quintuple system, the primary of which is generally typed as M1 V.}
\tablenotetext{f}{Also known as The Accident, this object does not yet have spectroscopic observations because it is too faint for ground-based spectroscopy. Its status as a subdwarf is assumed based on its highly unusual location on color-magnitude and color-color diagrams. See \cite{kirkpatrick2021b} for details.}
\tablenotetext{g}{The space motion of the primary in this system is used in lieu of an independent measurement for this object.}
\tablenotetext{h}{These objects are flagged by our analysis as belonging to the halo population, but their Gaia DR3 radial velocities are believed to be erroneous. For EGGR 290, the radial velocity measurement is likely incorrect due to the fact that this is a magnetic white dwarf and the fact that the Gaia DR3 radial velocity pipeline lacks white dwarf templates (\citealt{bailer-jones2022}). For UPM J0812$-$3529, its status as a DC white dwarf (\citealt{obrien2023}) means that there should not be any lines in the optical spectrum with which Gaia could measure a radial velocity, but this suspicious radial velocity measurement has nonetheless created much discussion regarding the object's possible close (future) fly-by of the Sun ({\citealt{bobylev2022,delafuentemarcos2022,bailer-jones2022}}). For G 188-48 B, the Gaia DR3 radial velocity is assumed to be spurious because the A component has a much more reasonable -- and better measured -- value of 20.06$\pm$0.14 km s$^{-1}$.}
\end{deluxetable*}

\begin{figure}
\includegraphics[scale=0.6,angle=0]{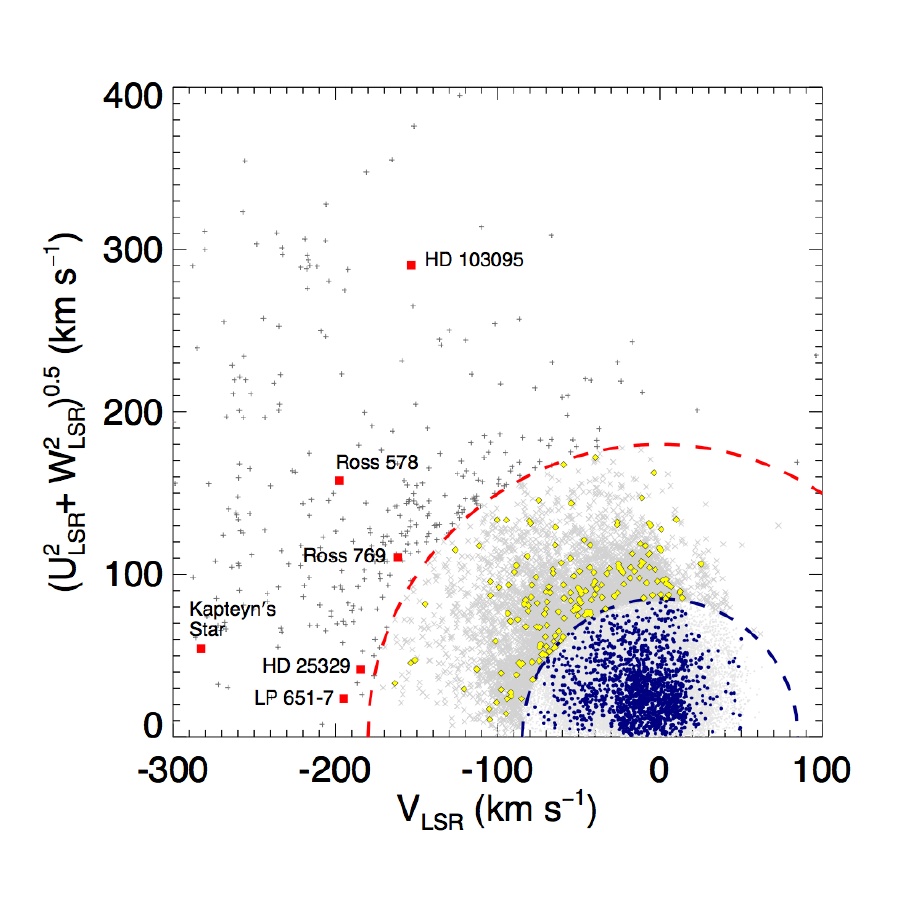}
\caption{Toomre diagram of $UVW$ space motions corrected to the Local Standard of Rest (LSR) for 74,066 {\it Gaia} DR2 stars within 100 pc of the Sun and having parallax errors $< 10\%$ (\citealt{kirkpatrick2021b}). Thin disk (light grey dots), thick disk (medium grey crosses), and halo (dark grey pluses) objects are marked, with halo stars falling outside the outer dashed circle (red) and thin disk objects falling inside the inner dashed circle (navy). Objects with measured radial velocities in Table~\ref{tab:20pc_census} or Table~\ref{tab:low_metallicity} are shown in navy if lying in the thin disk velocity zone, yellow for the thick disk zone, and red for the halo zone. The six halo members are highlighted with black labels. \label{fig:toomre_hot_stars}}
\end{figure}

To answer the first concern, we use the sky positions, parallaxes, and proper motions in Table~\ref{tab:20pc_census} along with published radial velocities in Table~\ref{tab:low_metallicity} to calculate the $U,V,W$ space velocities with respect to the Local Standard of Rest (LSR). We also calculate the $U, V, W$ values for all objects in Table~\ref{tab:20pc_census} with Gaia-based radial velocity measurements to see if any objects lacking low-metallicity spectral classifications are found to be halo members merely from their kinematics\footnote{Two such objects were found -- LP 651-7 and Ross 769 -- and for ease of reference, these have also been added to Table~\ref{tab:low_metallicity}.}. Figure~\ref{fig:toomre_hot_stars} shows the Toomre diagram for both sets of objects. Also shown for comparison are stars having radial velocity measurements in Gaia DR2 and lying within 100 pc of the Sun, color coded as thin disk ($V_{\rm tot} \le 85$ km s$^{-1}$), thick disk ($85 < V_{\rm tot} \le 180$ km s$^{-1}$), or halo ($V_{\rm tot} > 180$ km s$^{-1}$) in accordance with the kinematic criteria of \cite{nissen2004}. This comparison demonstrates that only six objects -- LP 651-7, Ross 578, HD 25329, Kapteyn's Star, HD 103095, and Ross 769 -- appear to belong to the kinematic halo population. All others most likely belong to the thin or thick disk populations.

\begin{figure}
\includegraphics[scale=0.55,angle=0]{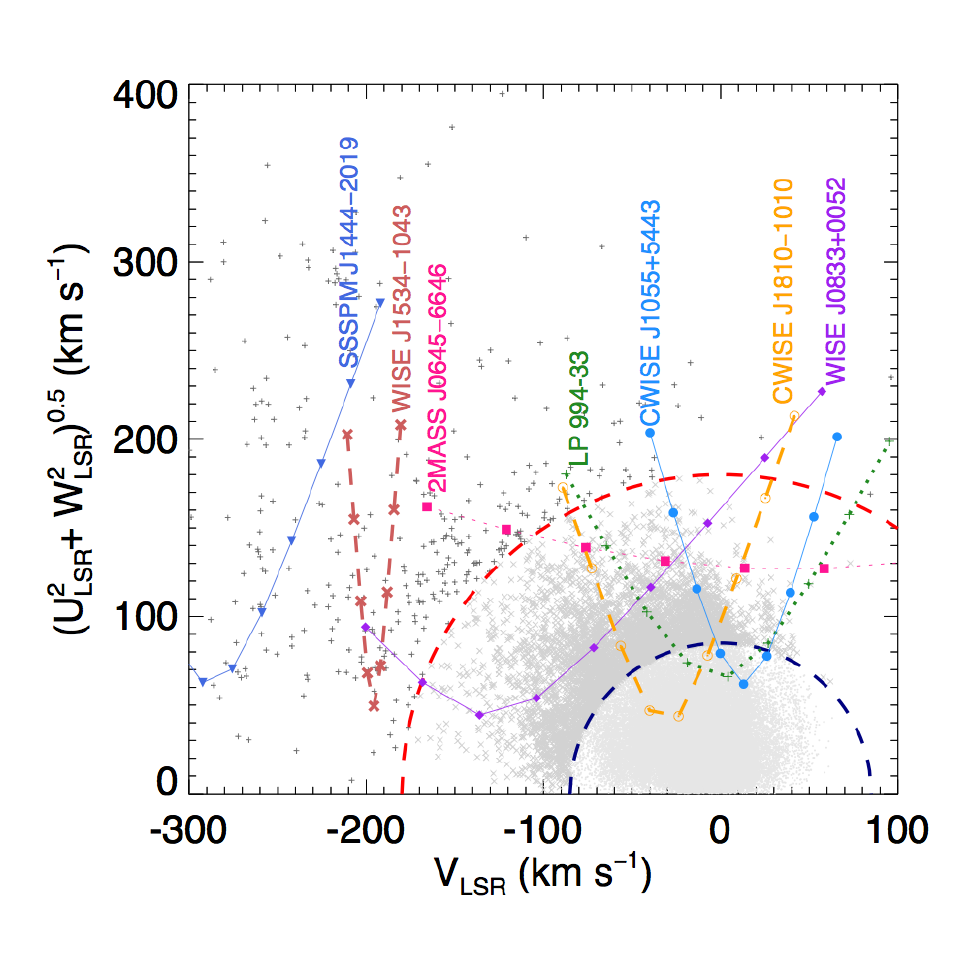}
\caption{Toomre diagram of the 100-pc sample from Figure~\ref{fig:toomre_hot_stars}, now overplotted with the seven objects (various colors and symbols) from Table~\ref{tab:low_metallicity} that lack radial velocity measurements. For these, results are shown for nine assumed radial velocities ranging from $-200$ to +200 km s$^{-1}$, in steps of 50 km s$^{-1}$. As in Figure~\ref{fig:toomre_hot_stars}, the demarcation of the thin disk, thick disk, and halo populations are shown by the dashed circles in red and navy.
\label{fig:toomre_cold_stars}}
\end{figure}

As stated in Section~\ref{sec:radial_velocities}, Gaia contains radial velocities only for those objects having $G_{RVS} \lesssim 14$ mag, which omits many of the M dwarfs and all of the L, T, and Y dwarfs within 20 pc. For these objects, we leverage spectroscopic indications of low metallicity to build a list of potential halo members, then we scour the literature for other published radial velocities. These objects are also listed in Table~\ref{tab:low_metallicity}. Many of these lack any radial velocity measurements, so assumed values from $-$200 to +200 km s$^{-1}$, in increments of 50 km s$^{-1}$, were used to calculate a range of possible $U, V, W$ velocities. These results, shown in Figure~\ref{fig:toomre_cold_stars}, suggest that only two of these colder objects -- SSSPM J1444$-$2019 and WISEA J153429.75$-$104303.3 (aka "The Accident") -- are likely to be true halo members. 

Figures~\ref{fig:toomre_hot_stars} and \ref{fig:toomre_cold_stars} taken together suggest that only eight objects (all of them believed to be single) out of 3,589 total in the 20-pc census, or $0.22\%$, are halo interlopers. Although this is slightly higher than the percentage of $0.15\%$ used in Table A of \cite{bensby2014} based on F and G stars alone, it nevertheless confirms that contamination by halo objects in the 20-pc census is extremely small. Although these objects will still be included in our mass function, any systematic offset imprinted upon their mass estimates can be ignored in subsequent analyses. 

The second concern is difficult to address, as very few low-metallicity objects have had their masses measured via direct methods. The coldest subdwarfs, for instance, have a multiplicity fraction of only ${\sim}1\%$ (\citealt{gonzalez-payo2021}); therefore, few such objects exist for dynamical analyses (e.g., \citealt{rebassa-mansergas2019}). Single subdwarfs are obvious targets for lensing-based mass measurements, as their high velocities increase the likelihood of "encounters" with background objects, but accurate whole-sky astrometry is just now advancing to the stage at which such measurements can be predicted and planned for (e.g. \citealt{sahu2020}). So, to address this concern, we instead note that only forty-two low-metallicity systems are known within the 20-pc census\footnote{It is worth noting that other low-luminosity, low-metallicity objects are likely to be found within this volume given the fact that parallaxes for both WISEA J153429.75$-$104303.3 ($d_{true} = 16.3^{+1.4}_{-1.2}$ pc vs.\ $d_{est} = 38.0^{+5.6}_{-4.9}$ pc; \citealt{kirkpatrick2021b,meisner2020a}) and CWISE J181005.77$-$101001.2 ($d_{true} = 8.9^{+0.7}_{-0.6}$ pc vs.\ $d_{est} = $ 14 - 64 pc; \citealt{lodieu2022,schneider2020}) placed them far closer to the Sun than original estimates predicted.} (Table~\ref{tab:low_metallicity}), which represents only $1.5\%$ of the total. Thus, if small biases are present in converting a subdwarf's spectral type, colors, or absolute magnitudes to masses, the bias in the overall 20-pc mass distribution will be negligible. 

The above logic on the scarcity of objects also holds for systems with {\it higher} metallicity than the Sun. This set of objects has a much smaller range in metallicity than the metal-poor objects above, and there are just a handful of examples. Only the higher-mass stars $\iota$ Hor AB (Fe+0.3), $\nu$ Phe (Fe+0.4), HD 176051 AB (slightly metal strong), and HD 207129 (Fe+0.4) have spectroscopic classifications that fall into this class. Another object, 14 Her, has a supersolar metallicity ([Fe/H] $\approx 0.4$; \citealt{rosenthal2021}) although its listed spectral type in Table~\ref{tab:20pc_census} gives no indication of this. Curiously, even though members of the Hyades Cluster have metallicities that are slightly supersolar ([Fe/H] $= 0.14{\pm}0.05$; \citealt{perryman1998}) and lie, on average, only 47.0$\pm$0.2 pc from the Sun (\citealt{lodieu2019}), there are no confirmed Hyads within the 20-pc volume (\citealt{smart2020,schneider2022}).

\subsubsection{Formation process\label{sec:exoplanets}}

Because we are interested in objects formed via the star formation process, we need a criterion to distinguish objects that may have formed via alternative formation mechanisms at the lowest masses.
When brown dwarfs were first theorized (\citealt{kumar1963,hayashi1963}), they were regarded as direct products of the star formation process -- ones that had insufficient mass to sustain prolonged thermonuclear fusion in their cores -- and as such represented the lower-mass extension of hydrogen-burning stars themselves. These could be contrasted with another low-mass formation product, planets, that were believed to be formed via a secondary process -- from a protoplanetary disk created around a newly formed protostar or brown dwarf. In the early 1960s, there were no known examples of brown dwarfs, and our own Solar System provided the only known examples of planets.

As brown dwarf and exoplanet discoveries began in earnest (see reviews by \citealt{kirkpatrick2005,winn2015}), it became clear that nature produces some low-mass products that are difficult to classify as either brown dwarf or exoplanet (e.g., 2MASSWJ 1207334$-$393254b, \citealt{chauvin2004}). The earlier definition based on formation was cumbersome to use in practice; unless an object was still in its infancy, its exact formation process would be difficult, if not impossible, to ascertain from observations. As an alternative, \cite{burrows1997} proposed another theoretically based definition. This alternative uses mass to distinguish between a brown dwarf and an (exo)planet, the dividing line being the somewhat arbitrarily chosen deuterium burning limit, which is $\sim13 M_{Jup}$ for solar metallicity. Somewhat surprisingly, this definition was thereafter widely (though not universally) adopted, in no small part because lower-mass discoveries that earlier would have been called "brown dwarfs" could now be referred to by a more attention-grabbing label of "exoplanet."

This alternative definition, however, came three and a half decades after the original brown dwarf definition, and the concept of planets having being born from a protoplanetary disk (the "nebular hypothesis") had been in the astronomical lexicon for over two centuries (\citealt{kant1755,laplace1796}). Thus, labeling an object below 13 $M_{Jup}$ as a planet often leads to confusion, as some readers -- and even researchers -- unwittingly apply both definitions in tandem. That is, they assume that a so-named "planet" (by the new, mass-based definition) must have formed via a protoplanetary disk (by the former, formation-based definition). It is difficult to divorce the term "planet" from its formation scenario.

In this paper, one of our goals is to define, or place limits on, the low-mass terminus of star formation. If we were to use the newer definition to include/exclude objects for the mass function analysis, our results would return a terminus of 13$M_{Jup}$, which merely reflects the dividing line chosen by the arbitrary definition. We must, therefore, more carefully consider whether the lowest mass objects in the 20-pc census should be counted as star-formation products or planetary-formation products.

As stated earlier, this definition also lacks easy observational verification. Nonetheless, some methods have been proposed to distinguish formation mechanisms. \cite{oberg2011} postulated that the carbon to oxygen ratio could be used as one tracer. Planets that formed close to a star would have a solar-like $C/O$ value, like brown dwarfs formed via gravitational collapse, whereas planets formed via accretion of ices beyond the water snowline would have a supersolar $C/O$ value. Those authors acknowledged, however, that measuring an accurate value of $C/O$ is fraught with difficulties (even within our own Solar System), and \cite{calamari2022} made a similar conclusion based on their analysis of the spectrum of the brown dwarf Gliese 229B. \cite{molliere2022} show that this simplified picture of the $C/O$ ratio is somewhat more complicated when disk chemical evolution and pebble accretion are taken into account, as well.

Similarly, \cite{morley2019} showed that the deuterium to hydrogen ratio could be used to distinguish between planets with solar $D/H$ values like Jupiter and Saturn, that formed directly from accretion of gas in the protostellar nebula, and planets with enhanced $D/H$ values like Neptune and Uranus, that presumably formed from accretion of ices. Both $C/O$ and $D/H$ thus have limitations: some objects formed via a protoplanetary disk have values indistinguishable from those of objects born via star formation. 

Another promising avenue is the overall metallicity. The giant planets of our Solar System have metal enhancements well above solar values (\citealt{wong2004}, \citealt{fletcher2009}), and exoplanets are preferentially found around metal-rich host stars (\citealt{fischer2005}, \citealt{wang2015}). These facts led \cite{fortney2008} to propose metallicity-based diagnostics that could distinguish between formation scenarios. Specifically for objects with $T_{\rm eff} < 1400K$, a strong 4.5 $\mu$m CO absorption band along with enhanced $H$- and $K$-band fluxes (from a relative lack of collision-induced absorption by $H_2$) are proposed as fingerprints of planet-like formation. However, these diagnostics are likely only useful when comparing populations of objects and not when establishing the formation pathway of individual objects. Metal enrichment is not unique to planet formation, as a collapsing metal-rich cloud can also produce low-mass objects.

\cite{bowler2023} note that the orientation between the spin axis of the star and the orbital plane of the companion shows promise as another marker of formation, as star-like formation shows a wide range of orientations, whereas planet-like formation prefers values near $90^\circ$. This is, however, another marker that can distinguish between populations but cannot be used on an individual object basis.

\cite{schlaufman2018} demonstrates that companions above $\sim10 M_{Jup}$ lack the tendency to fall primarily around metal-rich hosts that companions below $\sim4 M_{Jup}$ exhibit, which is taken as evidence of core accretion in the lower-mass set. \cite{hoch2023} likewise find a tentative difference in the trend of $C/O$ values at $\sim4 M_{Jup}$, which is taken as further evidence that those objects are primarily formed via core accretion, although, as stated above, $C/O$ ratios can be difficult to interpret.  Similarly, \cite{ribas2007} find differing radial velocity distributions above and below $M \sin(i)$ values of $\sim4 M_{Jup}$. \cite{schlaufman2018} states that planet-like formation appears to cease above $\sim4 M_{Jup}$, but not necessarily that star-like formation ceases below $\sim10 M_{Jup}$. There might still be a range in mass, below $\sim4 M_{Jup}$, where both processes contribute.

The methods addressed above require data that are so far lacking for most exoplanets or can be used only in comparing populations. Instead, for this paper, we propose a simple scheme whose purpose is merely to exclude objects with a high likelihood of having been formed via a protoplanetary disk while including all others as {\it possible} products of star formation. For our scheme, we require at least three bodies in a system because the only parameters available for two-body systems -- mass ratio, separation, etc.\ -- can lead to ambiguities when trying to distinguish between formation scenarios. 

As an example, \cite{bowler2020} have used twenty-seven long-period companions labeled as giant planets and brown dwarfs to search for differences in parameters. They find that the population of brown dwarfs has an eccentricity distribution peaking in the range $0.6 < e < 0.9$, whereas binaries with mass ratios significantly different from one have an eccentricity distribution peaking closer to $e \approx 0$. These results indicate that the star formation process tends to create binaries with large eccentricity, and the protoplanetary process tends to form binaries with near-zero eccentricity. To reiterate a point from above, while such trends may be indicative of a {\it population} of objects, eccentricity alone cannot be used on an object-by-object basis to distinguish between formation scenarios. The same is true for mass ratio, as doing so can bias our list of potential companions to only the higher mass ones, which could impact our ability to determine star formation's low-mass cutoff. (Similarly, {\it not} selecting on mass ratio can bias our results in the opposite direction, a point we address further in the next section.)

In triple (and higher-order) systems, however, we have other parameters available. Specifically, we note that the hierarchy of empirically observed triple {\it star} systems is such that the period of the outer component must be at least five times that of the inner pair (\citealt{tokovinin2004}). This is in good agreement with dynamical stability expectations for objects in circular orbits, and the period of the outer component must be even larger than five times the inner one when elliptical orbits are considered (\citealt{mardling2001}). Planets that have formed from a protoplanetary disk, on the other hand, can often arrange themselves in stable orbital configurations (e.g., in resonances with one another) that violate the above law. A multi-star system that formed via a collapsing cloud could, presumably, arrange itself in a similar manner if conditions were ideal, but such examples must be exceedingly rare. Therefore, we will use the ratios of orbital periods to identify "exoplanet" systems in the 20-pc census that most likely formed via a protoplanetary disk, and we retain all others for consideration as possible products of star formation.

To this end, Table~\ref{tab:20pc_exoplanets} lists all of the host objects from Table~\ref{tab:20pc_census} that were labeled as having one or more confirmed exoplanets in the NASA Exoplanet Archive as of 2022 Sep 01\footnote{One additional complication, as stated in Section~\ref{sec:multiplicity_and_exoplanets}, is that the NASA Exoplanet Archive uses a $30M_{Jup}$ dividing line, not $13M_{Jup}$, to distinguish between brown dwarfs and exoplanets. Hence, some of the brown dwarfs already listed in Table~\ref{tab:20pc_census} will be double counted; that is, they will have their own separate row in the table while also being listed under the "\#Planets" column. Such objects are flagged in Table~\ref{tab:20pc_exoplanets}.}. For systems in which any pair of "exoplanets" violate the $P_{outer} < 5P_{inner}$, we indicate the innermost pair that violates the rule and exclude all of the planets, thus including only the host star in later analysis. For all others, we have used the NASA Exoplanet Archive to compile their mass measurements. For objects identified only through radial velocity monitoring, we list the $M \sin(i)$ values, since the inclination of the system is not known. For other objects -- transiting systems, radial velocity systems with astrometric imaging, etc. -- we list the actual measured masses. Incorporating these objects in to the stellar mass function analysis will be discussed further in Section~\ref{sec:further_analysis}.

\startlongtable
\begin{deluxetable*}{llccllc}
\tabletypesize{\scriptsize}
\tablecaption{20-pc Objects Hosting "Planets"\label{tab:20pc_exoplanets}}
\tablehead{
\colhead{Name} &
\colhead{Coords} &
\colhead{\# of} &
\colhead{Note\tablenotemark{a}} &
\colhead{$M \sin(i)$} &
\colhead{$M$} &
\colhead{Mass} \\
\colhead{} &
\colhead{J2000} &
\colhead{Planets} &
\colhead{} &
\colhead{($M_{Jup}$)} &
\colhead{($M_{Jup}$)} &
\colhead{Ref.} \\
\colhead{(1)} &                          
\colhead{(2)} &
\colhead{(3)} &
\colhead{(4)} &
\colhead{(5)} &
\colhead{(6)} &
\colhead{(7)}
}
\startdata
Sun             &  \nodata    &  8   &   $P_{\rm Venus} < 5P_{\rm Mercury}$& \nodata & \nodata & \nodata\\
HD 1237 A       &  0016$-$7951&  1   &   consider & 3.37$\pm$0.09& \nodata&  1\\
GX And          &  0018+4401  &  2\tablenotemark{b}   
                                     &   consider & $0.11^{+0.08}_{-0.06}$& \nodata&  2\\
54 Psc AB       &  0039+2115  &  1   &   consider & 0.228$\pm$0.011&  \nodata& 3\\
HD 3765         &  0040+4011  &  1   &   consider & $0.173^{+0.014}_{-0.013}$& \nodata& 4\\
G 268-38        &  0044$-$1516&  2   &   consider & \nodata& 0.0201$\pm$0.0014&  5\\
\nodata         &  \nodata    &  -   &   consider & \nodata& $0.00554^{+0.00053}_{-0.00050}$&  5\\
BD+61 195       &  0102+6220  &  1   &   consider & 0.0177$\pm$0.0021& \nodata& 6\\
YZ Cet          &  0112$-$1659&  3   &   $P_c < 5P_b$ & \nodata& \nodata& \nodata\\
CD$-$54 269     &  0114$-$5356&  1   &   consider & 0.026$\pm$0.005& \nodata& 7\\
HD 7924         &  0121+7642  &  3   &   $P_c < 5P_b$ & \nodata& \nodata& \nodata\\
$\upsilon$ And A&  0136+4124  &  3   &   consider& 0.6876$\pm$0.0044& \nodata& 8\\
\nodata         &  \nodata    &  -   &   consider& 1.981$\pm$0.019& \nodata& 8\\
\nodata         &  \nodata    &  -   &   consider& 4.132$\pm$0.029& \nodata& 8\\
q01 Eri         &  0142$-$5344&  1   &   consider& 0.94$\pm$0.08& \nodata& 9 \\
$\tau$ Cet      &  0144$-$1556&  4   &   $P_h < 5P_g$& \nodata& \nodata& \nodata\\
TZ Ari          &  0200+1303  &  2\tablenotemark{c}   
                                     &   consider& 0.21$\pm$0.02& \nodata& 10\\
HD 13445 A      &  0210$-$5049&  1   &   consider& 4.42$\pm$0.20& \nodata& 11\\
BD+47 612       &  0222+4752  &  1   &   consider& $0.0619^{+0.0076}_{-0.0072}$& \nodata& 12 \\
$\iota$ Hor AB  &  0242$-$5048&  1\tablenotemark{d}  
                                     &   already included& 2.27±0.25& \nodata& 11 \\
Teegarden's Star&  0253+1652  &  2   &   $P_c < 5P_b$& \nodata& \nodata& \nodata \\
G 245-61        &  0257+7633  &  1   &   consider& \nodata& 0.00733$\pm$0.00063& 13\\
BD$-$17 588 A   &  0301$-$1635&  2   &   $P_b < 5P_c$& \nodata& \nodata& \nodata \\
CD Cet          &  0313+0446  &  1   &   consider& $0.0124^{+0.0013}_{-0.0014}$& \nodata& 14 \\
e Eri           &  0319$-$4304&  4   &   $P_c < 5P_b$& \nodata& \nodata& \nodata  \\
HD 21749        &  0326$-$6329&  2   &   $P_b < 5P_c$& \nodata& \nodata& \nodata  \\
$\epsilon$ Eri  &  0332$-$0927&  1   &   consider& \nodata& $0.66^{+0.12}_{-0.09}$& 15 \\
HD 22496        &  0335$-$4825&  1   &   consider& $0.0175^{+0.0023}_{-0.0021}$& \nodata& 16 \\
L 372-58        &  0335$-$4430&  3   &   $P_c < 5P_b$& \nodata& \nodata& \nodata  \\
L 229-91        &  0409$-$5322&  3   &   $P_c < 5P_b$& \nodata& \nodata& \nodata \\
$\omicron^2$ Eri A&0415$-$0739&  1   &   consider& 0.0266$\pm$0.0015& \nodata& 17 \\
$\epsilon$ Ret A&  0416$-$5918&  1   &   consider& 1.56$\pm$0.14& \nodata& 18 \\
L 375-2 AB      &  0432$-$3947&  2   &   $P_c < 5P_b$& \nodata& \nodata& \nodata \\
HD 285968       &  0442+1857  &  1   &   consider& 0.0285$\pm$0.0043& \nodata& 4 \\
Wolf 1539       &  0452+0628  &  1   &   consider& 0.82$\pm$0.07& \nodata&  19 \\
L 736-30        &  0453$-$1746&  3   &   $P_c < 5P_b$& \nodata& \nodata& \nodata \\
LP 656-38       &  0501$-$0656&  2   &   consider& $0.00636^{+0.00082}_{-0.00079}$& \nodata& 20 \\
\nodata         &  \nodata    &  -   &   consider& $0.0073^{+0.0016}_{-0.0015}$& \nodata& 20 \\
UCAC4 211-005570&  0505$-$4756&  1\tablenotemark{e}   
                                     &   consider& \nodata& \nodata& \nodata  \\
Kapteyn's Star  &  0511$-$4501&  1   &   consider& $0.022^{+0.004}_{-0.003}$& \nodata& 21  \\
$\pi$ Men       &  0537$-$8028&  3   &   consider& $0.0113^{+0.0015}_{-0.0014}$& \nodata&  22\\
\nodata         &  \nodata    &  -   &   consider& 0.0421$\pm$0.0043& \nodata& 23\\
\nodata         &  \nodata    &  -   &   consider& \nodata& 12.2$\pm$1.3& 22\\
$\beta$ Pic     &  0547$-$5103&  2   &   exclude\tablenotemark{f}& \nodata& \nodata& \nodata \\
HD 40307        &  0554$-$6001&  5   &   $P_c < 5P_b$& \nodata& \nodata& \nodata \\
Gl 229 A        &  0610$-$2151&  2   &   $P_b < 5P_c$& \nodata& \nodata& \nodata \\
HD 260655       &  0637+1733  &  2   &   $P_c < 5P_b$& \nodata& \nodata& \nodata \\
HD 265866       &  0654+3316  &  1   &   consider& 0.0126$\pm$0.0013& \nodata& 24 \\
BD+05 1668      &  0727+0513  &  2   &   $P_b < 5P_c$& \nodata& \nodata& \nodata \\
$\beta$ Gem     &  0745+2801  &  1   &   consider& 2.30$\pm$0.45& \nodata& 25 \\
L 34-26         &  0749$-$7642&  1\tablenotemark{g}  
                                     &   already included& \nodata& \nodata& \nodata  \\
CD$-$24 6144    &  0754$-$2518&  2   &   consider& 0.025$\pm$0.005& \nodata&  26\\
\nodata         &  \nodata    &  -   &   consider& 0.152$\pm$0.023& \nodata&  26\\
L 97-3 A        &  0806$-$6618&  1\tablenotemark{h}  
                                     &   already included& \nodata& \nodata& \nodata \\
L 98-59         &  0818$-$6818&  4   &   $P_c < 5P_b$& \nodata& \nodata& \nodata \\
HD 69830        &  0818$-$1237&  3   &   $P_c < 5P_b$& \nodata& \nodata& \nodata \\
L 675-81        &  0840$-$2327&  2   &   consider& 1.753$\pm$0.058& \nodata& 26\\
\nodata         &  \nodata    &  -   &   consider& 1.644$\pm$0.060& \nodata& 26\\
G 234-45        &  0841+5929  &  2   &   consider& $0.46^{+0.02}_{-0.01}$& \nodata& 27\\
\nodata         &  \nodata    &  -   &   consider& 0.20$\pm$0.01& \nodata& 27\\
$\rho^1$ Cnc A  &  0852+2819  &  5   &   $P_c < 5P_b$& \nodata& \nodata& \nodata  \\
HD 79211        &  0914+5241  &  1   &   consider& 0.0334$\pm$0.0038& \nodata& 28 \\
L 678-39        &  0936$-$2139&  3   &   $P_c < 5P_b$& \nodata& \nodata& \nodata  \\
CD$-$45 5378    &  0944$-$4546&  1   &   consider& \nodata& 0.00172$\pm$0.00025& 29 \\
HD 85512        &  0951$-$4330&  1   &   consider& 0.011$\pm$0.002& \nodata& 30 \\
BD+63 869       &  0956+6247  &  1   &   consider& 0.0721$\pm$0.0088& \nodata& 26 \\
BD+48 1829      &  1002+4805  &  1   &   consider& $0.0410^{+0.0064}_{-0.0063}$& \nodata& 31 \\
HD 87883        &  1008+3414  &  1   &   consider& \nodata& $5.37^{+0.51}_{-0.59}$& 22\\
L 320-124       &  1014$-$4709&  2   &   consider& \nodata& 0.00522$\pm$0.00072& 32\\
\nodata         &  \nodata    &  -   &   consider& 0.0083$\pm$0.0014& \nodata& 32\\
BD+01 2447      &  1028+0050  &  1   &   consider& 0.00538$\pm$0.00076& \nodata& 33 \\
47 UMa          &  1059+4025  &  3   &   $P_c < 5P_b$& \nodata& \nodata& \nodata \\
Lalande 21185   &  1103+3558  &  2   &   consider& $0.00846^{+0.00060}_{-0.00057}$& \nodata& 34\\
\nodata         &  \nodata    &  -   &   consider& $0.0428^{+0.0076}_{-0.0072}$& \nodata& 34\\
BD$-$18 3106    &  1107$-$1917&  2   &   $P_c < 5P_b$& \nodata& \nodata& \nodata \\
HD 97101 A      &  1111+3026  &  2   &   consider& $0.0239^{+0.0077}_{-0.0069}$& \nodata& 35\\
\nodata         &  \nodata    &  -   &   consider& $0.169^{+0.029}_{-0.027}$& \nodata& 35\\
HD 304043       &  1116$-$5732&  1   &   consider& 0.0348$\pm$0.0035& \nodata&  7\\
83 Leo B        &  1126+0300  &  1   &   consider& 0.070$\pm$0.050& \nodata& 11 \\
CD$-$31 9113    &  1135$-$3232&  3   &   $P_d < 5P_b$& \nodata& \nodata& \nodata \\
Ross 1003       &  1141+4245  &  2   &   consider& $0.3043^{+0.0044}_{-0.0032}$& \nodata& 36\\
\nodata         &  \nodata    &  -   &   consider& $0.214^{+0.015}_{-0.007}$& \nodata& 36\\
Ross 905        &  1142+2642  &  1   &   consider& \nodata& 0.070$\pm$0.007&  37\\
HD 102365 A     &  1146$-$4030&  1   &   consider& 0.050$\pm$0.008& \nodata& 38\\
Ross 128        &  1147+0048  &  1   &   consider& 0.00440$\pm$0.00066& \nodata&  39\\
HD 238090       &  1212+5429  &  1   &   consider& $0.0217^{+0.0029}_{-0.0030}$& \nodata&  24\\
Wolf 433        &  1238+1141  &  1   &   consider& 0.0415$\pm$0.0053& \nodata&  26\\
Wolf 437        &  1247+0945  &  1   &   consider& \nodata& $0.00887^{+0.00035}_{-0.00038}$&  40\\
BD+13 2618 AB   &  1300+1222  &  1\tablenotemark{i}  
                                     &   already included& \nodata& \nodata& \nodata \\
HD 113538       &  1304$-$5226&  2   &   $P_c < 5P_b$& \nodata& \nodata& \nodata \\
e Vir           &  1316+0925  &  1\tablenotemark{j}   
                                     &   already included& \nodata& \nodata& \nodata \\
61 Vir          &  1318$-$1818&  3   &   $P_d < 5P_c$& \nodata& \nodata& \nodata \\
Ross 1020       &  1322+2428  &  1   &   consider& 0.025$\pm$0.002& \nodata&  41\\
70 Vir          &  1328+1346  &  1   &   consider& 7.416$\pm$0.057& \nodata&  42 \\
BD+11 2576      &  1329+1022  &  1   &   consider& 0.016$\pm$0.003& \nodata& 43\\
$\tau$ Boo A    &  1347+1727  &  1   &   consider& 4.32$\pm$0.04& \nodata&  44\\
HD 122303       &  1401$-$0239&  1   &   consider& $0.0169^{+0.0022}_{-0.0020}$& \nodata& 45\\
Proxima Centauri&  1429$-$6240&  1   &   consider& 0.00337$\pm$0.00019& \nodata& 46 \\
HD 128311       &  1436+0944  &  2   &   $P_c < 5P_b$& \nodata& \nodata& \nodata \\
BD$-$07 4003    &  1519$-$0743&  3   &   $P_b < 5P_e$& \nodata& \nodata& \nodata \\
$\nu^2$ Lup     &  1521$-$4819&  3   &   $P_c < 5P_b$& \nodata& \nodata& \nodata \\
Ross 508        &  1523+1727  &  1   &   consider& 0.0126$\pm$0.0017& \nodata&  47\\
$\lambda$ Ser   &  1546+0721  &  1   &   consider& 0.0429$\pm$0.0046& \nodata&  4\\
G 180-18        &  1558+3524  &  2   &   consider& \nodata& 0.0055$\pm$0.0014& 48\\
\nodata         &  \nodata    &  -   &   consider& 0.0180$\pm$0.0030& \nodata&  48 \\
$\rho$ CrB      &  1601+3318  &  2\tablenotemark{k}   
                                     &   $P_c < 5P_b$& \nodata& \nodata& \nodata  \\
GJ 3942         &  1609+5256  &  1   &   consider& 0.0225$\pm$0.0019& \nodata& 49\\
14 Her          &  1610+4349  &  2   &   consider& \nodata& $8.1^{+1.6}_{-1.0}$& 22\\
\nodata         &  \nodata    &  -   &   consider& \nodata& $5.0^{+0.9}_{-1.0}$& 22\\
LP 804-27       &  1612$-$1852&  1   &   consider& 2.1& \nodata& 50 \\
HD 147379       &  1616+6714  &  1   &   consider& $0.0898^{+0.0047}_{-0.0046}$& \nodata& 12\\
HD 147513       &  1624$-$3911&  1   &   consider& 1.21& \nodata&  51\\
G 202-48        &  1625+5418  &  1   &   consider& 0.0089$\pm$0.0016& \nodata& 52\\
BD$-$12 4523    &  1630$-$1239&  3   &   $P_c < 5P_b$& \nodata& \nodata& \nodata \\
BD+25 3173      &  1658+2544  &  1   &   consider& 0.328$\pm$0.032& \nodata&  53\\
HD 154345       &  1702+4704  &  1   &   consider& 0.82$\pm$0.07& \nodata& 11\\
HD 154088       &  1704$-$2834&  1   &   consider& 0.021$\pm$0.003& \nodata& 54 \\
G 139-21        &  1715+0457  &  1   &   consider& \nodata& 0.0257$\pm$0.0014& 55 \\
BD+11 3149      &  1716+1103  &  2   &   consider& 0.00777$\pm$0.00085& \nodata& 56\\
\nodata         &  \nodata    &  -   &   consider& $0.0197^{+0.0025}_{-0.0024}$& \nodata&  56\\
HD 156384 C     &  1718$-$3459&  5   &   $P_c < 5P_b$& \nodata& \nodata& \nodata \\
CD$-$46 11540   &  1728$-$4653&  1   &   consider& 0.035& \nodata& 57\\
CD$-$51 10924   &  1730$-$5138&  4   &   consider& 0.0120$\pm$0.0010& \nodata& 22\\
\nodata         &  \nodata    &  -   &   consider& $0.0211^{+0.0022}_{-0.0020}$& \nodata& 22\\
\nodata         &  \nodata    &  -   &   consider& \nodata& $5.78^{+0.48}_{-0.46}$& 22\\
\nodata         &  \nodata    &  -   &   already included\tablenotemark{l}
                                     & \nodata& 13.43$\pm$1.1& 22\\
G 226-66        &  1735+6140  &  1   &   consider& $0.0283^{+0.0053}_{-0.0057}$& \nodata& 58\\
BD+68 946       &  1736+6820  &  2   &   consider& 0.0541$\pm$0.0031& \nodata& 26\\
\nodata         &  \nodata    &  -   &   consider& 0.050$\pm$0.013& \nodata& 26\\
CD$-$44 11909   &  1737$-$4419&  2   &   $P_c < 5P_b$& \nodata& \nodata& \nodata \\
BD+18 3421      &  1737+1835  &  1   &   consider& 0.0208$\pm$0.0014& \nodata&  59\\
$\mu$ Ara       &  1744$-$5150&  4   &   $P_b < 5P_e$& \nodata& \nodata& \nodata \\
BD+45 2743 A    &  1835+4544  &  1   &   consider& 0.0429$\pm$0.0025& \nodata& 60\\
HD 176029       &  1858+0554  &  1   &   consider& $0.0093^{+0.0016}_{-0.0015}$& \nodata& 61 \\
HD 177565       &  1906$-$3748&  1   &   consider& $0.048^{+0.020}_{-0.019}$& \nodata& 62\\
HD 180617       &  1916+0510  &  1   &   consider& 0.0384$\pm$0.0033& \nodata& 59\\
HD 189733       &  2000+2242  &  1   &   consider& \nodata& 1.13$\pm$0.08&  11\\
HD 190007       &  2002+0319  &  1   &   consider& 0.0518$\pm$0.0052& \nodata& 59\\
HD 190360       &  2003+2953  &  2   &   consider& 0.0600$\pm$0.0076& \nodata&  63\\
\nodata         &  \nodata    &  -   &   consider& \nodata& 1.8$\pm$0.2& 64\\
HD 189567       &  2005$-$6719&  2   &   $P_c < 5P_b$& \nodata& \nodata& \nodata \\
HD 192263       &  2013$-$0052&  1   &   consider& 0.56$\pm$0.05& \nodata&  11\\
HD 192310       &  2015$-$2701&  2   &   consider& 0.0532$\pm$0.0028& \nodata& 30\\
\nodata         &  \nodata    &  -   &   consider& 0.076$\pm$0.016& \nodata& 30\\
AU Mic          &  2045$-$3120&  2   &   $P_c < 5P_b$& \nodata& \nodata& \nodata \\
LSPM J2116+0234 &  2116+0234  &  1   &   consider& $0.0418^{+0.0031}_{-0.0035}$& \nodata& 65\\
LSPM J2122+2255 &  2122+2255  &  1   &   consider& 0.33$\pm$0.02& \nodata&  10\\
HD 204961       &  2133$-$4900&  2\tablenotemark{m}   
                                     &   consider& 0.68$\pm$0.09& \nodata& 66\\
HN Peg AB       &  2144+1446  &  1\tablenotemark{n} 
                                     &   already included& \nodata& \nodata& \nodata \\
G 264-12        &  2146+6648  &  2   &   $P_c < 5P_b$& \nodata& \nodata& \nodata \\
$\epsilon$ Ind A&  2203$-$5647&  1   &   consider& \nodata& $3.25^{+0.39}_{-0.65}$&  67\\
BD$-$05 5715    &  2209$-$0438&  2   &   $P_c < 5P_b$& \nodata& \nodata& \nodata \\
L 788-37        &  2213$-$1741&  1   &   consider& 0.023$\pm$0.002& \nodata& 41\\
HD 211970       &  2222$-$5433&  1   &   consider& 0.0409$\pm$0.0079& \nodata& 68\\
L 119-213       &  2241$-$6910&  1   &   consider\tablenotemark{o}
                                                     & \nodata& \nodata& \nodata \\
HD 216520       &  2247+8341  &  2   &   $P_c < 5P_b$& \nodata& \nodata& \nodata  \\
BD$-$15 6290 AB &  2253$-$1415&  4   &   $P_b < 5P_c$& \nodata& \nodata& \nodata \\
51 Peg          &  2257+2046  &  1   &   consider& 0.472$\pm$0.039& \nodata&  18\\
HD 217987       &  2305$-$3551&  2   &   $P_c < 5P_b$& \nodata& \nodata& \nodata \\
TRAPPIST-1      &  2306$-$0502&  7   &   $P_c < 5P_b$& \nodata& \nodata& \nodata \\
HD 219134       &  2313+5710  &  6   &   $P_c < 5P_b$& \nodata& \nodata& \nodata \\
$\gamma$ Cep AB &  2339+7737  &  1   &   consider& 1.85$\pm$0.16& \nodata&  69\\	
\enddata
\tablenotetext{a}{Indicates whether the "exoplanets" in this system should be considered in our stellar mass function analysis. Notes regarding period violations indicate objects that are excluded, as these are likely objects formed via a protoplanetary disk. In these cases, we list at least one example of component pairs ("b" vs.\ "c", "c" vs.\ "d", etc.) that violate the rule.}
\tablenotetext{b}{The "b" planet is currently considered controversial by the NASA Exoplanet Archive, so only the "c" planet is considered for inclusion in our mass function analysis.}
\tablenotetext{c}{This object was reported to have two exoplanets at the time of our original query to the NASA Exoplanet Archive, but it is now believed that the "b" component was a false positive.}
\tablenotetext{d}{This exoplanet, aka HR 810b, has a period from radial velocity variations of 302.8$\pm$2.3d (\citealt{stassun2017}) and is presumably the same object listed in Tabe~\ref{tab:20pc_census} as $\iota$ Hor B, the companion giving rise to the 331.7d orbital period in the Gaia DR3 non-single star table.}
\tablenotetext{e}{This transiting exoplanet, aka TOI-540 b, has no mass measurement.}
\tablenotetext{f}{The inner and outer planets do not violate the $P_{outer} < 5P_{inner}$ rule, but the imaging of this youthful system nonetheless shows the debris disk remaining from protoplanetary formation.}
\tablenotetext{g}{This "exoplanet" is the T dwarf WISEPA J075108.79$-$763449.6, already included as its own row in Table~\ref{tab:20pc_census}.}
\tablenotetext{h}{This "exoplanet" is the Y dwarf L 97-3 B (sometimes referred to as WD 0806$-$661 B), already included as its own row in Table~\ref{tab:20pc_census}.}
\tablenotetext{i}{This "exoplanet" is the T dwarf ULAS J130041.74+122114.7, already included as its own row in Table~\ref{tab:20pc_census}.}
\tablenotetext{j}{This "exoplanet" is the late-T/early-Y dwarf e Vir Ab, already included as its own row in Table~\ref{tab:20pc_census}.}
\tablenotetext{k}{This assumes that the exoplanet candidate $\rho$ CrB b is real.}
\tablenotetext{l}{In Table~\ref{tab:20pc_census}, we label the object causing the Gaia DR3 acceleration (from the non-single star list) as CD$-$51 10924 B. For the purposes of accounting, we will equate that object with this one, GJ 676 A c.}
\tablenotetext{m}{The "c" planet is currently considered controversial by the NASA Exoplanet Archive, so only the "b" planet is considered for inclusion in our mass function analysis.}
\tablenotetext{n}{This "exoplanet" is the T dwarf HN Peg B, already included as its own row in Table~\ref{tab:20pc_census}.}
\tablenotetext{o}{This object has no mass determination.}
\tablerefs{References for the mass measurements:
1 = \cite{naef2001},
2 = \cite{pinamonti2018},
3 = \cite{wittenmyer2019}, 
4 = \cite{rosenthal2021},
5 = \cite{lillo-box2020},
6 = \cite{perger2019},
7 = \cite{feng2020},
8 = \cite{curiel2011},
9 = \cite{marmier2013},
10 = \cite{quirrenbach2022},
11 = \cite{stassun2017},
12 = \cite{hobson2018},
13 = \cite{soto2021},
14 = \cite{bauer2020},
15 = \cite{llop-sayson2021},
16 = \cite{lillo-box2021},
17 = \cite{ma2018},
18 = \cite{butler2006},
19 = \cite{howard2010},
20 = \cite{astudillo-defru2017},
21 = \cite{anglada-escude2014},
22 = \cite{feng2022},
23 = \cite{hatzes2022},
24 = \cite{stock2020},
25 = \cite{hatzes2006},
26 = \cite{feng2020b},
27 = \cite{lopez-santiago2020},
28 = \cite{ditomasso2023},
29 = \cite{lam2021},
30 = \cite{pepe2011},
31 = \cite{hobson2019},
32 = \cite{bonfils2018},
33 = \cite{amado2021},
34 = \cite{hurt2022},
35 = \cite{dedrick2021},
36 = \cite{trifonov2018},
37 = \cite{maciejewski2014},
38 = \cite{tinney2011},
39 = \cite{bonfils2018b},
40 = \cite{trifonov2021},
41 = \cite{luque2018},
42 = \cite{luhn2019},
43 = \cite{damasso2022},
44 = \cite{borsa2015},
45 = \cite{suarez-mascareno2017},
46 = \cite{faria2022},
47 = \cite{harakawa2022},
48 = \cite{beard2022},
49 = \cite{perger2017},
50 = \cite{apps2010},
51 = \cite{mayor2004},
52 = \cite{suarez-mascareno2017b},
53 = \cite{johnson2010},
54 = \cite{unger2021},
55 = \cite{cloutier2021},
56 = \cite{affer2016},
57 = \cite{bonfils2007},
58 = \cite{pinamonti2019},
59 = \cite{burt2021},
60 = \cite{gonzalez-alvarez2021},
61 = \cite{toledo-padron2021},
62 = \cite{feng2017},
63 = \cite{wright2009},
64 = \cite{feng2021},
65 = \cite{lalitha2019},
66 = \cite{wittenmyer2014},
67 = \cite{feng2019},
68 = \cite{feng2019b},
69 = \cite{endl2011}.
}
\end{deluxetable*}

\clearpage

\subsection{Objects on the main sequence\label{sec:mass_estimates_main_sequence}}

Main sequence objects with directly measured masses can be used to calibrate relations of mass vs.\ absolute magnitude or mass vs.\ spectral type. Studies have shown that the relation with the smallest intrinsic scatter for K and M dwarfs is the one using absolute $K$-band magnitude (\citealt{delfosse2000}). The fact that the $K$-band relation shows the least scatter across the optical to near-infrared range is also predicted by model atmospheres, as this is the wavelength regime where competing physical effects modulated by metallicity variations largely cancel one another (\citealt{delfosse2000,mann2019}). More (and improved) dynamical mass measurements of binary stars\footnote{In reality, the total mass of the binary system was used, rather than the masses of the individual components, the latter of which are not generally known. The resulting relation is nonetheless applicable to individual objects, as sections 7.3 and 7.4 of \cite{mann2019} discuss in detail. See in particular their figure 15, which shows a direct comparison between directly measured individual masses and the resulting mass-$M_{Ks}$ relation.} along with improved Gaia parallaxes have enabled \cite{mann2019} to construct a mass vs.\ $M_{Ks}$ relation that results in estimated masses with only 2-3\% uncertainty. Specifically, $K_s$ is used because 2MASS provides all-sky coverage at this band. We use the \cite{mann2019} mass-$M_{Ks}$ relation (their equation 2) over the range $5.0 \le M_{Ks} \le 11.0$, roughly corresponding to spectral types from early-M to late-M\footnote{Their relation covers the range from $M_{Ks} \approx 4.0$ mag down to a spectral type of $\sim$L1. However, the bright end of the relation is poorly constrained, so for these objects we defer to the TIC mass estimates (see the next paragraph). At the faint end, we use the effective temperature analysis for objects cooler than M9.5, as further described in Section~\ref{sec:analysis_brown_dwarfs}.}. These estimates and their propagated uncertainties are listed in columns "EstMassMKs" and "EstMassMKsErr" of Table~\ref{tab:20pc_census}.

For other main sequence stars, we can use the methodology employed by \cite{stassun2019}. Using $\sim$20,000 (non-reddened) stars within 100 pc of the Sun with spectroscopically determined effective temperatures, they established a relation between $T_{\rm eff}$ and $G_{BP}-G_{RP}$ color. This is then mated with the results of \cite{torres2010} that relate $T_{\rm eff}$ to mass for stars with dynamically measured masses (\citealt{stassun2018b}). This gives mass estimates with uncertainties of ${\sim}6.4\%$ (\citealt{stassun2019}). We take mass estimates and their uncertainties directly from the revised TESS Input Catalog (TIC; \citealt{stassun2019}) for stars in our Table~\ref{tab:20pc_census}. These values are listed in columns "EstMassTIC" and "EstMassTICErr". We note, however, that the \cite{stassun2019} prescription for stars with $T_{\rm eff} \lesssim 4000$K (see their appendix A.1 along with \citealt{muirhead2018}) followed a different methodology. For these objects, masses were estimated using $K_s$ magnitudes, Gaia DR2 parallaxes, and the \cite{mann2019} mass-vs.-$M_{Ks}$ relation. 

\begin{figure}
\includegraphics[scale=0.325,angle=0]{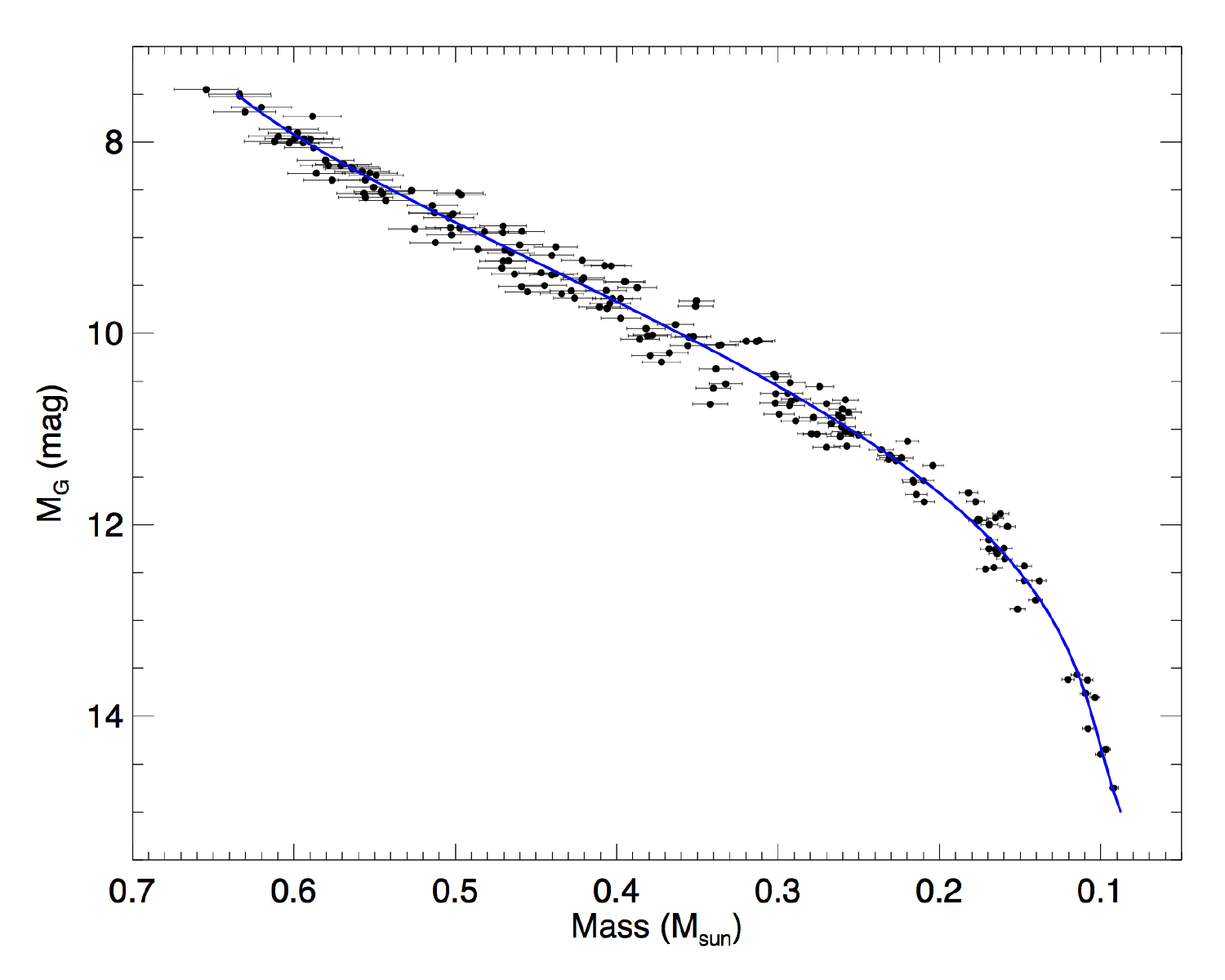}
\caption{Absolute G-band magnitude plotted against estimated mass for 180 well studied late-K and M dwarfs from \cite{mann2015}. The solid blue line shows our fitted relation from Equation~\ref{eqn:mass_vs_MG}. See text for details.
\label{fig:mass_vs_MG}}
\end{figure}

Some main sequence stars lack both $K_s$ magnitudes and an entry in the TESS Input Catalog. For these, we resort to two other estimation methods. The first is the mass vs.\ $M_G$ relation. \cite{chontos2021} took a list of well-studied late-K and M dwarfs (tables 5-7 from \citealt{mann2015}) and refined their mass estimates using more precise Gaia DR3 parallaxes and the \cite{mann2019} mass-vs.-$M_{Ks}$ relation from above. They derived a relation between this estimated mass and the absolute G-band magnitude. However, the coefficients in \cite{chontos2021} are published with insufficient accuracy to re-create the relation show in their figure 7, so we have re-derived them here. Our methodology is identical to theirs except that we exclude the sdM3 object L 750-42 (\citealt{gizis1997}) and do not incorporate a dependence on metallicity because the metallicity has not been measured for most of the M dwarfs within 20 pc. Using a functional form of 
\begin{equation}
    {\rm Mass} = \sum_{i=0}^{4}c_i(M_G-10.5)^i,
    \label{eqn:mass_vs_MG}
\end{equation} 
where $M_G$ is the G-band absolute magnitude in magnitudes and Mass is in units of $M_\odot$, we find best-fit coefficients of $c_0 = 0.30548$, $c_1 = -0.10588$, $c_2 = 0.011471$, $c_3 = 0.0021352$, and $c_4 = -0.00041023$. Our fit is illustrated in Figure~\ref{fig:mass_vs_MG}. The relation is valid from $7.5 \le M_G \le 15.0$ (spectral types from $\sim$K7 to $\sim$M8). For uncertainty propagation, we adopt the \cite{chontos2021} practice of a 2.2\% uncertainty added in quadrature to the $\sim$3\% uncertainty inherent to the \cite{mann2019} relation. This mass-$M_G$ relation is particularly useful for estimating masses of individual components of close double systems that are currently resolved only by Gaia. In Table~\ref{tab:20pc_census} we provide columns labeled "EstMassMG" and "EstMassMGErr" listing mass estimates for all objects for which these $M_G$-based estimates can be computed.

The second alternative estimation method is StarHorse (\citealt{anders2022}), which uses Gaia EDR3 data cross-matched to photometry from Pan-STARRS1, SkyMapper, 2MASS, and AllWISE to estimate stellar parameters from stellar isochrones (from PARSEC 1.2S; \citealt{marigo2017}) providing the closest match. When \cite{anders2022} mass estimates are available, these are listed in columns "EstMassSH" and "EstMassSHErr" of Table~\ref{tab:20pc_census}. These published mass uncertainties can be anomalously low compared to the other estimates discussed in this section because they pertain only to the internal model errors and do not include the systematic component coming from a model-to-truth comparison.

\begin{figure*}
\includegraphics[scale=0.5,angle=0]{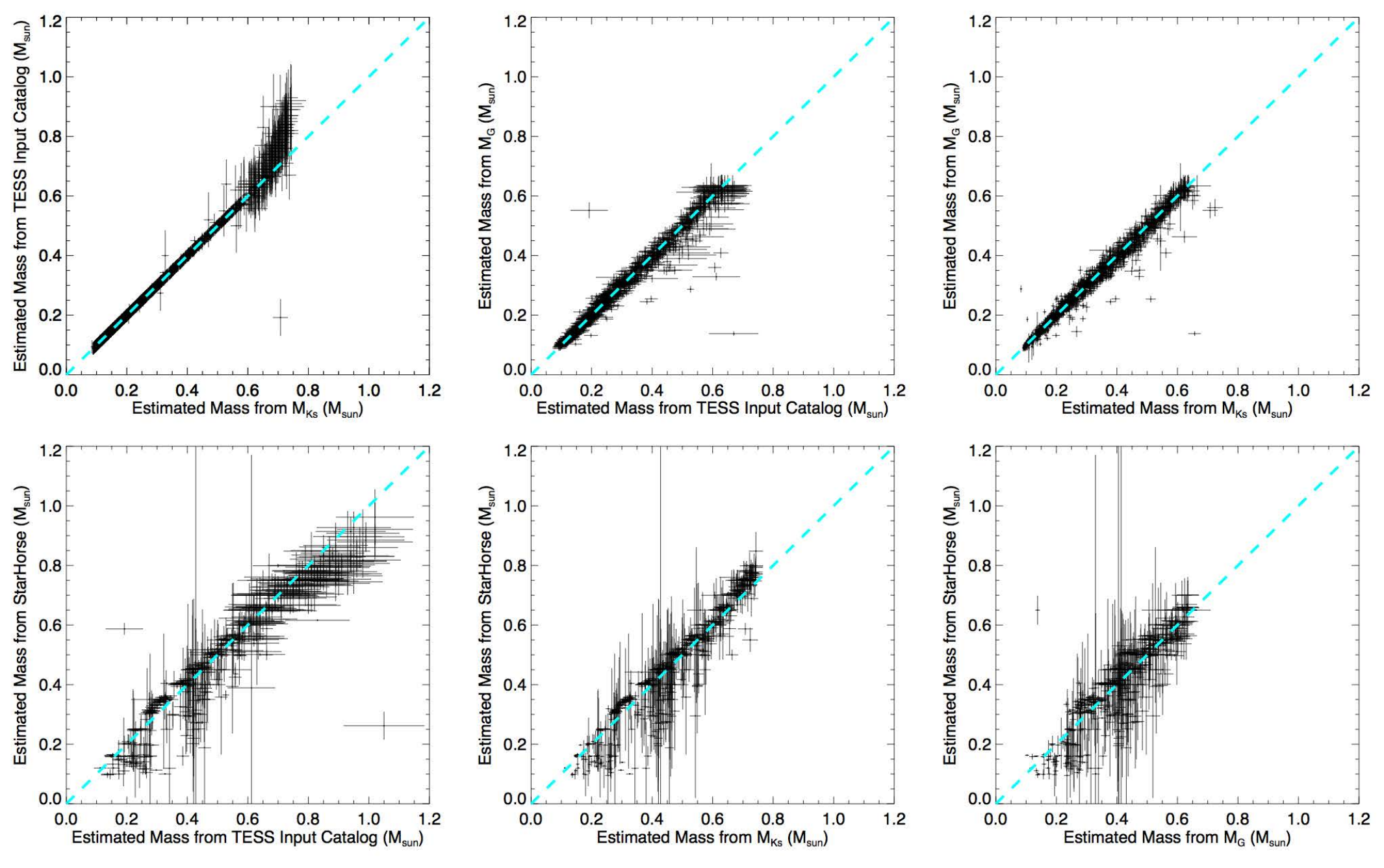}
\caption{Intercomparisons of results from our four mass estimation techniques. The line of one-to-one correspondence is shown by the blue dashes. See text for details.\label{fig:mass_estimate_intercomparisons}}
\end{figure*}

Figure~\ref{fig:mass_estimate_intercomparisons} shows the four estimation techniques compared to each other. The top three panels show the intercomparisons between the TESS Input Catalog estimates, the $M_{Ks}$ estimates, and the $M_G$ estimates. As these are all based on the same underlying mass vs.\ $M_{Ks}$ relation of \cite{mann2019}, the correspondence is generally excellent. (In fact, the correspondence between the TESS Input Catalog estimates and estimates from our $M_{Ks}$ technique are nearly perfect, differing only in the Gaia data release from which the parallax values were obtained.) The only deviation is for masses greater than $\sim0.65 M_\odot$ in the comparison between the TESS Input Catalog values and those derived from $M_{Ks}$, where the difference can be as large as 13\%.

The bottom three panels of Figure~\ref{fig:mass_estimate_intercomparisons} show small systematics between the three estimation techniques above and StarHorse. As stated above, masses from StarHorse are based on theoretical models, so such systematics might be expected between theory and observation. At masses of $\sim0.3 M_\odot$, StarHorse tends to overpredict (by $\sim$10\%) the mass relative to the other techniques, and at smaller masses may significantly underpredict (by $\sim$35\%). At masses near $0.8 M_\odot$, a small underprediction (by $<$5\%) relative to the TESS Input Catalog becomes an overprediction (by $\sim$5\%) relative to masses from the $M_{Ks}$ relation. At masses closer to $1.0 M_\odot$, StarHorse leads to underpredictions (by $\sim$10\%) relative to estimates from the TESS Input Catalog.

Given that systematic offsets of up to 15\% are seen even between the sets with empirical underpinnings, we are reluctant to apply corrections to offsets smaller than this value. The only exception to this is the $\sim$35\% offset seen for StarHorse estimates below StarHorse values of $\sim0.275 M_\odot$. In this case, rather than applying an offset, we will simply not use any StarHorse estimates below $0.275 M_\odot$.

\section{Further Analysis\label{sec:further_analysis}}

For each individual object ("\#CompsOnThisRow" = 1) in Table~\ref{tab:20pc_census}, we have adopted a mass and its uncertainty. These are listed in columns "AdoptedInitialMass" and "AdoptedInitialMassErr" along with an additional column, "AdoptedInitialMassNote", indicating the origin of the data from elsewhere in the table. These are labeled with the term "Initial" as a reminder that for white dwarfs, we need their initial masses on the main sequence; for all other objects, their current masses are assumed identical to their initial masses. The codes for "AdoptedInitialMassNote" are as follows, listed in their order of selection:

\begin{itemize}
    \item {\it wd IFMR, wd low, wd ultra-low}, or {\it wd conjecture}: The initial mass and its uncertainty have been computed via the initial-to-final mass relation or other means (see Table~\ref{tab:wd_masses}), if this object is a white dwarf.
    \item {\it measured}: Directly measured mass values from "Mass" and "MassErr" are used. The methodology used and its reference are listed in columns "MassMethod" and "MassRef". (For L, T, and Y dwarfs, directly measured masses are not retained because these are estimated in bulk through statistical means; see the Teff bullet, below.)
    \item {\it M\_Ks}: The mass and its uncertainty from the \cite{mann2019} $M_{Ks}$ relation ("EstMassMKs" and "EstMassMKsErr") are used. 
    \item {\it TIC}: The mass and its uncertainty from the TESS Input Catalog (\citealt{stassun2019}; "EstMassTIC" and "EstMassTICErr") are used.
    \item {\it M\_G}: The mass and its uncertainty from the $M_G$ relation of Equation~\ref{eqn:mass_vs_MG} ("EstMassMG" and "EstMassMGErr") are used.
    \item {\it SH}: The mass and its uncertainty from StarHorse (\citealt{anders2022}; "EstMassSH" and "EstMassSHErr") are used, unless that estimate falls below $0.275 M_\odot$ (see section~\ref{sec:mass_estimates_main_sequence}).
    \item {\it literature}: The mass and its uncertainty are taken from columns "EstMassLit" and "EstMassLitErr", the mass estimation method and reference for which are listed in "EstMassLitMethod" and "EstMassLitRef". (Literature values can supersede other values above if the object is listed as a giant or subgiant in Table~\ref{tab:giants_subgiants}.)
    \item {\it see GeneralNotes}: For objects with this code, the mass and its uncertainty were computed by us, as detailed in the "GeneralNotes" column of the table.
    \item {\it Teff}: For objects of type L, T, or Y, individual masses are not computed. These are handled statistically via the distribution of $T_{\rm eff}$ values and their uncertainties ("Teff" and "Teff\_unc"), as described in detail below.
\end{itemize}

For cases in which literature values did not list a mass uncertainty, a value of 10\% is  arbitrarily assumed. The quoted StarHorse uncertainty is also replaced with a 10\% uncertainty, based on the under- and over-predictions noted when comparing StarHorse values to other estimates (see discussion at end of Section~\ref{sec:mass_estimates_main_sequence}), unless the quoted StarHorse internal uncertainty is already larger, in which case we retain the published value.

\begin{figure}
\includegraphics[scale=0.45,angle=0]{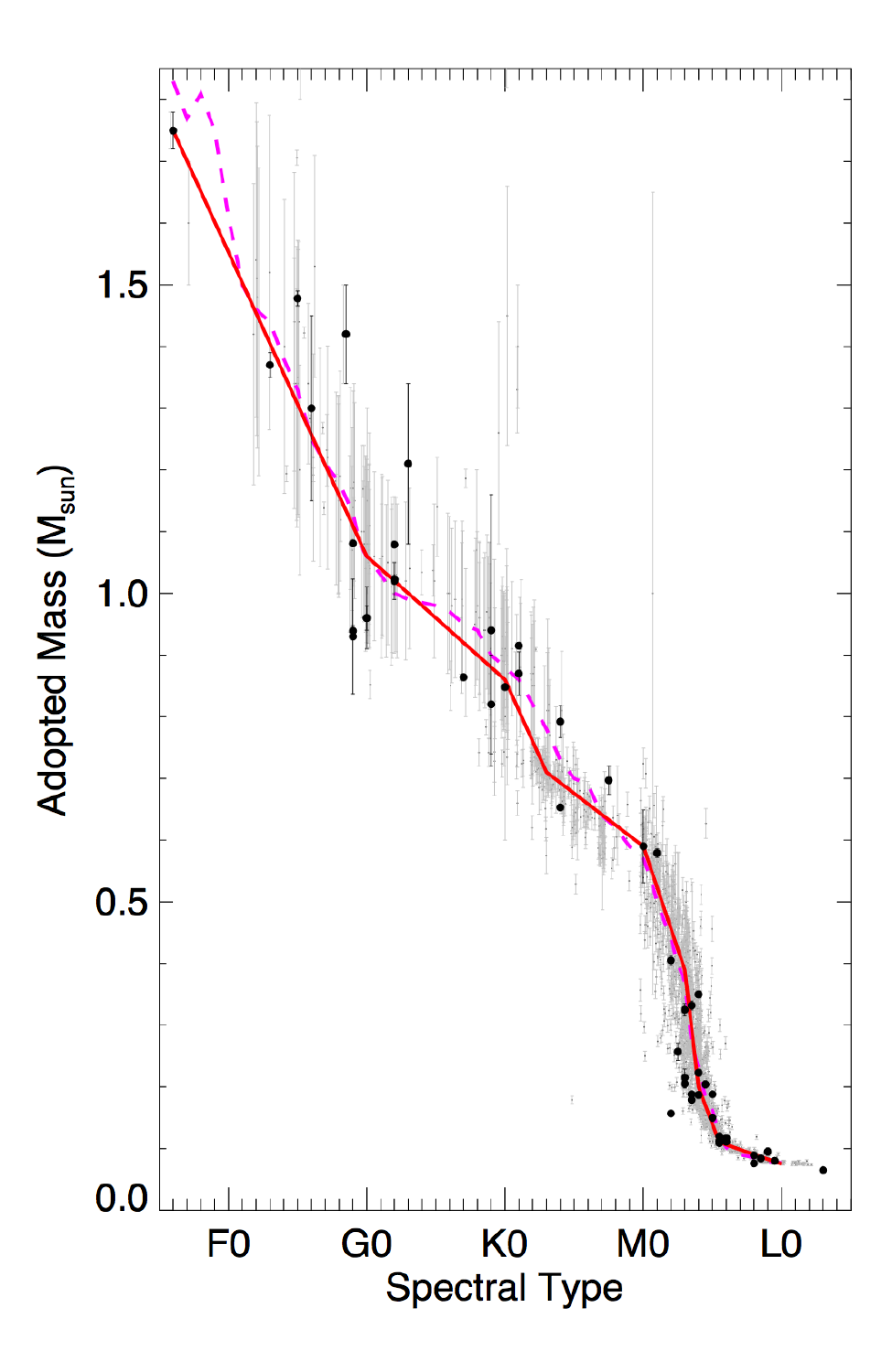}
\caption{Mass as a function of spectral type for 20-pc objects with measured (black points) or estimated (grey points) masses and optical spectral types in Table~\ref{tab:20pc_census}. The adopted initial mass (see text) is used for each object. For objects with estimated (not measured) masses, a random value between $-$0.25 and +0.25 has been added to the spectral type to better visualize otherwise overlapping data points. Our piecewise fit to the relation for dwarf stars is shown by the solid red line and is quantified in Table~\ref{tab:piecewise_mass_vs_spectype}. For comparison, we show the average mass per spectral type as tabulated in the 2022.04.16 version of \url{https://www.pas.rochester.edu/~emamajek/EEM_dwarf_UBVIJHK_colors_Teff.txt} (\citealt{pecaut2013}; magenta dashed line).\label{fig:mass_vs_spectype}}
\end{figure}

For cases where only a miscellaneous magnitude or delta magnitude of a companion were available, it is instructive to estimate a spectral type for the object in order to estimate its mass. Figure~\ref{fig:mass_vs_spectype} shows a comparison between masses and measured spectral types for those Table~\ref{tab:20pc_census} objects having mass estimates (or direct measures) from one of the other methods. The piecewise fit shown in the figure is the one we use to translate a dwarf spectral type estimate into a mass estimate. Other per-object details can be found in the General Notes column of Table~\ref{tab:20pc_census}.

\begin{deluxetable}{ccc}
\tabletypesize{\scriptsize}
\tablecaption{Piecewise Fit to Mass vs.\ Dwarf Spectral Type Relation\label{tab:piecewise_mass_vs_spectype}}
\tablehead{
\colhead{Spectral Type} &
\colhead{Spectral Index} &
\colhead{Mass ($M_\odot$)} \\
\colhead{(1)} &                          
\colhead{(2)} &
\colhead{(3)} 
}
\startdata
A6  &  6.0& 1.75\\
G0  & 20.0& 1.06\\
K0  & 30.0& 0.86\\
K3  & 33.0& 0.71\\
M0  & 40.0& 0.59\\
M3  & 43.0& 0.39\\
M4  & 44.0& 0.20\\
M5.5& 45.5& 0.11\\
L0  & 50.0& 0.075 \\
\enddata
\tablecomments{Each row in this table represents an inflection point in the red, piecewise fit of Figure~\ref{fig:mass_vs_spectype}.}
\end{deluxetable}

\subsection{Analysis of brown dwarfs\label{sec:analysis_brown_dwarfs}}

We use the methodology adopted by \cite{kirkpatrick2019,kirkpatrick2021} and Raghu et al.\ (submitted) to determine the mass function for L, T, and Y dwarfs, most of which are brown dwarfs lacking any color (or spectral type or absolute magnitude) to mass correlation. Specifically, the mass function for these objects is determined by comparing the distribution of present-day temperatures to predicted temperature distributions. Predictions are drawn from a grid of models with varying mass functions, birthrates, and low-mass cutoffs. For each point in the grid, we build a predicted mass/age distribution that is then passed through a set of evolutionary models to predict the current-day $T_{\rm eff}$ distribution. Using this grid of predictions allows us to find the combination of mass function, birthrate, and cutoff mass that best fits the observed temperature distribution.

\begin{figure*}
\includegraphics[scale=0.80,angle=0]{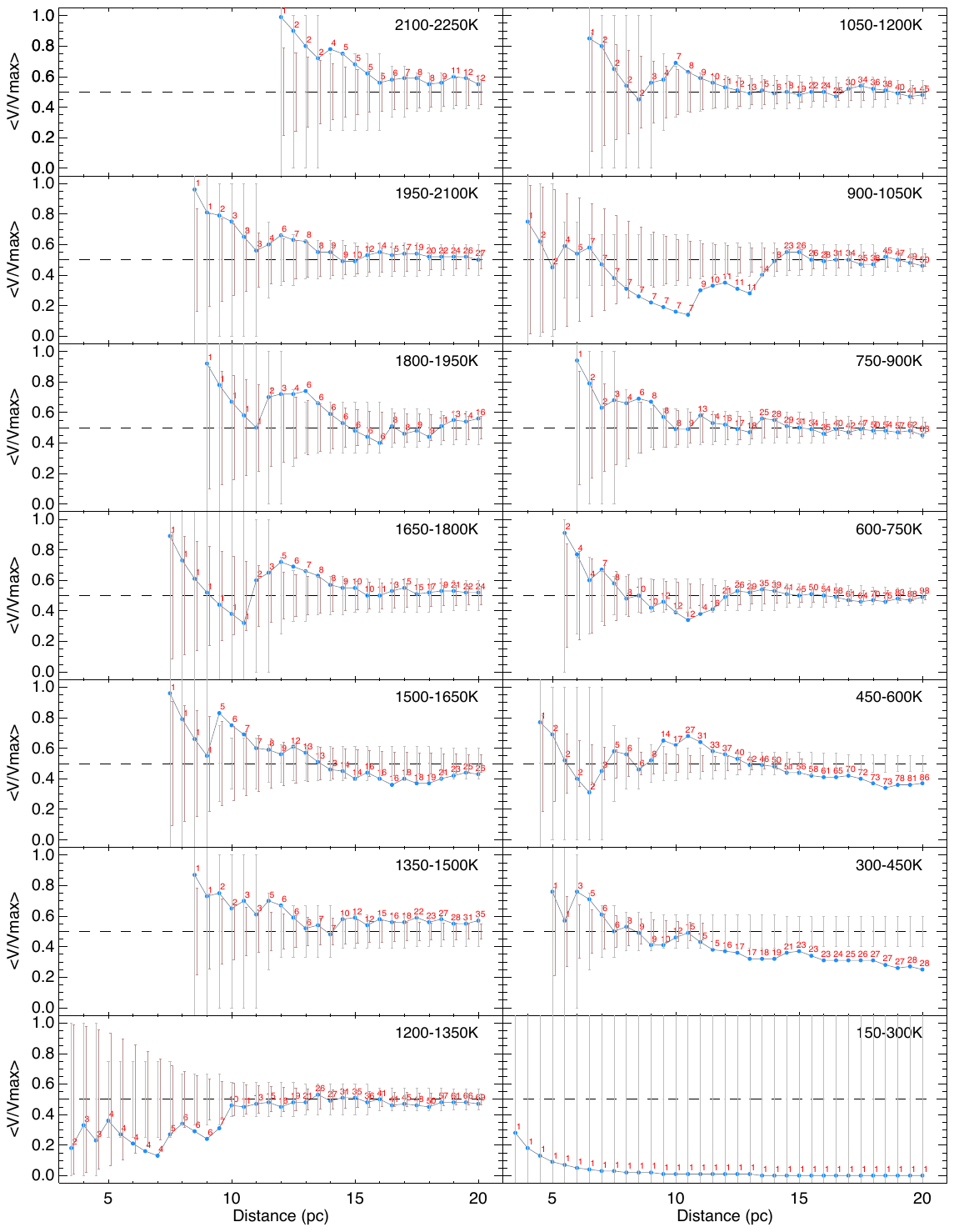}
\caption{The average $V/V_{\rm max}$ value in 0.5-pc intervals across fourteen 150-K bins encompassing L, T, and Y dwarfs. Blue dots show the empirical sample, and red labels denote the number of objects at each 0.5-pc computation. The black dashed line shows the $\langle{V}/{V_{max}}\rangle = 0.5$ level indicative of a complete sample. The grey error bars show the approximate 1$\sigma$ range that a sample of the size shown in red would exhibit, given random statistics. The brown error bars, offset by +0.05 pc from the grey error bars for clarity, show the 1$\sigma$ variation obtained by simulations using 10,000 Monte Carlo realizations having the number of objects and completeness limit listed in Table~\ref{tab:space_densities_brown_dwarfs}. See section 8.2 of \cite{kirkpatrick2021} for more details. 
\label{fig:v_over_vmax}}
\end{figure*}

For the empirical distribution, we estimate the $T_{\rm eff}$ value for each L, T, or Y dwarf (see Table~\ref{tab:brown_dwarf_additions_subtractions} in this paper and table 11 of \citealt{kirkpatrick2021}) and then calculate space densities as a function of $T_{\rm eff}$. To compute space densities, we need to determine the distances at which our brown dwarf subsamples are truly complete, as the coldest Y dwarfs are so intrinsically dim that we are unable to push their completeness to the 20-pc limit targeted in this paper. As described in \cite{kirkpatrick2021}, we determine completeness via the $V/V_{max}$ test (\citealt{schmidt1968}) using 150K bins and computing $\langle{V}/{V_{max}}\rangle$ at half-parsec steps within each bin. The computation starts with the first half-parsec step falling just larger than the distance of the bin's nearest object and advances in distance out to $d = 20$ pc. These results are shown in Figure~\ref{fig:v_over_vmax}.

A comparison of this figure to figure 23 of \cite{kirkpatrick2021} shows that, despite the many new discoveries (and many fewer retractions) noted in Table~\ref{tab:brown_dwarf_additions_subtractions}, each 150K bin has the same completeness limit as before. As one example, consider the bin with the largest change, 600-750K. In both \cite{kirkpatrick2021} and here, this bin is complete out to 20 pc, but the number of objects has nonetheless increased from eighty-three in \cite{kirkpatrick2021} to ninety-eight in this paper; see also Table~\ref{tab:space_densities_brown_dwarfs}. (The $V/V_{max}$ test is only as robust as the Poisson statistics allow, which is why both sets of numbers were deemed to be complete.) As another example, the number of objects interior to the completeness limit of 15.0 pc in the 450-600K bin has increased from fifty-three to fifty-six. 

As noted in \cite{kirkpatrick2021}, the $V/V_{max}$ test does not check for inhomogeneities in surface area, the most likely cause of which would be confusion along the Galactic plane that hinders our ability to find nearby brown dwarfs. Do the increased densities now reported in this paper indicate that these corrections can be reduced or dropped altogether?

\begin{figure*}
\includegraphics[scale=0.425,angle=0]{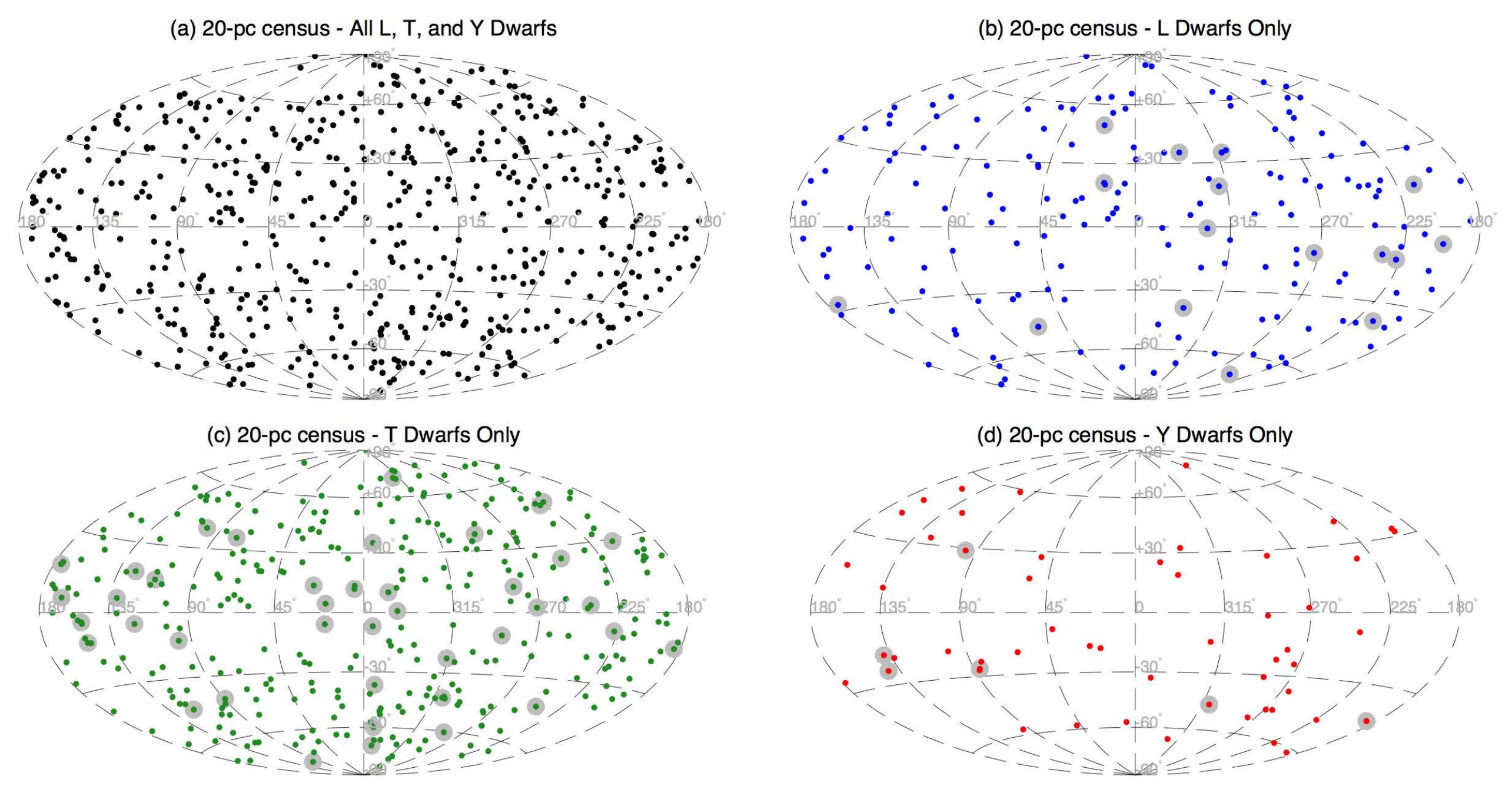}
\caption{Plots of the 20-pc L, T, and Y dwarf census in Galactic coordinates. The four panels display (a) the sample in its entirety (black), (b) only the L dwarfs (blue), (c) only the T dwarfs (green), and (d) only the Y dwarfs (red). New additions to the sample since \cite{kirkpatrick2021} are plotted with grey haloes in panels (b) through (d).
\label{fig:brown_dwarf_census_sky_plots}}
\end{figure*}

Figure~\ref{fig:brown_dwarf_census_sky_plots} shows the positions in Galactic coordinates of all 583 L, T and Y dwarfs in the 20-pc census. As was done in \cite{kirkpatrick2021}, we divide the sky into two zones: a zone along the Galactic plane ($|glat| < 14{\fdg}48$) and another ($|glat| \ge 14{\fdg}48$) well outside of the plane. This value of $|glat|$ was chosen so that the non-plane zone contains exactly three times the area of the plane zone. If there is no incompleteness along the Galactic plane, then the ratio of non-plane to plane objects should be three. For the volume-complete portions of our 20-pc census, we find that this ratio is $138/44 = 3.1$ for L dwarfs, $257/65 = 4.0$ for T dwarfs, and $31/4 = 7.8$ for Y dwarfs, suggesting that the Galactic plane does not introduce any significant incompleteness ($< 1$\%) for L dwarfs but does still impede the discovery of fainter T and Y dwarfs. In contrast, \cite{kirkpatrick2021} derived ratios of $137/34 = 4.0$, $234/34 = 6.9$, and $24/4 = 6.0$ for the L, T and Y dwarf samples, respectively. 

Incompleteness along the Galactic plane has improved in the current 20-pc census for the L and T dwarfs. For the Y dwarfs, the view is complicated by smaller number statistics. Taking the non-plane numbers of Y dwarfs as truth, then the number of plane Y dwarfs in the current sample should be $31/3 \pm (\sqrt{31})/3 = 10.3 \pm 1.9$, which is $3.3\sigma$ different from the value of 4 actually found. The same computation for the \cite{kirkpatrick2021} numbers gives a number of plane Y dwarfs that was only $2.5\sigma$ different. Hence, the under-density of Y dwarfs in the plane is now significantly worse, due to the fact that all new discoveries of Y dwarfs within the volume have been found outside of the plane zone.

\begin{deluxetable}{ccccccc}
\tablecaption{Space Densities for Early-L through Early-Y Dwarfs\label{tab:space_densities_brown_dwarfs}}
\tablehead{
\colhead{T$_{\rm eff}$} & 
\colhead{Complete-} &
\colhead{Raw} &
\colhead{Corr.} &
\colhead{Adjusted} &
\colhead{Adopted} \\
\colhead{Bin} & 
\colhead{ness Limit} &
\colhead{No.\ of} &
\colhead{Factor\tablenotemark{a}} &
\colhead{No.\ of} &
\colhead{Space Density\tablenotemark{b}} \\
\colhead{(K)} & 
\colhead{(pc)} &
\colhead{Objects} &
\colhead{} &
\colhead{Objects} &
\colhead{($\times$10$^{-3}$ pc$^{-3}$)} \\
\colhead{} &                          
\colhead{$d_{max}$} &
\colhead{$raw$} &
\colhead{$corr$} &
\colhead{$adj$} &
\colhead{$dens$} \\
\colhead{(1)} &                          
\colhead{(2)} &
\colhead{(3)} &
\colhead{(4)} &
\colhead{(5)} &
\colhead{(6)} \\
}
\startdata
2100-2250& 20.0\tablenotemark{c}& 
                  12& 1.00& 13.3$\pm$2.8& $>$0.36\\ 
1950-2100& 20.0& 27& 1.00& 21.2$\pm$3.4& 0.81$\pm$0.19\\
1800-1950& 20.0& 16& 1.00& 21.1$\pm$3.6& 0.48$\pm$0.16\\
1650-1800& 20.0& 24& 1.00& 23.0$\pm$3.8& 0.72$\pm$0.19\\
1500-1650& 20.0& 26& 1.00& 25.1$\pm$3.9& 0.78$\pm$0.19\\
1350-1500& 20.0& 35& 1.00& 34.2$\pm$4.7& 1.04$\pm$0.23\\
1200-1350& 20.0& 69& 1.03& 54.6$\pm$5.4& 2.12$\pm$0.30\\
1050-1200& 20.0& 45& 1.03& 44.3$\pm$5.3& 1.38$\pm$0.26\\
 900-1050& 20.0& 50& 1.06& 43.5$\pm$5.0& 1.58$\pm$0.27\\
  750-900& 20.0& 63& 1.06& 54.9$\pm$5.6& 1.99$\pm$0.31\\
  600-750& 20.0& 98& 1.06& 77.6$\pm$6.3& 3.10$\pm$0.37\\
  450-600& 15.0& 56& 1.06& 44.8$\pm$4.9& 4.20$\pm$0.67\\
  300-450& 11.0& 15& 1.15& 17.2$\pm$3.0& $>$3.09\tablenotemark{d}\\ 
  150-300& \nodata&  1&  \nodata& \nodata& \nodata\\
\enddata
\tablenotetext{a}{As the $T_{\rm eff}$ bins from 1050 to 1350K encompass both L and T dwarfs (see figure 22b of \citealt{kirkpatrick2021}), we average the correction factor for L dwarfs (1.00) and T dwarfs (1.06).}
\tablenotetext{b}{This value is computed via the equations $$dens = \Bigl(raw\Bigr)\Bigl(corr\Bigr)\bigg/\Bigl(\frac{4}{3}{\pi}{d_{max}}^3\Bigr)$$
and $$\sigma_{dens} = \sqrt{\bigl({\sigma_{raw}}^2 + {\sigma_{adj}}^2\bigr)}\Bigl(corr\Bigr)\bigg/\Bigl(\frac{4}{3}{\pi}{d_{max}}^3\Bigr)$$ where $\sigma_{raw} = \sqrt{raw}$.}
\tablenotetext{c}{This bin is complete only for its L dwarf complement. Since late-M dwarfs are also expected to populate this bin, the derived space density is considered to be a lower limit.}
\tablenotetext{d}{This temperature bin is not fully populated, as WISE sensitivity limits cannot probe to the quoted completeness distance for Y dwarfs below $T_{\rm eff} \approx 400$K.}
\end{deluxetable}

L dwarfs no longer show an underdensity in the plane, so no correction is needed for our derived L dwarf space densities. T dwarf space densities should, however, be multiplied by 1.06 to account for the observed incompleteness.  The Y dwarf incompleteness is harder to assess given the small number of Y dwarfs in the plane, but the raw numbers suggest a conservative correction factor of 1.15, slightly larger than the 1.13 factor adopted by \cite{kirkpatrick2021}. These factors are listed in Table~\ref{tab:space_densities_brown_dwarfs}.

\begin{figure}
\includegraphics[scale=0.45,angle=0]{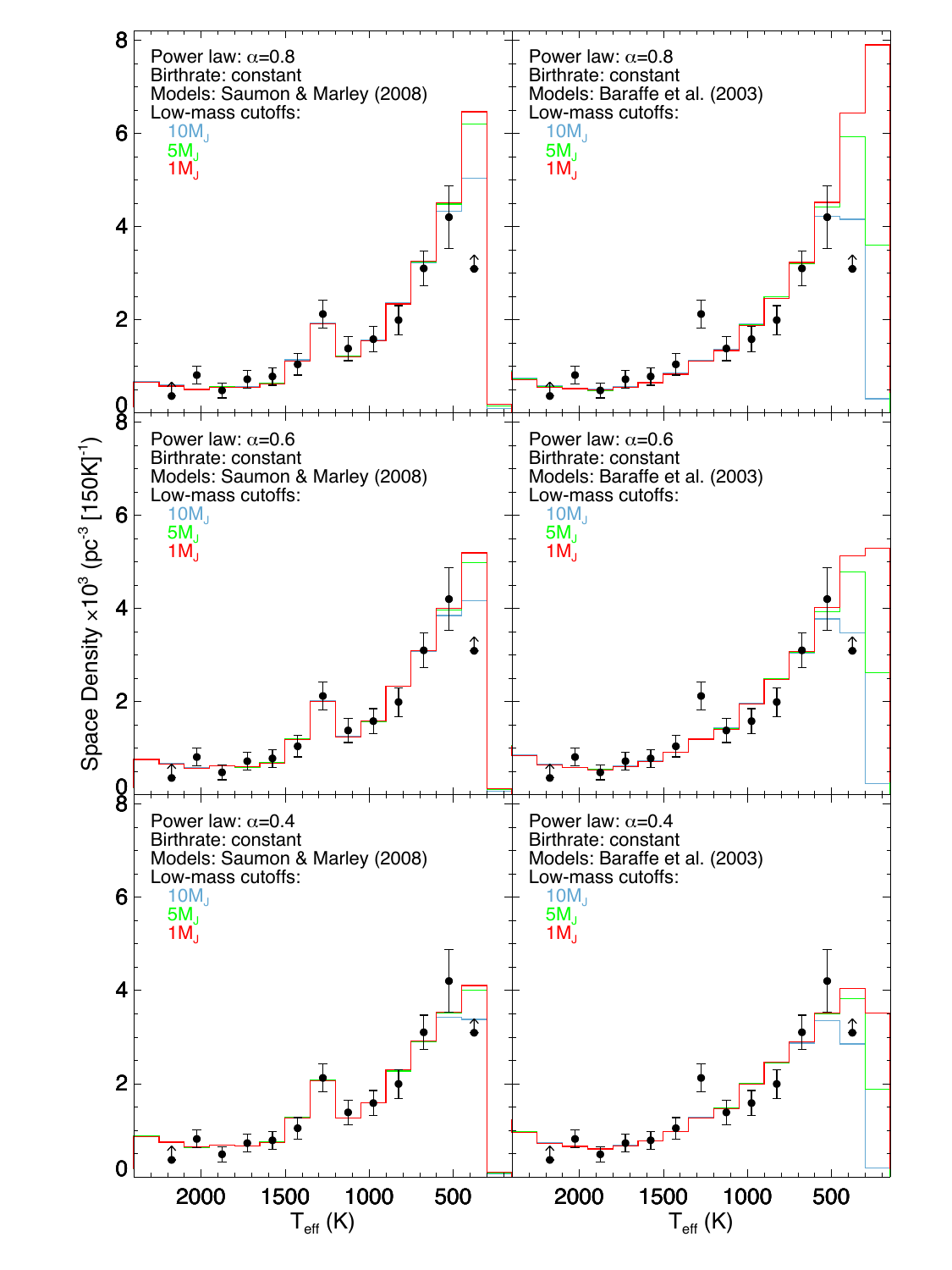}
\caption{Our measured L, T, and Y dwarf space densities from Table~\ref{tab:space_densities_brown_dwarfs} (black dots) as a function of effective temperature overplotted on different simulations from Raghu et al.\ (submitted). In all panels, simulations assuming a constant birthrate are shown, along with the results for three different low-mass cutoffs: 10 $M_{Jup}$ (light blue), 5 $M_{Jup}$ (green), and 1 $M_{Jup}$ (red). Panels in the left column use the \cite{saumon2008} evolutionary models, and panels in right column use \cite{baraffe2003}. The top row shows simulations with a power law of $\alpha = 0.8$, the middle row shows $\alpha = 0.6$, and the bottom row shows $\alpha = 0.4$. The best overall fits are those shown in the left panel in the middle row, using $\alpha = 0.6$ and the \cite{saumon2008} models.
\label{fig:space_densities_brown_dwarfs}}
\end{figure}

The final step in measuring the space densities of L, T, and Y dwarfs is assessing their measurement uncertainties. For this we adopt the same methodology used in \cite{kirkpatrick2021}. To summarize, our confidence in assigning an object to a $T_{\rm eff}$ bin is directly related to the measurement uncertainty on $T_{\rm eff}$, which is often comparable to the bin size itself. To estimate our confidence in the numbers of objects in each bin, we have run simulations with 10,000 Monte Carlo realizations wherein we take the uncertainty in $T_{\rm eff}$ and multiply it by a random value generated from a normal distribution having a mean of 0 and a standard deviation of 1. For each simulation, this uncertainty is added onto the measured value and the object (re-)assigned to the appropriate $T_{\rm eff}$ bin. The computed means and standard deviations across all 10,000 realizations are given in column 5 of Table~\ref{tab:space_densities_brown_dwarfs}. We use only these computed standard deviations in our adopted space densities, but not the adjusted means. As further explained in \cite{kirkpatrick2021}, the reason for this is that the number of objects is not preserved across the Monte Carlo simulations because some objects scatter into the hotter, incomplete bin at 2100-2250K and are lost, while objects at the other temperature extreme may be lost because they fall outside the completeness limit of the colder bin. This last loss is one-sided, however, as any colder objects scattering into the warmer bin would be necessarily retained. Hence, we compute our adopted space densities using the raw number counts, but including the uncertainties derived from our simulations, as shown in the footnote of Table~\ref{tab:space_densities_brown_dwarfs}. These densities are graphically illustrated in Figure~\ref{fig:space_densities_brown_dwarfs}.

The measured space densities can now be compared to the simulated $T_{\rm eff}$ distributions (Raghu et al., submitted) to infer the form of the mass function at this low-mass end. Following on the results of \cite{kirkpatrick2021}, which showed the best match to be a power law, $dN/dM \propto M^{-\alpha}$, with $\alpha \approx 0.6$, Raghu et al.\ (submitted) assume power-law functional forms with $\alpha$ values between 0.3 and 0.8 and, like \cite{kirkpatrick2021}, choose low-mass cutoffs of $\sim$1, 5, and 10 $M_{Jup}$. Unlike \cite{kirkpatrick2021}, however, they vary the birthrate to include not only a constant birthrate over the lifetime of the Milky Way, but also consider two other birthrates -- called inside-out and late-burst -- from \cite{johnson2021} that are constrained by new results from Gaia. The inside-out birthrate represents a declining birthrate over the 10 Gyr lifetime of the Galactic disk, and the late-burst birthrate is identical to  the inside-out form, except with an abrupt increase (by a factor of $\sim$3) in star formation $\sim$3-5 Gyr ago. 

Evolutionary models are used to infer the current $T_{\rm eff}$ value of each simulated object (using its mass and age). Raghu et al.\ (submitted) expand the model set used in \cite{kirkpatrick2021} by including the newer \cite{marley2021} predictions and show (again) that the only evolutionary models able to fit the bump in the L/T transition in the $T_{\rm eff}$ distribution are those of \cite{saumon2008}. 

It has been shown in \cite{kirkpatrick2021} and Raghu et al.\ (submitted) that the low-mass cutoff has little effect on the shape of the mass function at $T_{\rm eff}$ values above 450K, where our fitting is taking place. Therefore, we consider each $\alpha$ + birthrate pair and compute the median of the least squared values for the simulations across all three cutoff masses. The minimum is achieved for $\alpha = 0.6$ and a constant birthrate, identical to the findings in \cite{kirkpatrick2021}. The second best fit is achieved for $\alpha = 0.5$ and a constant birthrate. The third best fit is a tie among the $\alpha = 0.7$ + constant, the $\alpha = 0.4$ + late-burst, and $\alpha = 0.5$ + late-burst models. Use of either the late-burst or inside-out birthrates results in a slightly reduced $\alpha$ because those birthrates create a small overabundance, relative to the constant birthrate models, of older brown dwarfs that have already cooled to cooler temperatures.

In Figure~\ref{fig:space_densities_brown_dwarfs}, we show the fits for three values of $\alpha$ (0.4, 0.6, and 0.8) all paired with a constant birthrate. The panels in the left column of the figure show that the $\alpha = 0.6$ model with a constant birthrate and using the \cite{saumon2008} evolutionary models is an excellent representation of the empirical data. Can any new conclusions be gleaned regarding the low-mass cutoff? As figure 4 of Raghu et al.\ (submitted) illustrates, the \cite{saumon2008} models are incomplete below masses of $\sim$0.015$M_\odot$ ($\sim$16$M_{Jup}$), so they are a poor choice for determining what the low-mass cutoff might be. Instead, we revert back to the \cite{baraffe2003} models, that are complete down to $\sim$5$M_{Jup}$. As the rightmost panels in Figure~\ref{fig:space_densities_brown_dwarfs} illustrate, our ability to distinguish between low-mass cutoffs depends on measuring accurate space densities below 450K. Using the 20-pc census to say confidently that star formation's terminus is below 10$M_{Jup}$ or even 5$M_{Jup}$ depends on surveying the sky more deeply at the wavelengths of these objects' peak emission {\it and} obtaining the necessary astrometry to measure accurate distances. As the simulations using the \cite{baraffe2003} models show, measuring an accurate space density for the 300-450K bin will allow us to distinguish between the cutoff masses, and even a few more objects discovered in the 150-300K bin, which currently has only the 250K Y dwarf WISE J085510.83$-$071442.5 in it, will provide even tighter constraints.

\begin{figure}
\includegraphics[scale=0.35,angle=0]{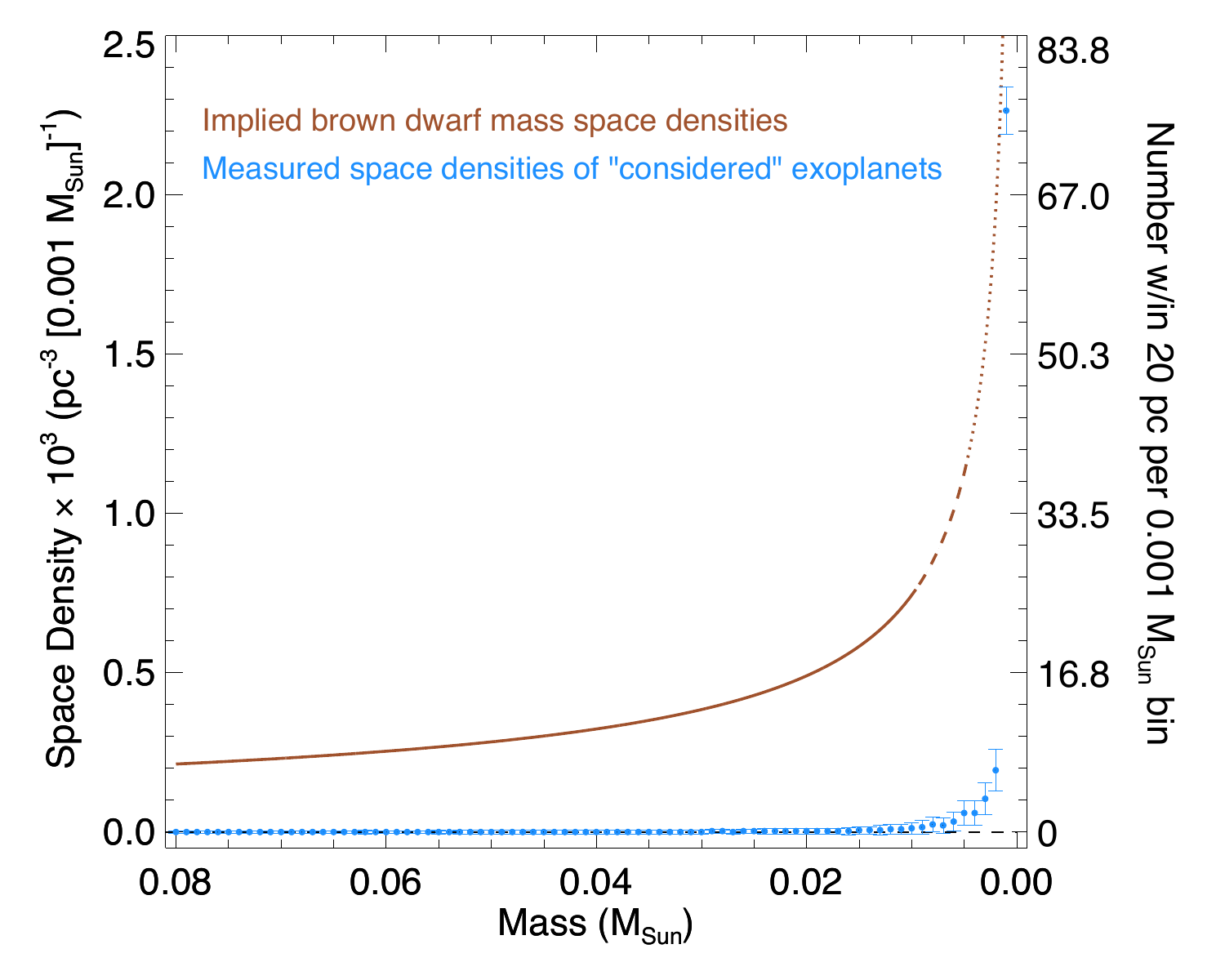}
\caption{Plots of the implied space densities of brown dwarfs (brown) in 0.001$M_\odot$ bins compared to the measured space densities of other possible low-mass star formation products from Table~\ref{tab:20pc_exoplanets} (blue). The brown dwarf space densities are divided into three mass zones -- $M > 10M_{Jup}$ (solid brown), $5M_{Jup} < M < 10M_{Jup}$ (dashed brown), and $1M_{Jup} < M < 5M_{Jup}$ (dotted brown). Note that the densities of the possible pseudo-exoplanets do not affect our measurement of the brown dwarf space densities, as their numbers only become appreciable at masses well below $5M_{Jup}$.
\label{fig:space_densities_exoplanets}}
\end{figure}

Have some of these ultra-low mass products of star formation already been identified, and are they masquerading in the literature as exoplanet discoveries to higher mass objects? We use our analysis in Table~\ref{tab:20pc_exoplanets} to see first if the omission of these objects has biased our derivation of the brown dwarf mass function above. With the exception of the two objects (the companions to UCAC4 211-005570 and L 119-213) lacking mass estimates, we take all objects labeled as "consider" in column 4 of Table~\ref{tab:20pc_exoplanets} and estimated their contribution to the overall mass function. For objects with $M \sin(i)$ measurements only, we pull a random number from a distribution of values uniformly distributed between 0 and 1 and multiply that number by 90 degrees to assign each an inclination, which we then use to assign an actual mass value. For all masses, whether or not they are true masses or adjusted $M \sin(i)$ measurements, we then pull a random number from a normal distribution with a mean of 0 and standard deviation of 1 and multiply that number by the uncertainty, which we then add back to the mass value. We perform this methodology over 10,000 Monte Carlo iterations and find the mean and standard deviation of the resulting space density, binned over 0.001-$M_\odot$ mass intervals, as illustrated in Figure~\ref{fig:space_densities_exoplanets}.

This figure shows that our derived space density of brown dwarfs, (which we find to be $\xi(M) = dN/dM = 0.0469 \times M^{-0.6}$ in units of \# $pc^{-3} [1M_\odot]^{-1}$, with $M$ in units of $M_\odot$; see Section~\ref{sec:discussion}), overwhelms the space density above $5M_{Jup}$ where our fitting took place. So, the omission of these objects has no impact on our derivation. However, the second question is whether any of these objects could be products of star formation itself rather than the secondary by-products of a proto-planetary disk. That question cannot be answered from this diagram, but it is a statistical certainty that at least a few of the objects on the high-mass tail of this distribution are star formation products. One striking result from Figure~\ref{fig:space_densities_exoplanets}, however, is the high space density of objects in the lowest mass bin, given that the census of such low-mass objects, whether resulting from star formation or proto-planetary disks, is still woefully incomplete. There is clearly no shortage of ultra-low mass objects in the Milky Way.

\subsection{Combined stellar and brown dwarf space densities\label{sec:combining_stellar_and_BD_densities}}

With the brown dwarf portion of the mass function now fitted, we can combine the stellar and brown dwarf portions to determine the shape of the overall mass function.

First we take the number counts across the stellar regime and perform a similar Monte Carlo analysis as was done on the brown dwarfs. Specifically, for each object we pull a random number from a normal distribution with a mean of 0 and a standard deviation of 1. We then multiply the mass measurement uncertainty by the random number and add that back to the mass value to get a true mass. We do this for each of the stars in our sample, and repeat the process 10,000 times to simulate 10,000 possible histograms. We then compute the mean value in each histogram bin along with its standard deviation. Because the 20-pc volume around the Sun is just one of many such volumes that can be taken as a sample of the Milky Way, we add the Poisson uncertainty and the standard deviation from above in quadrature to provide the final uncertainty per bin. (This parallels the brown dwarf space density analysis of Table~\ref{tab:space_densities_brown_dwarfs}.)

\begin{figure}
\includegraphics[scale=0.575,angle=0]{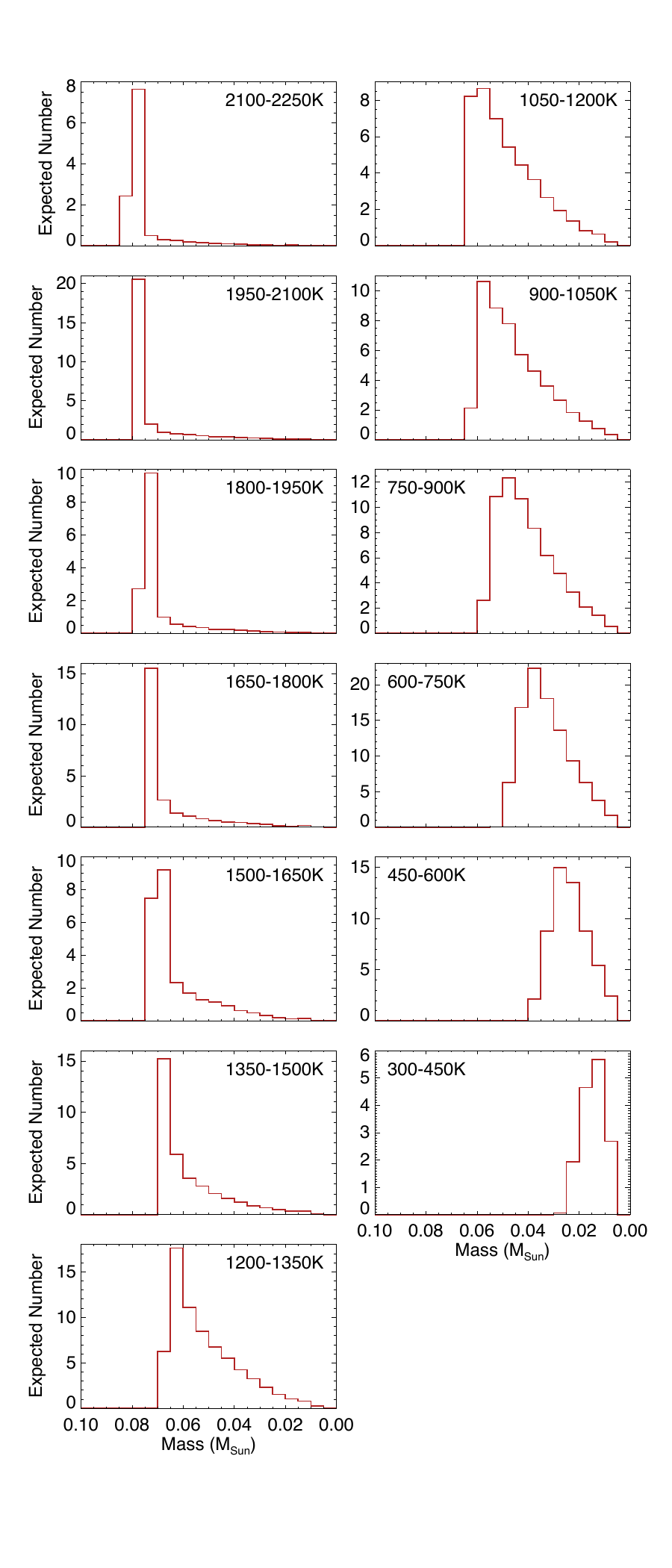}
\caption{The predicted distributions of brown dwarf masses in each of our 150K effective temperature bins based on the best-fit Raghu et al.\ (submitted) simulation to our measured L, T, and Y dwarf space densities (see text for details). Each histogram is scaled to match the total number of objects listed for that $T_{\rm eff}$ bin in Table~\ref{tab:space_densities_brown_dwarfs}.
\label{fig:brown_dwarf_mass_distribution_alpha06_constant_5MJup_SM08}}
\end{figure}

We can now append the substellar contribution onto this stellar distribution. To do this, we look at the predictions from the best fit Raghu et al.\ (submitted) model to the brown dwarf $T_{\rm eff}$ distribution from above, which is the $\alpha=0.6$ power law with a constant birthrate function and passed through the \cite{saumon2008} evolutionary models. We also choose a 0.005$M_\odot$ ($\sim5M_{Jup}$) cutoff to parallel the more detailed cutoff analysis from \cite{kirkpatrick2021}. This simulation gives the predicted mass distributions shown in Figure~\ref{fig:brown_dwarf_mass_distribution_alpha06_constant_5MJup_SM08}. Each histogram is scaled so that the total number of objects in each histogram matches the raw numbers of objects per bin listed in Table~\ref{tab:space_densities_brown_dwarfs}. As one example, the 27 objects in the 1950-2100K bin are predicted to fall almost exclusively in the 0.075-0.080 $M_\odot$ bin, and these predictions suggest that our 27 objects be apportioned as 20.5 objects in the 0.075-0.080 $M_\odot$ bin, 2.0 objects in the 0.070-0.075 $M_\odot$ bin, 1.0 object in the 0.065-0.070 $M_\odot$ bin, and fractional numbers of objects in bins of lower mass. As another example, the 63 objects in the 750-900K bin are spread over a wide range of masses from 0.005-0.060$M_\odot$ and are apportioned as 2.6 objects in the 0.055-0.060$M_\odot$ bin, 10.8 objects in the 0.050-0.055$M_\odot$ bin, 12.3 objects in the 0.045-0.050$M_\odot$ bin, 10.7 objects in the 0.040-0.045$M_\odot$ bin, 8.3 objects in the 0.035-0.040$M_\odot$ bin, etc.

We take the apportionment across all thirteen temperature bins and tally the results in each of the 0.005$M_\odot$-wide mass bins, after also applying the factor ({\it corr} in Table~\ref{tab:space_densities_brown_dwarfs}) to correct for losses of objects along the Galactic plane and extrapolating the numbers to the full 20-pc volume if that temperature bin was not complete to 20 pc. For example, the raw number counts in the 450-600K bin shown in both Table~\ref{tab:space_densities_brown_dwarfs} and Figure~\ref{fig:brown_dwarf_mass_distribution_alpha06_constant_5MJup_SM08} were multiplied by the 1.06 correction factor then multiplied by $(20/15)^3$ to extrapolate to the full volume. In Table~\ref{tab:space_densities_brown_dwarfs}, we find that our lowest temperature bin with a space density measurement, 300-450K, is considered to be incomplete, and the mass distribution for that bin in Figure~\ref{fig:brown_dwarf_mass_distribution_alpha06_constant_5MJup_SM08} suggests that that bin's objects fall exclusively below 0.025$M_\odot$. Therefore we consider any space density measurements below this mass value to be lower limits only.

\begin{figure*}
\includegraphics[scale=0.575,angle=0]{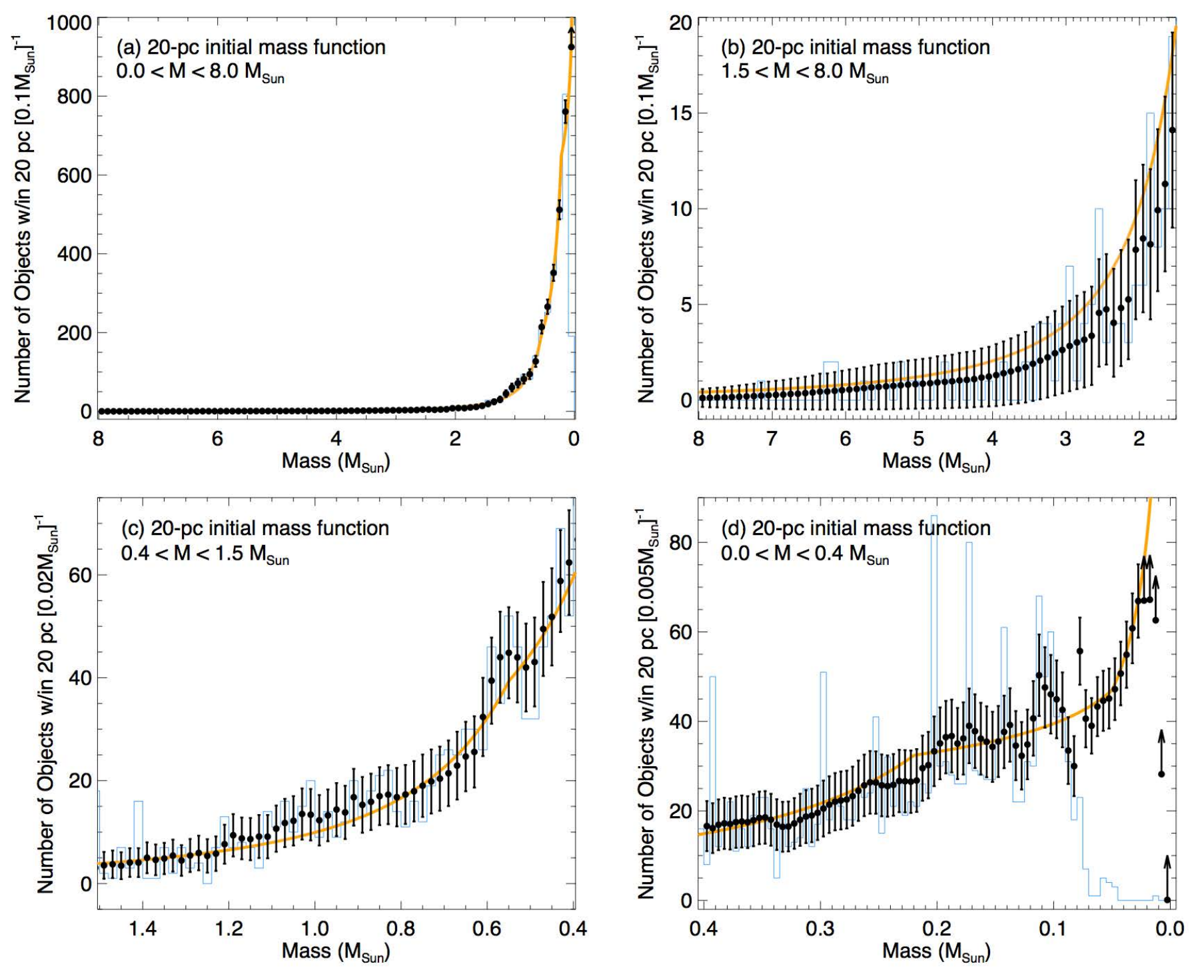}
\caption{The 20-pc initial mass function across all stellar and substellar masses. Our measured values and their uncertainties are shown in black. The raw number counts for stars of type M9.5 and earlier are shown by the blue histogram. (a) The full mass range, 0.0-8.0$M_\odot$, with 0.1$M_\odot$ binning; (b) A zoom-in of the high-mass end, from 1.5 to 8.0 $M_\odot$, with the same binning; (c) A zoom-in of the mid-range, from 0.4 to 1.5 $M_\odot$, with 0.02$M_\odot$ binning. (d) A zoom-in of the low-mass portion, 0.0-0.4$M_\odot$, with 0.005$M_\odot$ binning. Our fit to the mass function is shown by the orange line.
\label{fig:full_initial_mass_function}}
\end{figure*}

We now add these brown dwarf masses to the results of our Monte Carlo analysis of stellar masses above to produce a mass function across the entire mass range. This initial mass function is illustrated in Figure~\ref{fig:full_initial_mass_function}. Panel (a) shows the mass function across the full mass range from 0 to 8 $M_\odot$, binned in 0.1$M_\odot$ increments. The mass function rises with decreasing mass, and it continues to rise beyond our 0.025$M_\odot$ ($\sim26M_{Jup}$) completeness limit. Subsequent panels show details. Panel (b) shows the high-mass end of the initial mass function from 1.5 to 8.0 $M_\odot$, again with 0.1$M_\odot$ binning. The statistics above 3$M_\odot$ are poor but nonetheless show a steady increase from there down to 1.5$M_\odot$. Panel (c) shows the mid-mass range ($0.4 < M < 1.5 M_\odot$), now binned into smaller 0.02$M_\odot$ increments because the statistics here are richer.  Panel (d) zooms in on the smallest mass portion, below 0.4$M_\odot$, and chooses yet a smaller mass binning  of 0.005$M_\odot$. With the exception of a few small features (discussed below), the mass function is seen to rise monotonically from 1.5 to 0.025$M_\odot$. Mostly within the measurement errors (see more discussion below), the initial mass function is seen to continue rising well below our 0.025$M_\odot$ completeness limit and at least down to 0.015$M_\odot$.

The numbers on which Figure~\ref{fig:full_initial_mass_function} is based are given in Table~\ref{tab:space_densities_all}. For ease of reference, both the number of stars and the space density is given for each mass bin. Three mass binnings are tabulated, roughly paralleling what is shown in Figure~\ref{fig:full_initial_mass_function}: 0.1$M_\odot$ binning across the entire 0.0-8.0$M_\odot$ range (80 bins), 0.02$M_\odot$ binning across the range 0.0-1.6$M_\odot$ (80 bins), and 0.005$M_\odot$ binning across the range 0.0-0.4$M_\odot$ (80 bins). Mass bins with incomplete statistics are labeled as lower limits in the final column of the table.

\begin{deluxetable}{ccccccc}
\tablecaption{Number of Objects and Space Densities per Mass Bin for the 20-pc Census\label{tab:space_densities_all}}
\tablehead{
\colhead{Mass Bin} & 
\colhead{Total \#} &
\colhead{Space Density\tablenotemark{a}} &
\colhead{Lower Limit?} \\
\colhead{($M_\odot$)} & 
\colhead{} &
\colhead{(pc$^{-3}$ [$M$ bin]$^{-1}$)} &
\colhead{} \\
\colhead{(1)} &                          
\colhead{(2)} &
\colhead{(3)} &
\colhead{(4)} \\
}
\startdata
   0.00-0.10& 924.9$\pm$30.4& 0.0276005$\pm$0.0009072& yes \\
   0.10-0.20& 761.0$\pm$28.9& 0.0227094$\pm$0.0008624\\  
   0.20-0.30& 511.6$\pm$24.2& 0.0152669$\pm$0.0007222\\
   0.30-0.40& 351.7$\pm$20.5& 0.0104953$\pm$0.0006118\\
   0.40-0.50& 265.6$\pm$18.2& 0.0079259$\pm$0.0005431\\
   0.50-0.60& 214.3$\pm$16.8& 0.0063950$\pm$0.0005013\\
   0.60-0.70& 127.0$\pm$14.1& 0.0037899$\pm$0.0004208\\
   0.70-0.80&  94.5$\pm$12.6& 0.0028200$\pm$0.0003760\\
\enddata
\tablecomments{(This table is available in its entirety in machine-readable form.)}
\tablenotetext{a}{The [$M$ bin]$^{-1}$ portion of the units should be replaced with the bin size for that row. For example, for the first row of the table, the units will be pc$^{-3}$ [$0.10 M_\odot$]$^{-1}$ because that bin is $0.10 M_\odot$ wide.}
\end{deluxetable}

\begin{figure*}
\gridline{\fig{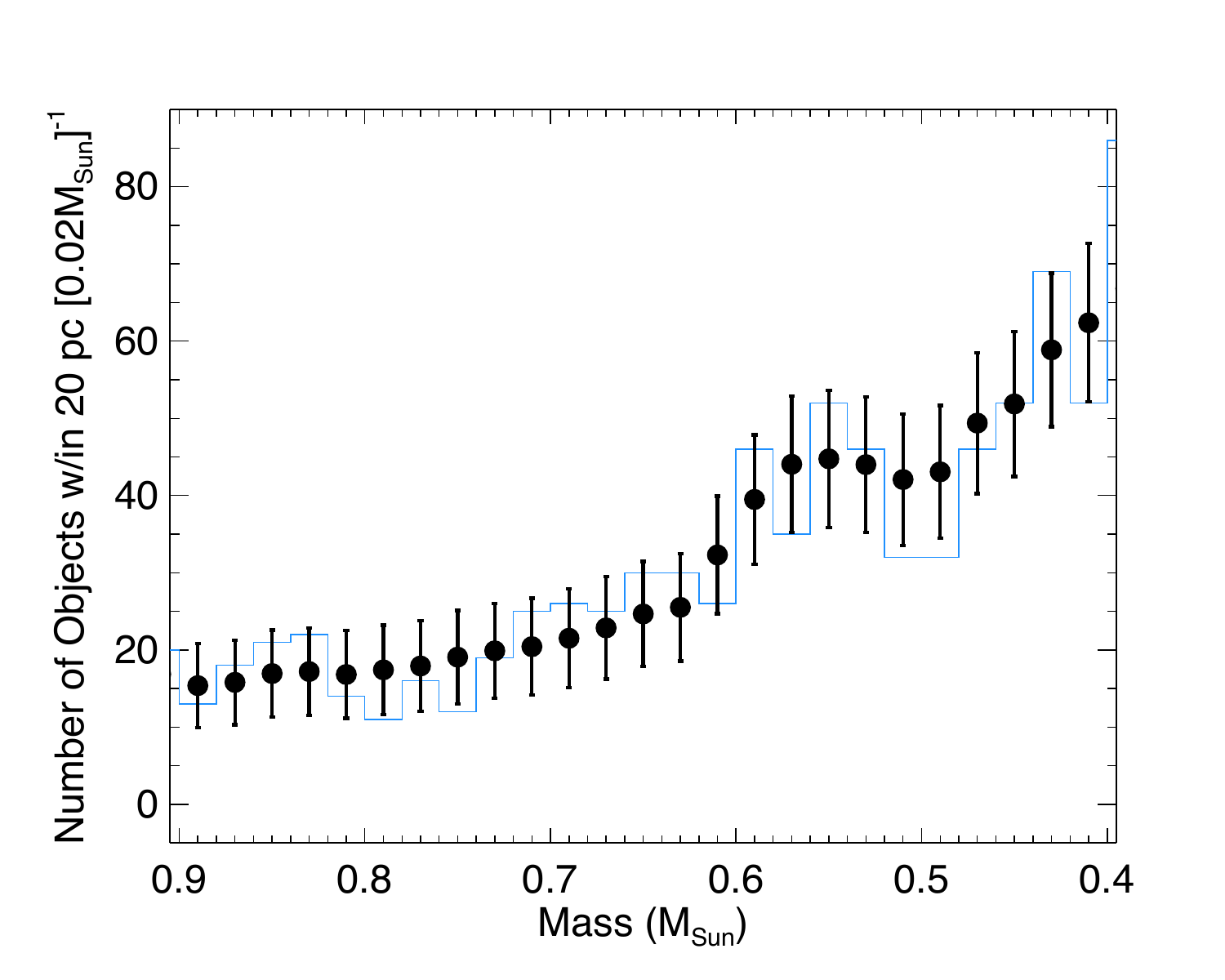}{0.3\textwidth}{(a)}
          \fig{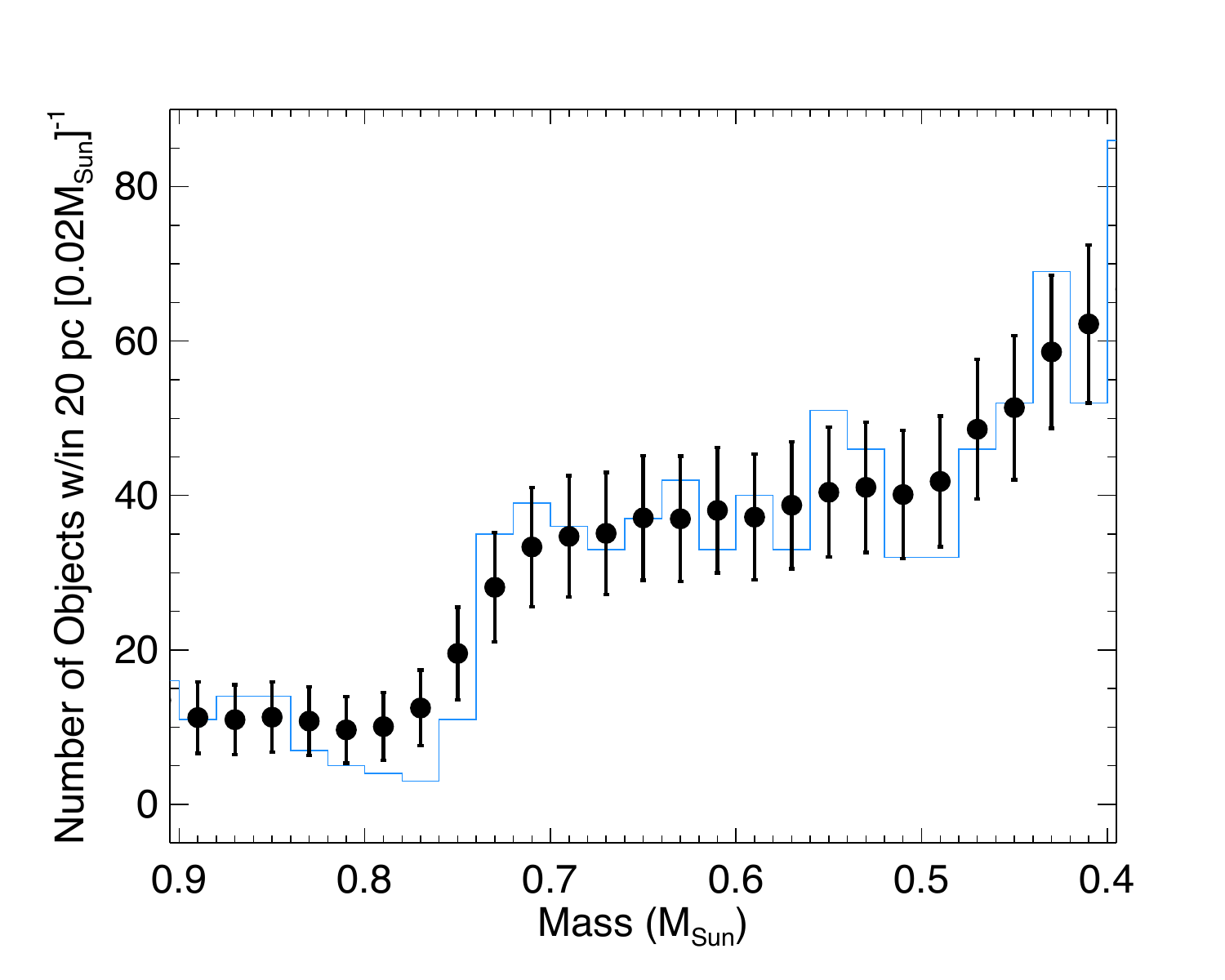}{0.3\textwidth}{(b)}
          \fig{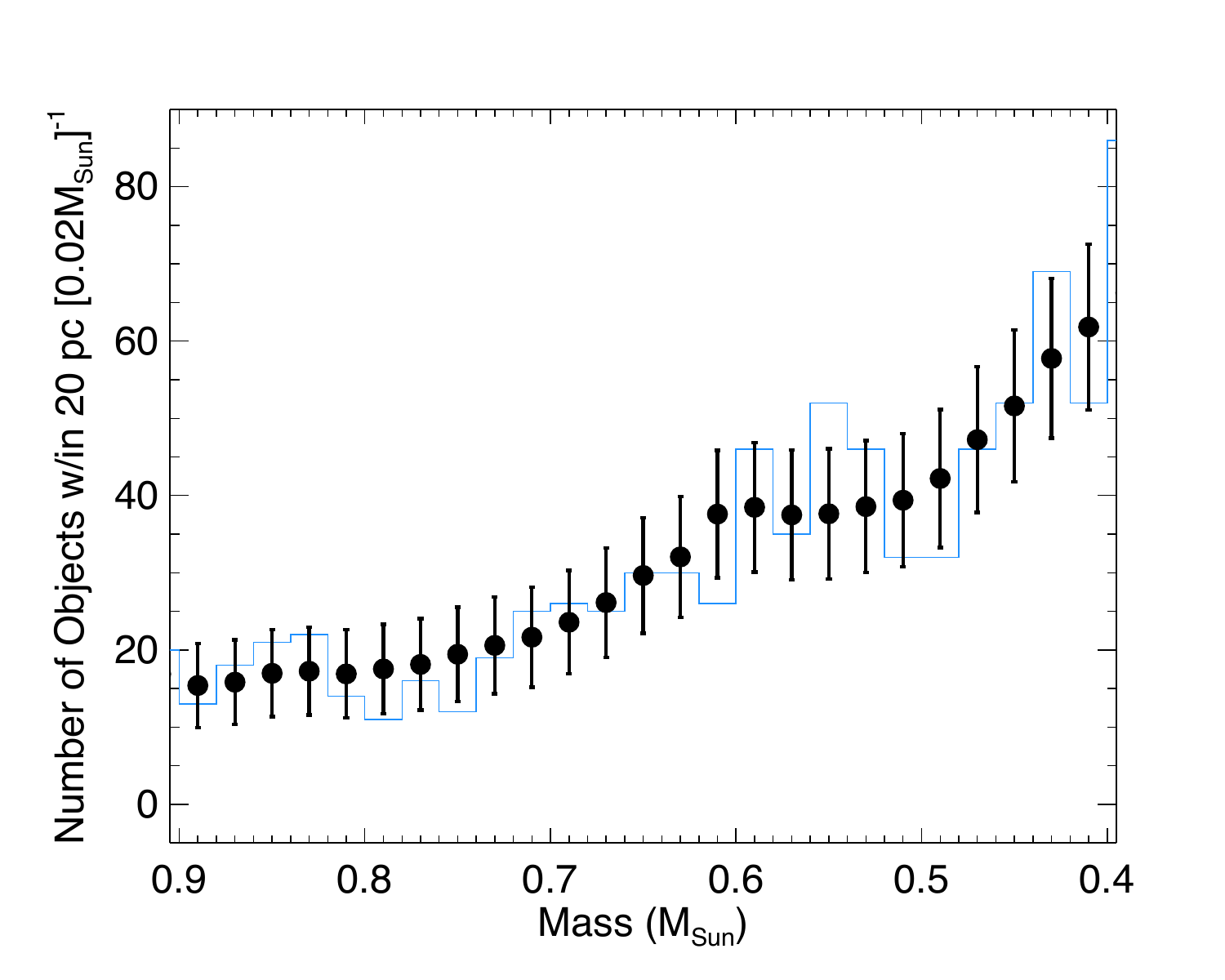}{0.3\textwidth}{(c)}}
\caption{Tests of the bump in the initial mass function seen near 0.55$M_\odot$ in Figure~\ref{fig:full_initial_mass_function}. (a) A zoom-in showing the bump in Figure~\ref{fig:full_initial_mass_function}(c). (b) The number counts over the same mass range but where we have moved the switchover in the mass estimate from $M_{Ks} = 5.0$ mag to $M_{Ks} = 4.0$ mag. (c) The number counts over the same mass range but where we have kept the $M_{Ks} = 5.0$ mag switchover point and inflated the uncertainties on the $M_{Ks}$-derived masses by a factor of four. For other details, see the caption to Figure~\ref{fig:full_initial_mass_function}.
\label{fig:mf_bump_tests}}
\end{figure*}

There are a few features in Figure~\ref{fig:full_initial_mass_function}(c) and (d) that warrant special attention. The first is the bump in the object counts near 0.55$M_\odot$ in panel (c). This falls near the point at which our mass estimation switches from that of the TIC relations of \cite{stassun2019} at higher masses to that of the $M_{Ks}$ relation of \cite{mann2019} at lower masses. Currently, we switch between these two relations at $M_{Ks} = 5.0$ mag, corresponding to a mass of $\sim0.6M_\odot$. As a test, if we change the switchover point to be at $M_{Ks} = 4.0$ mag ($M \approx 0.7 M_\odot$) instead, we find that this bump in the space densities moves to higher masses, with a deficit around $0.8 M_\odot$, as shown in Figure~\ref{fig:mf_bump_tests}. We also note that the uncertainties in the masses resulting from the \cite{mann2019} relation are three to four times smaller than those derived from the \cite{stassun2019} relation. As another test, we can artificially inflate the mass uncertainties on the \cite{mann2019}-derived masses while keeping the current switchover point at $M_{Ks} = 5.0$ mag. That result is also shown in Figure~\ref{fig:mf_bump_tests}. In this case, the bump is greatly diminished in the number counts, but an inflection point is still seen near 0.6$M_\odot$. Given that this feature in the number counts moves in response to the mass estimation used, we believe it is an artificial effect. Furthermore, given that the \cite{stassun2019} mass estimation relies on dynamically measured {\it individual} masses whereas the \cite{mann2019} relation uses Bayesian statistics to ferret out individual masses from binaries in which only the {\it total} system mass is measured, this likely indicates a small systematic offset that slightly deflates the \cite{mann2019}-derived masses relative to truth. In fact, an effect in this direction and representing a systematic offset of $\sim2\%$ is seen when comparing results of the $M_{Ks}$ relation to individually derived masses (figure 15 of \citealt{mann2019}). Obtaining more directly measured individual masses in this regime, corresponding to late-K and early-M dwarfs, would help to put this issue to rest.

Other features are seen in Figure~\ref{fig:full_initial_mass_function}(d). There is a small drop in the number counts near 0.13$M_\odot$ followed by a sudden rise near 0.11$M_\odot$. This feature is likely artificial, as the bump at 0.11$M_\odot$ is due primarily to components in multiple systems about which little information is known, and these were arbitrarily assigned masses of 0.11$M_\odot$ based on an anticipated spectral type of M5.5. This type lies at a very sharp inflection point (see Table~\ref{tab:piecewise_mass_vs_spectype}) in our mass vs. spectral type relation (Figure~\ref{fig:mass_vs_spectype}). 

The other feature is the rapidly changing number count between masses of 0.06 and 0.08 $M_\odot$, a mass range that straddles the stellar/substellar break. Some of the early- to mid-L dwarfs that we have included in our brown dwarf mass function analysis are likely very low-mass stars and not brown dwarfs themselves. As a consequence, these are assigned masses that are a bit too high (the extraneous high point in the 0.075-0.080$M_\odot$ bin), which likely leads to concomitant deficits in the next higher mass bins. In fact, Table~\ref{tab:20pc_census} lists three early-L dwarfs in the 20-pc census that have dynamical mass measurements, and one of these (LP 388-55 B; \citealt{dupuy2017}) has a mass just above the 0.075-0.080$M_\odot$ mass bin (0.083$\pm$0.03 $M_\odot$).

\section{Discussion\label{sec:discussion}}

How do our 20-pc results compare to other attempts in the literature to measure the initial mass function? Pioneering work by \cite{salpeter1955} found that a power law form $\xi(M) = dN/dM \propto M^{-\alpha}$ with $\alpha = 2.35$ best fit the initial mass function over the range $0.3 \le M \le 10 M\odot$. Subsequent work by \cite{miller1979} found $\alpha = 1.4$ for $0.1 \le M \le 1 M_\odot$ and $\alpha = 2.5$ for $1 \le M \le 10 M_\odot$. \cite{scalo1986} determined $\alpha = 2.7$ for $2 \le M \le 10 M_\odot$, and \cite{reid2002} found $\alpha = 1.3$ for $0.1 \le M \le 0.7 M_\odot$. 

As more accurate measurements of the initial mass function became possible, researchers realized the importance of distinguishing whether the masses used for the computation were that of the stellar system or of its individual components. For example, in an analysis of data from the Sloan Digital Sky Survey, \cite{bochanski2010} found $\alpha = 2.38$ for $0.32 < M < 0.80 M_\odot$ and $\alpha = 0.35$ for $0.10 < M < 0.32 M_\odot$ for the mass function of systems but $\alpha = 2.66$ and $\alpha = 0.98$ for the mass function of single stars over the same two mass regimes, respectively. Earlier, \cite{kroupa2001} had found that single-star initial mass functions resulting from the analysis of young star clusters generally gave values of $\alpha$ that were higher by $\sim$0.5 (for $0.1 < M < 1.0 M_\odot$) than the field initial mass function, for which systems were not resolved. \cite{reid2005IMF} cautions that unresolved multiplicity complicates interpretation of the initial mass function; the initial mass function of {\it systems} is more directly tied to the fragmentation of the original molecular cloud, but the initial mass function of {\it individual objects} give the mass distribution of the (sub)stellar bodies formed.

Our work on the 20-pc census has concentrated on analysis of the individual products of star formation, as we are curious to know how frequently this process produces, for example, very low-mass brown dwarfs compared to higher-mass objects. We will therefore restrict subsequent analysis here to the single-object initial mass function and leave analysis of the 20-pc census regarding multiplicity and the system mass function to those investigating how the formation of systems relates back to cloud fragmentation. 

\subsection{Comparison of Initial Mass Functions}

Here we compare two very well established forms of the initial mass function and compare their predictions to our results based on the 20-pc census.

\cite{chabrier2001,chabrier2003a, chabrier2003b} developed several functional forms for the initial mass function. The latest one relating to single objects is given by \cite{chabrier2003b} and is comprised of a power law at high masses and a log-normal form at lower masses:
\begin{equation}
\label{eqn:IMF_chabrier}
\begin{split}
  \xi(M) & = \frac{C_1}{\ln{10}}M^{-\alpha}, \: \textrm{for}\;  M > 1.0 M_\odot \\
  & = \frac{C_2}{M \ln{10}} e^{\frac{-(\log{M} - \log{M_c})^2}{2\sigma^2}}, \: \textrm{for}\; M \le 1.0 M_\odot
\end{split}
\end{equation}
where $C_1 = 0.0443$,
$\alpha = 2.3$, $C_2 = 0.158$, $M_c = 0.079 M_\odot$, and $\sigma = 0.69 M_\odot$. The value of $\xi(M)$ is in units of \# of objects per pc$^3$ per $M_\odot$. See equation 2 and table 1 of \cite{chabrier2003b} for the derivation shown above. We note that the values of $C_1$ and $C_2$ are set by \cite{chabrier2003b} to match an empirical space density measurement (at 1$M_\odot$ from \citealt{scalo1986}) of the initial mass function in the Milky Way's disk population.

Likewise, \cite{kroupa2013} found that a tripartite power-law form best describes the single-object initial mass function:
\begin{equation}
\label{eqn:IMF_kroupa}
\begin{split}
   \xi(M) & = C \left( {\frac{0.5}{0.07}} \right) ^{-\alpha_2} \left( {\frac{M}{0.5}} \right) ^{-\alpha_1}, \: \textrm{for}\; 0.5 < M < 150 M_\odot \\
   & = C \left( \frac{M}{0.07} \right) ^{-\alpha_2}, \: \textrm{for}\; 0.07 < M < 0.5 M_\odot \\
   & = \frac{C}{3} \left( \frac{M}{0.07} \right)^{-\alpha_3}, \: \textrm{for}\; 0.01 < M < 0.15 M_\odot \\
\end{split}
\end{equation}
where $\alpha_1 = 2.3$, $\alpha_2 = 1.3$, and $\alpha_3 = 0.3$. As above, the value of $\xi(M)$ is in units of \# of objects per pc$^3$ per $M_\odot$. Note that there are two components of this mass function that contribute to the mass range $0.15 < M < 0.07 M_\odot$. The value of $C$ is not specified by \cite{kroupa2013}. However, figure 4-24 of \cite{kroupa2013} provides a comparison of this initial mass function with the \cite{chabrier2003b} version in Equation~\ref{eqn:IMF_chabrier}, showing that they are identical at $\sim$0.85$M_\odot$, resulting in a value of $C \approx 1.15$.

\begin{figure*}
\includegraphics[scale=0.45,angle=0]{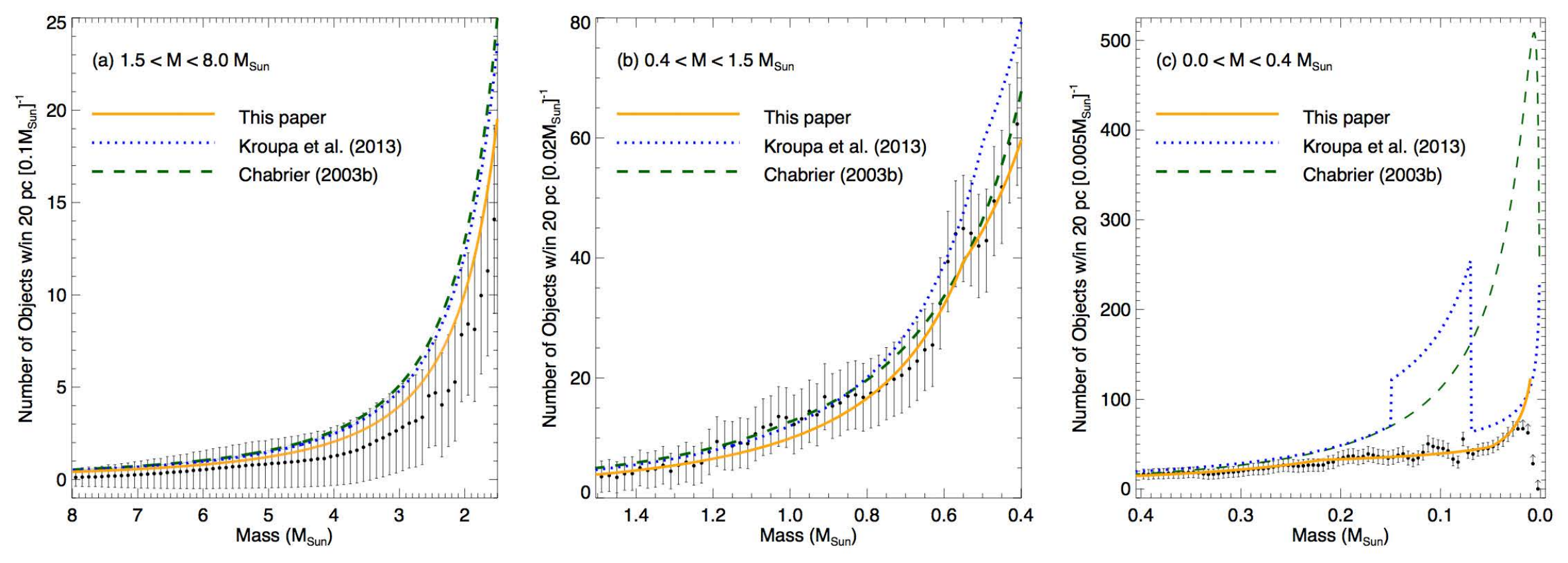}
\caption{A comparison of our 20-pc number counts (black points with error bars) and our fit of the initial mass function (solid orange line) to the functional forms of \cite{kroupa2013} (dotted blue line) and \cite{chabrier2003b} (dashed green line). Each panel shows a zoom-in of a different mass segment: (a) $1.5 < M < 8.0 M_\odot$, (b) $0.4 < M < 1.5 M_\odot$, (c) $0.0 < M < 0.4 M_\odot$.
\label{fig:full_initial_mass_function_comparison}}
\end{figure*}

As shown in Figure~\ref{fig:full_initial_mass_function_comparison}, neither of these parameterizations adequately describes the 20-pc initial mass function derived here. The main reason for this is that prior determinations were done pre-WISE, pre-Spitzer, and pre-Gaia before the L, T, and Y dwarf complement of the mass function was fully characterized and before exquisite parallax determinations became available for almost all objects in the 20-pc volume. As such, these prior works relied on poorer statistics with fits done in logarithmic scaling on both the $dN$ and $dM$ axes. Our careful accounting of objects within the 20-pc volume now allows for a more precise determination of the single-object initial mass function.

We thus provide a new multi-part power law parameterization that is bounded by the following caveats: (1) We assume $\alpha = 2.3$ at the high mass end, as has been determined from earlier studies, and we do this because our 20-pc census has few stars with $M > 2 M\odot$ to better constrain this. (2) We assume $\alpha = 1.3$ at intermediate masses, as this has also been established by earlier studies. (3) We take $\alpha = 0.6$ in the brown dwarf regime, as was determined in Section~\ref{sec:analysis_brown_dwarfs}. We do not constrain the stitch points in mass between the power law pieces nor do we limit the number of power law pieces to only three. We perform these fits by eye, keeping in mind the caveats from Section~\ref{sec:combining_stellar_and_BD_densities} concerning the non-physical bumps and troughs in the number counts as a function of mass. Given the constraints above, we find that a three-piece power law does not adequately describe the number counts in the mid- to late-M dwarf regime ($0.08 \lesssim M \lesssim 0.20 M_\odot$), but that a four-piece power law can. This best fit is given below and illustrated by the orange line in Figure~\ref{fig:full_initial_mass_function}:
\begin{equation}
\label{eqn:IMF_kirkpatrick}
\begin{split}
   \xi(M) & = C_1  (M)^{-\alpha_1}, \: \textrm{for}\; 0.55 < M < 8.0 M_\odot \\
   & = C_2 (M)^{-\alpha_2}, \: \textrm{for}\; 0.22 < M < 0.55 M_\odot \\
   & = C_3 (M)^{-\alpha_3}, \: \textrm{for}\; 0.05 < M < 0.22 M_\odot \\
   & = C_4 (M)^{-\alpha_4}, \: \textrm{for}\; 0.01 < M < 0.05 M_\odot \\
\end{split}
\end{equation}
where $C_1 = 0.0150$, $\alpha_1 = 2.3$, 
$C_2 = 0.0273$, $\alpha_2 = 1.3$, 
$C_3 = 0.134$, $\alpha_3 = 0.25$, 
$C_4 = 0.0469$, $\alpha_4 = 0.6$, 
As above, the value of $\xi(M)$ is in units of \# of objects per pc$^3$ per $M_\odot$. 

If we integrate under our best fit from 0.075 to 8.0 $M_\odot$, we find 3002 stars, which can be compared to the 3000 individual objects in the 20-pc census that have (measured or implied) types of M9.5 or earlier. The integration under our fit for 0.020 to 0.075 $M_\odot$ gives 789 brown dwarfs, compared to the 582 individual L, T, and Y dwarfs in Table~\ref{tab:20pc_census}. Most of the missing $\sim200$ brown dwarfs are ones with $T_{\rm eff}$ values 450-600K and distances of 15-20 pc or ones with 300-450K temperatures and 11-20 pc distances, where our current accounting is known to be incomplete (see Table~\ref{tab:space_densities_brown_dwarfs}). These results show that the number of stars relative to brown dwarfs is 3002/789, or $\sim$4. However, we believe that the brown dwarf mass function extends to at least 0.010 $M_\odot$, which would give a ratio of 3002/986 ($\sim$3) if the $\alpha = 0.6$ functional form continues to that mass. In the limiting case in which it continues to zero mass, we find a star-to-brown-dwarf ratio of 3002/1602, or $\sim2$. We note that as recently as a decade ago, this ratio in the Solar neighborhood was believed to be as high as 6:1 (\citealt{kirkpatrick2012}).

This decrease in the ratio of stars to brown dwarfs is not in tension with microlensing results, as analysis of OGLE data found that an even steeper power law in the brown dwarf regime ($\alpha = 0.8$ for $0.01 < M < 0.08 M_\odot$; \citealt{mroz2017}) best fits the observed distribution of short-timescale (low-mass) events. Furthermore, the possibility that the mass function extends into a regime significantly below 0.010 $M_\odot$ is bolstered by recent JWST observations of the Orion Nebula Complex that show a significant population of objects, down to at least 0.001$M_\odot$, that are apparently direct products of star formation (\citealt{mccaughrean2023,pearson2023}).

Finally, we note that our accounting of the mass of hydrogen-burning stars in the 20-pc census along with our best fit to the mass function of brown dwarfs allows us to calculate the average mass of an object in this sample. Integrating our mass function  down to a mass of 0.020 $M_\odot$, we find that value to be 0.41 $M_\odot$. There are likely many undiscovered brown dwarfs in the solar neighborhood too faint to be currently detected, so this average value could be pushed lower. Assuming there is no low-mass cutoff of star formation and the brown dwarf mass function continues to zero mass with a power law slope of $\alpha = 0.6$, we find that the average mass of objects in the 20-pc census would drop to 0.34 $M_\odot$. This can be considered as the limiting case unless the functional form at the lowest masses has an $\alpha$ value considerably greater than 0.6.

\section{Conclusions\label{sec:conclusions}}

In this paper our aim was to study the initial mass distribution of star formation's by-products, from the highest-mass progenitors of present-day white dwarfs to the lowest-mass brown dwarfs. For this, we have produced a volume-complete sample of stellar and substellar objects within 20 pc of the Sun. We have split multiple systems into their separate components and characterized each individual object to provide an accurate mass assignment.

Our main conclusions can be summarized as follows:

1) The initial mass function steadily increases as a function of descending mass. Its peak in (linear) mass space is not yet defined but is located below 0.020$M_\odot$ ($\sim20M_{Jup}$; Figure~\ref{fig:full_initial_mass_function}). We find that a quadripartite power-law ($\xi(M) = dN/dM \propto M^{-\alpha}$) fits the observed space densities well (Equation~\ref{eqn:IMF_kirkpatrick}). Going from high mass to low mass, we find exponents of $\alpha =$ 2.3, 1.3, 0.25, and 0.6, with stitch points between segments of 0.55, 0.22, and 0.05 $M_\odot$, respectively. Although the rate of ascent of the mass function is slowly retarded as a function of descending mass through the stellar and high-mass brown dwarf regimes, its ascent increases again for the lower-mass brown dwarfs.

2) This initial mass function agrees well with previous determinations in the high-mass regime (by design) but differs markedly from other established formalisms in the M, L, T, and Y dwarf regimes (Section~\ref{sec:discussion}). Functional forms proposed by \cite{chabrier2003b} and \cite{kroupa2013} overpredict the number of these lower-mass dwarfs relative to their more massive counterparts.

3) The 20-pc census currently consists of $\sim$3000 stars and $\sim$600 brown dwarfs (Table~\ref{tab:20pc_census}). At face value, this implies a stellar-to-substellar ratio of $\sim$5, but corrections for incompleteness for brown dwarfs with temperatures from 300-600K shows that the ratio is currently measured at $\sim$4. Incompletenesses at lower temperatures may yet bring this ratio as low as $\sim$3 (Section~\ref{sec:discussion}). The average mass of objects in the 20-pc census is currently measured as 0.41 $M_\odot$ but could drop as low as 0.34 $M_\odot$ if many colder brown dwarfs, yet to be discovered, actually exist (Section~\ref{sec:discussion}).

4) The 20-pc census of objects colder than 600K, corresponding to spectral type $\sim$T8.5, is still incomplete beyond 15 pc, and the completeness volume shrinks to 11 pc at 450K, corresponding to spectral type $\sim$Y0 (Table~\ref{tab:space_densities_brown_dwarfs}). Moreover, an additional source of incompleteness for objects as warm as 1350K is high backgrounds and confusion along the Galactic plane (Section~\ref{sec:analysis_brown_dwarfs}).

5) There are direct indications that many unrecognized companions still exist to already identified members of the 20-pc census. Acceleration (aka proper motion anomaly) has been used to flag many such systems (Section~\ref{sec:accelerators}), and large Gaia RUWE/LUWE values significantly higher than 1.0 flag many others (Section~\ref{sec:RUWE_LUWE}). Additional follow-up of these systems would help to better flesh out the 20-pc census itself while also providing much firmer statistics on multiplicity and the prevalence of hierarchical systems.

6) Our "complete" (see caveats \#4 and \#5, above) 20-pc census produced for this paper is available for additional uses (Table~\ref{tab:20pc_census}). As one example, this nearby sample is particularly useful as the hunting grounds for the closest habitable worlds to our own Solar System and is thus also available via the NASA Exoplanet Archive\footnote{\url{ https://exoplanetarchive.ipac.caltech.edu/docs/20pcCensus.html}}.

7) Except for white dwarfs (Section~\ref{sec:wd_masses}) and brown dwarfs (Section~\ref{sec:brown_dwarfs}), masses are used directly when they have been measured (Section~\ref{sec:mass_methods}). Most objects, though, lack actual mass determinations.  For these we use a variety of mass estimation methods and select the ones that provide the most reliable results, when comparison to truth is available (Section~\ref{sec:mass_estimates_main_sequence}). Nonetheless, our resulting space density computations binned by mass show some spurious features that appear to be caused by shortcomings in the estimation method. These are most obvious in the early-M dwarf region and in the regime from late-M to early-L dwarfs  (Section~\ref{sec:combining_stellar_and_BD_densities}). Dedicated programs, such as those by \cite{vrijmoet2022} and \cite{dupuy2022} directly determining masses of objects in these zones are clearly needed.

8) Dueling definitions in the literature of "brown dwarf" and "exoplanet" could bias our results if objects tagged as exoplanets are omitted from the initial mass function. We account for this and find that most of the objects labeled exoplanets (via the 13$M_{Jup}$-based definition) fall in a small-mass regime separate from the objects that have been more traditionally labelled as brown dwarfs (via the formation-based definition) and do not affect our conclusions regarding the mass function (Section~\ref{sec:exoplanets}). This having been stated, future studies of the initial mass function might wisely consider no such division, as planet formation can be thought of merely as a (delayed) secondary process resulting from star formation itself. Including planetary formation products as another branch of the initial mass function will, however, not be feasible until a statistically robust, volume-complete set of exoplanets can be reliably measured.

9) Gaia DR3 detections comprise only $\sim75\%$ of the volume-complete 20-pc census. Objects within 20 pc of the Sun can be missed by Gaia because they are too bright for Gaia observations, too faint for Gaia to detect, or are companions to Gaia-detected host stars that are (presently) inadequately characterized astrometrically (Section~\ref{sec:building_the_list_of_20pc_systems}).

10) Citizen science continues to produce new discoveries within 20 pc (Section~\ref{sec:BYW_discoveries}), even including a possible new Y dwarf with a "bonus" Spitzer parallax (Section~\ref{sec:appendix_astrometry}). Such discoveries are coming largely from WISE data sets, but these sets are being pushed to their sensitivity limits. Completing the 20-pc census in the 300-600K temperature range will require a deeper survey at $\sim$5 um, the best prospect for which is the NASA mission NEO Surveyor (\citealt{kirkpatrick2019b, mainzer2023}). Results from that mission, along with searches for cooler companions to known 20-pc hosts with JWST, represent our best short-term prospects for determining the occurrence rate of objects such as WISE J085510.83$-$071442.5 that reside below 300K.

11) The 20-pc census enables us to identify the nearest star or brown dwarf in each constellation (Table~\ref{tab:proximas}). Interestingly, six of the eighty-eight constellations have a Y dwarf as their nearest member despite the fact that Y dwarfs have not yet been fully mapped within this 20-pc volume.

NOTE ADDED IN PROOF: The nearby brown dwarf candidate CWISE J165909.91$-$351108.5 from Table~\ref{tab:poss_20pc_members_MLTY} has been confirmed as a late-L dwarf by \cite{robbins2023a}, but it likely falls outside the 20-pc census. Additionally, \cite{robbins2023b} find that CWISE J105512.11+544328.3, from Table~\ref{tab:low_metallicity}, is not a T subdwarf but rather an anomalously blue Y dwarf.

\clearpage

ACKNOWLEDGMENTS: Davy Kirkpatrick, Federico Marocco, and Chris Gelino acknowledge support from grant \#80NSSC20K0452 awarded for proposal 18-2ADAP18-0175 under the NASA Astrophysics Data Analysis Program. 
Data presented here are based on observations obtained at the Hale Telescope, Palomar Observatory as part of a continuing collaboration between the California Institute of Technology, NASA/JPL, Yale University, and the National Astronomical Observatories of China. Some of the data presented herein were obtained at the W. M. Keck Observatory, which is operated as a scientific partnership among the California Institute of Technology, the University of California and the National Aeronautics and Space Administration. The Observatory was made possible by the generous financial support of the W. M. Keck Foundation. The authors wish to recognize and acknowledge the very significant cultural role and reverence that the summit of Maunakea has always had within the indigenous Hawaiian community.  We are most fortunate to have the opportunity to conduct observations from this mountain.
This research has made use of the Keck Observatory Archive (KOA), which is operated by the W. M. Keck Observatory and the NASA Exoplanet Science Institute (NExScI), under contract with the National Aeronautics and Space Administration.
The first author would like to thank Patrick Shopbell at Caltech for resurrecting an Exabyte drive that successfully read raw CTIO data from 1997.
Part of this research was carried out at the Jet Propulsion Laboratory, California Institute of Technology, under a contract with the National Aeronautics and Space Administration (80NM0018D0004). He would also like to thank Mike Cushing for discussion of evolved star loci in the 20-pc color-magnitude diagrams.

One observation reported in this paper was obtained with the Southern African Large Telescope (SALT) under program 2021-2-SCI-027 (PI: Faherty).
Roberto Raddi acknowledges support from Grant RYC2021-030837-I funded by MCIN/AEI/ 10.13039/501100011033 and by “European Union NextGenerationEU/PRTR”,
and partial support by the AGAUR/Generalitat de Catalunya grant SGR-386/2021 and by the Spanish MINECO grant PID2020-117252GB-I00.
Eileen Gonzales acknowledges support from the Heising-Simons Foundation through a 51 Pegasi b Fellowship.
This publication makes use of VOSA, developed under the Spanish Virtual Observatory (\url {https://svo.cab.inta-csic.es}) project funded by MCIN/AEI/10.13039/501100011033/ through grant PID2020-112949GB-I00. This research made use of the Montreal Open Clusters and Associations (MOCA) database, operated at the Montr\'eal Plan\'etarium (J. Gagn\'e et al., in preparation).
Backyard Worlds research was supported by NASA grant 2017-ADAP17-0067 and by the NSF under grants AST- 2007068, AST-2009177, and AST-2009136. 
Johanna Vos acknowledges support from a Royal Society - Science Foundation Ireland University Research Fellowship (URF$\backslash$1$\backslash$221932).

\pagebreak

\facilities{
WISE, 
Gaia, 
Spitzer (IRAC), 
Hale (WIRC, DBSP, TSpec), 
Keck:I (MOSFIRE, NIRES), 
IRTF (SpeX),
Gemini:South (FLAMINGOS-2), 
Blanco (RCSPec, ARCoIRIS),
NTT (SOFI),
Magellan:Baade (FIRE),
Shane (Kast),
SALT (RSS),
SOAR (OSIRIS),
ARC (TSpec).
}

\software{
WiseView (\citealt{caselden2018}),
{\texttt{Spextool}} (\citealt{cushing2004,vacca2003}),
{\texttt{MOPEX/APEX}} (\citealt{makovoz2005a,makovoz2005b}),
{\texttt{IRAF}} (\citealt{tody1986,tody1993}),
{\texttt{kastredux}} (Burgasser et al., in prep.),
{\texttt{FIREHOSE/MASE}} (\citealt{bochanski2009}),
{\texttt{crowdsource}} (\citealt{schlafly2018}).
}

\appendix
\restartappendixnumbering

\section{Photometric, Spectroscopic, and Astrometric Follow-up\label{sec:appendix_phot_spec_astrom}}

\subsection{Photometry\label{sec:appendix_photometry}}

For possible M, L, T, and Y dwarf additions to the 20-pc census, we have searched for published near- and mid-infrared photometry using online surveys such as the Two Micron All Sky Survey (2MASS; \citealt{skrutskie2006}), the VISTA Hemisphere Survey (VHS; \citealt{mcmahon2013}), the UKIRT Hemisphere Survey (UHS; \citealt{dye2018}), and the Wide-field Infrared Survey Explorer (WISE; \citealt{wright2010}). These photometric measurements are listed in Table~\ref{tab:poss_20pc_members_MLTY}. In other cases, we have obtained our own ground-based follow-up or have searched the Spitzer Heritage Archive\footnote{\url {https://sha.ipac.caltech.edu/applications/Spitzer/SHA/}} for images with which to measure photometry. These results are also presented in Table~\ref{tab:poss_20pc_members_MLTY} but discussed further in the subsections below.

\subsubsection{Palomar/WIRC}

Eighteen objects were observed with the Wide-field Infrared Camera (WIRC; \citealt{wilson2003b}) at the Hale 5-m telescope on Palomar Mountain during the nights 2014 July 03, 2014 September 14, 2016 February 26, 2018 September 01, 2019 July 14, 2020 February 05, 2020 June 03, 2020 July 03, 2020 September 03, 2020 October 09, 2021 July 01, and 2021 August 10 UT. Data were acquired in the Maunakea Observatory filter system's $J$ and $H$ bands. Our standard observing technique, calibration strategy, and reduction methodology have been discussed in section 3.1.5 of \cite{kirkpatrick2011} with updates as discussed in section 9.2 of \cite{meisner2020a}.

\subsubsection{Keck/MOSFIRE}

Thirteen objects were observed with the Multi-Object Spectrometer For Infra-Red Exploration (MOSFIRE; \citealt{mclean2012}) at the 10-m W.\ M.\ Keck I telescope on Maunakea, Hawai'i, on the nights of 2021 August 27 and 2022 January 21 UT. Photometric acquisition and reductions followed the procedures described in section 3.1.1 of \cite{schneider2021}.

\subsubsection{Gemini-South/FLAMINGOS-2}

Nine objects were observed with the FLoridA Multi-object Imaging Near-infrared Grism Observational Spectrometer 2 (FLAMINGOS-2; \citealt{eikenberry2006}, \citealt{elston2003}, \citealt{jannuzi2004}) at the 8.1-m Gemini-South Observatory on Cerro Pach{\'o}n, Chile, on the nights of 2014 December 01, 2015 June 30, 2019 April 14, 2019 April 29, 2021 February 22, 2021 July 17, 2021 July 21, and 2021 October 18 UT. Photometric acquisition and reductions followed the procedures described in section 9.1 of \cite{meisner2020a}.

\subsubsection{SOAR/OSIRIS}

One object was observed with the Ohio State Infra-Red Imager/Spectrometer (OSIRIS) at the 4.1-m SOuthern Astrophysics Research (SOAR) Telescope on Cerro Pach{\'o}n, Chile, on the night of 2012 March 12 UT. Photometric acquisition and reduction of this lone $H$-band data point followed the methodology outlined in section 2.2.5 of \cite{kirkpatrick2012}.

\subsubsection{Spitzer/IRAC}

The Spitzer Heritage Archive was queried for directed or serendipitous observations of objects in our 20-pc candidate list (Table~\ref{tab:poss_20pc_members_MLTY}). The locations of thirty-one of these candidates were found to have Spitzer/IRAC observations in ch1 and/or ch2. The full list is shown in Table~\ref{tab:spitzer_photometry}. Although some of the Spitzer data were too shallow to detect our objects or were obtained at an epoch when our candidate was blended with a background source, most objects had measurable photometry. We used the MOPEX/APEX software (\citealt{makovoz2005a}, \citealt{makovoz2005b}) on each Astronomical Observation Request (AOR) to create mosaics, perform source detection, and then measure the photometry using the stack of individual frames. The output of the APEX code is the flux, in $\mu$Jy, for each detection using both aperture and PRF-fit techniques. The Spitzer photometry reported in Table~\ref{tab:poss_20pc_members_MLTY} is the PRF-fit photometry after converting to magnitudes using the correction factors listed in Table C.1 of the IRAC Handbook\footnote{\url {https://irsa.ipac.caltech.edu/data/SPITZER/docs/irac/iracinstrumenthandbook/home/}} along with the flux zeropoints in each band, as given in Table 4.1 of the Handbook. For objects having multiple AORs for a band, the reported photometry in that band is a weighted mean of the individual measurements in each AOR.

\startlongtable
\begin{deluxetable*}{lccccl}
\tabletypesize{\scriptsize}
\tablecaption{Ancillary Spitzer Data\label{tab:spitzer_photometry}}
\tablehead{
\colhead{Object} &
\colhead{AOR} &
\colhead{Bands\tablenotemark{a}} &                          
\colhead{Obs. Date} &                          
\colhead{Program}&
\colhead{PI}\\
\colhead{Name} &
\colhead{} &
\colhead{} &                          
\colhead{(UT)} &                          
\colhead{+ Phase\tablenotemark{b}}&
\colhead{Name}\\
\colhead{(1)} &                          
\colhead{(2)} &
\colhead{(3)} &
\colhead{(4)} &
\colhead{(5)} &
\colhead{(6)}
}
\startdata
CWISE J001322.53$-$114300.7& 60981248&  1,2&  2017-10-13&  13116PC&  Kelley        \\
         &   60981504&  1,2&  2017-10-16&  13116PC&  Kelley        \\
CWISE J025711.65$-$390626.9& 45160192&  1,2&  2012-03-20&  80109PC&  Kirkpatrick   \\
CWISE J041822.64+272958.8&   11233792&  1,2&  2005-02-20&   3584C &  Padgett       \\
         &   11237632&  1,2&  2005-02-21&   3584C &  Padgett       \\
         &   19030272&  1,2&  2007-03-30&  30816C &  Padgett       \\
         &   47116288&  1,2&  2013-11-07&  90071PC&  Kraus         \\
CWISE J042325.55+264045.0&    5074688&  1,2&  2004-10-05&    139C &  Evans         \\
         &    5075200&  1,2&  2004-10-07&    139C &  Evans         \\
         &   11233280&  1,2&  2005-02-21&   3584C &  Padgett       \\
         &   11237120&  1,2&  2005-02-21&   3584C &  Padgett       \\
         &   14604544&  1,2&  2005-09-16&  20386C &  Myers         \\
         &   14604800&  1,2&  2005-09-17&  20386C &  Myers         \\
         &   42296064&    1&  2011-11-04&  80053PC&  Paladini      \\
         &   47105280&  1,2&  2013-11-15&  90071PC&  Kraus         \\
CWISE J043227.67+260616.7&   11231744&  1,2&  2005-02-24&   3584C &  Padgett       \\
         &   11235584&  1,2&  2005-02-24&   3584C &  Padgett       \\
         &   47096320&  1  &  2013-11-10&  90071PC&  Kraus         \\
         &   47096832&  2  &  2013-11-10&  90071PC&  Kraus         \\
CWISE J044947.80$-$681745.6& 14351616&  1,2&  2005-07-18&  20203C &  Meixner       \\
         &   14364160&  1,2&  2005-10-25&  20203C &  Meixner       \\
CWISE J054921.19+264755.0&   38861568&  1,2&  2010-04-22&  61070PC&  Whitney       \\
         &   38873600&  2  &  2010-04-23&  61070PC&  Whitney       \\
CWISE J070314.97$-$062929.8& 38991360&  2  &  2010-05-08&  61071PC&  Whitney       \\
         &   38996992&  1  &  2010-05-07&  61071PC&  Whitney       \\
         &   39024896&  1,2&  2010-05-10&  61071PC&  Whitney       \\
         &   42089728&  1  &  2011-05-30&  61071PC&  Whitney       \\
         &   42133760&  1  &  2011-06-07&  61073PC&  Whitney       \\
CWISE J072900.97$-$742943.1& 41550592&  1,2&  2011-04-14&  70062PC&  Kirkpatrick   \\
CWISE J073748.86$-$252613.0& 39051520&  1,2&  2010-05-20&  61071PC&  Whitney       \\
CWISE J075744.48$-$300504.3& 39030528&  1,2&  2010-05-31&  61071PC&  Whitney       \\
         &   39032320&  2  &  2010-05-31&  61071PC&  Whitney       \\
CWISE J080940.43$-$372003.7&  6580224&  1,2&  2003-12-07&    104C &  Soifer        \\
         &   39328768&  1,2&  2010-06-13&  61072PC&  Whitney       \\
         &   39337984&  1  &  2010-06-13&  61072PC&  Whitney       \\
CWISE J091942.64$-$495243.7& 23707392\tablenotemark{d}&  1,2&  2008-03-04&  40791C &  Majewski      \\
CWISE J093035.01$-$743148.6& 44563200&  1,2&  2012-04-04&  80109PC&  Kirkpatrick   \\
CWISE J110238.85$-$775039.7& 19986432&  1,2&  2007-05-16&  30574C &  Allen         \\
         &   20006400&  1,2&  2007-05-16&  30574C &  Allen         \\
         &   43320064&  1  &  2011-07-23&  80053PC&  Paladini      \\
         &   47091712&  1,2&  2013-08-24&  90071PC&  Kraus         \\
CWISE J112106.36$-$623221.5& 23699712\tablenotemark{d}&  1,2&  2008-07-20&  40791C &  Majewski      \\
         &   42735360&  1,2&  2012-03-21&  80074PC&  Whitney       \\
CWISE J112440.19+663051.1&   70016512&  1,2&  2019-07-13&  14299PC&  Faherty       \\
CWISE J123455.88$-$641923.7& 42703104&  1,2&  2012-04-05&  80074PC&  Whitney       \\
CWISE J133427.46$-$625736.6&  9233664&  2  &  2004-03-10&    190C &  Churchwell    \\
         &    9234176&  1  &  2004-03-10&    190C &  Churchwell    \\
         &   11761920&  2  &  2004-07-21&    189C &  Churchwell    \\
         &   11768832&  1  &  2004-07-21&    189C &  Churchwell    \\
         &   45388800&  1  &  2012-04-10&  80074PC&  Whitney       \\
CWISE J135338.04+441017.6&   43424512&  2  &  2011-08-25&  80095PC&  Werner        \\
         &   45116416&  1,2&  2012-04-04&  80109PC&  Kirkpatrick   \\
VVV J165507.19$-$421755.5& 11955712\tablenotemark{c}&  1,2&  2004-09-06&  192C   &  Churchwell    \\
CWISE J181125.34+665806.4&   49728000&  1,2&  2014-05-09&  10147PC&  Bock          \\
         &   49728256&  1,2&  2014-05-09&  10147PC&  Bock          \\
         &   49728512&  1,2&  2014-05-09&  10147PC&  Bock          \\
         &   49739520&  2  &  2014-07-24&  10147PC&  Bock          \\
         &   49739776&  2  &  2014-07-24&  10147PC&  Bock          \\
         &   49740032&  1,2&  2014-07-25&  10147PC&  Bock          \\
         &   49752576&  1,2&  2014-09-14&  10147PC&  Bock          \\
         &   49752832&  1,2&  2014-09-14&  10147PC&  Bock          \\
         &   49753088&  2  &  2014-09-14&  10147PC&  Bock          \\
         &   62371328&  1  &  2017-02-19&  13153PC&  Capak         \\
         &   62377728&  2  &  2017-02-14&  13153PC&  Capak         \\
         &   62846464&  1  &  2017-07-05&  13153PC&  Capak         \\
         &   62921984&  1,2&  2018-01-04&  13153PC&  Capak         \\
         &   62937344&  2  &  2017-07-17&  13153PC&  Capak         \\
         &   62949120&  2  &  2017-11-20&  13153PC&  Capak         \\
         &   65073152&  2  &  2018-01-19&  13153PC&  Capak         \\
         &   65094400&  1,2&  2018-01-12&  13153PC&  Capak         \\
         &   65114624&  1  &  2018-06-09&  13153PC&  Capak         \\
         &   65124608&  1,2&  2018-06-04&  13153PC&  Capak         \\
         &   65133312&    2&  2018-06-03&  13153PC&  Capak         \\
         &   65190656&  1  &  2018-02-27&  13153PC&  Capak         \\
         &   65196288&  1,2&  2018-02-26&  13153PC&  Capak         \\
         &   65692160&  1  &  2019-01-05&  13153PC&  Capak         \\
         &   65718016&  2  &  2018-06-22&  13153PC&  Capak         \\
         &   65765632&  1,2&  2018-08-05&  13153PC&  Capak         \\
         &   65810176&  1,2&  2019-01-15&  13153PC&  Capak         \\
         &   65817088&  1  &  2018-06-26&  13153PC&  Capak         \\
         &   68615680&  1,2&  2019-02-27&  13153PC&  Capak         \\
         &   68631296&  2  &  2019-02-13&  13153PC&  Capak         \\
         &   68632576&  1  &  2019-02-09&  13153PC&  Capak         \\
CWISE J181429.08$-$202534.4& 21272832&  1  &  2007-05-16&  30570C &  Benjamin      \\
         &   21339392&  2  &  2007-05-16&  30570C &  Benjamin      \\
CWISE J185316.77$-$540658.0& 45091584&  1,2&  2012-05-25&  80109PC&  Kirkpatrick   \\
CWISE J190405.09$-$372616.9& 27041280&  1,2&  2008-05-10&  30574C &  Allen         \\
         &   47020032&  1,2&  2012-12-02&  90071PC&  Kraus         \\
CWISE J191118.88+085456.3&   11966976&  1,2&  2004-10-09&  187C   &  Churchwell    \\
CWISE J191839.52+441835.6&   50112512\tablenotemark{e}&  1,2&  2014-01-03&  10067PC&  Werner        \\
CWISE J192738.93$-$851335.6& 46936064&  1,2&  2012-12-03&  80077PC&  Leggett       \\
CWISE J204055.20+465148.0&   39489280\tablenotemark{c}&  1,2&  2010-07-15&  61072PC&  Whitney       \\
CWISE J205338.54$-$353922.5& 45070336&  1,2&  2012-06-23&  80109PC&  Kirkpatrick   \\
CWISE J213322.07$-$174151.5& 53349888&  1  &  2016-09-05&  11097PC&  Rozitis       \\
\enddata
\tablenotetext{a}{The bands refer to 1 = ch1 = 3.6 $\mu$m and 2 = ch2 = 4.5 $\mu$m.}
\tablenotetext{b}{The letter code after the program number refers to C = cryogenic data or PC = post-cryogenic data.}
\tablenotetext{c}{Source identification is uncertain or blended in both bands.}
\tablenotetext{d}{Source is undetected in both bands, as these data are very shallow.}
\tablenotetext{e}{Source is badly blended and not extracted in ch1.}
\end{deluxetable*}

\subsection{Spectroscopy\label{sec:appendix_spectroscopy}}

\begin{figure*}
\includegraphics[scale=0.475,angle=0]{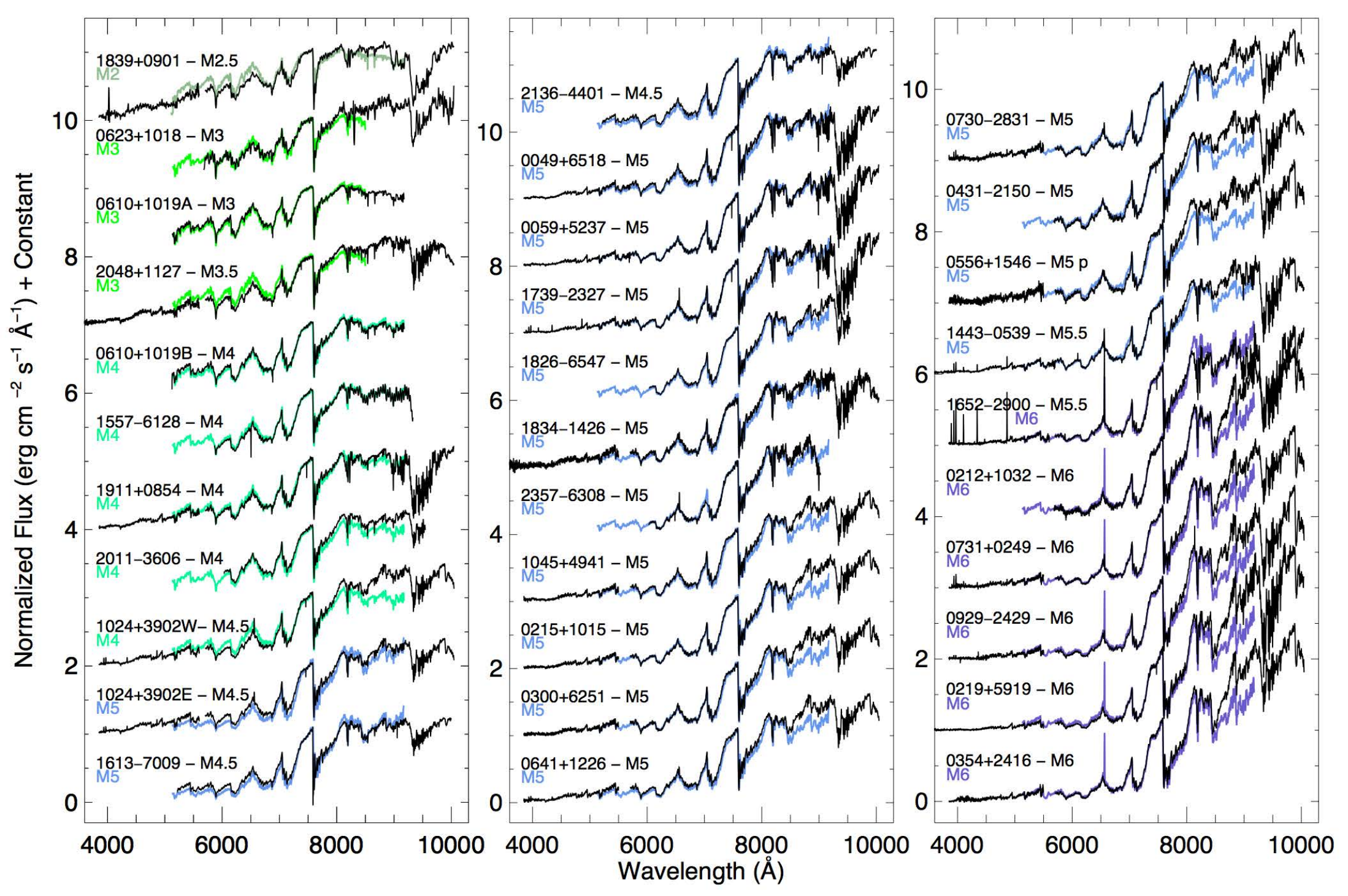}
\caption{Optical spectroscopic follow-up for objects classified as early-M through mid-M. Each target object (black) is normalized to one at 7500 \AA\ and overplotted (in other colors) with the spectral standard nearest the same spectral type. Integral offsets have been added to separate the spectra vertically. Target objects are labeled with brief RA/Dec (hhmm$\pm$ddmm) identifiers. The two target objects at upper left -- 1839+0901 and 0623+1018 -- have been smoothed to improve the signal-to-noise in each wavelength bin. Most spectra have not been corrected for earth's atmospheric absorption, so the contaminating B- and A-bands of O$_2$ at $\sim$6850-6900 and $\sim$7600-7700 \AA\ and telluric bands of H$_2$O at $\sim$7150-7300, 8150-8350, and 9000-9600 \AA\ remain.
\label{fig:opt_spec_panel1}}
\end{figure*}

\begin{figure*}
\includegraphics[scale=0.5,angle=0]{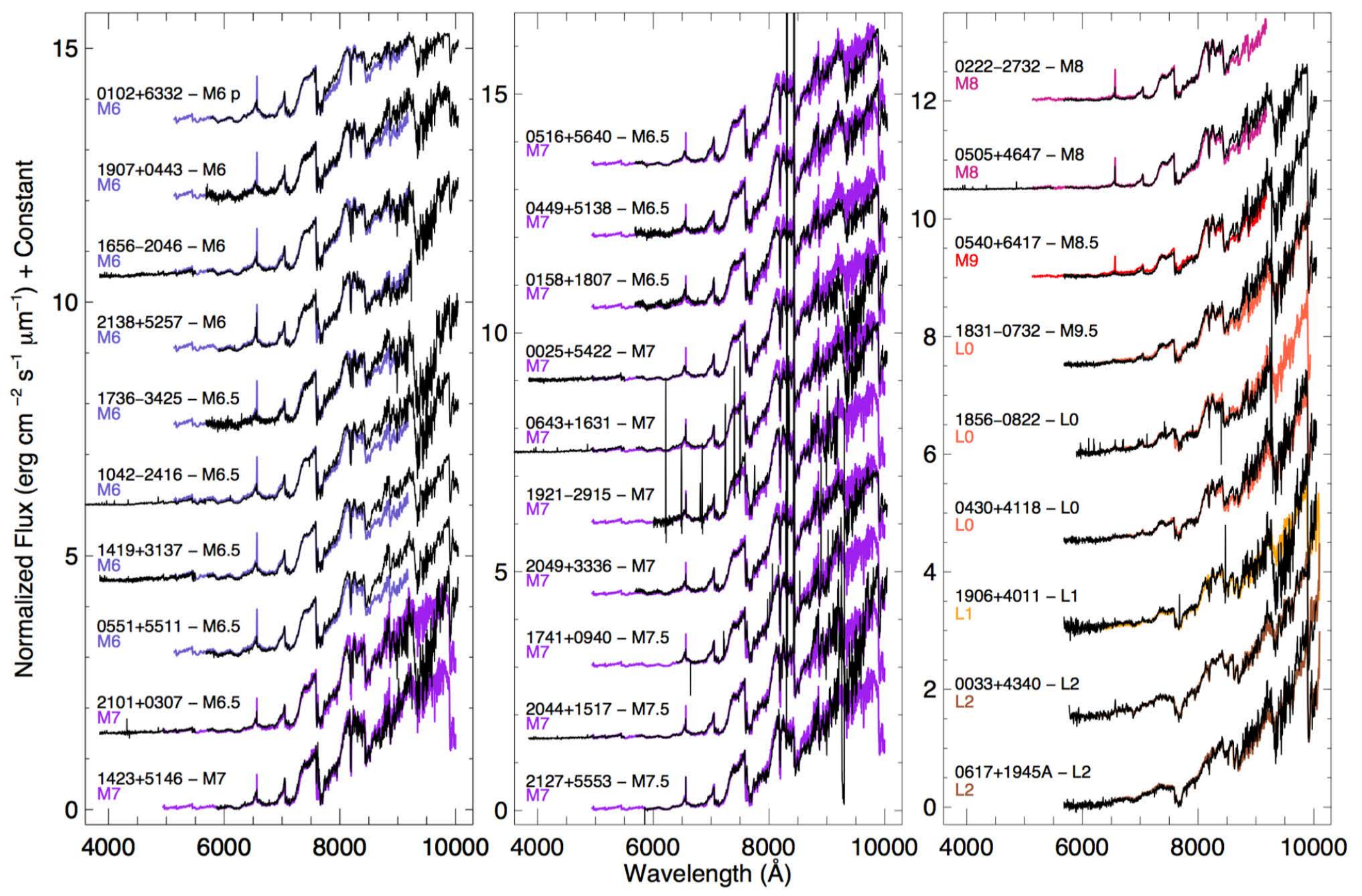}
\caption{Optical spectroscopic follow-up for objects classified as mid-M through early-L. Each target object (black) in the two left panels is normalized to one at 7500 \AA\ and overplotted (in other colors) with the spectral standard nearest the same spectral type. In the far right panel, this normalization is done instead at 8250 \AA. Offsets in steps of 1.5 have been added to separate the spectra vertically. A few target objects -- 1921$-$2915 (M7), 1906+4011 (L1), 0033+4340 (L2), and 0617+1945A (L2) -- have been smoothed to improve the signal-to-noise in each wavelength bin. The spectrum of 1921$-$2915 (M7) also suffers from residual cosmic ray hits. See the caption to Figure~\ref{fig:opt_spec_panel1} for more details.
\label{fig:opt_spec_panel2}}
\end{figure*}

\begin{figure*}
\includegraphics[scale=0.65,angle=0]{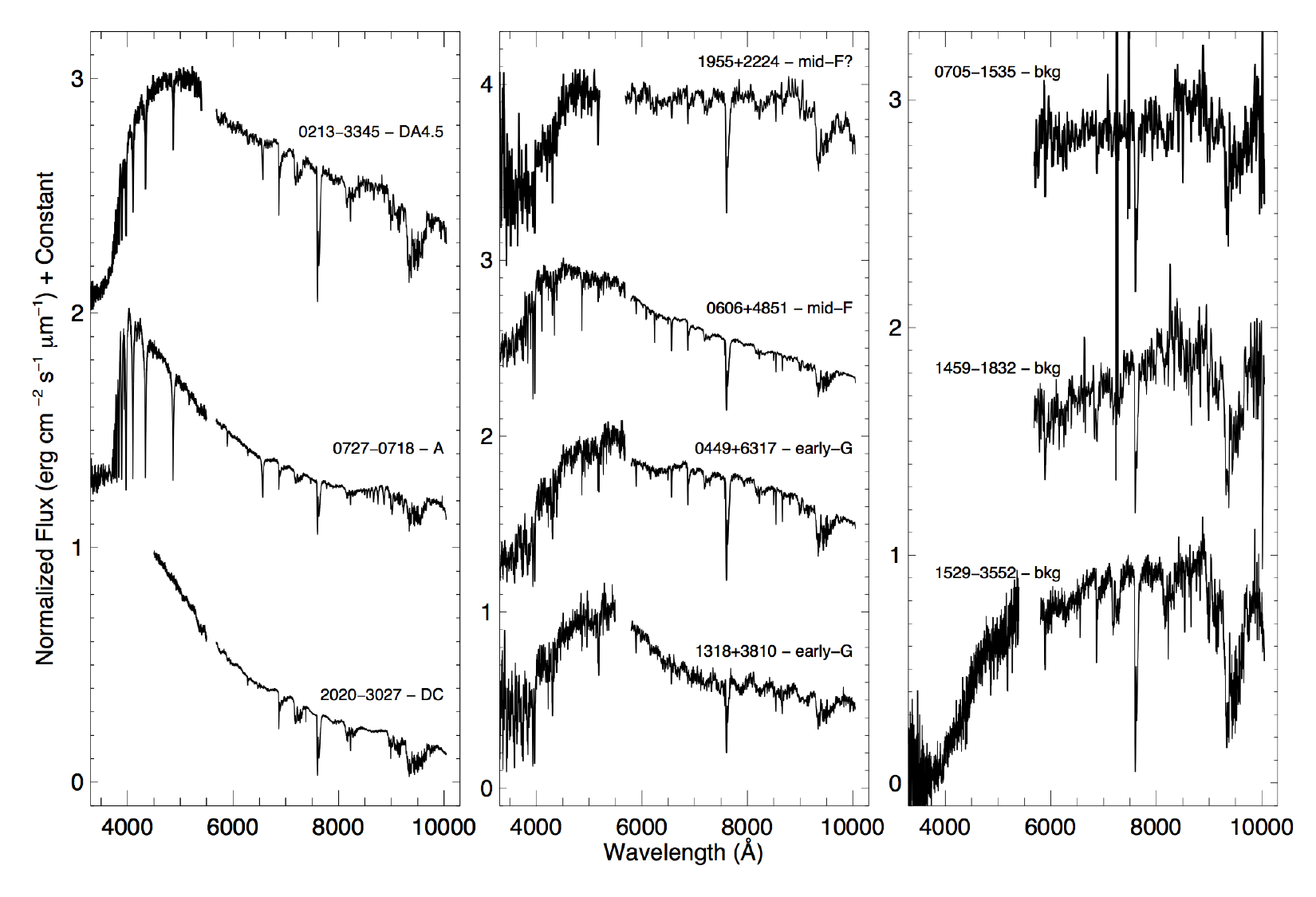}
\caption{Optical spectroscopic follow-up of objects not classified as M dwarfs or L dwarfs. Each target object is normalized to one at its peak flux. Objects in the far left panel are hot stars, and objects in the two right panels are colder stars or other background objects. Integral offsets have been added to separate the spectra vertically. A few spectra -- 0213$-$3345 (wd), 1955+2224 (mid-F?), 1318+3810 (early-G), and all of those in the rightmost panel -- have been smoothed to improve the signal-to-noise in each wavelength bin. See the caption to Figure~\ref{fig:opt_spec_panel1} for more details.
\label{fig:opt_spec_panel3}}
\end{figure*}

\begin{figure*}
\includegraphics[scale=0.525,angle=0]{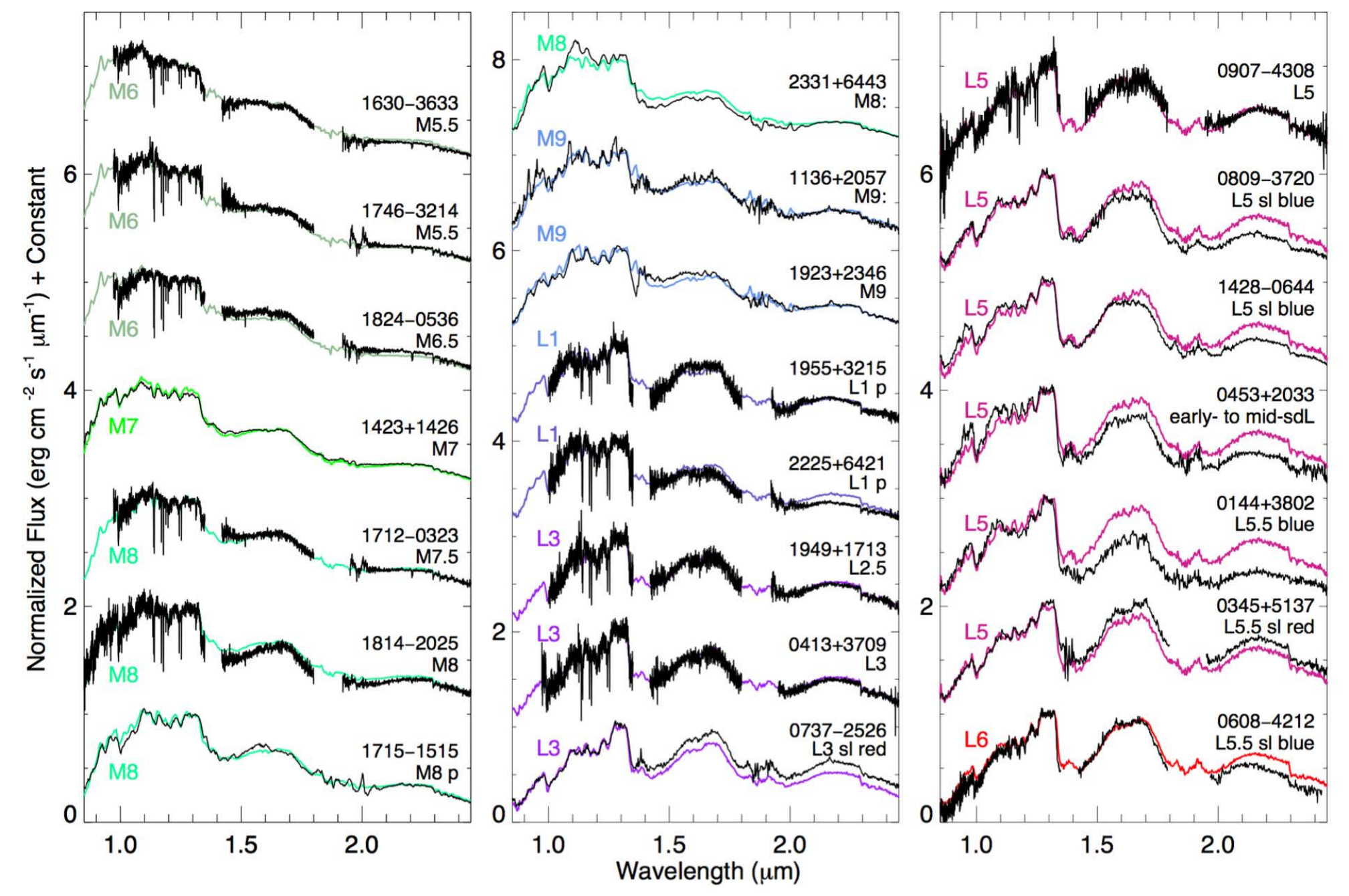}
\caption{Near-infrared spectroscopic follow-up of objects classified as mid-M through mid-L. Each target object (black) is normalized to one at 1.28 $\mu$m and overplotted (in other colors) with the spectral standard nearest the same spectral type. Integral offsets have been added to separate the spectra vertically. Target objects are labeled with brief RA/Dec (hhmm$\pm$ddmm) identifiers. One spectrum -- 0907$-$4308 (L5) -- has been smoothed to improve the signal-to-noise in each wavelength bin. Data deep within the telluric water bands near $\sim$1.4 and $\sim$1.75 $\mu$m are not displayed for some targets because of the poor signal-to-noise in those regions. 
\label{fig:nir_spec_panel1}}
\end{figure*}

\begin{figure*}
\includegraphics[scale=0.525,angle=0]{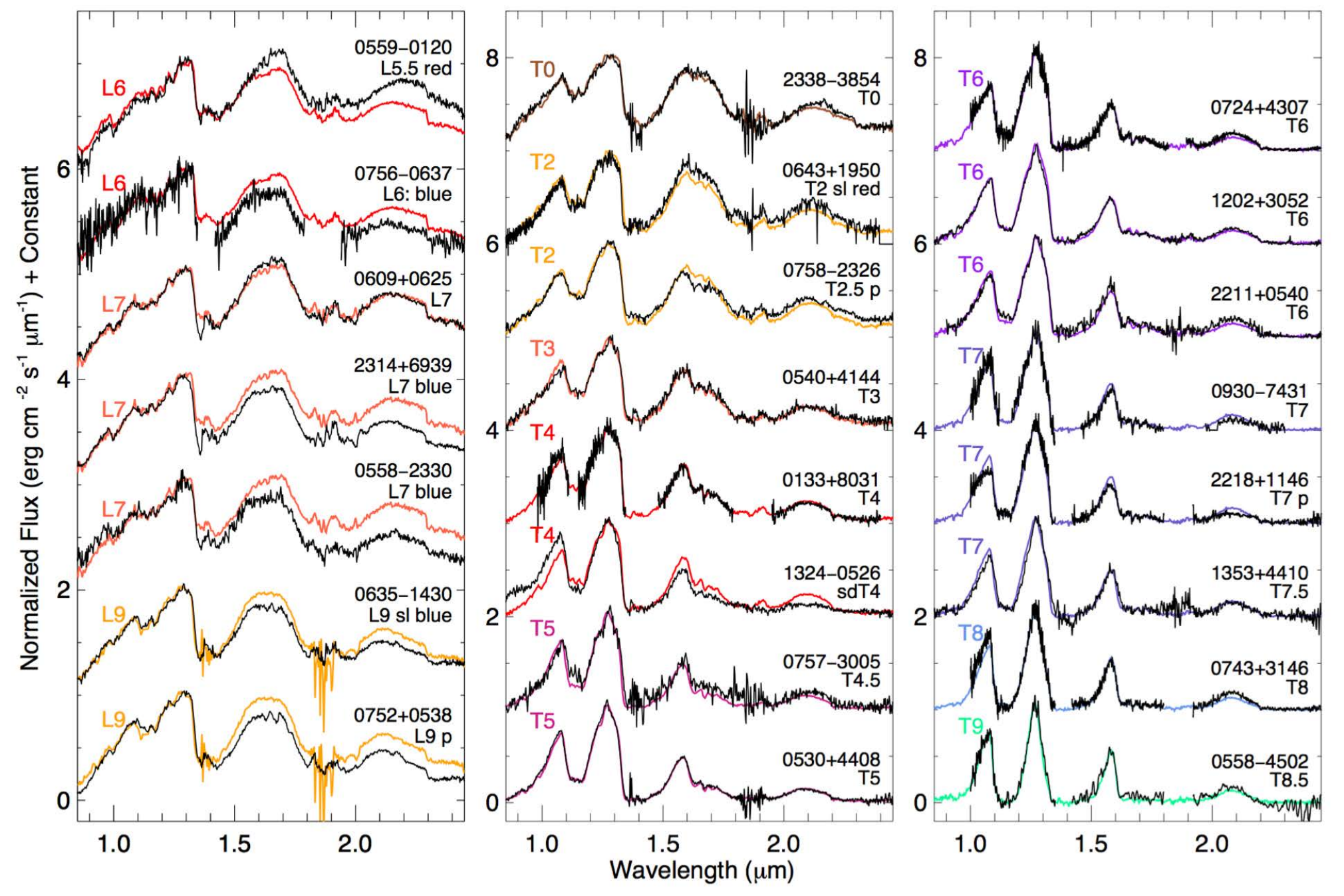}
\caption{Near-infrared spectroscopic follow-up of objects classified as mid-L through late-T. One spectrum -- 0133+8031 (T4) -- has been smoothed to improve the signal-to-noise in each wavelength bin. Data deep within the telluric water bands near $\sim$1.15, $\sim$1.4, and $\sim$1.75 $\mu$m are not displayed for some targets because of the poor signal-to-noise in those regions. See the caption to Figure~\ref{fig:nir_spec_panel1} for other details.
\label{fig:nir_spec_panel2}}
\end{figure*}

To aid in the characterization of objects, spectroscopy was acquired of 20-pc members discovered by Gaia, 20-pc suspects discovered by the Backyard Worlds: Planet 9 citizen science group, or 20-pc members lacking published spectral types. These reduced spectra are illustrated in Figures~\ref{fig:opt_spec_panel1} through \ref{fig:nir_spec_panel2}. Ten different instruments were used for this follow-up, as detailed below and summarized in Table~\ref{tab:spec_followup}.

\subsubsection{CTIO/RCSpec}

Optical follow-up of five objects was obtained on the UT dates of 1995 August 13-14, 1997 July 14, and 1997 July 16 at the Cerro Tololo Interamerican Observatory (CTIO) 4m telescope and on 1996 May 20 at the CTIO 1.5m telescope using the R-C Spectrograph with Folded Schmidt Camera (RCSpec). For the 1995 and 1996 observations, a 300 line mm$^{-1}$ grating with a GG 495 order-blocking filter was used with the 1024$\times$1024 CCD to cover a wavelength range from 6050 to 9550 \AA. For the 1997 observations, a 316 line mm$^{-1}$ grating with an OG 515 order-blocking filter was used with the Loral 3K CCD to cover a useable wavelength range from 5200 to 10000 \AA. In addition to the targets, standard calibrations -- biases, dome flats, arcs, and flux calibration standards -- were also obtained. Reductions were accomplished using the Image Reduction and Analysis Facility (\texttt{IRAF}; \citealt{tody1986,tody1993}), as described in \cite{kirkpatrick1997}.

\subsubsection{Lick/Kast}

The Kast Double Spectrograph at the Lick 3m Shane Telescope was used for optical follow-up of six objects on UT dates 2019 September 20,  2020 March 06, 2020 August 15-16, 2020 December 14, and 2022 July 02. The only data used were from the red arm, which employed a 600 line mm$^{-1}$ grating blazed at 7500 \AA\ to cover the wavelength range from $\sim$6000 to $\sim$9000 \AA. In addition to standard wavelength and flux calibrations, G2V and A0V stars were obtained near in time and on sky to the targets to correct for telluric absorption. The data were reduced using the \texttt{kastredux}\footnote{\url{https://github.com/aburgasser/kastredux}} package (Burgasser et al., in prep.), as further described in \cite{schneider2021}.

\subsubsection{Palomar/DBSP}

The Double Spectrograph (DBSP; \citealt{oke1982}) at the Hale 5m telescope on Palomar Mountain was used for sixty additional follow-up spectra. The UT nights of observation were 1995 December 02, 2021 December 06, 2022 January 02, 2022 February 03, 2022 February 07, 2022 May 24, 2022 May 30, 2022 June 07, and 2022 August 27. For the 1995 run, the D68 dichroic was used to split the light at $\sim$6800 \AA\ between the two arms. Gratings with 300 line mm$^{-1}$ blazed at 3990 \AA\ and with 316 line mm$^{-1}$ blazed at 7150 \AA\ were used in the blue and red arms, respectively, producing continuous wavelength coverage from $\sim$5100 to 9200 \AA. For the 2021 and 2022 runs, the D55 dichroic was used instead to split the light near 5500 \AA. A 600 line mm$^{-1}$ grating blazed at 3780 \AA\ was used in the blue arm and a 316 line mm$^{-1}$ blazed at 7150 \AA\ was used in the red arm, producing coverage from $\sim$3300 to 5500 \AA\ on the blue side and from $\sim$5700 to 10000 \AA\ on red side. (For the 2022 June run, fringing in the blue arm caused data shortward of 4500 \AA\ to be unusable.) Standard calibrations and \texttt{IRAF} data reductions were employed, as further described in \cite{kirkpatrick1991}.

\subsubsection{SALT/RSS}

An additional optical spectrum was acquired with the Robert Stobie Spectrograph (RSS; \citealt{burgh2003,kobulnicky2003}) on the 11.1$\times$9.8-m Southern African Large Telescope (SALT; \citealt{buckley2006}) on 2021 December 26 UT. The spectrograph was used in long slit mode using the PG0900 grating at an angle of 20$^\circ$, which produces coverage over the ranges 6033-7028, 7079-8045, and 8091-9023 \AA\ across the 3$\times$1 mini-mosaic. Our reductions began with the observatory-provided pre-processed data, for which gain correction, correction for cross-talk, and overscan subtraction had been applied. We then wavelength calibrated using neon arc lines obtained immediately after the target's spectroscopic data and flux calibrated using the \cite{hamuy1994} standard EG21 acquired with the same spectroscopic setup on 2023 January 24 UT.

\subsubsection{APO/TSpec}

TripleSpec (\citealt{wilson2004}) on the ARC 3.5m telescope at Apache Point Observatory was used for near-infrared follow-up of two objects on the nights of 2018 September 23 and 2019 October 08 UT. The spectrograph provides spectral coverage from 0.95 to 2.46 $\mu$m across five spectral orders. Data were taken with the conventional near-infrared technique of nodded pairs to perform background/bias subtraction, and standard calibrations were also acquired, including quartz lamps for flat fielding and A0 stars for telluric correction and flux calibration. Wavelength calibration was accomplished using night sky lines. Data reduction used \texttt{Tspectool}\footnote{\url{http://www.apo.nmsu.edu/arc35m/Instruments/TRIPLESPEC/\#7}}, a modified version of \texttt{Spextool} (\citealt{cushing2004}) rewritten specifically for APO/TripleSpec. 

\subsubsection{CTIO/ARCoIRIS}

Three objects were observed on the nights of 2018 April 02-03 UT using the Astronomy Research with the Cornell Infra Red Imaging Spectrograph (ARCoIRIS) at the 4m Victor Blanco telescope at CTIO. Spectra are acquired across six cross-dispersed orders covering 0.8 to 2.4 $\mu$m at a resolving power of $\sim$3500. Science exposures were taken with AB nod positions along the slit, which has a fixed width of 1$\farcs$1. Standard calibrations were acquired as discussed in \cite{greco2019}, and data were reduced using a modified version of the \texttt{Spextool} package (\citealt{cushing2004}), which utilizes A0 stars for telluric correction and flux calibration following the methodology of \cite{vacca2003}.

\subsubsection{IRTF/SpeX}

The SpeX instrument on the NASA Infrared Telescope Facility (IRTF) was used for near-infrared spectroscopy of twenty-seven objects over the nights of 2018 June 16, 2018 November 25, 2019 January 23, 2019 March 16-17, 2020 October 30, 2020 November 25, 2021 May 31, 2021 June 30, 2021 September 11, 2021 October 23, 2022 January 09, 2022 January 19, 2022 February 12, 2022 February 21, 2022 March 07, and 2022 March 11 UT. Two different setups were employed. The first, used mainly for brighter targets, was a cross-dispersed mode that provides spectra over the range 0.9-2.4 $\mu$m at a resolving power of $R \equiv {\lambda}/{\Delta}{\lambda} \approx 1200$. The second, used primarily for fainter targets, was the prism mode that provides spectra over the range 0.8-2.5 $\mu$m at a resolving power of $R \equiv {\lambda}/{\Delta}{\lambda} \approx 100-150$. As discussed in the subsections above, standard near-infrared calibrations  were obtained, and data were reduced using \texttt{Spextool} (\citealt{cushing2004}, \citealt{vacca2003}).

\subsubsection{Keck/NIRES}

Three objects were observed over the nights of 2019 February 14, 2020 July 07, and 2021 February 24 UT using the Near-Infrared Echellette Spectrometer (NIRES; e.g., \citealt{wilson2004}) at the 10m W.\ M.\ Keck II telescope. These data provided spectral coverage from 0.94 to 2.45 $\mu$m. Setup and calibrations were identical to those described in \cite{meisner2020b}, and reductions used a modified version of \texttt{Spextool} (\citealt{cushing2004}, \citealt{vacca2003}).

\subsubsection{Magellan/FIRE}

The Folded-port Infrared Echellette spectrograph (FIRE; \citealt{simcoe2013}) at the 6.5 m Walter Baade (Magellan I) telescope at Las Campanas Observatory was used to observe three objects over the nights of 2016 January 23, 2019 December 11, and 2020 February 18 UT. Observations were done in prism mode, which covers the range from 0.80 to 2.45 $\mu$m. Standard calibrations were acquired, and data were reduced using the \texttt{FIREHOSE} pipeline, which is based on the \texttt{MASE} (\citealt{bochanski2009}) and \texttt{Spextool} (\citealt{cushing2004}, \citealt{vacca2003}) packages.

\subsubsection{Palomar/TSpec}

Finally, seven near-infrared spectra were acquired with the Triple Spectrograph (TSpec; \citealt{herter2008}) at Palomar Mountain's 5m Hale telescope on the nights of 2018 April 28-29, 2018 October 17, and 2019 September 18 UT. Setup and calibrations were identical to those described in \citealt{kirkpatrick2011}. As with many of the other near-infrared spectroscopic data sets discussed above, TSpec data were also reduced with a modified version of \texttt{Spextool} (\citealt{cushing2004}, \citealt{vacca2003}).

\subsubsection{Analysis}

Spectral classification was accomplished by comparing spectra of the target objects to established on-sky anchors for each integral spectral type. For the optical spectra, these anchor points were taken from \cite{kirkpatrick1991} for objects of type mid-K through late-M and from \cite{kirkpatrick1999} for L dwarfs (Figures~\ref{fig:opt_spec_panel1}-\ref{fig:opt_spec_panel2}). Optical classifications of objects earlier than type K (Figure~\ref{fig:opt_spec_panel3}) used spectral anchors taken from \cite{gray2009}. Near-infrared classification (Figures~\ref{fig:nir_spec_panel1}-\ref{fig:nir_spec_panel2}) for M dwarfs, L dwarfs, and early-T dwarfs used the anchors described in \cite{kirkpatrick2010}, with the rest of the T dwarf anchors coming from \cite{burgasser2006}. For more on the methodology employed for both optical and near-infrared classifications, see \cite{kirkpatrick2010}.

These classification anchors are generally old disk objects with metallicities similar to the Sun. In a few cases, described below, the target object failed to match an anchor spectrum because of anomalous features attributable to extreme youth, lower metallicity, or other reasons including unresolved binarity. These special cases are addressed further below:

{\it CWISE J045334.34+203350.2:} At $J$-band, this object best matches the L5 standard, but there are clear discrepancies with the L5 standard across all wavelengths (Figure~\ref{fig:nir_spec_panel1}). The continuum of 0453+2033 is much flatter between 1.1 and 1.3 $\mu$m, the FeH band at 0.99 $\mu$m is much stronger, and the $H$- and $K$-band portions emit less flux relative to $J$-band than does the standard. We find that the $J$-band spectrum of 0453+2033 is a better match to 2MASS J17561080+2815238, which is typed in both the optical and the near-infrared as an sdL1 (\citealt{kirkpatrick2010}, \citealt{zhang2018}), in both the continuum shape and the strength of the FeH band. However, 0453+2033 has more flux at $H$ and $K$ relative to $J$ than does 2MASS J1756+2815, possibly indicating that the former is a slightly later subdwarf. Given that the set of anchors for the L subdwarf spectral sequence is still incomplete (\citealt{zhang2018}), we tentatively classify this object as an early- to mid-sdL.

{\it CWISE J055942.94$-$012002.4:} Of the spectra in Figures~\ref{fig:nir_spec_panel1}-\ref{fig:nir_spec_panel2} that have a "red" or "slightly red" classification, only 0559$-$0120 has the triangular-shaped $H$-band peak indicative of low gravity. Such low-gravity objects are necessarily young, as they have not yet contracted to their final radii. Using just the sky position and proper motion values (Table~\ref{tab:poss_20pc_members_MLTY}), as its parallax and radial velocity have not yet been measured, BANYAN $\Sigma$ (Bayesian Analysis for Nearby Young Associations; \citealt{gagne2018}) gives the object an 80\% chance of belonging to a known, young moving group -- either the AB Doradus group or, less likely, the $\beta$ Pictoris group. If an AB Dor member, BANYAN $\Sigma$ predicts 46$\pm$3 pc with a radial velocity of 22$\pm$2 km s$^{-1}$; if a $\beta$ Pic member, the predictions are d = 21$\pm$3 pc and RV = 19$\pm$2 km s$^{-1}$. Using solely an $M_{Ks}$ vs.\ spectral type relation (\citealt{dupuy2012}), as advocated for young L dwarfs in \cite{schneider2023}, we estimate a distance of $\sim$28.8 pc for this L5.5 dwarf, based on a value of $K_s = 14.34{\pm}0.09$ mag from the 2MASS All-Sky Point Source Catalog.

{\it CWISE J075227.38+053802.6:} We classify this object as L9 pec (Figure~\ref{fig:nir_spec_panel2}). The peculiarities stem from the two unusual absorption troughs at 1.63 and 1.67 $\mu$m within the $H$-band plateau and the unusual inflection near 2.21 $\mu$m at $K$-band. Such features are indicative of methane absorption, which should not be present shortward of 2.5 $\mu$m in an L9 dwarf. As previous papers such as \cite{burgasser2007b}, \cite{burgasser2010b}, and \cite{bardalez2014} have noted, such spectra may indicate the presence of an unresolved binary comprised of two morphologically distinct spectra -- a non-methane M or L dwarf and a methane-rich T dwarf. If 0752+0538 represents such an unresolved binary, modeling (see section 4.5 of \citealt{kirkpatrick2016}) suggests it is likely a late-L plus early-T composite system.

{\it CWISE J075853.12$-$232645.8:} We classify this object as T2.5 pec (Figure~\ref{fig:nir_spec_panel2}). Although its $J$-band peak matches both the T2 and T3 standards equally well, the $H$-band flux is suppressed and the $K$-band flux elevated relative to the standards. We find that a synthetic spectrum made up of components of types L8-L9 and T5-T6 fits the overall spectral shape slightly better than the single standards, suggesting perhaps that this object is an unresolved binary.

{\it CWISE J132403.81$-$052631.4:} The width of the $J$-band peak in this object best matches that of the T4 standard (Figure~\ref{fig:nir_spec_panel2}), but the fits at both shorter and longer wavelengths are much poorer. Specifically, the $H$- and $K$-band portions of 1324$-$0526 emit less flux relative to $J$-band than does the standard, and the $K$-band portion is notably flattened, an effect often ascribed to increased collision-induced absorption by H$_2$. Moreover, the $Y$-band portion emits more flux relative to $J$-band than does the standard. Elevated $Y$-band flux and suppressed $K$-band flux are seen in a comparison of the sdT5.5 dwarf HIP 73786B (Figure 1 of \citealt{zhang2019}) to standards of type T5 and T6, although the discrepancies are stronger in 1324$-$0526, and the latter also shows suppressed $H$-band flux. In the case of HIP 73786, the system has a measured subsolar metallicity of [Fe/H] = $-0.3{\pm}0.1$ ({\citealt{murray2011}}) from the K5 V primary, suggesting that the metallicity of 1324$-$0526 is somewhat lower still. We classify 1324$-$0526 as sdT4.

{\it CWISE J221859.41+114642.7:} The width of this object's $J$-band peak is most similar to the T7 standard, but its $H$- and $K$-band flux peaks are suppressed, with the latter being noticeably flattened. As with 1324$-$0526 above, such features are typical of subdwarfs, although the suppression of the $Y$-band peak in 2218+1146 runs contrary to the trend seen in T subdwarfs of slightly earlier type. In the sdT8 WISE J200520.38+542433.9, a wide companion in the low-metallicity ([Fe/H] = $-0.64{\pm}0.17$) Wolf 1130 system (\citealt{mace2013b}), the $Y$-band peak is shifted notably to the blue -- from 1.09 to 1.03 $\mu$m -- relative to the standards, an effect also seen in the isolated sdT6.5 dwarf ULAS J131610.28+075553.0 (\citealt{burningham2014}). Our spectrum is too noisy in this region to determine whether the same effect is present in 2218+1146, so we classify this object as T7 pec pending further confirmation of its subdwarf status.

\startlongtable
\begin{deluxetable*}{lcccccc}
\tabletypesize{\scriptsize}
\tablecaption{Spectroscopic Follow-up\label{tab:spec_followup}}
\tablehead{
\colhead{Object} &
\colhead{Coords} &
\colhead{Instrument} &                          
\colhead{Obs. Date} & 
\colhead{Observer}&
\colhead{Opt.}&
\colhead{NIR}\\
\colhead{} &
\colhead{(hhmm$\pm$ddmm)} &
\colhead{} &                          
\colhead{(UT)} &                          
\colhead{Code} &
\colhead{Spec. Type}&
\colhead{Spec. Type\tablenotemark{a}}\\
\colhead{(1)} &                          
\colhead{(2)} &
\colhead{(3)} &
\colhead{(4)} &
\colhead{(5)} &
\colhead{(6)} &
\colhead{(7)}
}
\startdata
2MASS J00251602+5422547&       0025+5422  &   Palomar/DBSP& 2022 Aug 27& S      & M7&      \nodata \\
CWISER J003350.99+434010.6 &   0033+4340  &   Palomar/DBSP& 2022 Jan 02& E,L,S,T& L2&      \nodata \\
2MASS J00492565+6518038&       0049+6518  &   Palomar/DBSP& 2022 Aug 27& S      & M5&      \nodata \\
WISE J005936.73+523719.0&      0059+5237  &   Palomar/DBSP& 2022 Aug 27& S      & M5&      \nodata \\
1RXS J010228.1+633256 &        0102+6332  &   Palomar/DBSP& 2022 Jan 02& E,L,S,T& M6 pec&  \nodata \\
CWISE J013343.58+803153.1 &    0133+8031  &   APO/TSpec   & 2019 Oct 08& A,Q    & \nodata& T4\\
CWISE J014407.64+380255.6 &    0144+3802  &   IRTF/SpeX   & 2022 Jan 09& C,F    & \nodata& L5.5 blue\\
WISEA J015815.65+180713.7 &    0158+1807  &   Palomar/DBSP& 2022 Feb 03& S      & M6.5&    \nodata \\
2MASS J02124635+1032546 &      0212+1032  &   Palomar/DBSP& 2022 Feb 03& S      & M6&      \nodata \\
LP 941-19 &                    0213$-$3345&   Palomar/DBSP& 2021 Dec 06& S      & DA4.5 &     \nodata \\
LP 469-205 &                   0215+1015  &   Palomar/DBSP& 2022 Feb 03& S      & M5&      \nodata \\
2MASS J02195603+5919273 &      0219+5919  &   Palomar/DBSP& 2021 Dec 06& S      & M6&      \nodata \\ 
2MASS J02224767$-$2732349 &    0222$-$2732&   Palomar/DBSP& 2021 Dec 06& S      & M8&      \nodata \\
2MASS J03000272+6251582 &      0300+6251  &   Palomar/DBSP& 2022 Feb 03& S      & M5&      \nodata \\
CWISE J034547.29+513716.0 &    0345+5137  &   IRTF/SpeX   & 2021 Nov 25& C,F,R  & \nodata& L5.5 sl.\ red\\
LP 357-56 &                    0354+2416  &   Palomar/DBSP& 2022 Feb 03& S      & M6&      \nodata \\
2MASS J04134574+3709087 &      0413+3709  &   Palomar/TSpec&2018 Oct 17& U,V,W  & \nodata& L3\\
Gaia EDR3 180116295441149824 & 0430+4118  &   Palomar/DBSP& 2021 Dec 06& S      & L0&      \nodata \\
LP 834-48 &                    0431$-$2150&   Palomar/DBSP& 2022 Feb 07& S      & M5&      \nodata \\
2MASS J04490464+5138412 &      0449+5138  &   Palomar/DBSP& 2022 Feb 03& S      & M6.5&    \nodata \\
UCAC4 767-032810 &             0449+6317  &   Palomar/DBSP& 2022 Jan 02& E,L,S,T& early-G& \nodata \\
CWISE J045334.34+203350.2 &    0453+2033  &   IRTF/SpeX   & 2018 Oct 01& K,Y,Z  & \nodata& early- to mid-sdL\\
2MASS J05053461+4648017 &      0505+4647  &   Palomar/DBSP& 2021 Dec 06& S      & M8&      \nodata \\
NLTT 14748 &                   0516+5640  &   Palomar/DBSP& 2022 Feb 03& S      & M6.5&    \nodata \\
CWISE J053046.20+440849.2 &    0530+4408  &   IRTF/SpeX   & 2020 Nov 25& C,F,R  & \nodata& T5\\
LSPM J0540+6417 &              0540+6417  &   Palomar/DBSP& 2021 Dec 06& S      & M8.5&    \nodata \\
CWISE J054034.89+414401.7 &    0540+4144  &   IRTF/SpeX   & 2022 Jan 09& C,F    & \nodata& T3\\
Gaia DR2 265201384281320448 &  0551+5511  &   Palomar/DBSP& 2022 Feb 03& S      & M6.5&    \nodata \\
WISEA J055600.48+154559.3 &    0556+1546  &   Palomar/DBSP& 2022 Feb 03& S      & M5 pec\tablenotemark{1}& \nodata \\
CWISE J055816.67$-$450233.4 &  0558$-$4502&  Magellan/FIRE& 2020 Feb 18& D      & \nodata& T8.5\\
CWISE J055829.92$-$233053.4&   0558$-$2330&   IRTF/SpeX   & 2019 Mar 16& F      & \nodata& L7 blue\\
CWISE J055942.94$-$012002.4&   0559$-$0120&   IRTF/SpeX   & 2022 Jan 09& F      & \nodata& L5.5 red\\
TYC 3382-603-1 &               0606+4851  &   Palomar/DBSP& 2022 Jan 02& E,L,S,T& mid-F&   \nodata \\
CWISE J060822.15$-$421244.7&   0608$-$4212&  Magellan/FIRE& 2016 Jan 23& M      & \nodata& L5.5 sl.\ blue\\
CWISE J060938.91+062513.2 &    0609+0625  &   IRTF/SpeX   & 2020 Oct 30& C,F,R,X& \nodata& L7\\
BD+10 1032A&                   0610+1019A &   Palomar/DBSP& 1995 Dec 02& 5      & M3&     \nodata \\
BD+10 1032B&                   0610+1019B &   Palomar/DBSP& 1995 Dec 02& 5      & M4&     \nodata \\
CWISE J061741.79+194512.8 A &  0617+1945A &   Palomar/DBSP& 2021 Dec 06& S      & L2&      \nodata \\
Gaia EDR3 3330473222213987072& 0623+1018  &   Palomar/DBSP& 2021 Dec 06& S      & M3&      \nodata \\
CWISE J063513.64$-$143029.4&   0635$-$1430&   IRTF/SpeX   & 2022 Jan 09& F      & \nodata& L9 sl.\ blue\\
UPM J0641+1226 &               0641+1226  &   Palomar/DBSP& 2022 Feb 03& S      & M5&      \nodata \\
WISEA J064313.95+163143.6 &    0643+1631  &   Palomar/DBSP& 2021 Dec 06& S      & M7&      \nodata \\
CWISE J064341.04+195039.3 &    0643+1950  &  Magellan/FIRE& 2019 Dec 11& G,K    & \nodata& T2 sl.\ red\\
Gaia EDR3 2936126887218756736& 0705$-$1535&   Palomar/DBSP& 2021 Dec 06& S      & bkg&     \nodata \\
CWISE J072418.16+430717.3 &    0724+4307  &   Keck/NIRES  & 2019 Feb 14& B,2    & \nodata& T6\\
UCAC4 414-032626 &             0727$-$0718&   Palomar/DBSP& 2021 Dec 06& S      & A\tablenotemark{2}&      \nodata \\
UPM J0730$-$2831 &             0730$-$2831&   Palomar/DBSP& 2022 Feb 03& S      & M5&      \nodata \\
2MASS J07312949+0249084 &      0731+0249  &   Palomar/DBSP& 2022 Feb 03& S      & M6&      \nodata \\
CWISE J073748.86$-$252613.0&   0737$-$2526&   IRTF/SpeX   & 2019 Jan 23& F,H    & \nodata& L3 sl.\ red\\
CWISE J074346.98+314603.4 &    0743+3146  &   Keck/NIRES  & 2021 Feb 24& B,J,2  & \nodata& T8\\
CWISE J075227.38+053802.6&     0752+0538  &   IRTF/SpeX   & 2022 Jan 09& F      & \nodata& L9 pec (composite)\\
CWISE J075628.41$-$063709.5 &  0756$-$0637&   CTIO/ARCoIRIS& 2018 Apr 03& F     & \nodata& L6: blue\\
CWISE J075744.48-300504.3&     0757$-$3005&   IRTF/SpeX   & 2022 Feb 12& F,N,O,R& \nodata& T4.5\\
CWISE J075853.12$-$232645.8&   0758$-$2326&   IRTF/SpeX   & 2022 Feb 21& F,O    & \nodata& T2.5 pec (composite?)\\
CWISE J080940.43$-$372003.7 &  0809$-$3720&   IRTF/SpeX   & 2019 Mar 17& F      & \nodata& L5 sl.\ blue\\
CWISER J090720.27$-$430856.7&  0907$-$4308&   CTIO/ARCoIRIS& 2018 Apr 02& F     & \nodata& L5\\
LP 846-7 &                     0929$-$2429&   Palomar/DBSP& 2022 Feb 03& S      & M6&      \nodata \\
CWISE J093035.01$-$743148.6&   0930$-$7431&   CTIO/ARCoIRIS& 2018 Apr 03& F     & \nodata& T7\\
LSPM J1024+3902W &             1024+3902W &   Palomar/DBSP& 2022 Feb 03& S      & M4.5&    \nodata \\
LSPM J1024+3902E &             1024+3902E &   Palomar/DBSP& 2022 Feb 03& S      & M4.5&    \nodata \\
LP 848-50 &                    1042$-$2416&   Palomar/DBSP& 2021 Dec 06& S      & M6.5&    \nodata \\
NLTT 25223 &                   1045+4941  &   Palomar/DBSP& 2022 Feb 03& S      & M5&      \nodata \\
CWISE J113646.36+205733.9 &    1136+2057  &   IRTF/SpeX   & 2019 Jan 23& F,H    & \nodata& M9:\\
CWISE J120258.26+305233.3 &    1202+3052  &   IRTF/SpeX   & 2022 Mar 07& C,O    & \nodata& T6\\
UCAC4 641-049451 &             1318+3810  &   Palomar/DBSP& 2022 Jan 02& E,L,S,T& early-G& \nodata \\
CWISE J132403.81$-$052631.4&   1324$-$0526&   IRTF/SpeX   & 2022 Jan 19& C,F    & \nodata&   sdT4  \\
CWISE J135338.04+441017.6 &    1353+4410  &   IRTF/SpeX   & 2022 Mar 11& C,N,O  & \nodata& T7.5\\
2MASS J14194617+3137094 &      1419+3137  &   Palomar/DBSP& 2022 Feb 03& S      & M6.5&    \nodata \\
NLTT 37185&                    1423+5146  &   Lick/Kast   & 2022 Jul 02& B,1,2  & M7\tablenotemark{3} &    \nodata\\
LP 440-17&                     1423+1426  &   IRTF/SpeX   & 2021 Jun 30& 3      & \nodata& M7\\
CWISE J142830.96$-$064435.5 &  1428$-$0644&   IRTF/SpeX   & 2019 Mar 16& F      & \nodata& L5 sl.\ blue\\
UCAC3 169-135909 &             1443$-$0539&   Palomar/DBSP& 2022 Feb 03& S      & M5.5&    \nodata \\
(ditto) &                      (ditto)    &   Palomar/DBSP& 2022 May 24& S      & M5.5&    \nodata \\
Gaia EDR3 6305165514134625024& 1459$-$1832&   Palomar/DBSP& 2022 May 24& S      & bkg&     \nodata \\
Gaia EDR3 6013647666939138688& 1529$-$3552&   Palomar/DBSP& 2022 May 24& S      & bkg&     \nodata \\
L 153-43&                      1557$-$6128&   CTIO 1.5m/RCSpec& 1996 May 20& 5  & M4&      \nodata \\
L 74-208&                      1613$-$7009&   CTIO 4m/RCSpec    & 1997 Jul 16& 5      & M4.5&     \nodata \\
SCR J1630$-$3633&              1630$-$3633&   Palomar/TSpec&2018 Apr 28& U,V    & \nodata& M5.5 \\
2MASS J16523515$-$2900186&     1652$-$2900&   Palomar/DBSP& 2022 Aug 27& S      & M5.5&    \nodata \\
SCR J1656$-$2046&              1656$-$2046&   Palomar/DBSP& 2022 Aug 27& S      & M6&      \nodata \\   
DENIS J171204.4$-$032328&      1712$-$0323&   Palomar/TSpec&2018 Apr 28& U,V    & \nodata& M7.5 \\
CWISER J171509.58$-$151534.6&  1715$-$1515&   IRTF/SpeX   & 2019 Mar 16& F      & \nodata& M8 pec \\
Gaia DR2 4053559111471124608&  1736$-$3425&   Palomar/DBSP& 2022 Aug 27& S      & M6.5&    \nodata \\   
2MASS J17392440$-$2327071&     1739$-$2327&   Palomar/DBSP& 2022 Aug 27& S      & M5&      \nodata \\
NLTT 45285&                    1741+0940  &   Lick/Kast   & 2020 Mar 06& B,J,2  & M7.5& \nodata\\
SCR J1746$-$3214&              1746$-$3214&   Palomar/TSpec&2018 Apr 29& U,V    & \nodata& M5.5 \\
CWISE J181429.08$-$202534.4&   1814$-$2025&   IRTF/SpeX   & 2021 May 31& 3      & \nodata& M8\\
WISEA J182423.61$-$053653.6&   1824$-$0536&   Palomar/TSpec&2018 Apr 28& U,V    & \nodata& M6.5 \\
WT 562&                        1826$-$6547&   CTIO 4m/RCSpec& 1995 Aug 14& 5    & M5 &     \nodata \\
Gaia DR2 4159791176135290752&  1831$-$0732&   Palomar/DBSP& 2022 May 30& P      & M9.5&    \nodata \\
UCAC4 378-124295&              1834$-$1426&   Palomar/DBSP& 2022 May 30& P      & M5&      \nodata \\
Gaia EDR3 4479498508613790464& 1839+0901  &   Palomar/DBSP& 2022 Aug 27& S      & M2.5&    \nodata \\ 
CWISE J185608.94$-$082257.6 &  1856$-$0822&   Lick/Kast   & 2019 Sep 20& J,1    & L0&      \nodata\\
WISEP J190648.47+401106.8&     1906+4011  &   Palomar/DBSP& 2022 Aug 27& S      & L1&      \nodata \\ 
NLTT 47423 &                   1907+0443  &   Palomar/DBSP& 2022 May 30& P      & M6&      \nodata \\
CWISE J191118.88+085456.3   &  1911+0854  &   Palomar/DBSP& 2022 Aug 27& S      & M4&      \nodata \\ 
2MASS J19212977$-$2915507&     1921$-$2915&   Lick/Kast   & 2020 Aug 16& B      & M7&     \nodata \\
CWISE J192351.88+234611.8 &    1923+2346  &   IRTF/SpeX   & 2021 Sep 11& C,F,R  & \nodata& M9\\
CWISE J194929.61+171301.3 &    1949+1713  &   APO/TSpec   & 2018 Sep 23& A      & \nodata& L2.5\\
Gaia DR2 2034222547248988032&  1955+3215  &   Palomar/TSpec&2019 Sep 18& U,V    & \nodata& L1 pec \\
UCAC4 563-099325 &             1955+2224  &   Palomar/DBSP& 2022 May 30& P      & mid-F?&   \nodata \\
HD 191408 B&                   2011$-$3606&   CTIO 4m/RCSpec& 1995 Aug 13& 5    & M4&      \nodata \\  
EC 20173$-$3036 &              2020$-$3027&   Palomar/DBSP& 2022 Jun 07& S      & DC&      \nodata \\
LSPM J2044+1517&               2044+1517  &   Palomar/DBSP& 2022 Aug 27& S      & M7.5&    \nodata \\ 
G 25-4 &                       2048+1127  &   Palomar/DBSP& 2022 May 30& P      & M3.5&    \nodata \\
2MASS J20492745+3336512&       2049+3336  &   Palomar/DBSP& 2022 May 30& P      & M7&      \nodata \\
MFL2000 J210104.18+030705.1&   2101+0307  &   Palomar/DBSP& 2022 Aug 27& S      & M6.5&    \nodata \\
2MASS J21272531+5553150&       2127+5553  &   Lick/Kast   & 2020 Aug 15& B,J,1  & M7.5&    \nodata \\
UCAC4 230-189452&              2136$-$4401&   CTIO 4m/RCSpec& 1997 Jul 14& 5      & M4.5&     \nodata \\
2MASS J21381698+5257188&       2138+5257  &   Lick/Kast   & 2020 Dec 14& B,I,J,1,2 & M6&      \nodata \\
CWISE J221113.55+054006.6 &    2211+0540  &   IRTF/SpeX   & 2021 Oct 23& C,F    & \nodata& T6\\
CWISE J221859.41+114642.7 &    2218+1146  &   Keck/NIRES  & 2020 Jul 07& B,I,J  & \nodata& T7 pec\\
Gaia DR2 2206265777300448768&  2225+6421  &   Palomar/TSpec&2019 Sep 18& U,V    & \nodata& L1 pec \\
CWISE J231403.13+693935.2 &    2314+6939  &   IRTF/SpeX   & 2018 Jun 16& F      & \nodata& L7 blue\\
CWISE J233135.66+644356.5 &    2331+6443  &   IRTF/SpeX   & 2018 Nov 25& F      & \nodata& M8:\\
CWISE J233819.49$-$385421.2&   2338$-$3854&   IRTF/SpeX   & 2021 Sep 11& C,F,R  & \nodata& T0\\
WISEA J235713.21$-$630827.6&   2357$-$6308&   SALT/RSS    & 2021 Dec 26& F      & M5&      \nodata \\
\enddata
\tablecomments{Observer code:
A = Katelyn Allers,
B = Adam Burgasser,
C = Emily Calamari,
D = Adam Schneider,
E = Peter Eisenhardt,
F = Jacqueline Faherty,
G = Daniella Bardalez Gagliuffi,
H = Eileen Gonzales,
I = Christopher Theissen,
J = Roman Gerasimov,
K = Rocio Kiman,
L = Guodong Li,
M = Jonathan Gagn\'e,
N = Dan Caselden,
O = Les Hamlet,
P = Thomas Connor,
Q = Blake Pantoja,
R = Austin Rothermich,
S = Daniel Stern,
T = Chao-Wei Tsai,
U = Eric Mamajek,
V = Federico Marocco,
W = Jon Rees
X = Jose I.\ Adorno,
Y = Johanna Vos,
Z = Mark Popinchalk,
1 = Christian Aganze,
2 = Chih-Chun Hsu,
3 = Richard Smart,
4 = Ryan Low,
5 = Davy Kirkpatrick.
}
\tablenotetext{a}{ Near-infrared spectral types of "sl.\ blue", "blue", "sl.\ red", and "red" refer to objects that are slightly ("sl.") or considerably bluer or redder in the $H$- and $K$-bands than the established spectral standard at that subtype, after the standard spectrum and target spectrum are overplotted at $J$-band. Those listed as "pec" (peculiar) have spectral morphologies that fail to match the established standards and cannot be characterized as easily. See section 5.2 of \cite{kirkpatrick2010} for more discussion.}
\tablenotetext{1}{0556+1546: Spectrum shows slightly stronger CaH bands relative to the M5 standard, though not enough to warrant a d/sdM5 designation.}
\tablenotetext{2}{0727$-$0718: See section 4.7 of \cite{tremblay2020} for an assessment of the spectral type. The Gaia DR2 parallax, which has a large uncertainty relative to other objects of similar magnitude, must be erroneous. No parallax is given in Gaia DR3. This star must actually fall well outside 20 pc.}
\tablenotetext{3}{1423+5146: Spectrum shows slightly stronger CaH bands relative to the M7 standard, though not enough to warrant a d/sdM7 designation.}
\end{deluxetable*}

\subsection{Astrometry\label{sec:appendix_astrometry}}

Additional parallaxes have been measured as part of an ongoing ground-based program and through serendipitous imaging data found in the Spitzer Heritage Archive. These results are discussed further below.

\subsubsection{NPARSEC results\label{sec:nparsec_astrometry}}

Nearby objects continue to be targeted as part of the NTT PARallaxes of Southern Extremely Cool objects (NPARSEC) project, a long-term program (186.C-0756 with R.\ Smart, PI; 105.C-0781, 108.21XQ.0001, and 108.21XQ.002 with E.\ Costa, PI) using the infrared spectrograph and imaging camera Son OF ISAAC (SOFI; \citealt{moorwood1998}) on the New Technology Telescope (NTT). The observational methodology and reduction procedures are identical to those discussed in \cite{smart2013}. For the eleven objects listed in Table~\ref{tab:nparsec_parallaxes}, the new NPARSEC preliminary values have smaller uncertainties than previously published parallaxes. The table gives the object names, J2000 coordinates, mean epoch of observation, the absolute parallax, the correction applied to the relative parallax to convert to absolute, the proper motion values in Right Ascension and Declination, the total time baseline of the NTT observations, the number of reference stars used, and the total number of separate observational epochs. 

Nine of the targets have absolute parallaxes greater than 50 mas, but for SDSS J163022.92+081822.0 and 2MASS J23312378$-$4718274, these better determined parallaxes have values below 50 mas, so we now exclude these three from the 20-pc census. We note, however, that these results are still considered preliminary and will be finalized once the NPARSEC program draws to a close.

\startlongtable
\begin{deluxetable*}{lrrcrcrrccc}
\tabletypesize{\footnotesize}
\tablecaption{Preliminary Parallax and Motion Fits for Objects on the NPARSEC Parallax Programs\label{tab:nparsec_parallaxes}}
\tablehead{
\colhead{Object} & 
\colhead{RA} &
\colhead{Dec} &
\colhead{Epoch} &
\colhead{$\pi_{abs}$} &
\colhead{Abs.} &
\colhead{$\mu_{RA}$} &
\colhead{$\mu_{Dec}$} &
\colhead{Baseline} &                          
\colhead{\# of} &
\colhead{\# of} \\
\colhead{Name} & 
\colhead{J2000} &
\colhead{J2000} &
\colhead{} &
\colhead{} &
\colhead{Corr.} &
\colhead{} &
\colhead{} &
\colhead{} &                          
\colhead{Ref.} &
\colhead{Obs.} \\
\colhead{} &
\colhead{(deg)} &
\colhead{(deg)} &
\colhead{} &
\colhead{(mas)} &
\colhead{(mas)} &
\colhead{(mas yr$^{-1}$)} &
\colhead{(mas yr$^{-1}$)} &
\colhead{(yr)} &                          
\colhead{Stars} &
\colhead{} \\
\colhead{(1)} &                          
\colhead{(2)} &  
\colhead{(3)} &  
\colhead{(4)} &
\colhead{(5)} &
\colhead{(6)} &
\colhead{(7)} &
\colhead{(8)} &
\colhead{(9)} &                          
\colhead{(10)} &
\colhead{(11)} \\
}
\startdata
2MASS J04070885+1514565     &  61.787883&  15.248440& 2019.80&    57.01$\pm$2.24& 0.73&   211.81$\pm$0.34&  -120.64$\pm$0.33& 11.1&     62&    20\\
2MASS J05103524$-$4208146   &  77.647251& -42.135453& 2011.12&    51.39$\pm$2.64& 0.66&    86.89$\pm$0.55&   588.34$\pm$0.39& 11.1&     87&    24\\
WISEPA J054231.26$-$162829.1&  85.629759& -16.473971& 2019.79&    63.26$\pm$1.65& 0.68&  -216.97$\pm$0.29&   297.02$\pm$0.35& 12.1&     93&    29\\
WISEPA J062720.07$-$111428.8&  96.840776& -11.246393& 2015.92&    77.02$\pm$2.86& 0.35&   -11.68$\pm$0.53&  -339.32$\pm$0.68& 12.1&    342&    27\\
2MASS J07290002$-$3954043   & 112.247022& -39.893729& 2013.23&   110.80$\pm$1.50& 0.36&  -564.46$\pm$0.33&  1694.63$\pm$0.31& 12.1&    394&    29\\
2MASS J09393548$-$2448279   & 144.900600& -24.812421& 2016.22&   189.80$\pm$2.68& 0.64&   569.65$\pm$0.60& -1040.60$\pm$0.63& 12.1&    124&    27\\
SDSS J163022.92+081822.0    & 247.595321&   8.305773& 2013.33&    41.76$\pm$2.79& 0.46&   -60.81$\pm$0.84&  -104.31$\pm$0.98&  7.0&    107&    19\\
2MASS J18283572$-$4849046   & 277.139127& -48.803065& 2013.33&    86.91$\pm$2.34& 0.39&   230.08$\pm$0.72&    88.50$\pm$0.58& 10.9&    503&    22\\
2MASS J22282889$-$4310262   & 337.120937& -43.175128& 2011.86&    95.60$\pm$2.23& 0.62&    97.98$\pm$0.42&  -306.91$\pm$0.46& 11.0&     31&    19\\
2MASS J23312378$-$4718274   & 352.849548& -47.307891& 2012.58&    49.00$\pm$4.23& 0.44&    73.01$\pm$0.70&   -64.37$\pm$0.69& 11.0&     18&    19\\
2MASSI J2356547$-$155310    & 359.226264& -15.888962& 2013.55&    64.78$\pm$2.26& 1.12&  -432.19$\pm$0.39&  -605.38$\pm$0.63& 11.1&     13&    19\\
\enddata
\end{deluxetable*}

\subsubsection{Spitzer results\label{sec:white_bear}}

\begin{figure*}
\includegraphics[scale=0.41,angle=0]{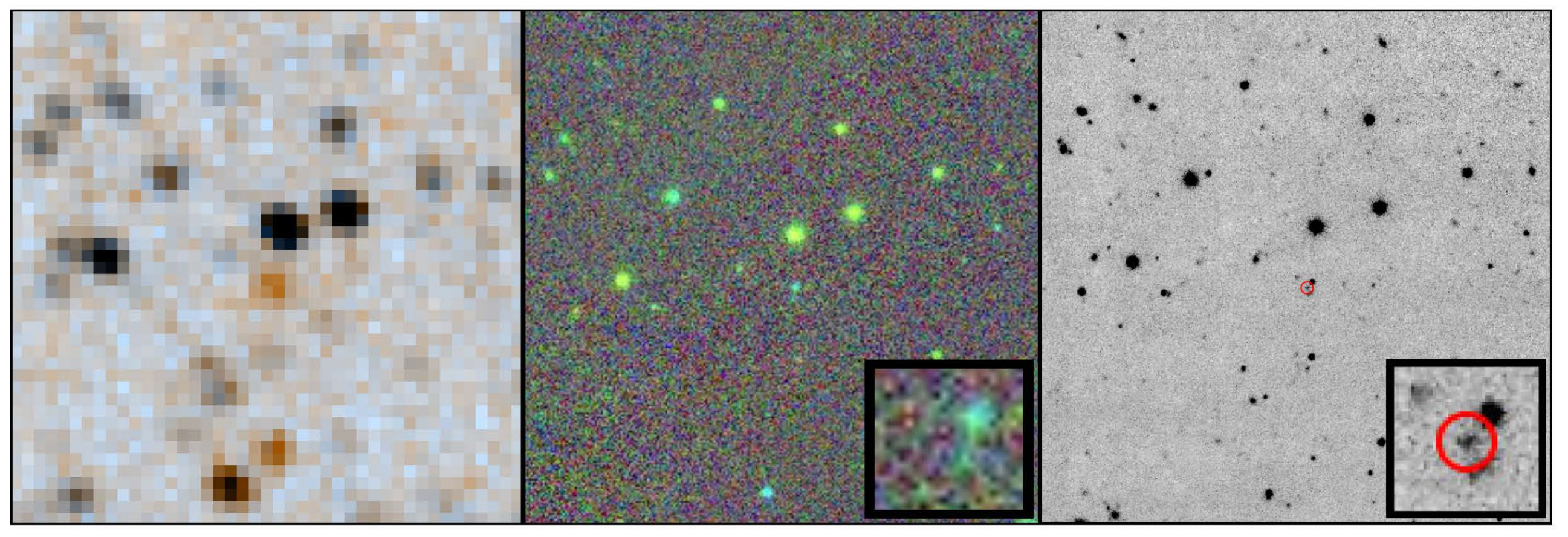}
\caption{Cutout images, 120$\arcsec$ on a side with north up and east to left, of 1811+6658. (Left) WISE data at epoch 2021.6. The separate W1 and W2 bands have been mapped into a color scheme in which objects appearing at roughly equal brightness in each will appear black, and those appearing primarily in W2 will appear orange (\citealt{caselden2018}). The brown dwarf 1811+6658 is the orange object at the center of the field. (Center) PanSTARRS data. Bands $y/i/g$ have been mapped into red/green/blue. Note the two blue background sources lying near the center of the field, which is shown in the zoomed inset. (Right) Keck/MOSFIRE data at epoch 2021.7. The detection of 1811+6658 is marked with a red circle (matched to the size of the aperture used in our photometric reductions) and is sandwiched between the two blue background sources seen in the PanSTARRS view. The inset shows a zoom of the field center.
\label{fig:1811p6658_wiseview}}
\end{figure*}

CWISE J181125.34+665806.4 (hereafter 1811+6658; see Figure~\ref{fig:1811p6658_wiseview}) is located only 1.2 degrees from the north ecliptic pole (NEP), and the area around the NEP was routinely observed by the Spitzer Space Telescope. As shown in Table~\ref{tab:spitzer_photometry}, the location of 1811+6658 was observed repeatedly in post-cryogenic programs 10147 (PI: Bock) and 13153 (PI: Capak) in an attempt to explore the genesis of fluctuations in the extragalactic background light and to provide IRAC/ch1 and IRAC/ch2 data on touchstone fields that will be used by Euclid, Roman, and JWST to study galaxy growth during the epoch of reionization. The data in program 10147 cover the timeframe from May 2014 to Sep 2014, and those in program 13153 cover Feb 2017 to Feb 2019.

To extract astrometry from these data sets, we searched for blocks of ch2 coverage that had sufficient depth and redundancy to provide a similar per-epoch astrometric accuracy to that obtained in our own Spitzer parallax programs (\citealt{kirkpatrick2019, kirkpatrick2021}). (This cold brown dwarf is much brighter at ch2, 15.95 mag, than at ch1, 18.23 mag, so only the longer-wavelength band would provide sufficient signal-to-noise for our astrometric needs.) Program 10147 used 30s exposures per frame, and the position of 1811+6658 was observed at four or fewer epochs. The data from program 13153, on the other hand, used 100s exposures per frame and had more coverage at each sky position.

We pared this data set down to include only those frames for which the location of 1811+6658 was far enough from the frame edge to provide a reasonable number of Gaia DR3 reference stars surrounding the target's location. Specifically, we retained only those ch2 frames that imaged all six of our pre-selected Gaia DR3 astrometric reference objects encircling a 60${\arcsec}$ zone centered on the location of 1811+6658. For program 10147, this left only two or three frames per epoch; this lack of redundancy combined with the short exposure time means that these data are unsuitable for astrometric analysis. For program 13153, however, we are left with four to six redundant, longer exposure frames per epoch (defined here to be per AOR), which is suitable for our reduction techniques. Of those program 13153 AORs listed in Table~\ref{tab:spitzer_photometry}, only the ch2 data in 62377728, 65133312, 68615680, and 68631296 lacked sufficient redundancy. The time span covered by the remaining data sets is Jul 2017 to Jan 2019. These data were extracted and astrometrically calibrated to the Gaia DR3 reference frame as described in (\citealt{kirkpatrick2021}).

Given that the usable Spitzer data only cover a year and a half, we turned to WISE astrometry to provide the additional baseline needed to disentangle parallax from proper motion. The NEP is within the boresight of the WISE spacecraft on every orbit, but given the 47$^{\prime}$-wide field of view, 1811+6658 is not within the continuous viewing zone. However, that location {\it is} viewed by WISE during a span of 50+ days every six months as the scan pattern rotates around the ecliptic pole. As such, there are several weeks of coverage twice per year covering its location.

We ran the {\texttt{crowdsource}} detection software (\citealt{schlafly2018}) on time-resolved unWISE coadds (\citealt{meisner2018}) for all ten-day epochal mosaics covering the position of 1811+6658. We retained those source lists for which the frame coverage depth at the location of 1811+6658 was 40 or greater. This was done in an effort to assure that the area surrounding the target's location also had sufficient coverage, as this area is needed for the astrometric calibrators. The measured positions of these surrounding astrometric standards were used to place the measured position of 1811+6658 onto the same Gaia DR3 reference frame used for the Spitzer observations.

\begin{figure*}
\includegraphics[scale=0.88,angle=0]{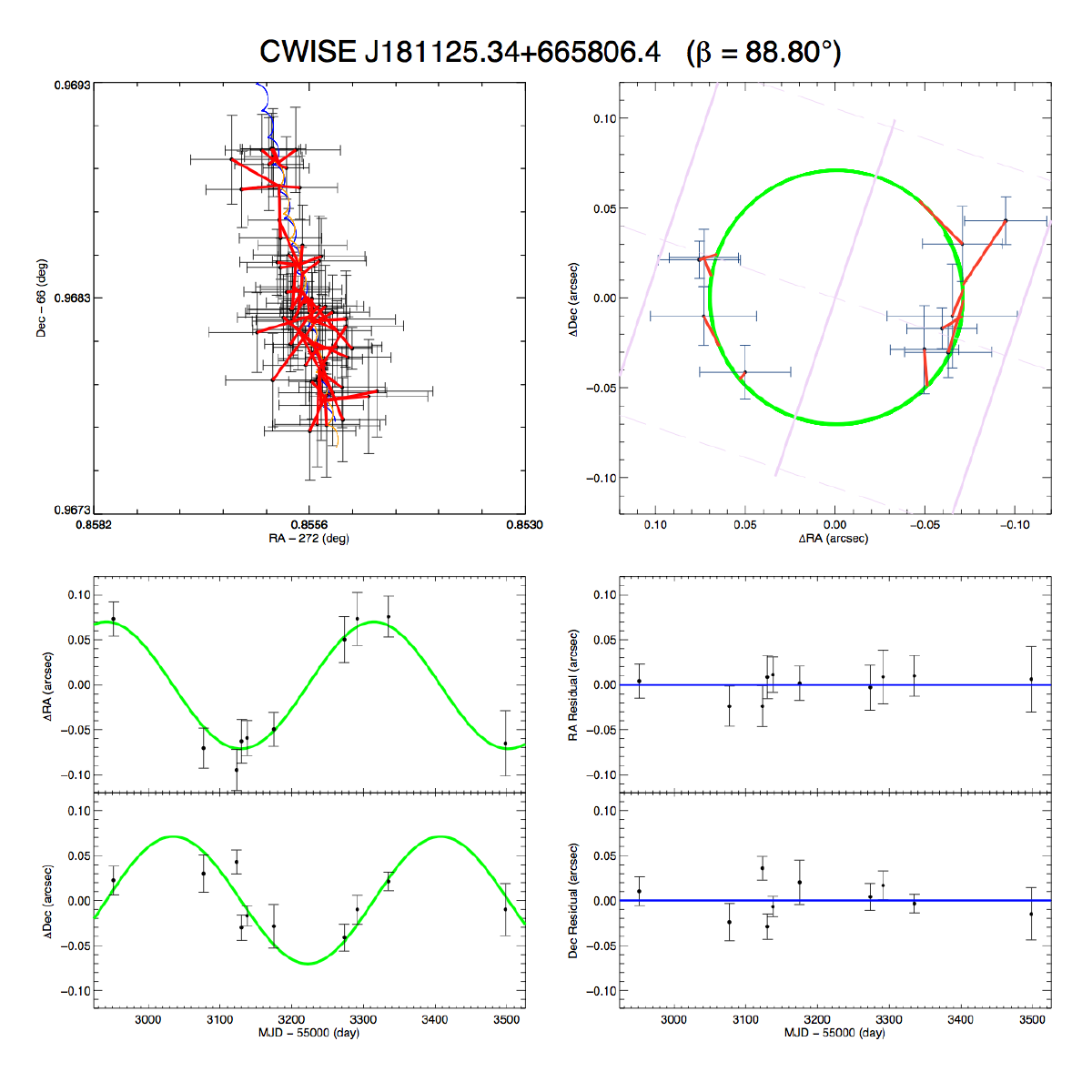}
\caption{Best fit to the parallax and proper motion of 1811+6658. (Upper left) Sky plot showing the track of the object along the sky. Black points with large uncertainties are the 57 individual unWISE time-resolved measurements. The orange curve shows the best fit as seen from the vantage point of WISE, and the blue curve shows this same fit from the vantage point of Spitzer. Red lines connect each observation to its predicted point along the best-fit curve. (Upper right) The parallax solution (green) with the proper motion component subtracted out. For clarity, only the 10 Spitzer data points are shown. Red lines connect the times of the Spitzer observations to their predicted points on the curve. (Lower left) The parallactic fit (green) as a function of time in RA and Dec, along with the measured Spitzer astrometry. (Lower right) Residuals around the parallactic fit as function of time in both RA and Dec. Blue lines mark residuals of zero. For additional info on this plot, see Figure 2 of \cite{kirkpatrick2021}.
\label{fig:1811p6658_plx_plots}}
\end{figure*}

This astrometry from Spitzer and WISE is listed in Table~\ref{tab:1811p6658_astrometric_pts}. A fit to the parallax and proper motion was performed on the combined astrometry using the methodology outlined in \cite{kirkpatrick2021}, resulting in the values shown in Table~\ref{tab:1811p6658_astrometric_soln}. The results of this fit are shown graphically in Figure~\ref{fig:1811p6658_plx_plots}. We find that the object has a distance of 14.3$^{+1.6}_{-1.2}$ pc and a value of $M_{ch2} = 15.16{\pm}0.21$ mag. A comparison to Figure 16d of \cite{kirkpatrick2021} suggests a spectral type of early-Y for this absolute magnitude. We further note that the measured colors -- $J_{MKO} -$ch2 = 5.66$\pm$0.04 mag, W1$-$W2 = 3.04$\pm$0.09 mag, and ch1$-$ch2 = 2.28$\pm$0.02 mag -- suggest a slightly earlier spectral type of around T9-T9.5 based on Figures 16e, 16g, and 16h of \cite{kirkpatrick2021}. A comparison of our Keck/MOSFIRE $J_{MKO}$, and WISE W1+W2 images with data from PanSTARRS (Figure~\ref{fig:1811p6658_wiseview}) shows that 1811+6658 is passing near two blue PanSTARRS sources. Given the low spatial resolution of the WISE (and Spitzer) data, we believe that our measurements of W1 (and ch1) are contaminated by these background objects. The higher resolution of the Keck/MOSFIRE data allows us to separate all three components, but our aperture photometry at $J_{MKO}$ is likely still compromised given our aperture radius of 6 pixels ($1{\farcs}1$). Therefore, our measured 
$J_{MKO} -$ ch2, W1$-$W2, and ch1$-$ch2 color are all likely bluer than their true values, supporting our assertion of a Y dwarf spectral type. 

\begin{deluxetable*}{rrrrccrrr}
\tabletypesize{\scriptsize}
\tablecaption{Astrometry on the {\it Gaia} DR3 Reference Frame for 1811+6658\label{tab:1811p6658_astrometric_pts}}
\tablehead{
\colhead{RA} &
\colhead{Dec} &     
\colhead{$\sigma_{\rm RA}$}  &
\colhead{$\sigma_{\rm Dec}$} &
\colhead{Band} &
\colhead{MJD} &
\colhead{$X$} &
\colhead{$Y$} &
\colhead{$Z$} \\
\colhead{(deg)} &
\colhead{(deg)} &     
\colhead{(asec)}  &
\colhead{(asec)} &
\colhead{} &
\colhead{(day)} &
\colhead{(km)} &
\colhead{(km)} &
\colhead{(km)} \\
\colhead{(1)} &                          
\colhead{(2)} &  
\colhead{(3)} &     
\colhead{(4)} &
\colhead{(5)} &  
\colhead{(6)} &                          
\colhead{(7)} &  
\colhead{(8)} &
\colhead{(9)}
}
\startdata
 272.8555557&  66.9680473& 0.01926& 0.01576& ch2&  57951.3599612&     -146795664.394331&     -29137217.001001&     -15336117.968022\\
 272.8554312&  66.9680074& 0.02229& 0.02085& ch2&  58077.0419038&      105457481.330533&    -102108810.219802&     -44432690.607376\\
 272.8554058&  66.9679956& 0.02281& 0.01313& ch2&  58122.9951224&      153046661.423622&      -8618952.216674&      -1585273.269887\\
 272.8554273&  66.9679729& 0.02411& 0.01427& ch2&  58130.1216015&      152890751.809545&       7809001.342473&       5815875.561150\\
 272.8554284&  66.9679740& 0.01951& 0.01134& ch2&  58137.7633120&      150349854.612387&      25297184.572655&      13659004.961915\\
 272.8554287&  66.9679583& 0.01908& 0.02450& ch2&  58175.0395945&      105086559.337856&      99767940.252168&      46540371.773800\\
\enddata
\tablecomments{(This table is available in its entirety in a machine-readable form in the online journal. A portion is shown here for guidance regarding its form and content.)}
\end{deluxetable*}

\begin{deluxetable}{lr}
\tabletypesize{\footnotesize}
\tablecaption{Parallax and Motion Fit for 1811+6658\label{tab:1811p6658_astrometric_soln}}
\tablehead{
\colhead{Parameter} &
\colhead{Value} \\
\colhead{(1)} &
\colhead{(2)} \\
}
\startdata
RA at t$_0$&       272.855621 deg $\pm$ 33.5 mas\\
Dec at t$_0$ &      66.968261 deg $\pm$ 28.7 mas\\
t$_0$ (MJD)&        57293.81 \\
$\varpi_{abs}$ &    69.7$\pm$6.8 mas \\
$\mu_{RA}$ &        $-$91.7$\pm$13.9 mas yr$^{-1}$\\
$\mu_{Dec}$ &       $-$439.8$\pm$11.6 mas yr$^{-1}$\\
$\chi^2$ &          65.903\\                
$\nu$ &             129 \\
$\chi_{\nu}^2$ &    0.511 \\
\#$_{Spitzer}$ &    10 \\
\#$_{WISE}$ &       57 \\
\#$_{Gaia}$ &        6 \\
\enddata
\tablecomments{The RA and Dec values are listed on the ICRS coordinate system. The last three rows represent the number of {\it Spitzer} ch2 epochs (\#$_{Spitzer}$) and the number of unWISE W2 epochs (\#$_{WISE}$) used in the fits, along with the number of five-parameter Gaia DR3 stars used for the astrometric re-registration (\#$_{Gaia}$).}
\end{deluxetable}

\pagebreak

\section{The List of Proximal Systems\label{sec:appendix_proximas}}

Despite recent WISE-based discoveries of the L+T dwarf binary system WISE J104915.57$-$531906.1 AB (\citealt{luhman2013}; 1.99 pc distant) and the Y dwarf WISE J085510.83$-$071442.5 (\citealt{luhman2014}; 2.28 pc distant) adding to our knowledge of the Sun's immediate neighbors, Proxima Centauri (\citealt{innes1915}; 1.30 pc distant) remains the closest. It is often just referred to as "Proxima", Latin for "nearest", because it is {\it the} nearest to the Sun. Yet, its full name translates to "Nearest of Centaurus". This has led some curious individuals to wonder what are the nearest stars -- i.e., the other proximal objects -- of each of the other eighty-seven official constellations.

Table~\ref{tab:20pc_census} allows us to answer this question, given our current knowledge of the 20-pc census. Proxima Centauri and its primary, $\alpha$ Centauri AB, represent a rare multi-object system for which the parallaxes of the individual components are so accurate that we can determine the far-flung companion to be closer to us than its host binary. For other multi-object systems, discerning the closest component may be far more difficult. Using a short-period binary as an example, determining the closest object in the system depends upon the orbital period and orientation with respect to the Sun, as one component may be the closer one at some epochs and the more distant one at others. Hence, we will identify only the proximal {\it systems} in each constellation when a multi-object system arises.

Table~\ref{tab:proximas} lists these proximal systems. As examples, the closest system in Canis Major is the binary Sirius AB, and the closest in Delphinus is the Y dwarf WISEPC J205628.90+145953.3. For several constellations, the proximal system is still ambiguous, given current uncertainties in the measured trigonometric parallaxes of the closest candidates. Constellations having two objects within 1$\sigma$ of the same closest value are indicated by footnotes. Note that all constellations have a proximal system within the 20-pc limit of Table~\ref{tab:20pc_census}, the most distant being the K dwarf HD 200779 at 15.05 pc, the closest known object in Equuleus.

\startlongtable
\begin{deluxetable}{lllr}
\tabletypesize{\scriptsize}
\tablecaption{The Proximal Systems for Each Constellation\label{tab:proximas}}
\tablehead{
\colhead{Constellation} &
\colhead{Proximal System} &                          
\colhead{Spectral Class} &                          
\colhead{Parallax (mas)}  \\
\colhead{(1)} &                          
\colhead{(2)} &
\colhead{(3)} &
\colhead{(4)}
}
\startdata
And& Ross 248                & M& 316.48 \\
Ant& DENIS J104814.6$-$395606& M& 247.21 \\
Aps& L 43-72 AB              & M+M&  85.71 \\
Aql& Altair ($\alpha$ Aql)   & A& 194.95 \\
Aqr& EZ Aqr ABC              & M+M+M& 289.50 \\
Ara& CD$-$46 11540           & M& 219.64 \\
Ari& Teegarden's Star        & M& 260.98 \\
Aur& QY Aur AB               & M+M& 165.21 \\
Boo& HD 119850               & M& 183.99 \\
Cae& L 374-6                 & M&  70.06 \\
Cam& G 254-29                & M& 190.32 \\
Cap& LP 816-60               & M& 177.93 \\
Car& L 143-23                & M& 206.96 \\
Cas& Achird ($\eta$ Cas AB)  & F+K& 168.83 \\
Cen& $\alpha$ Cen AB + Proxima Cen& G+K+M& 768.06 \\
Cep& HD 239960 AB            & M+M& 249.96 \\
Cet& G 272-61 AB             & M+M& 373.84 \\
Cha& SCR J1138$-$7721        & M& 119.34 \\
Cir& DENIS-P J1454078$-$660447& L& 93.94 \\
CMa& Sirius ($\alpha$ CMa AB)& A+wd& 79.21 \\
CMi& Procyon ($\alpha$ CMi AB)& F+wd& 284.56 \\
Cnc& G 51-15                 & M& 279.24 \\
Col& AP Col                  & M& 115.39 \\
Com& $\beta$ Com             & F& 108.72 \\
CrA& L 489-58                & M&  80.41 \\
CrB& LSPM J1524+2925         & M&  76.46 \\
Crt& CD$-$23 9765            & M&  92.80 \\
Cru& L 194-11                & M&  76.35 \\
Crv& Ross 695                & M& 112.67 \\
CVn& Chara ($\beta$ CVn)     & G& 118.02 \\
Cyg& 61 Cyg AB               & K+K& 286.00 \\
Del& WISEPC J205628.90+145953.3& Y& 140.80 \\
Dor& L 230-188               & M& 140.69 \\
Dra& HD 173739 + HD 173740   & M+M& 283.84 \\
Equ& HD 200779               & K&  66.46 \\
Eri& Ran ($\epsilon$ Eri)    & K& 310.57 \\
For& LP 944-20               & M& 155.59 \\
Gem& HD 265866               & M& 179.06 \\
Gru& HD 204961               & M& 201.32 \\
Her& WISEPA J174124.26+255319.5& T& 214.30 \\
Hor& L 372-58                & M& 272.16 \\
Hya& WISE J085510.83$-$071442.5& Y& 439.00 \\
Hyi& $\beta$ Hyi             & G& 133.71 \\
Ind& $\epsilon$ Ind ABC      & K+T+T& 274.84 \\
Lac& EV Lac                  & M& 197.95 \\
Leo& Wolf 359                & M& 415.17 \\
Lep& Gl 229 AB               & M+T& 173.69 \\
Lib& Gl 570 ABCD             & K+M+M+T& 170.01 \\
LMi& G 119-36                & M& 102.75 \\
Lup& CD$-$40 9712            & M& 168.99 \\
Lyn& G 111-47                & M& 112.99 \\
Lyr& 2MASSI J1835379+325954  & M& 175.79 \\
Men& L 32-8 + L 32-9         & M+M& 113.13 \\
Mic& AX Mic                  & M& 251.91 \\
Mon& Ross 614 AB\tablenotemark{a}& M+M& 242.96 \\
Mus& LAWD 37                 & wd& 215.67 \\
Nor& WISEA J154045.67$-$510139.3& M& 187.72 \\
Oct& L 49-19                 & M& 116.31 \\
Oph& Barnard's Star          & M& 546.97 \\
Ori& G 99-49                 & M& 192.01 \\
Pav& SCR J1845$-$6357 AB     & M+T& 249.91 \\
Peg& WISE J220905.73+271143.9\tablenotemark{b}& Y& 161.70 \\
Per& 2MASS J04195212+4233304\tablenotemark{c} & M&  97.44 \\
Phe& LAWD 96                 & wd& 120.01 \\
Pic& Kapteyn's Star          & M& 254.19 \\
PsA& HD 217987               & M& 304.13 \\
Psc& Wolf 28                 & wd& 231.78 \\
Pup& 2MASS J07290002$-$3954043\tablenotemark{d}& T& 126.30 \\
Pyx& CD$-$32 5613            & wd& 117.39 \\
Ret& WISE J035000.32$-$565830.2& Y& 176.40 \\
Scl& HD 225213               & M& 230.09 \\
Sco& CD$-$44 11909           & M& 199.69 \\
Sct& WISEA J182423.61$-$053653.6& M& 75.67 \\
Ser& HD 165222               & M& 129.21 \\
Sex& LP 731-58               & M& 219.33 \\
Sge& HD 349726 + Ross 730    & M+M& 113.25 \\
Sgr& Ross 154                & M& 336.02 \\
Tau& WISEPA J041022.71+150248.5& Y& 151.30 \\
Tel& L 347-14                & M& 169.23 \\
TrA& WISE J163940.86$-$684744.6& Y& 219.60 \\
Tri& LP 245-10               & M&  96.73 \\
Tuc& CD$-$68 47              & M& 121.44 \\
UMa& Lalande 21185           & M& 392.75 \\
UMi& WISEPC J150649.97+702736.0& T& 193.94 \\
Vel& WISE J104915.57$-$531906.1AB& L+T& 501.55 \\
Vir& Ross 128                & M& 296.30 \\
Vol& LAWD 26                 & wd& 122.41 \\
Vul& WISE J192841.35+235604.9 & T& 146.40 \\
\enddata
\tablenotetext{a}{The T dwarf UGPS J072227.51$-$054031.2 is equidistant with this source within the measurement uncertainties.}
\tablenotetext{b}{The M dwarf pair BD+19 5116AB is equidistant with this source within the measurement uncertainties.}
\tablenotetext{c}{The T dwarf WISE J043052.92+463331.6 is equidistant with this source within the measurement uncertainties.}
\tablenotetext{d}{The M dwarf SCR J0740$-$4257 is equidistant with this source within the measurement uncertainties.}
\end{deluxetable}

It is worth noting the prevalence of brown dwarfs in Table~\ref{tab:proximas}. There are four T dwarf companions residing in proximal systems with K or M dwarf primaries, one L+T binary as its own proximal system, and eleven solivagant L, T, or Y dwarfs that are proximal objects in their own right. In this latter group, over half (six) of the these are Y dwarfs, a spectral type that is not yet fully sampled near the Sun.

\end{document}